\documentclass[reqno,12pt,a4paper]{article}

\usepackage{amsfonts,amssymb,amsthm, 
amsmath,amscd, bbm, bm, mathabx, 
mathrsfs,delarray,subfigure}
\usepackage[dvipsnames]{xcolor}
\usepackage{hyperref}
\usepackage{cite}
\usepackage{xypic}
\usepackage{sectsty}

\DeclareSymbolFont{sfletters}{OML}{cmbrm}{m}{it}  

\DeclareMathSymbol{\sfGamma}{\mathord}{sfletters}{"00}
\DeclareMathSymbol{\sfDelta}{\mathord}{sfletters}{"01}
\DeclareMathSymbol{\sfTheta}{\mathord}{sfletters}{"02}
\DeclareMathSymbol{\sfLambda}{\mathord}{sfletters}{"03}
\DeclareMathSymbol{\sfXi}{\mathord}{sfletters}{"04}
\DeclareMathSymbol{\sfPi}{\mathord}{sfletters}{"05}
\DeclareMathSymbol{\sfSigma}{\mathord}{sfletters}{"06}
\DeclareMathSymbol{\sfUpsilon}{\mathord}{sfletters}{"07}
\DeclareMathSymbol{\sfPhi}{\mathord}{sfletters}{"08}
\DeclareMathSymbol{\sfPsi}{\mathord}{sfletters}{"09}
\DeclareMathSymbol{\sfOmega}{\mathord}{sfletters}{"0A}
\DeclareMathSymbol{\sfalpha}{\mathord}{sfletters}{"0B}
\DeclareMathSymbol{\sfbeta}{\mathord}{sfletters}{"0C}
\DeclareMathSymbol{\sfgamma}{\mathord}{sfletters}{"0D}
\DeclareMathSymbol{\sfdelta}{\mathord}{sfletters}{"0E}
\DeclareMathSymbol{\sfepsilon}{\mathord}{sfletters}{"0F}
\DeclareMathSymbol{\sfzeta}{\mathord}{sfletters}{"10}
\DeclareMathSymbol{\sfeta}{\mathord}{sfletters}{"11}
\DeclareMathSymbol{\sftheta}{\mathord}{sfletters}{"12}
\DeclareMathSymbol{\sfiota}{\mathord}{sfletters}{"13}
\DeclareMathSymbol{\sfkappa}{\mathord}{sfletters}{"14}
\DeclareMathSymbol{\sflambda}{\mathord}{sfletters}{"15}
\DeclareMathSymbol{\sfmu}{\mathord}{sfletters}{"16}
\DeclareMathSymbol{\sfnu}{\mathord}{sfletters}{"17}
\DeclareMathSymbol{\sfxi}{\mathord}{sfletters}{"18}
\DeclareMathSymbol{\sfpi}{\mathord}{sfletters}{"19}
\DeclareMathSymbol{\sfrho}{\mathord}{sfletters}{"1A}
\DeclareMathSymbol{\sfsigma}{\mathord}{sfletters}{"1B}
\DeclareMathSymbol{\sftau}{\mathord}{sfletters}{"1C}
\DeclareMathSymbol{\sfupsilon}{\mathord}{sfletters}{"1D}
\DeclareMathSymbol{\sfphi}{\mathord}{sfletters}{"1E}
\DeclareMathSymbol{\sfchi}{\mathord}{sfletters}{"1F}
\DeclareMathSymbol{\sfpsi}{\mathord}{sfletters}{"20}
\DeclareMathSymbol{\sfomega}{\mathord}{sfletters}{"21}
\DeclareMathSymbol{\sfvarepsilon}{\mathord}{sfletters}{"22}
\DeclareMathSymbol{\sfvartheta}{\mathord}{sfletters}{"23}
\DeclareMathSymbol{\sfvarpi}{\mathord}{sfletters}{"24}
\DeclareMathSymbol{\sfvarrho}{\mathord}{sfletters}{"25}
\DeclareMathSymbol{\sfvarsigma}{\mathord}{sfletters}{"26}
\DeclareMathSymbol{\sfvarphi}{\mathord}{sfletters}{"27}

\DeclareMathSymbol{\spartial}{\mathord}{sfletters}{"40}

\DeclareMathSymbol{\sfA}{\mathord}{sfletters}{"41}
\DeclareMathSymbol{\sfB}{\mathord}{sfletters}{"42}
\DeclareMathSymbol{\sfC}{\mathord}{sfletters}{"43}
\DeclareMathSymbol{\sfD}{\mathord}{sfletters}{"44}
\DeclareMathSymbol{\sfE}{\mathord}{sfletters}{"45}
\DeclareMathSymbol{\sfF}{\mathord}{sfletters}{"46}
\DeclareMathSymbol{\sfG}{\mathord}{sfletters}{"47}
\DeclareMathSymbol{\sfH}{\mathord}{sfletters}{"48}
\DeclareMathSymbol{\sfI}{\mathord}{sfletters}{"49}
\DeclareMathSymbol{\sfJ}{\mathord}{sfletters}{"4A}
\DeclareMathSymbol{\sfK}{\mathord}{sfletters}{"4B}
\DeclareMathSymbol{\sfL}{\mathord}{sfletters}{"4C}
\DeclareMathSymbol{\sfM}{\mathord}{sfletters}{"4D}
\DeclareMathSymbol{\sfN}{\mathord}{sfletters}{"4E}
\DeclareMathSymbol{\sfO}{\mathord}{sfletters}{"4F}
\DeclareMathSymbol{\sfP}{\mathord}{sfletters}{"50}
\DeclareMathSymbol{\sfQ}{\mathord}{sfletters}{"51}
\DeclareMathSymbol{\sfR}{\mathord}{sfletters}{"52}
\DeclareMathSymbol{\sfS}{\mathord}{sfletters}{"53}
\DeclareMathSymbol{\sfT}{\mathord}{sfletters}{"54}
\DeclareMathSymbol{\sfU}{\mathord}{sfletters}{"55}
\DeclareMathSymbol{\sfV}{\mathord}{sfletters}{"56}
\DeclareMathSymbol{\sfW}{\mathord}{sfletters}{"57}
\DeclareMathSymbol{\sfX}{\mathord}{sfletters}{"58}
\DeclareMathSymbol{\sfY}{\mathord}{sfletters}{"59}
\DeclareMathSymbol{\sfZ}{\mathord}{sfletters}{"5A}
\DeclareMathSymbol{\sfa}{\mathord}{sfletters}{"61}
\DeclareMathSymbol{\sfb}{\mathord}{sfletters}{"62}
\DeclareMathSymbol{\sfc}{\mathord}{sfletters}{"63}
\DeclareMathSymbol{\sfd}{\mathord}{sfletters}{"64}
\DeclareMathSymbol{\sfe}{\mathord}{sfletters}{"65}
\DeclareMathSymbol{\sff}{\mathord}{sfletters}{"66}
\DeclareMathSymbol{\sfg}{\mathord}{sfletters}{"67}
\DeclareMathSymbol{\sfh}{\mathord}{sfletters}{"68}
\DeclareMathSymbol{\sfi}{\mathord}{sfletters}{"69}
\DeclareMathSymbol{\sfj}{\mathord}{sfletters}{"6A}
\DeclareMathSymbol{\sfk}{\mathord}{sfletters}{"6B}
\DeclareMathSymbol{\sfl}{\mathord}{sfletters}{"6C}
\DeclareMathSymbol{\sfm}{\mathord}{sfletters}{"6D}
\DeclareMathSymbol{\sfn}{\mathord}{sfletters}{"6E}
\DeclareMathSymbol{\sfo}{\mathord}{sfletters}{"6F}
\DeclareMathSymbol{\sfp}{\mathord}{sfletters}{"70}
\DeclareMathSymbol{\sfq}{\mathord}{sfletters}{"71}
\DeclareMathSymbol{\sfr}{\mathord}{sfletters}{"72}
\DeclareMathSymbol{\sfs}{\mathord}{sfletters}{"73}
\DeclareMathSymbol{\sft}{\mathord}{sfletters}{"74}
\DeclareMathSymbol{\sfu}{\mathord}{sfletters}{"75}
\DeclareMathSymbol{\sfv}{\mathord}{sfletters}{"76}
\DeclareMathSymbol{\sfw}{\mathord}{sfletters}{"77}
\DeclareMathSymbol{\sfx}{\mathord}{sfletters}{"78}
\DeclareMathSymbol{\sfy}{\mathord}{sfletters}{"79}
\DeclareMathSymbol{\sfz}{\mathord}{sfletters}{"7A}

\usepackage[LGR,T1]{fontenc}
\usepackage{amsmath,etoolbox}

\newcommand{\declarebsfgreek}[2]{%
  \protected\csdef{bsf#1}{\mathord{\text{\bsfgreekfont#2}}}%
}
\newcommand{\bsfgreekfont}{\usefont{LGR}{cmss}{bx}{it}}

\declarebsfgreek{alpha}{a}
\declarebsfgreek{beta}{b}
\declarebsfgreek{gamma}{g}
\declarebsfgreek{delta}{d}
\declarebsfgreek{epsilon}{e}
\declarebsfgreek{zeta}{z}
\declarebsfgreek{eta}{h}
\declarebsfgreek{theta}{j}
\declarebsfgreek{iota}{i}
\declarebsfgreek{kappa}{k}
\declarebsfgreek{lambda}{l}
\declarebsfgreek{mu}{m}
\declarebsfgreek{nu}{n}
\declarebsfgreek{xi}{x}
\declarebsfgreek{omicron}{o}
\declarebsfgreek{pi}{p}
\declarebsfgreek{rho}{r}
\declarebsfgreek{sigma}{s}
\declarebsfgreek{tau}{t}
\declarebsfgreek{upsilon}{u}
\declarebsfgreek{phi}{f}
\declarebsfgreek{chi}{q}
\declarebsfgreek{psi}{y}
\declarebsfgreek{omega}{w}
\declarebsfgreek{varsigma}{c}

\declarebsfgreek{Gamma}{G}
\declarebsfgreek{Delta}{D}
\declarebsfgreek{Upsilon}{U}
\declarebsfgreek{Omega}{W}
\declarebsfgreek{Theta}{J}
\declarebsfgreek{Lambda}{L}
\declarebsfgreek{Xi}{X}
\declarebsfgreek{Pi}{P}
\declarebsfgreek{Sigma}{S}
\declarebsfgreek{Phi}{F}
\declarebsfgreek{Psi}{Y}

\newcommand{\declarebsfitalic}[2]{%
  \protected\csdef{bsf#1}{\mathord{\text{\bsfitalicfont#2}}}%
}
\newcommand{\bsfitalicfont}{\usefont{T1}{cmss}{bx}{it}}

\declarebsfitalic{a}{a}
\declarebsfitalic{b}{b}
\declarebsfitalic{c}{c}
\declarebsfitalic{d}{d}
\declarebsfitalic{e}{e}
\declarebsfitalic{f}{f}
\declarebsfitalic{g}{g}
\declarebsfitalic{h}{h}
\declarebsfitalic{i}{i}
\declarebsfitalic{j}{j}
\declarebsfitalic{k}{k}
\declarebsfitalic{l}{l}
\declarebsfitalic{m}{m}
\declarebsfitalic{n}{n}
\declarebsfitalic{o}{o}
\declarebsfitalic{p}{p}
\declarebsfitalic{q}{q}
\declarebsfitalic{r}{r}
\declarebsfitalic{s}{s}
\declarebsfitalic{t}{t}
\declarebsfitalic{u}{u}
\declarebsfitalic{v}{v}
\declarebsfitalic{x}{x}
\declarebsfitalic{y}{y}
\declarebsfitalic{z}{z}

\declarebsfitalic{A}{A}
\declarebsfitalic{B}{B}
\declarebsfitalic{C}{C}
\declarebsfitalic{D}{D}
\declarebsfitalic{E}{E}
\declarebsfitalic{F}{F}
\declarebsfitalic{G}{G}
\declarebsfitalic{H}{H}
\declarebsfitalic{I}{I}
\declarebsfitalic{J}{J}
\declarebsfitalic{K}{K}
\declarebsfitalic{L}{L}
\declarebsfitalic{M}{M}
\declarebsfitalic{N}{N}
\declarebsfitalic{O}{O}
\declarebsfitalic{P}{P}
\declarebsfitalic{Q}{Q}
\declarebsfitalic{R}{R}
\declarebsfitalic{S}{S}
\declarebsfitalic{T}{T}
\declarebsfitalic{U}{U}
\declarebsfitalic{V}{V}
\declarebsfitalic{X}{X}
\declarebsfitalic{Y}{Y}
\declarebsfitalic{Z}{Z}

\usepackage{accents}

\sectionfont{\fontsize{12}{15}\selectfont}
\subsectionfont{\fontsize{12}{15}\selectfont}

\usepackage{enumitem}

\makeatletter 
\newcommand\raggedtop{%
  \topskip=1\topskip plus 10pt} 

\makeatother 
\raggedtop

\DeclareMathOperator{\ad}{ad}
\DeclareMathOperator{\Ad}{Ad}

\DeclareMathOperator{\ran}{ran} 
\DeclareMathOperator{\id}{id}

\DeclareMathOperator{\Fun}{Fun}

\DeclareMathOperator{\Map}{Map}

\DeclareMathOperator{\Vect}{Vect}

\DeclareMathOperator{\Aut}{Aut}

\DeclareMathOperator{\INN}{INN}
\DeclareMathOperator{\Der}{Der}

\DeclareMathOperator{\Lie}{Lie}

\DeclareMathOperator{\tr}{tr}

\DeclareMathOperator{\ee}{e}

\DeclareMathOperator{\AD}{AD}
\DeclareMathOperator{\CC}{C\hspace{-1pt}}
\DeclareMathOperator{\RR}{R\hspace{-1pt}}

\DeclareMathOperator{\GL}{GL}

\DeclareMathOperator{\SO}{SO}
\DeclareMathOperator{\SU}{SU}

\DeclareMathOperator{\NN}{N\hspace{-1pt}}
\DeclareMathOperator{\ON}{ON\hspace{-1pt}}
\DeclareMathOperator{\TT}{T\hspace{-1pt}}
\DeclareMathOperator{\WW}{W\hspace{-1pt}}
\DeclareMathOperator{\OW}{OW\hspace{-1pt}}
\DeclareMathOperator{\ZZ}{Z\hspace{-1pt}}
\DeclareMathOperator{\DD}{D\hspace{-1pt}}

\numberwithin{equation}{subsection} 
\numberwithin{subsection}{section} 

\newcommand{\ceqref}[1]{{\textcolor{blue}{\eqref{#1}}}}
\newcommand{\cref}[1]{{\textcolor{blue}{\ref{#1}}}}
\newcommand{\ccite}[1]{{\textcolor{blue}{\!\cite{#1}}}}

\newcommand{\sss}{{\hbox{\large $\sum$}}}

\newcommand{\ddd}{{\hbox{\large $\bigoplus$}}}

\newcommand{\ul}[1]{{\underline{#1}}}
\newcommand{\bfs}[1]{{\boldsymbol{#1}}}

\newcommand{\sdot}{\hspace{.5pt}\dot{}\hspace{.5pt}}
\newcommand{\hfpt}{\hspace{.75pt}}
\newcommand{\mhfpt}{\hspace{-.75pt}}
\newcommand{\dd}{\text{d}}

\newcommand{\mathsans}[1]{{{\sf #1}}}

\font\euler=eusm10 at 12.8 truept
\font\scripteuler=eusm7
\font\scriptscripteuler=eusm5 
\def\eul{\fam=12}
\newcommand{\matheul}[1]{{{\eul #1}}}

\newtheorem{defi}{{\sf Definition}}[section]
\newtheorem{prop}{{\sf Proposition}}[section]

\newtheorem{lemma}{{\sf Lemma}}[section]

\DeclareMathSymbol{*}{\mathbin}{symbols}{"03} 

\begin{document}

\vskip1.5cm
\begin{large}
{\flushleft\textcolor{blue}{\sffamily\bfseries 4-d Chern--Simons Theory:}}  
{\flushleft\textcolor{blue}{\sffamily\bfseries Higher Gauge Symmetry and Holographic Aspects}}  
\end{large}
\vskip1.cm
\hrule height 1.5pt
\vskip1.cm
{\flushleft{\sffamily \bfseries Roberto Zucchini}\\
\it Dipartimento di Fisica ed Astronomia,\\
Universit\`a di Bologna,\\
I.N.F.N., sezione di Bologna,\\
viale Berti Pichat, 6/2\\
Bologna, Italy\\
Email: \textcolor{blue}{\tt \href{mailto:roberto.zucchini@unibo.it}{roberto.zucchini@unibo.it}}, 
\textcolor{blue}{\tt \href{mailto:zucchinir@bo.infn.it}{zucchinir@bo.infn.it}}}


\vskip.5cm
\vskip.4cm 
{\flushleft\sc
Abstract:} 
We present and study a 4--d Chern--Simons (CS) model whose gauge symmetry
is encoded in a balanced Lie group crossed module.  Using the 
derived formal set--up recently found, the model can be formulated in a way that in many
respects closely parallels that of the familiar 3--d CS one.
In spite of these formal resemblance, the gauge invariance properties of the 4--d CS model differ considerably.
The 4–d CS action is fully gauge invariant if the underlying base 4–fold has no boundary.
When it does, the action is gauge variant, the gauge variation being a boundary term.
If certain boundary
conditions are imposed on the gauge fields and gauge transformations, level quantization can then occur.
In the canonical formulation of the theory, it is found that, depending again on boundary conditions,
the 4--d CS model is characterized by surface charges obeying a non trivial Poisson bracket algebra. This is
a higher counterpart of the familiar WZNW current algebra arising in the 3–d model.
4--d CS theory thus exhibits rich holographic properties.
The covariant Schroedinger quantization of the 4--d CS model is performed.
A preliminary analysis of 4--d CS edge field theory is also provided. 
The toric and Abelian projected models are described in some detail.

\vspace{2mm}
\par\noindent


\vfil\eject
\tableofcontents

\vfil\eject

\section{\textcolor{blue}{\sffamily Introduction}}\label{sec:intro}

There are several reasons why the formulation and study of 4-dimensional
higher Chern--Simons (CS) theory is an interesting and worthwhile endeavour.
Some have to do with physics, other with mathematics.

4--dimensional BF theory is an instance of 4--dimensional CS theory. Though as a topological quantum
field theory (TQFT)
it involves no metric and possesses no local degrees of freedom, it yields however
general relativity when the basic fields are suitably expressed in terms of or related to
metric data. This has been done in two independent ways \ccite{MacDowell:1977jt,Plebanski:1977zz}
which differ by the choice of the gauge group. A 4-dimensional gravitational CS model was
worked out in ref. \ccite{Morales:2017zjw} by Kaluza--Klein compactification of a 5-dimensional (anti)
de Sitter gravitational CS model. Specific instances of 4--dimensional CS theory have appeared 
as topological sectors of CS modified gravity \ccite{Carroll:1989vb,Grumiller:2007rv} and 
string based cosmological models \ccite{Shiu:2015xda} describing axion--like fields and their coupling
to gauge fields. 

In refs. \ccite{Costello:2013zra,Costello:2013sla,Costello:2017dso,Costello:2018gyb}
the three types of solutions of the Yang--Baxter equation, rational, trigonometric and elliptic,
and their properties were obtained from a 4--dimensional CS model compactified down to 2 dimensions.
The model is defined on a 4--fold of the form $\varSigma\times C$, where $\varSigma$ is a 2--fold and
$C$ is a Riemann surface, and involves a background meromorphic (1,0)--form on $C$. So, it is only
partially topological, the 4-dimensional diffeomorphism symmetry being broken to the 2--dimensional one. 

There exist two main quite different approaches to the construction of CS type TQFTs:
the algebraic--topological approach and the differential--geometrical approach. These frameworks are 
related. It is possible to identify a TQFT defined in the former combinatorial approach and another TQFT
defined in the latter continuum one if the partition functions of these TQFTs can be proven to be equal
for appropriate assignments of input data. 
This correspondence is well understood in three dimensions. The Turaev--Viro--Barrett--Westbury model
\ccite{Turaev:1992hq,Barrett:1993ab} is a combinatorial model of 3--dimensional quantum gravity
with cosmological constant known to be equivalent to 3--dimensional BF theory with cosmological term
\ccite{Horowitz:1989ng,Blau:1989bq,Cattaneo:1995tw} when their underlying Lie groups are 
the same and the quantum group parameter and cosmological constant are properly related.
The Reshetikhin--Turaev model \ccite{Reshetikhin:1991tc}
is a 3--dimensional combinatorial TQFT believed to be equivalent 
3--dimensional CS TQFT \ccite{Witten:1988hf,Frohlich:1989gr} under similar conditions.
The Dijkgraaf--Witten model \ccite{Dijkgraaf:1989pz} is another combinatorial TQFT that can be related
to a 3--dimensional Abelian BF theory (being in fact viewable as a special case of a general
Turaev--Viro--Barrett-Westbury construction). The correspondence is not as well understood in its details
in four dimensions. The 4--dimensional counterpart of the Turaev--Viro--Barrett--Westbury model
is the Crane--Yetter--Broda model \ccite{Crane:1993cra,Broda:1993bu,Crane:1994ji}, which 
is identified with 4--dimensional BF theory with cosmological term 
for equal underlying Lie groups and related parameters  \ccite{Baez:1995ph}.
Likewise the Yetter model \ccite{Yetter:1993dh}, a 4-dimensional higher analogue of the 
Dijkgraaf--Witten model, can be related to a 4-dimensional Abelian BF theory.  
All 4--dimensional geometrically defined continuum TQFT mentioned above 
are again instances of 4--dimensional CS models.

At low energy, topologically ordered phases of
matter are described by TQFTs. In 3--dimensional spacetime, many
fractional quantum Hall states as well as lattice
models such as Kitaev's toric code model can be
explained by suitable CS models \ccite{Kitaev:2003fta,Levin:2004mi,Walker:2011mda}.
In this cases, fractional braiding statistics between quasiparticles emerges through the
correlation functions of a pair of Wilson loops forming a Hopf link. 
It is expected that fractional braiding statistics has a  4--dimensional spacetime analog.
Since particle--like excitations do not braid and have only ordinary bosonic/fermionic
statistics in this case, fractional statistics can only arise from the braiding of
either a point--like and a loop--like or two loop--like excitations. This has been adequately described
through BF type TQFTs \ccite{Balachandran:1992qg,Bergeron:1994ym,Szabo:1998ej} through the correlation
functions of Wilson loops and surfaces, pointing again to 4--dimensional CS theory.

Wilson lines \ccite{Wilson:1974sk} are relevant in the study of confinement in quantum chromodynamics, 
loop quantum gravity, symmetry breaking in string theory and, as we have just seen, condensed 
matter physics. They depend on the topology of the underlying knots
and, as shown in Witten's foundational work \ccite{Witten:1988hf}, 
they can be used to study knot topology in 3-dimensional CS theory
using basic techniques of quantum field theory.
CS correlators of Wilson line operators provide knot and link invariants. 
The 4-dimensional counterpart of Wilson lines, Wilson surfaces, 
are expected to be relevant in the analysis of non perturbative features of higher form 
gauge theory and quantum gravity. They also should be a basic element of any 
field theoretic approach to 4-dimensional 2--knot topology \ccite{CottaRamusino:1994ez}.
Based on Witten's paradigm, 
it should be possible to study surface knot topology in 4--dimensions computing correlators
of Wilson surfaces in an appropriate 4--dimensional version of CS theory
using again techniques of quantum field theory
\ccite{Soncini:2014zra,Zucchini:2015wba,Zucchini:2015xba,Alekseev:2015hda,Chekeres:2018kmh,Zucchini:2019mbz}.  

The holographic principle \ccite{tHooft:1993dmi,Susskind:1994vu}
has emerged as one of the most far reaching theoretical ideas in the last two decades. 
The first known occurrence of holography in quantum field theory is the correspondence
discovered by Witten \ccite{Witten:1988hf} between 3--dimensional
CS theory with gauge group $G$ as the bulk field theory and the 2--dimensional Wess--Zumino--Novikov--Witten (WZNW)
model \ccite{Wess:1971yu,Witten:1983tw} with target group $G$ as the boundary field theory, manifesting itself as
an equivalence between the space of quantum states of the CS model on a 2--fold $S$ and the space of conformal blocks
of the WZNW model on $S$. The holographic principle has also allowed for an effective 3--dimensional
CS description of edge states is the fractional quantum Hall effect \ccite{Fujita:2009kw,Takayanagi:2013uya}. 
Little is known, to the best of our knowledge on the holographic features of 4--dimensional CS theory.

\subsection{\textcolor{blue}{\sffamily Outline of the paper}}\label{subsec:outline}

The above considerations support our claim that the study of 4--dimensional
CS theory is a worthy undertaking.
In this paper, following the differential--geometric approach to TQFT, we shall 
formulate a 4--dimensional CS theory as a certain kind of strict higher gauge theory.
In fact, as ordinary CS theory exists only in odd dimensional
manifolds, 4--dimensional CS theory must be necessarily built in the framework
of higher gauge theory (see \ccite{Baez:2010ya} for a review). Such theory has been already considered in refs.
\ccite{Zucchini:2011aa,Soncini:2014ara,Zucchini:2015ohw} also in the semistrict case
on 4-folds without boundary. Here, we shall study it on 4-folds with boundary.
As we shall see, it is precisely in this case that the theory exhibits its most interesting
holographic features. Below, we outline briefly the construction of the 4--dimensional CS model we have carried out. 

The higher gauge symmetry of our 4--dimensional CS model is 
encoded by a Lie group crossed module. Various relevant topics of crossed module theory
are discussed in sect. \cref{sec:liecrmod}. Here 
it is enough to recall that a Lie group crossed module $\mathsans{M}$ consists of two Lie groups $\mathsans{E}$, $\mathsans{G}$ 
together with a Lie group action $\mu:\mathsans{G}\times\mathsans{E}\rightarrow\mathsans{E}$
of $\mathsans{G}$ on $\mathsans{E}$ by automorphisms and an equivariant target 
map $\tau:\mathsans{E}\rightarrow\mathsans{G}$ satisfying a certain identity \ccite{Baez5,Baez:2003fs}.

On general grounds, in order to construct the kinetic term of the Lagrangian of a field theory,
a non singular bilinear pairing invariant under the symmetries of the theory is required.
In the 4--dimensional CS model, a crossed module $\mathsans{M}$ with invariant pairing
$\langle\cdot,\cdot\rangle:\mathfrak{g}\times\mathfrak{e}\rightarrow\mathbb{R}$ 
carries out this task. In fact,
invariant pairings in higher gauge theory play a role similar to that of invariant traces in ordinary
one. An invariant pairing selects further isotropic crossed submodules of $\mathsans{M}$
as a distinguished subclass of submodules. 
These correspond to standard choices of linear boundary conditions for CS gauge fields and gauge transformations. 


In sect. \cref{sec:derform}, we introduce the formal set--up of derived Lie groups
and algebras originally worked out in refs. \ccite{Zucchini:2019rpp,Zucchini:2019pbv}.
This is essentially a superfield formalism providing an elegant and convenient way
of handling certain structural elements of a Lie group crossed module by organizing them
as functions of a formal odd variable $\bar\alpha\in\mathbb{R}[-1]$.
The derived Lie group $\DD\mathsans{M}$ of a crossed module $\mathsans{M}$ consists
of superfields of the form $\mathrm{P}(\bar\alpha)=\ee^{\bar\alpha P}p$,
where $p\in\mathsans{G}$, $P\in\mathfrak{e}[1]$, 
with certain group operations determined by $\mathsans{M}$. Its Lie algebra $\DD\mathfrak{m}$,
the derived Lie algebra,  
consists similarly of superfields of the form $\mathrm{U}(\bar\alpha)=u+\bar\alpha U$, 
where $u\in\mathfrak{g}$, $U\in\mathfrak{e}[1]$,  with the associated Lie bracket. 

The relevant fields of the 4--dimensional CS model are based on a 4--fold $M$.
They are crossed module valued inhomogeneous form fields.
More formally, they are maps from the shifted tangent bundle $T[1]M$ of $M$
into either the derived group $\DD\mathsans{M}$ or the derived algebra $\DD\mathfrak{m}$
or a shifted version of this.  They are thus representable as derived superfields,
functions of a formal odd variable $\alpha\in\mathbb{R}[1]$.

The derived superfield formalism is particularly suited for our 4--dimensional CS model because
of its compactness and capability of presenting it as an ordinary CS theory with an
exotic graded gauge group, the derived group. Indeed, by making evident the close relationship of higher to ordinary gauge
theory, it allows importing many ideas and techniques of the latter to the former. 
In particular, the gauge fields and the gauge transformations of the 4--dimensional CS model
can be treated in this fashion in the derived set--up. 

In sect. \cref{sec:4dchern}, we present the 4--dimensional CS model as a strict higher gauge theoretic model
using a the derived superfield formalism. 
In this way, if $\mathsans{M}$ is the gauge crossed module, the gauge field of the model is a superfield of the form
\begin{equation}
\Omega(\alpha)=\omega-\alpha\varOmega, 
\label{higau1/intr}
\end{equation}
where $\omega\in\Map(T[1]M,\mathfrak{g}[1])$, $\varOmega\in\Map(T[1]M,\mathfrak{e}[2])$.
$\omega$, $\varOmega$ represent the usual 1-- and 2--form components of a gauge 2--connection
in higher gauge theory. The 4--dimensional CS action reads as 
\begin{equation}
\sfC\hspace{-.75pt}\sfS(\Omega)
=\frac{k}{4\pi}\int_{T[1]M}\varrho_M\!\left(\Omega,\dd\Omega+\tfrac{1}{3}[\Omega,\Omega]\right)\!,
\label{4dchern1/intr}
\end{equation}
where $k$ is a constant, $(\cdot,\cdot)$ is a certain degree 1 invariant pairing derived by the invariant pairing
$\langle\cdot,\cdot\rangle$ and $\dd$ is a certain degree 1 nilpotent differential extending the de
Rham differential $d$. $\sfC\hspace{-.75pt}\sfS$ is formally identical to the 3--dimensional
CS action. It is 4--dimensional however because of the 1 unit of degree provided by the pairing $(\cdot,\cdot)$. 
$\sfC\hspace{-.75pt}\sfS$ can be expressed explicitly in components as 
\begin{multline}
\sfC\hspace{-.75pt}\sfS(\omega,\varOmega)=\frac{k}{2\pi}\int_{T[1]M}\varrho_M
\left\langle d\omega+\tfrac{1}{2}[\omega,\omega]-\tfrac{1}{2}\dot\tau(\varOmega),\varOmega\right\rangle
\\
-\frac{k}{4\pi}\int_{T[1]\partial M}\varrho_{\partial M} \left\langle\omega,\varOmega\right\rangle.
\label{4dchern2/intr}
\end{multline}
This can be described as a generalized BF theory with boundary term and cosmological term determined by the Lie
differential $\dot\tau$ of the target map $\tau$ of the crossed module $\mathsans{M}$. 

The higher gauge symmetry of the 4--dimensional CS model can also be described in the derived set--up. 
Analogously to a gauge field,  \pagebreak a gauge transformation is a superfield expression of the form \hphantom{xxx}
\begin{equation}
\mathrm{U}(\alpha)=\ee^{\alpha U}u,
\label{higau11/intr}
\end{equation}
where $u\in\Map(T[1]M,\mathsans{G})$, $U\in\Map(T[1]M,\mathfrak{e}[1])$.
$u$, $U$ represent the usual 0-- and 1--form components of a 1--gauge transformation
in higher gauge theory. $\mathrm{U}$ acts on the higher gauge field $\Omega$ as
\begin{equation}
\Omega^{\mathrm{U}}=\Ad\mathrm{U}^{-1}(\Omega)+\mathrm{U}^{-1}\mathrm{dU}.
\label{higau9/intr}
\end{equation}
This gauge transformation action is formally identical to that of ordinary gauge theory, but it yields when
expressed explicitly in components the usual gauge transformation relations of higher gauge
theory. 

In spite of the formal resemblance of 4-- and 3--dimensional CS theory when the derived
formulation is used, the invariance properties of the higher CS model differ in several important aspects
from those of the ordinary CS one, especially in relation to the effect of a boundary in the base manifold.
Unlike its 3--dimensional counterpart, the 4--dimensional CS action $\sfC\hspace{-.75pt}\sfS$
is fully gauge invariant if the 4--fold $M$
has no boundary. When $M$ does have a boundary, the action is no longer invariant, but the gauge variation 
is just a boundary term. The gauge invariance of the 4--dimensional theory depend therefore in a decisive way
on the kind of boundary conditions which are imposed on the gauge fields $\Omega$ and
gauge transformations $U$. These are discussed in sect. \cref{sec:4dchern}.

To quantize higher CS theory, proceeding as in the ordinary case, 
one should presumably allow for the broadest gauge symmetry
leaving the Boltzmann weight $\exp(i\hfpt\sfC\hspace{-.75pt}\sfS)$ invariant possibly
restricting the value of the CS level $k$ as appropriate.
On a 4--fold $M$ with no boundary, the theory is fully gauge invariant and so there are no restrictions
on either the gauge symmetry or the level. 
On a 4--fold $M$ with boundary, one should impose
on the relevant higher gauge fields $\Omega$ and transformations $\mathrm{U}$ the weakest possible boundary
conditions capable of rendering the gauge variation an integer multiple of $2\pi$.
These are also discussed in sect. \cref{sec:4dchern}. Depending also on crossed module $\mathsans{M}$
and the invariant pairing $\langle\cdot,\cdot\rangle$, level quantization can occur. $\vphantom{\ul{\ul{\ul{g}}}}$

In sect. \cref{sec:4dchern}, a canonical analysis a la Dirac of the 4--dimensional CS model
is carried out. 
The close relationship of the canonical formulations of 4-- and 3--dimensional CS theory is again
especially evident in the derived framework. 
The results of the ordinary theory generalize to the higher one, but in a non trivial way.
The Hamiltonian generators of the gauge symmetry which through their weak vanishing define the physical
phase space of flat gauge fields of the theory are determined. They are first class only if appropriate
boundary conditions are obeyed by the gauge fields and the gauge transformations. 
It is found that, depending again on these boundary conditions, 
there exist generically surface charges obeying a non trivial Poisson bracket algebra.
This is a higher counterpart of the familiar WZNW current algebra arising
in the corresponding canonical analysis of 3--dimensional CS model.

Gauge theories defined on manifolds with boundaries may exhibit emergent boundary degrees
of freedom called edge modes. In fact, boundaries normally break gauge
invariance transforming in this way gauge degrees of freedom into physical ones. 
In sect. \cref{sec:4dchern}, we outline a canonical theory of the edge modes of 4--dimensional
CS theory and their physical symmetries, extending the corresponding analysis of the 3-dimensional theory
\ccite{Wen:1992vi,Carlip:2005zn,Afshar:2017okz,Donnelly:2016auv,Geiller:2019bti}.  

In sect. \cref{sec:4dchern}, we finally show how the covariant Schroedinger quantization of 4--dimensional
CS theory can be carried out on the lines of the 3--dimensional case \ccite{Axelrod:1989xt}.
Among other things, we obtain the higher analogue of the WZNW Ward identities obeyed by the
wave functionals and an expression of the higher WZNW action, which turns out to be fully
topological. 




In sect. \cref{sec:aplc}, we finally illustrate a few field theoretic models which are interesting
non trivial instances of 
4--dimensional CS theory, in particular the toric and the Abelian projection models.


\subsection{\textcolor{blue}{\sffamily Outlook}}\label{subsec:outlook}

There remain to ascertain to what extent the 4--dimensional CS theory presented in this paper
is capable to reproduce various 'disguised' CS model that have appeared in the literature and
yield new interesting ones.

Our analysis of the edge sector of 4--dimensional CS theory
is still incomplete. Although we have identified the edge fields and
the physical edge symmetry group, there remain
basic problems to be solved such as the lack of a Lagrangian and Hamiltonian
description of the dynamics of edge fields and the identification of the edge modes
observables. A viable edge field theory of the 4--dimensional  
CS model is a topic certainly deserving an in depth study.

In this paper, we have not discussed the incorporation of Wilson surfaces in the theory.
This can be done in the standard framework of strict higher gauge theory along the lines of refs. 
\ccite{Zucchini:2015wba,Zucchini:2015xba}. There is however another so far unexplored rout to
dealing with this problem. In a series of papers
\ccite{Balachandran:1977ub,Alekseev:1988vx,Diakonov:1989fc,Diakonov:1996zu}
(see also ref. \ccite{Beasley:2009mb} for a review)
a geometrical action capable to compute Wilson lines in 3--dimensional CS theory
was obtained and studied. Given the formal similarity of 4-- and 3--dimensional CS in the
derived formulation, it is conceivable that a formally analogous geometrical action may
be found that is capable of computing Wilson surfaces in 4--dimensional CS theory. This is
a promising line of inquiry deserving to be pursued. 

All the above matters are left for future work \cite{Zucchini:2021inp}.

\vfil\eject

\section{\textcolor{blue}{\sffamily Lie crossed modules and invariant pairings}}\label{sec:liecrmod}

Lie group and algebra crossed modules constitute the type of algebraic structure on which the higher
CS theory elaborated in this paper rests. In this section, we review this topic,
with no pretence of mathematical
rigour or completeness, dwelling only on those points which are relevant in the following analysis.
The theory of crossed modules is best formulated in a categorical framework. However, we shall not
insist on the categorical features of these algebraic structures. 
See ref. \ccite{Baez5,Baez:2003fs} for an exhaustive exposition of this subject.

The subject matter covered in this section is disparate and has been gathered
mainly for later reference. 
Subsect. \cref{subsec:liecrmod} is a review of the theory of Lie group and algebra crossed modules
serving also the purpose of set the notation used throughout this paper. The material of subsect.
\cref{subsec:crmodinv} is required mostly from subsect. \cref{subsec:4dchern} onwards.
The material of subsect. \cref{subsec:crosumo} is used primarily in subsects.
\cref{subsec:levelchern}, \cref{subsec:levelquant}, \cref{subsec:surfchern}.


\subsection{\textcolor{blue}{\sffamily Lie group and algebra crossed modules}}\label{subsec:liecrmod}

Crossed modules encode the symmetry of higher gauge theory both at the finite and the infinitesimal level.
Our path toward 4--dimensional CS theory must necessarily start with them. In this subsection, we
review the theory of Lie group and algebra crossed modules and module morphisms.
The precise definitions and properties of crossed modules are collected in app. \cref{app:def}. 

\vspace{2.5mm}

\noindent
{\it Lie group crossed modules}

The structure of finite Lie crossed module 
abstracts and extends the set--up consisting of a Lie group $\mathsans{G}$ and 
a normal Lie subgroup $\mathsans{E}$ of $\mathsans{G}$ acted upon by $\mathsans{G}$ by conjugation.
A Lie group crossed module $\mathsans{M}$ consists indeed of two Lie groups $\mathsans{E}$, $\mathsans{G}$ 
together with a Lie group action $\mu:\mathsans{G}\times\mathsans{E}\rightarrow\mathsans{E}$
of $\mathsans{G}$ on $\mathsans{E}$ by automorphisms and an equivariant Lie group 
map $\tau:\mathsans{E}\rightarrow\mathsans{G}$
rendering $\mu$ compatible with adjoint action of $\mathsans{E}$
(cf. eqs. \ceqref{liecrmod1}, \ceqref{liecrmod2}). $\mathsans{E}$, $\mathsans{G}$ and 
$\tau$, $\mu$ are called the source and target groups and the target and action structure maps of
$\mathsans{M}$, respectively. Below, we shall write $\mathsans{M}=(\mathsans{E},\mathsans{G},\tau,\mu)$
to specify the crossed module through its data. 

A morphism of finite Lie crossed modules is a map of crossed modules preserving the module structure
expressing a relationship of sameness or likeness of the modules involved. 
More explicitly, a morphism $\beta:\mathsans{M}'\rightarrow\mathsans{M}$ of Lie group crossed modules
consists of two Lie group morphisms $\phi:\mathsans{G}'\rightarrow\mathsans{G}$ and 
$\varPhi:\mathsans{E}'\rightarrow\mathsans{E}$ intertwining in the appropriate sense the structure 
maps $\tau'$, $\mu'$, $\tau$, $\mu$ (cf. eqs. \ceqref{liecrmod3}, \ceqref{liecrmod4}).
We shall normally write $\beta:\mathsans{M}'\rightarrow\mathsans{M}=(\varPhi,\phi)$
to indicate constituent morphisms of the crossed module morphism.

Taking the direct product of the relevant constituent data in the Lie group category, 
it is possible to construct the direct product $\mathsans{M}_1\times\mathsans{M}_2$ 
of two Lie group crossed modules $\mathsans{M}_1$, $\mathsans{M}_2$ and the direct product $\beta_1\times\beta_2$
of two Lie group crossed module morphisms $\beta_1$, $\beta_2$ in straightforward fashion.
Complicated crossed modules and module morphisms can sometimes be analyzed by factorizing them
into direct products of simpler modules and module morphisms. 

There exist many examples of Lie group crossed modules and crossed module morphisms.
In particular, Lie groups and automorphisms, representations and central extensions of Lie groups can be
described as instances of Lie group crossed modules. Lie group morphisms can be employed to
construct morphisms of such crossed modules. There are two basic model crossed modules to which a
broad range of crossed modules entering in the formulation of higher CS theory can be related to.
They are defined for any Lie group $\mathsans{G}$. 
The first is the inner automorphism crossed module of $\mathsans{G}$,
$\INN\mathsans{G}=(\mathsans{G},\mathsans{G},\id_{\mathsans{G}},\Ad_{\,\mathsans{G}})$.
The second is the (finite) coadjoint action crossed module of $\mathsans{G}$,
$\AD^*\mathsans{G}=(\mathfrak{g}^*,\mathsans{G},1_{\mathsans{G}},\Ad_{\mathsans{G}}{}^*)$, where 
$\mathfrak{g}$ is the Lie algebra of $\mathsans{G}$ and its dual space 
$\mathfrak{g}^*$ is viewed as an Abelian group.
A crossed module morphism
$\rho:\INN\mathsans{G}'\rightarrow\INN\mathsans{G}$ reduces to a
group morphism $\chi:\mathsans{G}'\rightarrow\mathsans{G}$. A 
crossed module morphism
$\alpha:\AD^*\mathsans{G}'\rightarrow\AD^*\mathsans{G}$ is specified by 
a group morphism $\lambda:\mathsans{G}'\rightarrow\mathsans{G}$ and an intertwiner
$\varLambda:\mathfrak{g}'^*\rightarrow\mathfrak{g}^*$ of $\Ad_{\mathsans{G}'}{}^*$ to
$\Ad_{\mathsans{G}}{}^*\circ\lambda$.

\vspace{2.5mm}

\noindent
{\it Lie algebra crossed modules}

The structure of infinitesimal Lie crossed module 
axiomatizes likewise the set--up consisting of a Lie algebra $\mathfrak{g}$ and 
a Lie ideal $\mathfrak{e}$ of $\mathfrak{g}$ equipped with the adjoint action of $\mathfrak{g}$.
It is therefore the differential version of that of finite Lie crossed module. A Lie algebra
crossed module $\mathfrak{m}$ consists so of two Lie algebras $\mathfrak{e}$, $\mathfrak{g}$ 
together with a Lie algebra action $m:\mathfrak{g}\times\mathfrak{e}\rightarrow\mathfrak{e}$
of $\mathfrak{g}$ on $\mathfrak{e}$ by derivations and an equivariant Lie algebra 
map $t:\mathfrak{e}\rightarrow\mathfrak{g}$ making $m$ compatible with adjoint action of $\mathfrak{e}$
(cf. eqs. \ceqref{liecrmod5}, \ceqref{liecrmod6}). $\mathfrak{e}$, $\mathfrak{g}$ and
$t$, $m$ are called the source and target algebras and the target and action structure maps of $\mathfrak{m}$,
respectively. Below, we shall write $\mathfrak{m}=(\mathfrak{e},\mathfrak{g},t,m)$
to specify the crossed module through its data. 

A morphism of infinitesimal Lie crossed modules is a map of crossed modules preserving the 
module structure describing a way such crossed modules are congruent. 
It is therefore the differential version of that of morphism of finite Lie crossed module. 
More explicitly, a morphism $p:\mathfrak{m}'\rightarrow\mathfrak{m}$ of Lie algebra crossed modules
consists of two Lie algebra morphisms $h:\mathfrak{g}'\rightarrow\mathfrak{g}$ and 
$H:\mathfrak{e}'\rightarrow\mathfrak{e}$ intertwining in the appropriate sense the structure 
maps $t'$, $m'$, $t$, $m$ (cf. eqs. \ceqref{liecrmod7}, \ceqref{liecrmod8}).
We shall use often the notation $p:\mathfrak{m}'\rightarrow\mathfrak{m}=(H,h)$
to indicate constituent morphisms of the crossed module morphism. 

Similarly to the Lie group case, taking the direct sum of the relevant
constituent data in the Lie algebra category, it is possible to define 
the direct sum $\mathfrak{m}_1\oplus\mathfrak{m}_2$ 
of two Lie algebra crossed modules $\mathfrak{m}_1$, $\mathfrak{m}_2$ and direct sum $p_1\oplus p_2$
of two Lie algebra crossed module morphisms $p_1$, $p_2$ in obvious fashion.
These notions answer at the differential level to those of direct products
of finite crossed modules and module morphisms. They allow to analyze
crossed modules and module morphisms by decomposing them
as direct sums of more elementary modules and module morphisms as we shall see in particular in
subsect. \cref{subsec:crmodinv} below. 

Many examples of Lie algebra crossed modules and crossed module morphisms
are also available. They pair with the basic examples of
Lie group crossed modules and crossed module morphisms recalled above.
Ordinary Lie algebras and derivations, representations and central extensions of Lie algebras can
be described as instances of Lie algebra crossed modules and Lie algebra morphisms can be assembled
variously to construct morphisms of such crossed modules. 
In particular, there are two basic model crossed modules defined for any Lie algebra
$\mathfrak{g}$ corresponding to the inner
automorphism and coadjoint action crossed modules introduced above. 
The first is the inner derivation crossed module of $\mathfrak{g}$,
$\INN\mathfrak{g}=(\mathfrak{g},\mathfrak{g},\id_{\mathfrak{g}},\ad_{\,\mathfrak{g}})$.
The second is the (infinitesimal) coadjoint action crossed module of $\mathfrak{g}$,
$\AD^*\mathfrak{g}=(\mathfrak{g}^*,\mathfrak{g},0_{\mathfrak{g}},\ad_{\mathfrak{g}}{}^*)$,
where $\mathfrak{g}^*$ is regarded as an Abelian algebra.
A crossed module morphism
$r:\INN\mathfrak{g}'\rightarrow\INN\mathfrak{g}$ reduces to an
algebra morphism $x:\mathfrak{g}'\rightarrow\mathfrak{g}$. A 
crossed module morphism
$a:\AD^*\mathfrak{g}'\rightarrow\AD^*\mathfrak{g}$ is specified by 
an algebra morphism $l:\mathfrak{g}'\rightarrow\mathfrak{g}$ and an intertwiner
$L:\mathfrak{g}'^*\rightarrow\mathfrak{g}^*$ of $\ad_{\mathfrak{g}'}{}^*$ to
$\ad_{\mathfrak{g}}{}^*\circ l$. 


Lie differentiation plays the same important role in Lie crossed module theory
as it does in Lie group theory.
With any Lie group crossed module $\mathsans{M}=(\mathsans{E},\mathsans{G},\tau,\mu)$
there is associated the Lie algebra crossed module
$\mathfrak{m}=(\mathfrak{e},\mathfrak{g},\dot\tau,\sdot\mu{}\sdot\hfpt)$, where 
$\mathfrak{e}$, $\mathfrak{g}$ are the Lie algebras of Lie groups $\mathsans{E}$, $\mathsans{G}$ respectively
and the dot notation $\dot{}$ denotes Lie differentiation along the relevant Lie group 
(cf. app. \cref{app:ident}) for more details),
much as a Lie algebra is associated with a Lie group. Similarly, with any Lie group crossed module morphism
$\beta:\mathsans{M}'\rightarrow\mathsans{M}=(\varPhi,\phi)$ there is associated the Lie algebra crossed module morphism
$\dot\beta:\mathfrak{m}'\rightarrow\mathfrak{m}=(\dot\varPhi,\dot\phi)$, 
just as a Lie algebra morphism is associated with a Lie group morphism.

As examples, we mention that the Lie algebra crossed modules of the Lie group crossed modules
$\INN\mathsans{G}$ and $\AD^*\mathsans{G}$ we introduced above for any Lie group $\mathsans{G}$ 
are precisely $\INN\mathfrak{g}$ and $\AD^*\mathfrak{g}$, respectively, as expected.


\subsection{\textcolor{blue}{\sffamily Crossed modules with invariant pairing}}\label{subsec:crmodinv}

Crossed modules with invariant pairing are an essential ingredient of the construction of
4--dimensional CS actions. 
Indeed, invariant pairings in higher gauge theory
play a role similar to that of invariant traces in ordinary gauge theory. 
We introduce and discuss this topic in this subsection.

On general grounds, in order to construct the kinetic term of the Lagrangian of a field theory,
a non singular bilinear pairing is required. Further, when the field theory is characterized by
certain symmetries, the same symmetries must be enjoyed by the pairing, which so is in addition
invariant.

The field content of 4--dimensional 
CS gauge theory whose symmetry is described infinitesimally by a Lie
algebra crossed module $\mathfrak{m}=(\mathfrak{e},\mathfrak{g},t,m)$ comprises a 
$\mathfrak{g}$--valued 1--form gauge field $\omega$ and  an $\mathfrak{e}$--valued 2--form
gauge field $\varOmega$. 
The bilinear pairing entering in the kinetic term of the gauge fields, thus,  
must be defined on either $\mathfrak{g}\times\mathfrak{g}$ or 
$\mathfrak{g}\times\mathfrak{e}$ or $\mathfrak{e}\times\mathfrak{e}$.
The CS kinetic term, which  must be of derivative order $1$ in order the field equations to be
equivalent to the flatness conditions for the gauge fields, can take thus three forms
\begin{equation}
K_3=\langle d\omega,\omega\rangle, \qquad  K_4=\langle d\omega,\varOmega\rangle,
\qquad  K_5=\langle d\varOmega,\varOmega\rangle, 
\label{kin1}
\end{equation}
These are a 3--, 4--, 5--form yielding 
a 3--, 4-- and 5--dimensional CS model respectively. As we are interested in a 4--dimensional one,
it is the second form of the pairing that is relevant for us. So, the pairing will be a non singular bilinear form
$\langle\cdot,\cdot\rangle:\mathfrak{g}\times\mathfrak{e}\rightarrow\mathbb{R}$.

The infinitesimal higher gauge symmetry described by $\mathfrak{m}$ is ultimately codified
in the adjoint action of $\mathfrak{g}$ on itself and the module action $m$ of $\mathfrak{g}$
on $\mathfrak{e}$. The kinetic term will so have the required invariance properties if
the pairing $\langle\cdot,\cdot\rangle$ obeys
\begin{equation}
\langle\ad z(x),X\rangle+\langle x,m(z,X)\rangle=0
\label{crmodinv2/rep}
\end{equation}
for $z,x\in\mathfrak{g}$, $X\in\mathfrak{e}$.
It remains to clarify how the pairing $\langle\cdot,\cdot\rangle$ behaves with regard to the
module target map $t$. This boils down to find an appropriate requirement for the difference 
$\langle t(X),Y\rangle-\langle t(Y),X\rangle$ with for $X,Y\in\mathfrak{e}$. The minimal choice
avoiding introducing further structures consists in demanding that 
\begin{equation}
\langle t(X),Y\rangle=\langle t(Y),X\rangle. 
\label{crmodinv1/rep}
\end{equation}

\vspace{2.5mm}

\noindent
{\it Lie algebra crossed modules with invariant pairing}

A Lie algebra crossed module with invariant pairing is 
Lie algebra crossed module $\mathfrak{m}=(\mathfrak{e},\mathfrak{g},t,m)$
equipped with a non singular bilinear form
$\langle\cdot,\cdot\rangle:\mathfrak{g}\times\mathfrak{e}\rightarrow\mathbb{R}$ 
enjoying properties \ceqref{crmodinv2/rep}, \ceqref{crmodinv1/rep}.

A crossed module $\mathfrak{m}$
with invariant pairing is balanced, which means that $\dim\mathfrak{g}=\dim\mathfrak{e}$,
because of the non singularity of the pairing. This is not as strong a restriction as it may appear
at first sight. It can be shown that any Lie algebra
crossed module $\mathfrak{m}$ 
can always be trivially  extended to a balanced crossed module
$\mathfrak{m}'$,  
for which depending on cases one has either 
$\mathfrak{e}'=\mathfrak{e}\oplus\mathfrak{p}$ and $\mathfrak{g}'=\mathfrak{g}$
or $\mathfrak{e}'=\mathfrak{e}$ and $\mathfrak{g}'=\mathfrak{g}\oplus\mathfrak{q}$ 
for suitable Abelian Lie algebras $\mathfrak{p}$, $\mathfrak{q}$. 

A morphism $p:\mathfrak{m}'\rightarrow\mathfrak{m}=(H,h)$ of Lie algebra crossed modules with
invariant pairings $\langle\cdot,\cdot\rangle'$, $\langle\cdot,\cdot\rangle$ is naturally
defined as a crossed module morphism that preserves the pairings (cf. eq. \ceqref{crmodinv3}).
Such a morphism describes a stronger form of sameness or likeness of the crossed modules
concerning not only their algebraic structures but
involving also to their invariant pairings. 

We shall now explore the implications
of having an invariant pairing structure attached to the crossed module.

\vspace{2.5mm}

\noindent
{\it Core and residue of a crossed module with invariant pairing}

If $\mathfrak{m}$ is a Lie algebra crossed module, then $\ker t$
is a central ideal of $\mathfrak{e}$ and $\ran t$ is an ideal of $\mathfrak{g}$. 
Using these properties,
one can show that with $\mathfrak{m}$ there are canonically associated two further Lie algebra
crossed modules.

The first one, which we shall call the core of $\mathfrak{m}$ in the following, is
the crossed module $\CC\mathfrak{m}=(\mathfrak{e}/\ker t,\ran t,
t_{\hspace{.33pt}\CC\hspace{.66pt}},m_{\hspace{.33pt}\CC\hspace{.66pt}})$, where
\begin{align}
&t_{\hspace{.33pt}\CC\hspace{.66pt}}(X+\ker t)=t(X),
\vphantom{\Big]}
\label{crmodinv5}
\\
&m_{\hspace{.33pt}\CC\hspace{.66pt}}(x,X+\ker t)=m(x,X)+\ker t
\vphantom{\Big]}
\label{crmodinv6}
\end{align}
for $x\in\ran t$, $X\in\mathfrak{e}$. It can be verified that the structure maps
$t_{\hspace{.33pt}\CC\hspace{.66pt}}$ and $m_{\hspace{.33pt}\CC\hspace{.66pt}}$
are well defined and satisfy the required properties \ceqref{liecrmod5}, \ceqref{liecrmod6}.
The core crossed module $\CC\mathfrak{m}$ is characterizing by the invertibility
of $t_{\hspace{.33pt}\CC\hspace{.66pt}}$. 

The second one, which we shall call the residue of $\mathfrak{m}$, is
the crossed module $\RR\mathfrak{m}=(\ker t,\mathfrak{g}/\ran t,t_{\hspace{.33pt}\RR\hspace{.66pt}},
m_{\hspace{.33pt}\RR\hspace{.66pt}})$, where
\begin{align}
&t_{\hspace{.33pt}\RR\hspace{.66pt}}(X)=\ran t,
\vphantom{\Big]}
\label{crmodinv7}
\\
&m_{\hspace{.33pt}\RR\hspace{.66pt}}(x+\ran t,X)=m(x,X) 
\vphantom{\Big]}
\label{crmodinv8}
\end{align}   
for $x\in\mathfrak{g}$, $X\in\ker t$. Again, it can be verified that the structure maps $t_{\hspace{.33pt}\RR\hspace{.66pt}}$
and $m_{\hspace{.33pt}\RR\hspace{.66pt}}$
are well defined and satisfy the properties \ceqref{liecrmod5}, \ceqref{liecrmod6}.
The characterizing property of the residue crossed module $\RR\mathfrak{m}$ is the vanishing 
of $t_{\hspace{.33pt}\RR\hspace{.66pt}}$. 


If $\mathfrak{m}$ is in addition equipped with
an invariant pairing $\langle\cdot,\cdot\rangle$, then $\CC\mathfrak{m}$ and $\RR\mathfrak{m}$
are equipped with induced invariant pairing $\langle\cdot,\cdot\rangle_{\hspace{.33pt}\CC\hspace{.66pt}}$
and $\langle\cdot,\cdot\rangle_{\hspace{.33pt}\RR\hspace{.66pt}}$.
For $\CC\mathfrak{m}$, we have 
\begin{equation}
\langle x,X+\ker t\rangle_{\hspace{.33pt}\CC\hspace{.66pt}}=\langle x,X\rangle
\label{crmodinv9}
\end{equation}
where $x\in\ran t$, $X\in\mathfrak{e}$. \pagebreak It is straightforward to check that
$\langle\cdot,\cdot\rangle_{\hspace{.33pt}\CC\hspace{.66pt}}$ is well defined and obeys conditions
\ceqref{crmodinv2}, \ceqref{crmodinv1}.
For $\RR\mathfrak{m}$, we have similarly
\begin{equation}
\langle x+\ran t,X\rangle_{\hspace{.33pt}\RR\hspace{.66pt}}=\langle x,X\rangle
\label{crmodinv10}
\end{equation}
where $x\in\mathfrak{g}$, $X\in\ker t$. Again, $\langle\cdot,\cdot\rangle_{\hspace{.33pt}\RR\hspace{.66pt}}$
is well defined and satisfies properties \ceqref{crmodinv2}, \ceqref{crmodinv1}. 

In subsect. \cref{subsec:liecrmod}, we introduced two basic model Lie algebra crossed modules, the inner
derivation crossed module $\INN\mathfrak{g}$ and the infinitesimal coadjoint action crossed module 
$\AD^*\mathfrak{g}$ of a Lie algebra $\mathfrak{g}$. They are evidently both balanced and they can both be
equipped with invariant pairings, as we shall show momentarily. They are indeed prototypical crossed modules
with these properties. 


The inner derivation crossed module of $\mathfrak{g}$ is
$\INN\mathfrak{g}=(\mathfrak{g},\mathfrak{g},\id_{\mathfrak{g}},\ad_{\mathfrak{g}})$. $\INN\mathfrak{g}$
carries no canonical invariant pairing, but any $\ad_{\mathfrak{g}}$ invariant symmetric non singular pairing
of $\mathfrak{g}$ can be used as one. 
$\INN\mathfrak{g}$ is characterized by the following property.
If $\mathfrak{m}=(\mathfrak{e},\mathfrak{g},t,m)$ is a Lie algebra crossed module with invariant
pairing such that $t$ is invertible, then there is an invariant pairing on $\INN\mathfrak{g}$
such that $\mathfrak{m}$ is isomorphic to $\INN\mathfrak{g}$, the isomorphism $i$ of $\mathfrak{m}$ onto
$\INN\mathfrak{g}$ being given by the pair $(t,\id_{\mathfrak{g}})$ and the invariant pairing on
$\INN\mathfrak{g}$ being related to that of $\mathfrak{m}$ by
$\langle x,t^{-1}(X)\rangle_{\INN}=\langle x,X\rangle$.

The infinitesimal coadjoint action crossed module of $\mathfrak{g}$ we consider next is  
$\AD^*\mathfrak{g}=(\mathfrak{g}^*,\mathfrak{g},0_{\mathfrak{g}},\ad_{\mathfrak{g}}{}^*)$. 
Unlike the inner derivation crossed module discussed above, $\AD^*\mathfrak{g}$
carries a natural invariant pairing, the duality pairing of $\mathfrak{g}$ and $\mathfrak{g}^*$. 
$\AD^*\mathfrak{g}$ enjoys the following property.
If $\mathfrak{m}=(\mathfrak{e},\mathfrak{g},t,m)$ is a Lie algebra crossed module with invariant pairing 
with $t$ vanishing, then $\mathfrak{m}$ is isomorphic to $\AD^*\mathfrak{g}$, 
the isomorphism $j$ of $\INN\mathfrak{g}$ onto $\mathfrak{m}$ 
being given by the pair $(J,\id_{\mathfrak{g}})$, where $J$ 
is the linear isomorphism of $\mathfrak{g}^*$ onto $\mathfrak{e}$ such that
$\langle x,J(X)\rangle=\langle x,X\rangle_{\AD^*}$ with $\langle\cdot,\cdot\rangle_{\AD^*}$
the duality pairing of $\mathfrak{g}$ and $\mathfrak{g}^*$. 

\vspace{2.75mm}

\noindent
{\it The decomposition theorem}

Consider again a generic Lie algebra crossed module $\mathfrak{m}=(\mathfrak{e},\mathfrak{g},t,m)$
with invariant pairing, The characterizing property of $\CC\mathfrak{m}$
is $t_{\hspace{.33pt}\CC\hspace{.66pt}}$ being a linear isomorphism. This makes $\CC\mathfrak{m}$ isomorphic to the 
crossed module $\INN\ran t$ equipped with a suitable invariant pairing.
Likewise, the characterizing property of $\RR\mathfrak{m}$ is $t_{\hspace{.33pt}\RR\hspace{.66pt}}$ vanishing. 
$\RR\mathfrak{m}$ is in this way isomorphic to the crossed module $\AD^*(\mathfrak{g}/\ran t)$
with the canonical invariant pairing. 

The following decomposition theorem is key to understanding relevant aspects of the gauge
symmetry of 4--dimensional CS theory studied later on.
Consider a balanced crossed module $\mathfrak{m}=(\mathfrak{e},\mathfrak{g},t,m)$ with invariant pairing.
Suppose that there exists an ideal $\mathfrak{h}$ of $\mathfrak{g}$ such that $\ran t\cap\mathfrak{h}=0$
and \hphantom{xxxxxxxxxxxx}
\begin{equation}
\mathfrak{g}\simeq\ran t\oplus\mathfrak{h}. 
\label{crmodinv11}
\end{equation}
Then, $\mathfrak{m}$ decomposes as \hphantom{xxxxxxxxxxx} 
\begin{equation}
\mathfrak{m}\simeq\CC\mathfrak{m}\oplus\RR\mathfrak{m}.
\label{crmodinv12}
\end{equation}
in the category of crossed modules with invariant pairing.
The proof of the theorem requires in an essential way the use of the 
invariant pairing (cf. app. \cref{app:coreres}). By the isomorphisms noticed earlier, we could write
\ceqref{crmodinv12} as
\begin{equation}
\mathfrak{m}\simeq\INN\ran t\oplus\AD^*(\mathfrak{g}/\ran t).
\label{crmodinv13}
\end{equation}
This means that in the analysis of the following sections, under weak conditions, we can assume 
that the relevant Lie algebra crossed module $\mathfrak{m}$ with invariant pairing
is of the form $\INN\mathfrak{g}_c\oplus\AD^*\mathfrak{g}_r$ for certain
Lie algebras $\mathfrak{g}_c$, $\mathfrak{g}_r$ with the direct summands equipped
respectively with the appropriate and the canonical invariant pairings.



\vspace{2.25mm}

\noindent
{\it Lie group crossed modules with invariant pairing}

The subtlest features of 4--dimensional CS theory emerge when the underlying higher gauge symmetry is
considered at the finite level through the appropriate Lie group crossed module. 
Invariant pairings are naturally defined only on Lie algebra crossed modules. It is possible however to attach
an invariant pairing also to a Lie group crossed module by endowing the associated Lie algebra crossed module
with one. However, upon doing so, it is necessary to strengthen the invariance condition of the pairing 
by requiring invariance to hold at the finite and not only infinitesimal level. 

We shall thus \pagebreak define a Lie group crossed module with invariant pairing as a 
crossed module $\mathsans{M}=(\mathsans{E},\mathsans{G},\tau,\mu)$ such that the associated
Lie algebra crossed module $\mathfrak{m}=(\mathfrak{e},\mathfrak{g},\dot\tau,\sdot\mu\sdot)$
(cf. app. \cref{app:ident}) is a 
crossed module with invariant pairing $\langle\cdot,\cdot\rangle$ 
enjoying the property that \hphantom{xxxxxxxx}
\begin{equation}
\langle\Ad a(x),\mu\sdot(a,X)\rangle=\langle x,X\rangle
\label{crmodinv14/rep}
\end{equation}
for $a\in\mathsans{G}$, $x\in\mathfrak{g}$, $X\in\mathfrak{e}$
(cf. apps. \cref{app:ident}, \cref{app:crmodinv})
Notice that \ceqref{crmodinv14/rep} implies \ceqref{crmodinv2/rep} with $m=\sdot\mu\sdot$ 
via Lie differentiation with respect to $a$, while \ceqref{crmodinv1/rep} holds with $t=\dot\tau$.

Again, the non singularity of $\langle\cdot,\cdot\rangle$ implies that
$\mathsans{M}$ is balanced, $\dim\mathsans{E}=\dim\mathsans{G}$. 
Analogously to the Lie algebra case, it
can be shown that any Lie group crossed module $\mathsans{M}$ 
can always be trivially extended to a balanced crossed module $\mathsans{M}'$, 
with either $\mathsans{E}'=\mathsans{E}\times\mathsans{P}$
and $\mathsans{G}'=\mathsans{G}$ 
or $\mathsans{E}'=\mathsans{E}$ and $\mathsans{G}'=\mathsans{G}\times\mathsans{Q}$ 
for suitable Abelian Lie groups $\mathsans{P}$, $\mathsans{Q}$, depending on cases.

For a Lie group crossed module $\mathsans{M}=(\mathsans{E},\mathsans{G},\tau,\mu)$
with invariant pairing $\langle\cdot,\cdot\rangle$, 
identity \ceqref{crmodinv14/rep} implies the relation 
\begin{equation}
\langle x,\sdot\mu(y,A)\rangle=\langle y,\sdot\mu(x,A^{-1})\rangle,
\label{crmodinv15}
\end{equation}
where $x,y\in\mathfrak{g}$ and $A\in\mathsans{E}$, under mild assumptions on the Lie group
$\mathsans{E}$. Specifical\-ly, \ceqref{crmodinv15} holds when $\mathsans{E}$
is connected and also when  $\mathsans{E}$ is not connected
in the connected component of the identity of $\mathsans{E}$ and in any connected component
of $\mathsans{E}$ where it holds for at least one element.  \ceqref{crmodinv15} holds also when $\tau$
is invertible with no restrictions on $\mathsans{E}$.
Property \ceqref{crmodinv15} in a sense completes \ceqref{crmodinv14/rep}.
We shall call the crossed module $\mathsans{M}$ fine if \ceqref{crmodinv15} holds. 
The seemingly technical condition plays in fact an important role in the analysis
of the gauge invariance of 4--dimensional CS theory, as we shall see in due course. 

If $\mathsans{M}=(\mathsans{E},\mathsans{G},\tau,\mu)$ is a Lie group crossed module 
with invariant pairing 
such that a direct sum decomposition of $\mathfrak{g}$ of the form \ceqref{crmodinv11}
is available, then  
the direct sum decomposition \ceqref{crmodinv12} of the associated Lie algebra crossed module
$\mathfrak{m}=(\mathfrak{e},\mathfrak{g},\dot\tau,\sdot\mu\sdot)$ into its
core and residue $\CC\mathfrak{m}$, $\RR\mathfrak{m}$ holds.
One does not expect a corresponding direct product factorization of the module $\mathsans{M}$ to
occur in analogy to what happens in the akin setting of
Lie group theory.$\vphantom{\ul{\ul{\ul{g}}}}$
\pagebreak It is nevertheless instructive to examine this issue is some detail.


Similarly to the Lie algebra case, it is possible to canonically associate with $\mathsans{M}$ two
Lie group crossed modules with invariant pairing, its core and residue crossed modules,
relying on the properties that $\ker\tau$ is a central normal subgroup $\mathsans{E}$ and
$\ran\tau$ is a normal subgroup of $\mathsans{G}$, in analogy with the Lie algebra case. 

The core of $\mathsans{M}$ is the crossed module $\CC\mathsans{M}=(\mathsans{E}/\ker\tau,\ran\tau,
\tau_{\hspace{.33pt}\CC\hspace{.66pt}},\mu_{\hspace{.33pt}\CC\hspace{.66pt}})$, where
\begin{align}
&\tau_{\hspace{.33pt}\CC\hspace{.66pt}}(A\ker\tau)=\tau(A),
\vphantom{\Big]}
\label{crmodinv16}
\\
&\mu_{\hspace{.33pt}\CC\hspace{.66pt}}(a,A\ker\tau)=\mu(a,A)\ker\tau 
\vphantom{\Big]}
\label{crmodinv17}
\end{align}
for $a\in\ran\tau$, $A\in\mathsans{E}$. It can be verified that the structure maps $\tau_{\hspace{.33pt}\CC\hspace{.66pt}}$
and $\mu_{\hspace{.33pt}\CC\hspace{.66pt}}$
are well defined and satisfy the required properties \ceqref{liecrmod1}, \ceqref{liecrmod2}.
The core crossed module $\CC\mathsans{M}$ is characterized by the invertibility of $\tau_{\hspace{.33pt}\CC\hspace{.66pt}}$. 
The Lie algebra crossed module associated with $\CC\mathsans{M}$ is precisely
the core $\CC\mathfrak{m}$ of $\mathfrak{m}$ defined in \ceqref{crmodinv7}, \ceqref{crmodinv8}.
The invariant pairing $\langle\cdot,\cdot\rangle$ of $\mathfrak{m}$ provides $\CC\mathfrak{m}$ with
the invariant pairing $\langle\cdot,\cdot\rangle_{\hspace{.33pt}\CC\hspace{.66pt}}$ defined by eq.
\ceqref{crmodinv9}. $\langle\cdot,\cdot\rangle_{\hspace{.33pt}\CC\hspace{.66pt}}$ in turn 
satisfies property \ceqref{crmodinv14} as a consequence of $\langle\cdot,\cdot\rangle$
doing so. $\CC\mathsans{M}$ is in this way a crossed module with invariant pairing. $\CC\mathsans{M}$
is fine, even if $\mathsans{M}$ is not,  as $\tau_{\hspace{.33pt}\CC\hspace{.66pt}}$ is invertible. 

The residue of $\mathsans{M}$ is
the crossed module $\RR\mathsans{M}=(\ker\tau,\mathsans{G}/\ran\tau,\tau_{\hspace{.33pt}\RR\hspace{.66pt}},
\mu_{\hspace{.33pt}\RR\hspace{.66pt}})$, where
\begin{align}
&\tau_{\hspace{.33pt}\RR\hspace{.66pt}}(A)=\ran\tau,
\vphantom{\Big]}
\label{crmodinv18}
\\
&\mu_{\hspace{.33pt}\RR\hspace{.66pt}}(a\ran\tau,A)=\mu(a,A)  
\vphantom{\Big]} 
\label{crmodinv19}
\end{align}   
for $a\in\mathsans{G}$, $A\in\ker\tau$. Again,
it can be verified that the structure maps $\tau_{\hspace{.33pt}\RR\hspace{.66pt}}$ and $\mu_{\hspace{.33pt}\RR\hspace{.66pt}}$
are well defined and satisfy the properties \ceqref{liecrmod1}, \ceqref{liecrmod2}. The residue crossed module 
$\RR\mathsans{M}$ is characterized by the vanishing of $\tau_{\hspace{.33pt}\RR\hspace{.66pt}}$. 
The Lie algebra crossed module associated with $\RR\mathsans{M}$ is precisely
the residue $\RR\mathfrak{m}$ of $\mathfrak{m}$ 
defined in \ceqref{crmodinv9}, \ceqref{crmodinv10}.
The invariant pairing $\langle\cdot,\cdot\rangle$ of $\mathfrak{m}$ provides $\RR\mathfrak{m}$ with
the invariant pairing $\langle\cdot,\cdot\rangle_{\hspace{.33pt}\RR\hspace{.66pt}}$ defined by eq.
\ceqref{crmodinv10} and $\langle\cdot,\cdot\rangle_{\hspace{.33pt}\RR\hspace{.66pt}}$
obeys \ceqref{crmodinv14} since $\langle\cdot,\cdot\rangle$ does.
So, $\RR\mathsans{M}$ too is a crossed module with invariant pairing. $\RR\mathsans{M}$ is not fine in general, but
it is fine if $\mathsans{M}$ is.

In subsect. \cref{subsec:liecrmod},$\vphantom{\ul{\ul{\ul{\ul{\ul{g}}}}}}$
\pagebreak we introduced two basic model Lie groups crossed modules, the inner
automorphism crossed module $\INN\mathsans{G}$ and the finite coadjoint action crossed module 
$\AD^*\mathsans{G}$ of a Lie group $\mathsans{G}$. Since their associated Lie algebra crossed modules
are respectively $\INN\mathfrak{g}$ and $\AD^*\mathfrak{g}$, they are both balanced and they can both
presumably be equipped with the invariant pairings of these latter. 


The inner automorphism crossed module of $\mathsans{G}$ is 
$\INN\mathsans{G}=(\mathsans{G},\mathsans{G},\id_{\mathsans{G}},\Ad_{\,\mathsans{G}})$.
The Lie algebra crossed module associated with $\INN\mathsans{G}$ is the inner
derivation crossed module of $\INN\mathfrak{g}$ which we discussed earlier. If $\mathfrak{g}$ is equipped with a
$\Ad_{\mathsans{G}}$--invariant symmetric non singular pairing, $\INN\mathfrak{g}$ gets endowed with an invariant
pairing obeying \ceqref{crmodinv14}. In this way, $\INN\mathsans{G}$ 
becomes a crossed module with invariant pairing. Since $\id_{\mathsans{G}}$ is trivially invertible,
$\INN\mathsans{G}$ is fine. 
$\INN\mathsans{G}$ is characterized by the following property. 
If $\mathsans{M}=(\mathsans{E},\mathsans{G},\tau,\mu)$ is a crossed module with invariant
pairing such that $\tau$ is invertible, then there is an invariant pairing on $\INN\mathsans{G}$
such that $\mathsans{M}$ is isomorphic to $\INN\mathsans{G}$. The isomorphism $\xi$ of $\mathsans{M}$ onto
$\INN\mathsans{G}$ is given by the pair $(\tau,\id_{\mathsans{G}})$. Its Lie differential $\dot\xi$
is precisely the isomorphism $i$ of $\mathfrak{m}$ onto $\INN\mathfrak{g}$, which
we described in the Lie algebra case. Note that if $\mathsans{M}$ is a crossed module with generic $\tau$,
then its core $\CC\mathsans{M}$ is isomorphic to $\INN\ran\tau$. 

The finite coadjoint action crossed module of $\mathsans{G}$ is 
$\AD^*\mathsans{G}=(\mathfrak{g}^*,\mathsans{G},1_{\mathsans{G}},\Ad_{\mathsans{G}}{}^*)$.
The Lie algebra crossed module associated with $\AD^*\mathsans{G}$ is the infinitesimal coadjoint action
crossed module of $\AD^*\mathfrak{g}$ which we also discussed above. 
$\AD^*\mathsans{G}$ is endowed with a natural invariant pairing obeying \ceqref{crmodinv14},
since $\AD^*\mathfrak{g}$ is equipped with the $\Ad_{\mathsans{G}}$--invariant duality pairing
of $\mathfrak{g}$ and $\mathfrak{g}^*$. $\AD^*\mathsans{G}$ is fine, since $\mathfrak{g}^*$ is
clearly connected. Contrary to what one may expect, 
if $\mathsans{M}=(\mathsans{E},\mathsans{G},\tau,\mu)$ is a crossed module with invariant pairing 
with $\tau$ vanishing, then $\mathsans{M}$ is not necessarily isomorphic to $\AD^*\mathsans{G}$.
The isomorphism $j$ of $\INN\mathfrak{g}$ onto $\mathfrak{m}$ we defined above,
given by the pair $(J,\id_{\mathsans{G}})$
with $J$ the linear isomorphism of $\mathfrak{g}^*$ onto $\mathfrak{e}$ such that
$\langle x,J(X)\rangle=\langle x,X\rangle_{\AD^*}$, does not integrate in general to an isomorphism 
$\eta$ of $\INN\mathsans{G}$ onto $\mathsans{M}$, because $J$ does not integrate to an isomorphism 
of $\mathfrak{g}^*$ onto $\mathsans{E}$. As a consequence, when $\mathsans{M}$ is a crossed module with generic $\tau$,
its residue $\RR\mathsans{M}$ is generally not isomorphic to $\AD^*(\mathsans{G}/\ran\tau)$. 

From the above discussion, it is now not surprising that 
$\mathsans{M}\nsimeq\CC\mathsans{M}\times\RR\mathsans{M}$ in general
for the reason that the Lie algebra isomorphisms
$h:\mathfrak{g}\rightarrow\ran\dot\tau\oplus(\mathfrak{g}/\ran\dot\tau)$ \linebreak and 
$H:\mathfrak{e}\rightarrow(\mathfrak{e}/\ker\dot\tau)\oplus\ker\dot\tau$ 
underlying the crossed module isomorphism \ceqref{crmodinv12}  (cf. app. \cref{app:coreres})
in general cannot be lifted to corresponding Lie 
group isomorphisms $\phi:\mathsans{G}\rightarrow\ran\tau\times(\mathsans{G}/\ran\tau)$ and 
$\varPhi:\mathsans{E}\rightarrow(\mathsans{E}/\ker\tau)\times\ker\tau$ 
unless all Lie groups involved are connected and simply connected.


\subsection{\textcolor{blue}{\sffamily Crossed submodules 
and isotropy}}\label{subsec:crosumo}

The concept of crossed submodule of a Lie group crossed module answers to the familiar concept
of subgroup of a Lie group and similarly in the Lie algebra case. 
In this subsection, we shall define and study submodules of a given 
crossed module and their associated normalizer and Weyl crossed modules.
When the crossed module is equipped with
an invariant pairing, isotropic crossed submodules can be considered and constitute
a distinguished subclass of submodules with special features. 
Before proceeding to detailing the properties of these substructures, however, we shall
explain in simple elementary terms why they are relevant in the construction of 4--dimensional CS theory.

In 4--dimensional CS theory, the higher gauge field $\Omega$ and infinitesimal gauge transformations $\Theta$
are non homogeneous forms on the underlying 4--fold valued in the Lie algebra crossed
module $\mathfrak{m}$ of the symmetry Lie group crossed module $\mathsans{M}$ (cf. subsect. \cref{subsec:higau}).

When the base manifold has a boundary, as we generally assume in 4--dimen\-sional CS theory, 
boundary conditions must be imposed on the gauge field $\Omega$
and infinitesimal gauge transformations $\Theta$ ensuring that the variational problem is well--posed on one
hand (cf. subsect. \cref{subsec:4dchern}) and that gauge invariance is preserved on the other
(cf. subsect. \cref{subsec:levelchern}). The boundary condition on $\Omega$ must be such to make
the boundary integral yielded by the variation of the CS action vanish. Since the boundary integrand  
involves the invariant pairing $\langle\cdot,\cdot\rangle$ of the crossed module $\mathfrak{m}$, this 
can be achieved by demanding the field $\Omega$ to be
valued on the boundary in a crossed submodule $\mathfrak{m}'$ of $\mathfrak{m}$
isotropic with respect to $\langle\cdot,\cdot\rangle$. 
Requiring this boundary condition to be gauge invariant 
forces also the transformations $\Theta$ to be
valued in $\mathfrak{m}'$ on the boundary.
Such particularly simple choice of boundary conditions, which we shall generally adopt in the following,
is called isotropic linear for evident reasons.


Infinitesimal gauge transformations form a Lie algebra. In particular, the Lie bracket $[\Theta,\Theta']$
is defined for any gauge transformation pair $\Theta$, $\Theta'$. 
In the canonical formulation of 4--dimensional CS theory (cf. subsect. \cref{subsec:holchern}),
the action of the gauge transformations $\Theta$ is generated by Hamiltonian 
functionals $\sfQ(\Theta)$ of the gauge field $\Omega$ (not explicitly shown here).
The functionals $\sfQ(\Theta)$ 
span under Poisson bracketing a centrally extended representation of the
gauge transformation Lie algebra. Explicitly, 
\begin{equation}
\{\sfQ(\Theta),\sfQ(\Theta')\}=\sfQ([\Theta,\Theta'])+\kappa\cdot\sfc(\Theta,\Theta'),
\label{crosumo1}
\end{equation}
where $\sfc$ is a certain 2--cocycle of the gauge transformation Lie algebra.

The physical higher gauge field phase space is defined by the constraints
\begin{equation}
\sfQ(\Theta)\approx 0,
\label{crosumo2}
\end{equation}
with $\Theta$ any gauge transformation, according to naive Dirac theory. 
However, the $\sfQ(\Theta)$ do not form a first class set of phase space functionals because of
the central term appearing in the right hand side of \ceqref{crosumo1}.
Furthermore, the vanishing of the $\sfQ(\Theta)$ by itself does not select the functional
submanifold of flat gauge fields $\Omega$, as one would like to in CS theory.

Both the 2--cocycle $\sfc(\Theta,\Theta')$ in \ceqref{crosumo1} and 
the obstructing term of $\sfQ(\Theta)$ preventing \ceqref{crosumo2}
from singling out the flat gauge field space are given by certain boundary integrals. 
The above problems can therefore be solved by imposing
suitable boundary conditions on the gauge field $\Omega$ and gauge transformations 
$\Theta$ making those integral expressions vanish. 
Since the integrands involve the invariant pairing
$\langle\cdot,\cdot\rangle$ of the crossed module $\mathfrak{m}$, this 
can be achieved by requiring both $\Omega$ and the $\Theta$ to be
valued on the boundary in a crossed submodule $\mathfrak{m}'$ of $\mathfrak{m}$ isotropic
with respect to $\langle\cdot,\cdot\rangle$, i.e. 
by imposing again isotropic linear boundary conditions. When $\Omega$ and the $\Theta$  
obey these conditions,  as we suppose, \ceqref{crosumo2} becomes a set of genuine flatness enforcing
first class constraints. 

If $\Theta$ is a gauge transformation such that 
\begin{equation}
\{\sfQ(\Theta),\sfQ(\Theta')\}\approx 0
\label{crosumo3}
\end{equation}
for any gauge transformation $\Theta'$ obeying the isotropic linear boundary condition, 
then $\sfQ(\Theta)$ represents a physical symmetry surface charge. 
Comparing \ceqref{crosumo3} to \ceqref{crosumo1}, we see that for this to be the case,
$\Theta$ must itself obey a boundary condition ensuring that $[\Theta,\Theta']$
satisfies the isotropic linear boundary condition and $\sfc(\Theta,\Theta')=0$
for any such $\Theta'$. 
This condition consists so in requiring $\Theta$ to be valued
on the boundary in the orthogonal normalizer crossed module $\ON\mathfrak{m}'$ of $\mathfrak{m}'$,
that is the maximal crossed submodule of $\mathfrak{m}$ that 
normalizes $\mathfrak{m}'$ in the appropriate sense and is orthogonal to $\mathfrak{m}'$ 
with respect to the invariant pairing $\langle\cdot,\cdot\rangle$. However, as
$\sfQ(\Theta)$ is weakly invariant under the shift
$\Theta\to\Theta+\Theta'$ with $\Theta'$ obeying the isotropic linear boundary condition
by virtue of \ceqref{crosumo2}, $\Theta$ is effectively valued on the orthogonal Weyl crossed module 
$\OW\mathfrak{m}'=\ON\mathfrak{m}'/\mathfrak{m}'$ of $\mathfrak{m}$. The boundary condition
is therefore called orthogonal Weyl linear.  

Having motivated the consideration of crossed submodules, in particular iso\-tropic ones,
in 4--dimensional CS theory, we now proceed to describe these structures in more precise
terms. Albeit the above discussion has been carried out mostly at the infinitesimal level,
we shall examine first the finite case.

\vspace{2.5mm}

\noindent
{\it Lie group crossed submodules and their normalizer and Weyl modules}

The notion of crossed submodule of a finite Lie crossed module is analogous to that of subgroup
of a Lie group. A crossed submodule $\mathsans{M}'$ of a Lie group crossed module $\mathsans{M}$
is indeed a substructure of $\mathsans{M}$ which is itself a Lie group crossed module.  More formally,
given two Lie group crossed modules $\mathsans{M}=(\mathsans{E},\mathsans{G},\tau,\mu)$,
$\mathsans{M}'=(\mathsans{E}',\mathsans{G}',\tau',\mu')$, $\mathsans{M}'$ is a submodule of $\mathsans{M}$
if $\mathsans{E}'$, $\mathsans{G}'$ are Lie subgroups of $\mathsans{E}$, $\mathsans{G}$ and $\tau'$, $\mu'$
are the restrictions of $\tau$, $\mu$ to $\mathsans{E}'$, $\mathsans{G}'\times\mathsans{E}'$, respectively.

The concept of normalizer crossed module of a submodule of a finite Lie crossed module
corresponds in turn to that of normalizer group of a subgroup of a Lie group.
The normalizer $\NN\mathsans{M}'$ of a crossed submodule $\mathsans{M}'$
of a Lie group crossed module $\mathsans{M}$ is the largest crossed submodule of $\mathsans{M}$
normalizing $\mathsans{M}'$ (cf. app. \cref{app:def}). $\NN\mathsans{M}'$ can be described
rather explicitly as follows. Let $\mathsans{M}=(\mathsans{E},\mathsans{G},\tau,\mu)$,
$\mathsans{M}'=(\mathsans{E}',\mathsans{G}',\tau',\mu')$.
The normalizer of $\mathsans{G}'$ in $\mathsans{G}$, $\NN\mathsans{G}'$, is the set of all elements
$a\in\mathsans{G}$ such that $aba^{-1},a^{-1}ba\in\mathsans{G}'$ for $b\in\mathsans{G}'$.
Similarly, the $\mu$--normalizer of $\mathsans{E}'$ in $\mathsans{G}$, $\mu\NN\mathsans{E}'$,
is the set of all elements $a\in\mathsans{G}$ such that 
$\mu(a,B),\mu(a^{-1},B)\in\mathsans{E}'$ for $B\in\mathsans{E}'$. 
The $\mu$--transporter of $\mathsans{G}'$ into $\mathsans{E}'$ in $\mathsans{E}$,
$\mu\TT_{\mathsans{G}'}\mathsans{E}'$ is the set of all elements $A\in\mathsans{E}$
such that $\mu(b,A)A^{-1},A^{-1}\mu(b,A)\in\mathsans{E}'$ for $b\in\mathsans{G}'$.
$\NN\mathsans{G}'$ and $\mu\NN\mathsans{E}'$ turn out to be Lie subgroups of $\mathsans{G}$
and likewise $\mu\TT_{\mathsans{G}'}\mathsans{E}'$ a Lie subgroup of $\mathsans{E}$.
Then, $\NN\mathsans{M}'=(\mu\TT_{\mathsans{G}'}\mathsans{E}',
\NN\mathsans{G}'\cap\mu\NN\mathsans{E}',\tau'{}_{\NN},\mu'{}_{\NN})$, where
$\tau'{}_{\NN}$, $\mu'{}_{\NN}$ are the restrictions of $\tau$, $\mu$ to
$\mu\TT_{\mathsans{G}'}\mathsans{E}'$, $\NN\mathsans{G}'\cap\mu\NN\mathsans{E}'\times\mu\TT_{\mathsans{G}'}\mathsans{E}'$,
respectively. It can be verified that the structure maps
$\tau'{}_{\NN}$, $\mu'{}_{\NN}$ 
are well defined and satisfy the required properties \ceqref{liecrmod1}, \ceqref{liecrmod2}.
$\NN\mathsans{M}'$ is so a Lie group crossed module. By construction, $\NN\mathsans{M}'$ 
is a crossed submodule of $\mathsans{M}$ containing $\mathsans{M}'$ as a crossed submodule
and normalizing it and is maximal with these properties. 

The Weyl crossed module of a submodule of a finite Lie crossed module
is much like the Weyl group of a subgroup of a Lie group.
For a crossed submodule $\mathsans{M}'$ of a Lie group crossed module $\mathsans{M}$, this 
is just the quotient $\WW\mathsans{M}'=\NN\mathsans{M}'/\mathsans{M}'$ removing from the normalizer 
crossed module $\NN\mathsans{M}'$ of $\mathsans{M}'$ the trivially normalizing submodule $\mathsans{M}'$.
Specifically,  let again $\mathsans{M}=(\mathsans{E},\mathsans{G},\tau,\mu)$,
$\mathsans{M}'=(\mathsans{E}',\mathsans{G}',\tau',\mu')$. It can be shown that
$\mathsans{G}'$, $\mathsans{E}'$ are normal Lie subgroups of $\NN\mathsans{G}'\cap\mu\NN\mathsans{E}'$,
$\mu\TT_{\mathsans{G}'}\mathsans{E}'$, respectively. We have then
$\WW\mathsans{M}'=(\mu\TT_{\mathsans{G}'}\mathsans{E}'/\mathsans{E}',
\NN\mathsans{G}'\cap\mu\NN\mathsans{E}'/\mathsans{G}',\tau'{}_{\WW},\mu'{}_{\WW})$,
where the structure maps $\tau'{}_{\WW}$, $\mu'{}_{\WW}$, are given by
\begin{align}
&\tau'{}_{\WW}(A\mathsans{E}')=\tau(A)\mathsans{G}',
\vphantom{\Big]}
\label{}
\\
&\mu'{}_{\WW}(a\mathsans{G}',A\mathsans{E}')=\mu(a,A)\mathsans{E}'
\vphantom{\Big]}
\label{}
\end{align}
for $a\in\NN\mathsans{G}'\cap\mu\NN\mathsans{E}'$, $A\in\mu\TT_{\mathsans{G}'}\mathsans{E}'$.
It can be verified that the structure maps $\tau'{}_{\WW}$, $\mu'{}_{\WW}$ 
are well defined and obey relations \ceqref{liecrmod1}, \ceqref{liecrmod2}.
$\WW\mathsans{M}'$ is therefore a Lie group crossed module as required.

There are plenty of examples of crossed submodules of Lie group crossed modules, in particular of the model ones 
described in subsect. \cref{subsec:liecrmod}.
They are illustrated next, but before doing so we introduce some notation. 
Let $\mathsans{G}$ be a fixed Lie group $\mathsans{H}$, $\mathsans{K}$ be Lie subgroups of 
$\mathsans{G}$. We denote by $[\mathsans{H},\mathsans{K}]$ the commutator subgroup
of $\mathsans{H}$, $\mathsans{K}$ in $\mathsans{G}$. We denote further by
$\TT_{\mathsans{H}}\mathsans{K}$ the transporter of $\mathsans{H}$ into
$\mathsans{K}$, the largest subgroup $\mathsans{L}$ of $\mathsans{G}$
such that $[\mathsans{H},\mathsans{L}]$ is contained in $\mathsans{K}$. This coincides
with the $\Ad$--transporter $\Ad\TT_{\mathsans{H}}\mathsans{K}$ as previously defined. 

Let $\mathsans{G}$ be a Lie group, $\mathsans{H}$ be a subgroup of $\mathsans{G}$
and $\mathsans{K}$ be a normal subgroup of $\mathsans{H}$. Then, $\INN_{\mathsans{H}}\mathsans{K}
=(\mathsans{K},\mathsans{H},\id_{\mathsans{H}}|_{\mathsans{K}},\Ad_{\mathsans{H}}|_{\mathsans{H}\times\mathsans{K}})$
is a Lie group crossed module, called inner $\mathsans{H}$--automorphism crossed module of $\mathsans{K}$,
and is a crossed submodule of $\INN\mathsans{G}=(\mathsans{G},\mathsans{G},\id_{\mathsans{G}},\Ad_{\,\mathsans{G}})$,
the inner automorphism crossed module of $\mathsans{G}$.
The normalizer crossed module of $\INN_{\mathsans{H}}\mathsans{K}$ is a crossed module of the same kind, 
viz $\NN\INN_{\mathsans{H}}\mathsans{K}=\INN_{\NN\mathsans{H}\cap\NN\mathsans{K}}\TT_{\mathsans{H}}\mathsans{K}$.
It is easily verified that $\TT_{\mathsans{H}}\mathsans{K}$ is a normal subgroup of
$\NN\mathsans{H}\cap\NN\mathsans{K}$ as required.
The Weyl crossed module of $\INN_{\mathsans{H}}\mathsans{K}$ is therefore 
$\WW\INN_{\mathsans{H}}\mathsans{K}=(\mathsans{T}_{\mathsans{K},\mathsans{H}},
\mathsans{N}_{\mathsans{K},\mathsans{H}},
\id_{\mathsans{N}_{\mathsans{K},\mathsans{H}}}|_{\mathsans{T}_{\mathsans{K},\mathsans{H}}},
\Ad_{\mathsans{N}_{\mathsans{K},\mathsans{H}}}|_{\mathsans{N}_{\mathsans{K},\mathsans{H}}\times\mathsans{T}_{\mathsans{K},\mathsans{H}}})$,
where $\mathsans{T}_{\mathsans{K},\mathsans{H}}=\TT_{\mathsans{H}}\mathsans{K}/\mathsans{K}$,
$\mathsans{N}_{\mathsans{K},\mathsans{H}}=\NN\mathsans{H}\cap\NN\mathsans{K}/\mathsans{H}$.
(Here, $\id_{\mathsans{N}_{\mathsans{K},\mathsans{H}}}|_{\mathsans{T}_{\mathsans{K},\mathsans{H}}}$,
$\Ad_{\mathsans{N}_{\mathsans{K},\mathsans{H}}}|_{\mathsans{N}_{\mathsans{K},\mathsans{H}}\times\mathsans{T}_{\mathsans{K},\mathsans{H}}}$
denote abusively the natural map 
$\mathsans{T}_{\mathsans{K},\mathsans{H}}\rightarrow\mathsans{N}_{\mathsans{K},\mathsans{H}}$
and the maps induced by  $\Ad_{\mathsans{N}_{\mathsans{K},\mathsans{H}}}$ upon composition with it, respectively.) 

Let $\mathsans{G}$ be a Lie group and $\mathsans{H}$, $\mathsans{K}$ be subgroups of $\mathsans{G}$
with $\mathsans{H}\subseteq\NN\mathsans{K}$. 
Then, $\AD_{\mathsans{H}}{}^*\mathsans{K}
=(\mathfrak{k}^0,\mathsans{H},1_{\mathsans{H}}|_{\mathfrak{k}^0},\Ad_{\mathsans{H}}{}^*|_{\mathsans{H}\times\mathfrak{k}^0})$,
$\mathfrak{k}^0\subseteq\mathfrak{g}^*$ being the annihilator of $\mathfrak{k}$, 
is a Lie group crossed module, called the finite coadjoint $\mathsans{H}$--action crossed module of $\mathsans{K}$,
and is a crossed submodule of $\AD^*\mathsans{G}=(\mathfrak{g}^*,\mathsans{G},1_{\mathsans{G}},\Ad_{\,\mathsans{G}}{}^*)$,
the finite coadjoint action crossed module of $\mathsans{G}$. 
If $\mathsans{H}$, $\mathsans{K}$ are connected, the normalizer crossed module of $\AD_{\mathsans{H}}{}^*\mathsans{K}$ is
a crossed module of the same kind, as
$\NN\AD_{\mathsans{H}}{}^*\mathsans{K}=\AD_{\NN\mathsans{H}\cap\NN\mathsans{K}}{}^*[\mathsans{H},\mathsans{K}]$.
Again, it can be straightforwardly verified that 
$\NN\mathsans{H}\cap\NN\mathsans{K}\subseteq\NN[\mathsans{H},\mathsans{K}]$ as required. 
Above, connectedness is assumed only for the sake of simplicity. 
The Weyl crossed module of $\AD_{\mathsans{H}}{}^*\mathsans{K}$ then is 
$\WW\AD_{\mathsans{H}}{}^*\mathsans{K}=(\mathfrak{t}_{{\mathsans{K},\mathsans{H}}},
\mathsans{N}_{\mathsans{K},\mathsans{H}}, 
1_{\mathsans{N}_{\mathsans{K},\mathsans{H}}}|_{\mathfrak{t}_{{\mathsans{K},\mathsans{H}}}},
\Ad_{\mathsans{N}_{\mathsans{K},\mathsans{H}}}{}^*|_{\mathsans{N}_{\mathsans{K},\mathsans{H}}\times\mathfrak{t}_{{\mathsans{K},\mathsans{H}}}})$,
where $\mathfrak{t}_{{\mathsans{K},\mathsans{H}}}=[\mathfrak{h},\mathfrak{k}]^0/\mathfrak{k}^0$,
$\mathsans{N}_{\mathsans{K},\mathsans{H}}=\NN\mathsans{H}\cap\NN\mathsans{K}/\mathsans{H}$.

\vspace{2.5mm}

\noindent
{\it Lie algebra crossed submodules and their normalizer and Weyl modules}

As it might be expected, there are infinitesimal counterparts of the Lie group theoretic notions 
introduced above. Albeit it is not difficult to guess them, we report them below for 
completeness.

The concept of crossed submodule of a infinitesimal Lie crossed module is inspired by that of subalgebra
of a Lie algebra. A crossed submodule $\mathfrak{m}'$ of a Lie algebra crossed module $\mathfrak{m}$
is indeed a substructure of $\mathfrak{m}$ which is itself a Lie algebra crossed module.  Specifically, 
given two Lie algebra crossed modules $\mathfrak{m}=(\mathfrak{e},\mathfrak{g},t,m)$,
$\mathfrak{m}'=(\mathfrak{e}',\mathfrak{g}',t',m')$, $\mathfrak{m}'$ is a submodule of $\mathfrak{m}$
if $\mathfrak{e}'$, $\mathfrak{g}'$ are Lie subalgebras of $\mathfrak{e}$, $\mathfrak{g}$ and $t'$, $m'$
are the restrictions of $t$, $m$ to $\mathfrak{e}'$, $\mathfrak{g}'\times\mathfrak{e}'$, respectively.

The notion of normalizer crossed module of a submodule of a infinitesimal Lie crossed module
correlates with that of normalizer algebra of a subalgebra of a Lie algebra as expected.
The normalizer $\NN\mathfrak{m}'$ of a crossed submodule $\mathfrak{m}'$
of a Lie algebra crossed module $\mathfrak{m}$ is the largest crossed submodule of $\mathfrak{m}$
normalizing $\mathfrak{m}'$ (cf. app. \cref{app:def}). Explicitly, $\NN\mathfrak{m}'$ can be specified 
as follows. Let $\mathfrak{m}=(\mathfrak{e},\mathfrak{g},t,m)$,
$\mathfrak{m}'=(\mathfrak{e}',\mathfrak{g}',t',m')$.
The normalizer of $\mathfrak{g}'$ in $\mathfrak{g}$, $\NN\mathfrak{g}'$,
is the set of all elements
$u\in\mathfrak{g}$ such that $[u,v]\in\mathfrak{g}'$ for $v\in\mathfrak{g}'$.
Similarly, the $m$--normalizer of $\mathfrak{e}'$ in $\mathfrak{g}$, $m\NN\mathfrak{e}'$,
is the set of all elements $u\in\mathfrak{g}$ such that 
$m(u,V)\in\mathfrak{e}'$ for $V\in\mathfrak{e}'$. 
The $m$--transporter of $\mathfrak{g}'$ into $\mathfrak{e}'$ in $\mathfrak{e}$,
$m\TT_{\mathfrak{g}'}\mathfrak{e}'$ is the set of all elements $U\in\mathfrak{e}$
such that $m(v,U)\in\mathfrak{e}'$ for $v\in\mathfrak{g}'$. 
$\NN\mathfrak{g}'$ and $m\NN\mathfrak{e}'$ turn out to be Lie subalgebras of $\mathfrak{g}$
and likewise $m\TT_{\mathfrak{g}'}\mathfrak{e}'$ a Lie subalgebra of $\mathfrak{e}$.
Then, $\NN\mathfrak{m}'=(m\TT_{\mathfrak{g}'}\mathfrak{e}',
\NN\mathfrak{g}'\cap m\NN\mathfrak{e}',t'{}_{\NN},m'{}_{\NN})$, where
$t'{}_{\NN}$, $m'{}_{\NN}$ are the restrictions of $t$, $m$ to
$m\TT_{\mathfrak{g}'}\mathfrak{e}'$, $\NN\mathfrak{g}'\cap m\NN\mathfrak{e}'\times m\TT_{\mathfrak{g}'}\mathfrak{e}'$,
respectively. It can be verified that the structure maps
$t'{}_{\NN}$, $m'{}_{\NN}$ 
are well defined and satisfy the required properties \ceqref{liecrmod5}, \ceqref{liecrmod6}.
$\NN\mathfrak{m}'$ so turns out to be a Lie algebra crossed module. By the way we have designed it,
$\NN\mathfrak{m}'$ is a crossed submodule of $\mathfrak{m}$ containing $\mathfrak{m}'$ as a crossed submodule
and normalizing it and is maximal with these properties. 

The Weyl crossed module of a submodule of a infinitesimal Lie crossed module
is now conceived similarly to the Weyl algebra of a subalgebra of a Lie algebra.
For a crossed submodule $\mathfrak{m}'$ of a Lie algebra crossed module $\mathfrak{m}$, this 
is just the quotient $\WW\mathfrak{m}'=\NN\mathfrak{m}'/\mathfrak{m}'$ removing from the normalizer 
crossed module $\NN\mathfrak{m}'$ of $\mathfrak{m}'$ the trivially normalizing submodule $\mathfrak{m}'$.
Concretely,  let again $\mathfrak{m}=(\mathfrak{e},\mathfrak{g},t,m)$,
$\mathfrak{m}'=(\mathfrak{e}',\mathfrak{g}',t',m')$. It can be shown that
$\mathfrak{g}'$, $\mathfrak{e}'$ are Lie ideals of $\NN\mathfrak{g}'\cap m\NN\mathfrak{e}'$,
$m\TT_{\mathfrak{g}'}\mathfrak{e}'$, respectively. We have then
$\WW\mathfrak{m}'=(m\TT_{\mathfrak{g}'}\mathfrak{e}'/\mathfrak{e}',
\NN\mathfrak{g}'\cap m\NN\mathfrak{e}'/\mathfrak{g}',t'{}_{\WW},m'{}_{\WW})$,
where the structure maps $t'{}_{\WW}$, $m'{}_{\WW}$, are given by
\begin{align}
&t'{}_{\WW}(U+\mathfrak{e}')=t(U)+\mathfrak{g}',
\vphantom{\Big]}
\label{}
\\
&m'{}_{\WW}(u+\mathfrak{g}',U+\mathfrak{e}')=m(u,U)+\mathfrak{e}'
\vphantom{\Big]}
\label{}
\end{align}
for $u\in\NN\mathfrak{g}'\cap m\NN\mathfrak{e}'$, $U\in m\TT_{\mathfrak{g}'}\mathfrak{e}'$.
It can be verified that the structure maps $t'{}_{\WW}$, $m'{}_{\WW}$ 
are well defined and obey relations \ceqref{liecrmod1}, \ceqref{liecrmod2}.
$\WW\mathfrak{m}'$ is therefore a Lie algebra crossed module as required.

We present next a class of examples of crossed submodules of the model Lie algebra
crossed modules of subsect. \cref{subsec:liecrmod}, matching those introduced
above in the finite case, after recalling a few basic notions of Lie theory.
Let $\mathfrak{g}$ be a Lie algebra and $\mathfrak{h}$, $\mathfrak{k}$ be Lie subalgebras of 
$\mathfrak{g}$. We denote by $[\mathfrak{h},\mathfrak{k}]$ the commutator subalgebra of $\mathfrak{h}$, $\mathfrak{k}$
in $\mathfrak{g}$. We denote further by $\TT_{\mathfrak{h}}\mathfrak{k}$ the transporter of $\mathfrak{h}$ into
$\mathfrak{k}$, that is the largest subalgebra $\mathfrak{l}$ of $\mathfrak{g}$
such that $[\mathfrak{h},\mathfrak{l}]$ is contained in $\mathfrak{k}$. This is just the $\ad$--transporter
$\ad\TT_{\mathfrak{h}}\mathfrak{k}$ we defined earlier. 

Let $\mathfrak{g}$ be a Lie algebra, $\mathfrak{h}$ be a subalgebra of $\mathfrak{g}$
and $\mathfrak{k}$ be an ideal of $\mathfrak{h}$. Then, $\INN_{\mathfrak{h}}\mathfrak{k}
=(\mathfrak{k},\mathfrak{h},\id_{\mathfrak{h}}|_{\mathfrak{k}},\ad_{\mathfrak{h}}|_{\mathfrak{h}\times\mathfrak{k}})$
turns out to be a Lie algebra crossed module, called the inner
$\mathfrak{h}$--derivation crossed module of $\mathfrak{k}$,
and is a crossed submodule of $\INN\mathfrak{g}$ $=(\mathfrak{g},\mathfrak{g},\id_{\mathfrak{g}},\ad_{\,\mathfrak{g}})$,
the inner derivation crossed module of $\mathfrak{g}$.
The normalizer crossed module of $\INN_{\mathfrak{h}}\mathfrak{k}$ is a crossed module of the same kind as it, since
we have $\NN\INN_{\mathfrak{h}}\mathfrak{k}=\INN_{\NN\mathfrak{h}\cap\NN\mathfrak{k}}\TT_{\mathfrak{h}}\mathfrak{k}$,
$\TT_{\mathfrak{h}}\mathfrak{k}$ being an ideal of $\NN\mathfrak{h}\cap\NN\mathfrak{k}$.
The Weyl crossed module of $\INN_{\mathfrak{h}}\mathfrak{k}$ is hence 
$\WW\INN_{\mathfrak{h}}\mathfrak{k}=(\mathfrak{t}_{\mathfrak{k},\mathfrak{h}},
\mathfrak{n}_{\mathfrak{k},\mathfrak{h}},
\id_{\mathfrak{n}_{\mathfrak{k},\mathfrak{h}}}|_{\mathfrak{t}_{\mathfrak{k},\mathfrak{h}}},
\ad_{\mathfrak{n}_{\mathfrak{k},\mathfrak{h}}}|_{\mathfrak{n}_{\mathfrak{k},\mathfrak{h}}\times\mathfrak{t}_{\mathfrak{k},\mathfrak{h}}})$,
where $\mathfrak{t}_{\mathfrak{k},\mathfrak{h}}=\TT_{\mathfrak{h}}\mathfrak{k}/\mathfrak{k}$, 
$\mathfrak{n}_{\mathfrak{k},\mathfrak{h}}=\NN\mathfrak{h}\cap\NN\mathfrak{k}/\mathfrak{h}$.
(Here, $\id_{\mathfrak{n}_{\mathfrak{k},\mathfrak{h}}}|_{\mathfrak{t}_{\mathfrak{k},\mathfrak{h}}}$,
$\ad_{\mathfrak{n}_{\mathfrak{k},\mathfrak{h}}}|_{\mathfrak{n}_{\mathfrak{k},\mathfrak{h}}\times\mathfrak{t}_{\mathfrak{k},\mathfrak{h}}}$
denote abusively as before the natural map $\mathfrak{t}_{\mathfrak{k},\mathfrak{h}}\rightarrow
\mathfrak{n}_{\mathfrak{k},\mathfrak{h}}$ and the map induced by 
$\ad_{\mathfrak{n}_{\mathfrak{k},\mathfrak{h}}}$ upon composition with it, respectively.)

Next, let $\mathfrak{g}$ be a Lie algebra and $\mathfrak{h}$, $\mathfrak{k}$ be subalgebras of $\mathfrak{g}$
such that $\mathfrak{h}\subseteq\NN\mathfrak{k}$.  
Then, $\AD_{\mathfrak{h}}{}^*\mathfrak{k}
=(\mathfrak{k}^0,\mathfrak{h},0_{\mathfrak{h}}|_{\mathfrak{k}^0},\ad_{\mathfrak{h}}{}^*|_{\mathfrak{h}\times\mathfrak{k}^0})$,
$\mathfrak{k}^0\subseteq\mathfrak{g}^*$ being again the annihilator of $\mathfrak{k}$,
is a Lie algebra crossed module, the infinitesimal coadjoint $\mathfrak{h}$--action crossed module of $\mathfrak{k}$,
and is a crossed submodule of $\AD^*\mathfrak{g}=(\mathfrak{g}^*,\mathfrak{g},0_{\mathfrak{g}},\ad_{\,\mathfrak{g}}{}^*)$,
the infinitesimal coadjoint action crossed module of $\mathfrak{g}$.
The normalizer crossed module of $\AD_{\mathfrak{h}}{}^*\mathfrak{k}$ is again a crossed module of the same kind, 
$\NN\AD_{\mathfrak{h}}{}^*\mathfrak{k}=\AD_{\NN\mathfrak{h}\cap\NN\mathfrak{k}}{}^*[\mathfrak{h},\mathfrak{k}]$,
where one has $\NN\mathfrak{h}\cap\NN\mathfrak{k}\subseteq\NN[\mathfrak{h},\mathfrak{k}]$. 
The Weyl crossed module of $\AD_{\mathfrak{h}}{}^*\mathfrak{k}$ is in this way found to be 
$\WW\AD_{\mathfrak{h}}{}^*\mathfrak{k}=(\mathfrak{t}_{{\mathfrak{k},\mathfrak{h}}},
\mathfrak{n}_{\mathfrak{k},\mathfrak{h}}, 
0_{\mathfrak{n}_{\mathfrak{k},\mathfrak{h}}}|_{\mathfrak{t}_{{\mathfrak{k},\mathfrak{h}}}},
\ad_{\mathfrak{n}_{\mathfrak{k},\mathfrak{h}}}{}^*|_{\mathfrak{n}_{\mathfrak{k},\mathfrak{h}}\times\mathfrak{t}_{{\mathfrak{k},\mathfrak{h}}}})$,
where $\mathfrak{t}_{{\mathfrak{k},\mathfrak{h}}}=[\mathfrak{h},\mathfrak{k}]^0/\mathfrak{k}^0$,
$\mathfrak{n}_{\mathfrak{k},\mathfrak{h}}=\NN\mathfrak{h}\cap\NN\mathfrak{k}/\mathfrak{h}$.

The above constructions are of course designed to be compatible with Lie differentiation. 
If $\mathsans{M}$, $\mathsans{M}'$ are Lie group crossed modules with associated Lie algebra crossed modules
$\mathfrak{m}$, $\mathfrak{m}'$ and $\mathsans{M}'$ is a crossed submodule of $\mathsans{M}$, then 
$\mathfrak{m}'$ is a crossed submodule of $\mathfrak{m}$. Further, the Lie algebra crossed modules
of the normalizer and Weyl crossed modules $\NN\mathsans{M}'$ and $\WW\mathsans{M}'$
of $\mathsans{M}'$ are precisely the normalizer and Weyl crossed modules $\NN\mathfrak{m}'$
and $\WW\mathfrak{m}'$ of $\mathfrak{m}'$.

Reconsider the model examples $\INN_{\mathsans{H}}\mathsans{K}\subseteq\INN\mathsans{G}$,
$\AD_{\mathsans{H}}{}^*\mathsans{K}\subset\AD^*\mathsans{G}$
of crossed submodules described earlier in this subsection
defined for a Lie group $\mathsans{G}$ and subgroups $\mathsans{H}$, $\mathsans{K}$ of $\mathsans{G}$
satisfying the stated conditions. Then, the Lie algebra crossed module
of $\INN_{\mathsans{H}}\mathsans{K}$, $\AD_{\mathsans{H}}{}^*\mathsans{K}$ 
are $\INN_{\mathfrak{h}}\mathfrak{k}$, $\AD_{\mathfrak{h}}{}^*\mathfrak{k}$, respectively, as it might be expected. 


\vspace{2.5mm}

\noindent
{\it The orthogonal case}

We now consider the case where an ambient Lie algebra crossed module is equipped with an
invariant pairing (cf. subsect. \cref{subsec:crmodinv}). This allows us to consider isotropic crossed submodules.

Let $\mathfrak{m}=(\mathfrak{e},\mathfrak{g},t,m)$ be a Lie algebra crossed module with invariant
pairing $\langle\cdot,\cdot\rangle$ (cf. subsect. \cref{subsec:crmodinv})
and let $\mathfrak{m}'=(\mathfrak{e}',\mathfrak{g}',t',m')$ be a crossed
submodule of $\mathfrak{m}$. $\mathfrak{m}'$ is said to be isotropic if $\mathfrak{e}'\subseteq\mathfrak{g}'{}^\perp$,
where ${}^\perp$ denotes orthogonal complement with respect to $\langle\cdot,\cdot\rangle$.
$\mathfrak{m}'$ is said to be Lagrangian if it is maximally isotropic, i.e. if $\mathfrak{e}'=\mathfrak{g}'{}^\perp$.
When $\mathfrak{m}'$ is isotropic, $\dim\mathfrak{g}'+\dim\mathfrak{e}'\leq\dim\mathfrak{g}=\dim\mathfrak{e}$,
the bound being saturated when $\mathfrak{m}'$ is Lagrangian.

When $\mathfrak{m}'$ is an isotropic submodule, it is possible to define the orthogonal normalizer $\ON\mathfrak{m}'$
of $\mathfrak{m}'$ refining the normalizer $\NN\mathfrak{m}'$.
$\ON\mathfrak{m}'$ is the orthogonal complement of $\mathfrak{m}'$ in $\NN\mathfrak{m}'$
with respect to the pairing $\langle\cdot,\cdot\rangle$ in the appropriate sense. More formally, 
$\ON\mathfrak{m}'=(m\TT_{\mathfrak{g}'}\mathfrak{e}'\cap\mathfrak{g}'^\perp,
\NN\mathfrak{g}'\cap m\NN\mathfrak{e}'\cap\mathfrak{e}'^\perp,t_{\ON},m_{\ON})$, where
$t_{\ON}$, $m_{\ON}$ are the restrictions of $t$, $m$ to
$m\TT_{\mathfrak{g}'}\mathfrak{e}'\cap\mathfrak{g}'^\perp$,
$\NN\mathfrak{g}'\cap m\NN\mathfrak{e}'\cap\mathfrak{e}'^\perp\times
m\TT_{\mathfrak{g}'}\mathfrak{e}'\cap\mathfrak{g}'^\perp$, respectively.
$\ON\mathfrak{m}'$ turns out to be a crossed submodule of $\NN\mathfrak{m}'$
and so $\mathfrak{m}$ itself. Further, $\ON\mathfrak{m}'$ contains $\mathfrak{m}'$ as a submodule.
The orthogonal Weyl crossed module is the quotient module $\OW\mathfrak{m}'=\ON\mathfrak{m}'/\mathfrak{m}'$. 
$\ON\mathfrak{m}'$ reduces to $\mathfrak{m}'$ and $\OW\mathfrak{m}'$ vanishes
when $\mathfrak{m}'$ is Lagrangian.

Suppose that $\mathsans{M}$ is a Lie group crossed module with invariant pairing $\langle\cdot,\cdot\rangle$
(cf. subsect. \cref{subsec:crmodinv})
and that $\mathsans{M}'$ is a Lie group crossed submodule of $\mathsans{M}$
and that $\mathfrak{m}$, $\mathfrak{m}'$ are the Lie algebra crossed modules of $\mathsans{M}$,
$\mathsans{M}'$, respectively. $\mathsans{M}'$ is said to be an isotropic submodule of $\mathsans{M}$
if $\mathfrak{m}'$ is an isotropic submodule of $\mathfrak{m}$ in the sense defined
above and similarly in the Lagrangian case. 

When $\mathsans{M}'$ is an isotropic crossed submodule of $\mathsans{M}$, it is possible
under certain conditions to define an orthogonal normalizer $\ON\mathsans{M}'$ of $\mathsans{M}'$
refining the normalizer $\NN\mathsans{M}'$ introduced above. $\ON\mathsans{M}'$ is a 
crossed submodule of $\NN\mathsans{M}'$
having $\ON\mathfrak{m}'$ as associated Lie algebra crossed module and containing
$\mathsans{M}'$ as a crossed submodule.
Because of its defining properties, $\ON\mathsans{M}'$
normalizes $\mathsans{M}'$ and so the orthogonal Weyl crossed
module $\OW\mathsans{M}'=\ON\mathsans{M}'/\mathsans{M}'$ of $\mathsans{M}'$ can be defined. 
We shall not investigate here the precise conditions ensuring the existence 
of $\ON\mathsans{M}'$. In the following we shall tacitly 
assume that they are satisfied. For instance, if $\mathsans{G}'$, $\mathsans{E}'$ are connected
a choice of $\ON\mathsans{M}'$ exists. 

We now examine whether the model crossed submodules introduced above 
turn out to be isotropic in the presence of an invariant pairing. 

Let $\mathfrak{g}$ be a Lie algebra equipped with an invariant symmetric non singular bilinear form
$\langle\cdot,\cdot\rangle$. With this extra datum, the inner derivation crossed module 
$\INN\mathfrak{g}$ $=(\mathfrak{g},\mathfrak{g},\id_{\mathfrak{g}},\ad_{\,\mathfrak{g}})$ of $\mathfrak{g}$
is a Lie algebra crossed module with invariant pairing (cf. subsect. \cref{subsec:crmodinv}). 
Let $\mathfrak{h}$ be a subalgebra of $\mathfrak{g}$
and $\mathfrak{k}$ be an ideal of $\mathfrak{h}$ such that \linebreak $\mathfrak{k}\subseteq\mathfrak{h}^\perp$. 
Then, the inner $\mathfrak{h}$--derivation crossed module $\INN_{\mathfrak{h}}\mathfrak{k}
=(\mathfrak{k},\mathfrak{h},\id_{\mathfrak{h}}|_{\mathfrak{k}},\ad_{\mathfrak{h}}|_{\mathfrak{h}\times\mathfrak{k}})$
of $\mathfrak{k}$ which we defined above is an isotropic submodule of $\INN\mathfrak{g}$.
The orthogonal normalizer crossed module of $\INN_{\mathfrak{h}}\mathfrak{k}$ is 
$\ON\INN_{\mathfrak{h}}\mathfrak{k}=\INN_{\NN\mathfrak{h}\cap\NN\mathfrak{k}\cap\mathfrak{k}^\perp}
(\TT_{\mathfrak{h}}\mathfrak{k}\cap\mathfrak{h}^\perp)$ and is once more of the same kind.
The orthogonal Weyl crossed module of $\INN_{\mathfrak{h}}\mathfrak{k}$ is 
$\OW\INN_{\mathfrak{h}}\mathfrak{k}=(\mathfrak{ot}_{\mathfrak{k},\mathfrak{h}},
\mathfrak{on}_{\mathfrak{k},\mathfrak{h}},
\id_{\mathfrak{on}_{\mathfrak{k},\mathfrak{h}}}|_{\mathfrak{ot}_{\mathfrak{k},\mathfrak{h}}},
\ad_{\mathfrak{on}_{\mathfrak{k},\mathfrak{h}}}|_{\mathfrak{on}_{\mathfrak{k},\mathfrak{h}}\times\mathfrak{ot}_{\mathfrak{k},\mathfrak{h}}})$,
where $\mathfrak{ot}_{\mathfrak{k},\mathfrak{h}}=\TT_{\mathfrak{h}}\mathfrak{k}\cap\mathfrak{h}^\perp/\mathfrak{k}$, 
$\mathfrak{on}_{\mathfrak{k},\mathfrak{h}}=\NN\mathfrak{h}\cap\NN\mathfrak{k}\cap\mathfrak{k}^\perp/\mathfrak{h}$.
We remark here that under the stated hypotheses the subalgebra $\mathfrak{h}$ is isotropic in
$\mathfrak{g}$. We also note that $\INN_{\mathfrak{h}}\mathfrak{k}$ is Lagrangian precisely when
$\mathfrak{k}=\mathfrak{h}^\perp$.

Next, let $\mathfrak{g}$ be a Lie algebra. Then, the duality pairing
$\langle\cdot,\cdot\rangle$ of $\mathfrak{g}$, $\mathfrak{g}^*$
renders the infinitesimal coadjoint action crossed module 
$\AD^*\mathfrak{g}=(\mathfrak{g}^*,\mathfrak{g},0_{\mathfrak{g}},\ad_{\,\mathfrak{g}}{}^*)$ of $\mathfrak{g}$
a Lie algebra crossed module with invariant pairing (cf. subsect. \cref{subsec:crmodinv}).
Let $\mathfrak{h}$, $\mathfrak{k}$ be subalgebras of $\mathfrak{g}$ with $\mathfrak{h}\subseteq\mathfrak{k}$.
Then, $\mathfrak{h}\subseteq\NN\mathfrak{k}$ and
the infinitesimal coadjoint $\mathfrak{h}$--action crossed module 
$\AD_{\mathfrak{h}}{}^*\mathfrak{k}
=(\mathfrak{k}^0,\mathfrak{h},0_{\mathfrak{h}}|_{\mathfrak{k}^0},\ad_{\mathfrak{h}}{}^*|_{\mathfrak{h}\times\mathfrak{k}^0})$
of $\mathfrak{k}$ which we defined above is an isotropic submodule of $\AD^*\mathfrak{g}$.
The orthogonal normalizer crossed module
of $\AD_{\mathfrak{h}}{}^*\mathfrak{k}$ is 
$\ON\AD_{\mathfrak{h}}{}^*\mathfrak{k}
=\AD_{\NN\mathfrak{h}\cap\mathfrak{k}}{}^*([\mathfrak{h},\mathfrak{k}]+\mathfrak{h})$,
and so is of the same kind too.
The orthogonal Weyl crossed module of $\AD_{\mathfrak{h}}{}^*\mathfrak{k}$ is 
$\OW\AD_{\mathfrak{h}}{}^*\mathfrak{k}=(\mathfrak{ot}_{{\mathfrak{k},\mathfrak{h}}},
\mathfrak{on}_{\mathfrak{k},\mathfrak{h}}, 
0_{\mathfrak{on}_{\mathfrak{k},\mathfrak{h}}}|_{\mathfrak{ot}_{{\mathfrak{k},\mathfrak{h}}}},
\ad_{\mathfrak{on}_{\mathfrak{k},\mathfrak{h}}}{}^*|_{\mathfrak{on}_{\mathfrak{k},\mathfrak{h}}\times\mathfrak{ot}_{{\mathfrak{k},\mathfrak{h}}}})$
with $\mathfrak{ot}_{{\mathfrak{k},\mathfrak{h}}}=([\mathfrak{h},\mathfrak{k}]+\mathfrak{h})^0/\mathfrak{k}^0$,
$\mathfrak{on}_{\mathfrak{k},\mathfrak{h}}=\NN\mathfrak{h}\cap\mathfrak{k}/\mathfrak{h}$.
We note that $\AD_{\mathfrak{h}}{}^*\mathfrak{k}$ is Lagrangian precisely when $\mathfrak{k}=\mathfrak{h}$.

\vfil\eject

\section{\textcolor{blue}{\sffamily Higher gauge theory in the derived formulation}}\label{sec:derform}

In this section, we formulate higher gauge theory in a novel derived formal framework
worked out in refs. \ccite{Zucchini:2019rpp,Zucchini:2019pbv},
that we shall adopt in the construction of 4-- dimensional CS theory. The framework has the distinguished merit of showing
that the relationship of higher to ordinary gauge theory is much closer than it was hitherto thought.
It also provides an elegant graded geometric set--up for the manipulation of crossed module valued non homogeneous forms.
There are two versions of the derived set up, the ordinary and the internal 
\footnote{$\vphantom{\dot{\dot{\dot{a}}}}$We recall briefly the difference between ordinary and internal maps.
Consider a pair $X$, $Y$ of graded manifolds and a map $\varphi$ from $X$ to $Y$. 
When expressed in terms of local 
body and soul coordinates $x^a$ and $\xi^r$ of $X$, the components of one such function 
with respect to local body and soul coordinates $y^i$ and $\eta^h$ of $Y$ are polynomials
in the $\xi^r$ with coefficients which are smooth functions of the $x^a$.
If $\varphi$ is an ordinary map, these coefficient have degree $0$ 
If instead $\varphi$ is internal, the may have non zero degree.}.
The former, suitable for the conventional formulation of 4--dimensional CS theory
studied in this paper, is illustrated in this section. The latter, 
required in the Batalin--Vilkovisky (BV) \ccite{Batalin:1981jr,Batalin:1984jr}
formulation, will be presented elsewhere.  


\subsection{\textcolor{blue}{\sffamily Derived Lie groups and algebras}}\label{subsec:dergralg}

The derived Lie group of a Lie group crossed module and the corresponding infinitesimal notion of
derived Lie algebra of a Lie algebra crossed module were originally introduced in refs.
\ccite{Zucchini:2019rpp,Zucchini:2019pbv}.
In the 4--dimensional CS theory we present, they pay a role analogous to that of the gauge
group of ordinary gauge theory.

Before proceeding to the illustration of this topic a few introductory comments are useful.
The formal set--up of derived Lie groups and algebras is an elegant and convenient way
of handling certain structural elements of the Lie group and algebra crossed modules entering
in the formulation of higher gauge theory.
In practice, it is a kind of superfield formalism not dissimilar to the analogous
formalisms broadly used in supersymmetric field theories. 
It is particularly suited for applications to 4--dimensional CS theory because
of its compactness and capability of presenting it as an ordinary CS theory
with an exotic graded gauge group or algebra.

It must be made clear that the derived Lie group of a Lie group crossed module does not fully
encode this latter, but it only describes an approximation of it in the sense of synthetic geometry.
In fact, the target map of the crossed module is not involved in the definition of the derived group, 
nor could it be because, roughly speaking, the approximation is such to push
the range of the target map away out of reach.
Similar considerations apply to the derived Lie algebra of a Lie algebra crossed module.
The reader is referred to ref. \ccite{Zucchini:2019rpp} for a more precise formal elaboration of this point. 

The deep reasons why the derived set--up allows for such a natural formulation of 4--dimensional CS theory
are still not completely clear. Future work will perhaps shed light on this point.

Consider a Lie group crossed module $\mathsans{M}=(\mathsans{E},\mathsans{G},\tau,\mu)$.
The derived Lie group $\DD\mathsans{M}$ of $\mathsans{M}$ consists of the internal maps
from $\mathbb{R}[-1]$ to the semidirect product $\mathsans{E}\rtimes_\mu\mathsans{G}$ of the Lie groups 
$\mathsans{E}$ and $\mathsans{G}$ with respect to the $\mathsans{G}$--action $\mu$ of the form 
\begin{equation}
\mathrm{P}(\bar\alpha)=\ee^{\bar\alpha P}p   
\label{liecmx1}
\end{equation}
with $\bar\alpha\in\mathbb{R}[-1]$, where $p\in\mathsans{G}$, $P\in\mathfrak{e}[1]$
with the following operations. 
For any $\mathrm{P},\mathrm{Q}\in\DD\mathsans{M}$ with
$\mathrm{P}(\bar\alpha)=\ee^{\bar\alpha P}p$, $\mathrm{Q}(\bar\alpha)=\ee^{\bar\alpha Q}q$, one has 
\begin{align}
&\mathrm{PQ}(\bar\alpha)=\ee^{\bar\alpha(P+\mu\sdot(p,Q))}pq,
\vphantom{\Big]}
\label{liecmx2}
\\
&\mathrm{P}^{-1}(\bar\alpha)=\ee^{-\bar\alpha\mu\sdot(p^{-1},P)}p^{-1}.
\vphantom{\Big]}
\label{liecmx3}
\end{align}
$\DD\mathsans{M}$ is a graded Lie group. The graded Lie group isomorphism 
\begin{equation}
\DD\mathsans{M}\simeq\mathfrak{e}[1]\rtimes_{\mu\sdot}\mathsans{G} 
\label{liecm13}
\end{equation}
holds, where $\mathfrak{e}$ is regarded as an Abelian Lie group 
and $\mathfrak{e}[1]\rtimes_{\mu\sdot}\mathsans{G}$
denotes the semidirect product of the Lie groups $\mathfrak{e}[1]$ and $\mathsans{G}$ 
with respect to the $\mathsans{G}$--action $\mu\sdot$.
The operator $\DD$ has nice functorial properties. 
A morphism $\beta:\mathsans{M}'\rightarrow\mathsans{M}$
of Lie group crossed modules induces through its constituent Lie group morphisms
$\varPhi:\mathsans{E}'\rightarrow\mathsans{E}$, $\phi:\mathsans{G}'\rightarrow\mathsans{G}$
a Lie group morphism $\DD\beta:\DD\mathsans{M}'\rightarrow\DD\mathsans{M}$.
Further, if $\mathsans{M}_1$, $\mathsans{M}_2$ are Lie group crossed modules, then
$\DD(\mathsans{M}_1\times\mathsans{M}_2)=\DD\mathsans{M}_1\times\DD\mathsans{M}_2$. 

The notion of derived Lie group has an infinitesimal counterpart.
Consider a Lie algebra crossed module $\mathfrak{m}=(\mathfrak{e},\mathfrak{g},t,m)$.
The derived Lie algebra $\DD\mathfrak{m}$ of $\mathfrak{m}$ consists of the internal maps
from $\mathbb{R}[-1]$ to the semidirect product $\mathfrak{e}\rtimes_m\mathfrak{g}$ of
the Lie algebras $\mathfrak{e}$ and $\mathfrak{g}$ with respect to the $\mathfrak{g}$--action $m$
of the form 
\begin{equation}
\mathrm{U}(\bar\alpha)=u+\bar\alpha U, 
\label{liecmx5}
\end{equation}
with $\bar\alpha\in\mathbb{R}[-1]$, where $u\in\mathfrak{g}$, $U\in\mathfrak{e}[1]$
with the obvious linear structure and the following Lie bracket.
For any $\mathrm{U},\mathrm{V}\in\DD\mathfrak{m}$ such that
$\mathrm{U}(\bar\alpha)=u+\bar\alpha U$, $\mathrm{V}(\bar\alpha)=v+\bar\alpha V$, one has 
\begin{equation}
[\mathrm{U},\mathrm{V}](\bar\alpha)=[u,v]+\bar\alpha(m(u,V)-m(v,U)). 
\label{liecmx6}
\end{equation}
$\DD\mathfrak{m}$ is a graded Lie algebra. The graded Lie algebra isomorphism 
\begin{equation}
\DD\mathfrak{m}\simeq\mathfrak{e}[1]\rtimes_m\mathfrak{g}
\label{liecm14}
\end{equation}
holds, where 
$\mathfrak{e}$ is regarded as an Abelian Lie algebra and  $\mathfrak{e}[1]\rtimes_m\mathfrak{g}$
denotes the semidirect product of the Lie algebras
$\mathfrak{e}[1]$ and $\mathfrak{g}$ with respect to the $\mathfrak{g}$--action $m$.
As in the finite case above, the operator $\DD$ has functorial properties. 
A morphism $p:\mathfrak{m}'\rightarrow\mathfrak{m}$
of Lie algebra crossed modules induces via its underlying Lie algebra morphisms
$H:\mathfrak{e}'\rightarrow\mathfrak{e}$, $h:\mathfrak{g}'\rightarrow\mathfrak{g}$, 
a Lie algebra morphism $\DD p:\DD\mathfrak{m}'\rightarrow\DD\mathfrak{m}$.
Further, if $\mathfrak{m}_1$, $\mathfrak{m}_2$ are Lie algebra crossed modules, 
$\DD(\mathfrak{m}_1\oplus\mathfrak{m}_2)=\DD\mathfrak{m}_1\oplus\DD\mathfrak{m}_2$. 

The derived set--up introduced above is fully compatible with Lie differentiation. 
If $\mathsans{M}=(\mathsans{E},\mathsans{G},\tau,\mu)$ is a Lie group crossed module and 
$\mathfrak{m}=(\mathfrak{e},\mathfrak{g},\dot\tau,{}\dot{}\mu{}\dot{})$ 
is its associated Lie algebra crossed module (cf. subsect. \cref{subsec:liecrmod}), 
then $\DD\mathfrak{m}$ is the Lie algebra of $\DD\mathsans{M}$.
Further, if $\beta:\mathsans{M}'\rightarrow\mathsans{M}$ is a Lie group crossed module morphism
and $\dot\beta:\mathfrak{m}'\rightarrow\mathfrak{m}$ is the corresponding Lie algebra
crossed module morphism, then $\dot\DD\beta=\DD\dot\beta$. 

The derived set--up is also consistent with the submodule structure of the underlying crossed module
(cf. subsect. \cref{subsec:crosumo}). 
If $\mathsans{M}$ is a Lie group crossed module and $\mathsans{M}'$ is a crossed submodule of $\mathsans{M}$, 
then $\DD\mathsans{M}'$ is Lie subgroup of $\DD\mathsans{M}$. Further, if the Lie group $\mathsans{E}'$
of $\mathsans{M}'$ is connected, then the normalizer and Weyl crossed modules$\vphantom{\ul{\ul{\ul{\ul{g}}}}}$
$\NN\mathsans{M}'$ and $\WW\mathsans{M}'$ of $\mathsans{M}'$ satisfy $\DD\NN\mathsans{M}'=\NN\DD\mathsans{M}'$
and $\DD\WW\mathsans{M}'=\WW\DD\mathsans{M}'$, where $\NN\DD\mathsans{M}'$
and $\WW\DD\mathsans{M}'=\NN\DD\mathsans{M}'/\DD\mathsans{M}'$ are the normalizer and Weyl Lie groups of 
$\DD\mathsans{M}'$. The reason why $\mathsans{E}'$ is required to be connected is that
$\NN\DD\mathsans{M}'$ normalizes only the Lie algebra $\mathfrak{e}'$ and so 
only the connected component of $1_{\mathsans{E}}$ in $\mathsans{E}'$ upon Lie integration. 
Likewise, if $\mathfrak{m}$ is a Lie algebra crossed module and $\mathfrak{m}'$ is a crossed submodule of $\mathfrak{m}$, 
then $\DD\mathfrak{m}'$ is Lie subalgebra of $\DD\mathfrak{m}$. Further,
the normalizer and Weyl crossed modules $\NN\mathfrak{m}'$ and
$\WW\mathfrak{m}'$ of $\mathfrak{m}'$ have the property that $\DD\NN\mathfrak{m}'=\NN\DD\mathfrak{m}'$ and 
$\DD\WW\mathfrak{m}'=\WW\DD\mathfrak{m}'$, where $\NN\DD\mathfrak{m}'$
and $\WW\DD\mathfrak{m}'=\NN\DD\mathfrak{m}'/\DD\mathfrak{m}'$ are the normalizer and Weyl Lie algebras of 
$\DD\mathfrak{m}'$.



Suppose that $\mathfrak{m}$ is a Lie algebra crossed module with invariant pairing $\langle\cdot,\cdot\rangle$
(cf. subsect. \cref{subsec:crmodinv}.
Then, $\DD\mathfrak{m}$ is equipped with an induced symmetric non singular bilinear form
$(\cdot,\cdot):\DD\mathfrak{m}\times\DD\mathfrak{m}\rightarrow\mathbb{R}[-1]$ defined by
\begin{equation}
(\mathrm{U},\mathrm{V})=\langle u,\sfsigma V\rangle+\langle v,\sfsigma U\rangle
\label{isotrop1}
\end{equation}
for any $\mathrm{U},\mathrm{V}\in\DD\mathfrak{m}$, where
$\sfsigma:\mathfrak{e}[1]\rightarrow\mathbb{R}[-1]\otimes\mathfrak{e}$
is the natural $2$--fold suspension map. 
$\sfsigma$ serves the purpose of identifying $\DD\mathfrak{m}$ with the internal map space
$\bfs{{\Map}}(*,\DD\mathfrak{m})$, where $*$ is the singleton,  
since $\bfs{{\Map}}(*,\mathfrak{v}[p])$ is isomorphic to $\mathbb{R}[-p]\otimes\mathfrak{v}$ for any
vector space $\mathfrak{v}$. The identification in turn is necessary for the right hand side
of \ceqref{isotrop1} to make sense. The pairing $(\cdot,\cdot)$ is invariant as
\begin{equation}
([\mathrm{W},\mathrm{U}],\mathrm{V})+(\mathrm{U},[\mathrm{W},\mathrm{V}])=0
\label{isotrop0}
\end{equation}
for $\mathrm{U},\mathrm{V},\mathrm{W}\in\DD\mathfrak{m}$. 

Let $\mathfrak{m}'$ be a crossed submodule of $\mathfrak{m}$.
Then, $(\cdot,\cdot)$ restricts to a symmetric bilinear form
$(\cdot,\cdot)':\DD\mathfrak{m}'\times\DD\mathfrak{m}'\rightarrow\mathbb{R}[-1]$, which however is not non singular
any longer in general. $\mathfrak{m}'$ is an isotropic crossed submodule of $\mathfrak{m}$
(cf. subsect. \cref{subsec:crosumo}) precisely when 
$\DD\mathfrak{m}'$ is an isotropic subalgebra of $\DD\mathfrak{m}$. 
In that case, 
$\dim\DD\mathfrak{m}'$ $\leq\frac{1}{2}\dim\DD\mathfrak{m}
=\dim\mathfrak{g}=\dim\mathfrak{e}$, the bound being saturated when
$\mathfrak{m}'$ is Lagrangian.

If $\mathfrak{m}'$ is an isotropic crossed submodule of $\mathfrak{m}$, 
the orthogonal normalizer and Weyl crossed modules $\ON\mathfrak{m}'$ and
$\OW\mathfrak{m}'$ of $\mathfrak{m}'$ turn out to be $\DD\ON\mathfrak{m}'=\ON\DD\mathfrak{m}'$ and 
$\DD\OW\mathfrak{m}'=\OW\DD\mathfrak{m}'$ respectively, where
$\ON\DD\mathfrak{m}'=\NN\DD\mathfrak{m}'\cap\DD\mathfrak{m}'^\perp$
and $\OW\DD\mathfrak{m}'=\ON\DD\mathfrak{m}'/\DD\mathfrak{m}'$ are the orthogonal normalizer
and Weyl Lie algebras of $\DD\mathfrak{m}'$, $\DD\mathfrak{m}'^\perp$ denoting the
orthogonal complement of $\DD\mathfrak{m}'$. $\vphantom{\ul{\ul{g}}}$

\vfil\eject

\subsection{\textcolor{blue}{\sffamily Derived superfield formulation}}\label{subsec:superfield}

In this subsection, we shall survey the main spaces of Lie group and algebra crossed module valued fields
using a derived superfield formulation. This allows for a very compact geometrically transparent
formulation of 4--dimensional CS theory studied in later sections. 

We assume that the fields propagate on a general manifold $X$. Later, we shall add
the restriction that $X$ is orientable and compact, possibly with boundary.
To include also differential forms, using however a convenient graded geometric description,
the fields will be maps from the shifted tangent bundle $T[1]X$ of $X$ into some graded target manifold
$T$. Below, we denote by $\Map(T[1]X,T)$ the space of ordinary maps from $T[1]X$ into $T$.
The broader space 
of internal maps from $T[1]X$ to $T$ required to incorporate ghost-like fields
in a BV set--up can also be considered, though we shall not do so in this paper. 


The fields we shall consider will be valued either in the derived Lie group $\DD\mathsans{M}$ of
a Lie group crossed module $\mathsans{M}=(\mathsans{E},\mathsans{G},\tau,\mu)$
or in the derived Lie algebra $\DD\mathfrak{m}$ of the associated Lie algebra crossed module
$\mathfrak{m}=(\mathfrak{e},\mathfrak{g},\dot\tau,\sdot\mu{}\sdot\hfpt)$ (cf. subsects. \cref{subsec:liecrmod})
and \cref{subsec:dergralg}). A more comprehensive treatment of this kind of fields is provided in ref. \ccite{Zucchini:2019rpp}.

We consider first $\DD\mathsans{M}$--valued fields. Fields of this kind are elements of the mapping space
$\Map(T[1]X,\DD\mathsans{M})$. If $\mathrm{U}\in\Map(T[1]X,\DD\mathsans{M})$, then 
\begin{equation}
\mathrm{U}(\alpha)=\ee^{\alpha U}u
\label{superfield1}
\end{equation}
with $\alpha\in\mathbb{R}[1]$, where $u\in\Map(T[1]X,\mathsans{G})$, $U\in\Map(T[1]X,\mathfrak{e}[1])$.
$u$, $U$ are the components of $\mathrm{U}$. 
$\Map(T[1]X,\DD\mathsans{M})$ has a Lie 
group structure induced by that of $\DD\mathsans{M}$: if $\mathrm{U}\in\Map(T[1]X,\DD\mathsans{M})$,
$\mathrm{V}\in\Map(T[1]X,\DD\mathsans{M})$, then
\begin{equation}
\mathrm{UV}(\alpha)=\ee^{\alpha(U+\mu\sdot(u,V))}uv, \qquad \mathrm{U}^{-1}(\alpha)
=\ee^{-\alpha\mu\sdot(u^{-1},U))}u^{-1}. 
\label{superfield2}
\end{equation}

Next, we consider first $\DD\mathfrak{m}$--valued fields. Fields of this kind are elements of the mapping space
$\Map(T[1]X,\DD\mathfrak{m})$. If $\Phi\in\Map(T[1]X,\DD\mathfrak{m})$, then 
\begin{equation}
\Phi(\alpha)=\phi+\alpha\varPhi  
\label{superfield3}
\end{equation}
with $\alpha\in\mathbb{R}[1]$, where $\phi\in\Map(T[1]X,\mathfrak{g})$, $\varPhi\in\Map(T[1]X,\mathfrak{e}[1])$. 
Again, $\phi$, $\varPhi$ are the components of $\Phi$. $\Map(T[1]X,\DD\mathfrak{m})$ has a Lie 
algebra structure induced by that of $\DD\mathfrak{m}$: if 
$\Phi\in\Map(T[1]X,\DD\mathfrak{m})$, $\Psi\in\Map(T[1]X,\DD\mathfrak{m})$, then 
\begin{equation}
[\Phi,\Psi](\alpha)=[\phi,\psi]
+\alpha\!\left(\hfpt\sdot\mu\sdot(\phi,\varPsi)-\sdot\mu\sdot(\psi,\varPhi)\right). 
\label{superfield4}
\end{equation}
$\Map(T[1]X,\DD\mathfrak{m})$ is the virtual Lie algebra of $\Map(T[1]X,\DD\mathsans{M})$. (For an explanation
of this terminology, see ref. \ccite{Zucchini:2019rpp}). 

As it turns out, the $\DD\mathfrak{m}$--valued fields introduced above are not enough for our proposes.
One also needs to incorporate fields that are valued in the degree shift\-ed linear spaces
$\DD\mathfrak{m}[p]$ with $p$ an integer. If $\Phi\in\Map(T[1]X,\DD\mathfrak{m}[p])$, then 
\begin{equation}
\Phi(\alpha)
=\phi+(-1)^p\alpha\varPhi  
\label{superfield5}
\end{equation}
with components $\phi\in\Map(T[1]X,\mathfrak{g}[p])$, $\varPhi\in\Map(T[1]X,\mathfrak{e}[p+1])$. 
There is a bilinear bracket that associates with a pair of fields 
$\Phi\in\Map(T[1]X,\DD\mathfrak{m}[p])$, $\Psi\in\Map(T[1]X,\DD\mathfrak{m}[q])$
a field $[\Phi,\Psi]\in\Map(T[1]X,\DD\mathfrak{m}[p+q])$ given by 
\begin{equation}
[\Phi,\Psi](\alpha)=[\phi,\psi]
+(-1)^{p+q}\alpha\!\left(\hfpt\sdot\mu\sdot(\phi,\varPsi)
-(-1)^{pq}\hfpt\sdot\mu\sdot(\psi,\varPhi)\right)
\label{superfield6}
\end{equation}
Setting $\ZZ\DD\mathfrak{m}=\bigoplus_{p=-\infty}^\infty\DD\mathfrak{m}[p]$, 
$\Map(T[1]X,\ZZ\DD\mathfrak{m})$ is a graded Lie algebra. This contains
the Lie algebra $\Map(T[1]X,\DD\mathfrak{m})$ as its degree $0$ subalgebra. 

An adjoint action of $\Map(T[1]X,\DD\mathsans{M})$ on the Lie algebra $\Map(T[1]X,\DD\mathfrak{m})$
and more generally on the graded Lie algebra $\Map(T[1]X,\ZZ\DD\mathfrak{m})$ is defined. For 
$\mathrm{U}\in\Map(T[1]X,\DD\mathsans{M})$, $\Phi\in\Map(T[1]X,\DD\mathfrak{m}[p])$, one has 
\begin{align}
&\Ad\mathrm{U}(\Phi)(\alpha)
=\Ad u(\phi)+(-1)^p\alpha(\mu\sdot(u,\varPhi)-\sdot\mu\sdot(\Ad u(\phi),U)),
\vphantom{\Big]}
\label{superfield7}
\\
&\Ad\mathrm{U}^{-1}(\Phi)(\alpha)=\Ad u^{-1}(\phi)+(-1)^p\alpha\mu\sdot(u^{-1},\varPhi+\sdot\mu\sdot(\phi,U)).
\vphantom{\Big]}
\label{superfield8}
\end{align}
The adjoint action preserves Lie brackets as in ordinary Lie theory. Indeed, for 
$\mathrm{U}\in\Map(T[1]X,\DD\mathsans{M})$, $\Phi\in\Map(T[1]X,\DD\mathfrak{m}[p])$,
$\Psi\in\Map(T[1]X,\DD\mathfrak{m}[q])$, 
\begin{equation}
[\Ad\mathrm{U}(\Phi),\Ad\mathrm{U}(\Psi)]=\Ad\mathrm{U}([\Phi,\Psi]).
\label{superfield9}
\end{equation}

As is well--known, \pagebreak in the graded geometric formulation we adopt, the nilpotent de Rham differential
$d$ is a degree $1$ homological vector field on $T[1]X$, $d^2=0$. $d$ induces a natural degree $1$ 
derived differential $\dd$ on the graded vector space $\Map(T[1]X,\ZZ\mathfrak{m})$.
Concisely, $\dd=d+\dot\tau d/d\alpha$. In more ore explicit terms, for 
$\Phi\in\Map(T[1]X,\DD\mathfrak{m}[p])$, the field $\dd\Phi\in\Map(T[1]X,\DD\mathfrak{m}[p+1])$
reads as 
\begin{equation}
\dd\Phi(\alpha)=d\phi+(-1)^p\dot\tau(\varPhi)+(-1)^{p+1}\alpha d\varPhi.
\label{superfield10}
\end{equation}
It can be straightforwardly verified that shown that 
\begin{equation}
\dd [\Phi,\Psi]=[\dd \Phi,\Psi]+(-1)^p[\Phi,\dd \Psi]
\label{superfield11}
\end{equation}
for $\Phi\in\Map(T[1]X,\DD\mathfrak{m}[p])$, $\Psi\in\Map(T[1]X,\DD\mathfrak{m}[q])$
and that
\begin{equation}
\dd^2=0.
\label{superfield12}
\end{equation}
In this way, $\Map(T[1]X,\ZZ\DD\mathfrak{m})$ becomes a differential graded Lie algebra. 

On several occasions, the pull--back $\mathrm{dUU}^{-1},\mathrm{U}^{-1}\mathrm{dU}\in\Map(T[1]X,\DD\mathfrak{m}[1])$
of the Maurer--Cartan forms of $\DD\mathsans{M}$ by a
$\DD\mathsans{M}$ field $\mathrm{U}\in\Map(T[1]X,\DD\mathsans{M})$ will enter our considerations.  
For these, there exits explicit expressions, 
\begin{align}
&\mathrm{dUU}^{-1}(\alpha)=duu^{-1}+\dot\tau(U)
\vphantom{\Big]}
\nonumber
\\
&\hspace{3.5cm}
-\alpha\!\left(dU+\tfrac{1}{2}[U,U]-\sdot\mu\sdot(duu^{-1}+\dot\tau(U),U)\right),
\vphantom{\Big]}
\label{superfield13}
\\
&\mathrm{U}^{-1}\mathrm{dU}(\alpha)=\Ad u^{-1}\!\left(duu^{-1}+\dot\tau(U)\right)
-\alpha\mu\sdot\!\left(u^{-1},dU+\tfrac{1}{2}[U,U]\right). 
\vphantom{\Big]}
\label{superfield14}
\end{align}
By the relation $\dd=d+\dot\tau d/d\alpha$, 
\ceqref{superfield13}, \ceqref{superfield14} follow from \ceqref{superfield1} and 
the variational identities 
$\delta\ee^{\alpha X}\ee^{-\alpha X}=\frac{\exp(\alpha\ad X)-1}{\alpha\ad X}\delta(\alpha X)$,
$\ee^{-\alpha X}\delta\ee^{\alpha X}=\frac{1-\exp(-\alpha\ad X)}{\alpha\ad X}\delta(\alpha X)$ 
with $\delta=\dot\tau d/d\alpha$, owing to the nilpotence of $\alpha$. 

Next, we assume that the Lie group crossed module $\mathsans{M}$ is equipped with
an invariant pairing $\langle\cdot,\cdot\rangle$. A pairing on the graded Lie algebra
$\Map(T[1]X,\ZZ\mathfrak{m})$ is induced in this way: for 
$\Phi\in\Map(T[1]X,\DD\mathfrak{m}[p])$, $\Psi\in\Map(T[1]X,\DD\mathfrak{m}[q])$
\begin{equation}
(\Phi,\Psi)=\langle\phi,\varPsi\rangle+(-1)^{pq}\langle\psi,\varPhi\rangle. 
\label{superfield15}
\end{equation}
Note that $(\Phi,\Psi)\in\Map(T[1]X,\mathbb{R}[p+q+1])$. The field pairing $(\cdot,\cdot)$
therefore has degree $1$. $(\cdot,\cdot)$ is bilinear. More generally, when scalars with non trivial
grading are involved, the left and right brackets $($ and $)$ behave as if they had respectively
degree $0$ and $1$. For instance, $(c\Phi,\Psi)=c(\Phi,\Psi)$ whilst $(\Phi,\Psi c)=(-1)^k(\Phi,\Psi)c$
if the scalar $c$ has degree $k$. $(\cdot,\cdot)$ is further graded symmetric, 
\begin{equation}
(\Phi,\Psi)=(-1)^{pq}(\Psi,\Phi).
\label{superfield16}
\end{equation}
$(\cdot,\cdot)$ is also non singular. 

The field pairing $(\cdot,\cdot)$ has several other properties which make it a very natural ingredient in the field
theoretic constructions of later sections. 
First, $(\cdot,\cdot)$ is $\DD\mathsans{M}$ invariant.
If $\Phi\in\Map(T[1]X,\DD\mathfrak{m}[p])$, $\Psi\in\Map(T[1]X,\DD\mathfrak{m}[q])$, we have 
\begin{equation}
\left(\Ad\mathrm{U}(\Phi),\Ad\mathrm{U}(\Psi)\right)=(\Phi,\Psi)
\label{superfield19}
\end{equation}
for $\mathrm{U}\in\Map(T[1]X,\DD\mathsans{M})$. By Lie differentiation, $(\cdot,\cdot)$
enjoys also $\DD\mathfrak{m}$ invariance. This latter, however, admits a graded extension, because of which 
\begin{equation}
([\Xi,\Phi],\Psi)+(-1)^{pr}(\Phi,[\Xi,\Psi])=0,
\label{superfield17}
\end{equation}
for $\Xi\in\Map(T[1]M,\DD\mathfrak{m}[r])$. 

Second, $(\cdot,\cdot)$ is compatible 
with the derived differential $\dd$, i. e. 
the de Rham vector field $d$ differentiates $(\cdot,\cdot)$ through $\dd$, 
\begin{equation}
d(\Phi,\Psi)=(\dd \Phi,\Psi)+(-1)^p(\Phi,\dd \Psi). 
\label{superfield18}
\end{equation}

Let $\mathsans{M}$, $\mathsans{M}'$ be Lie group crossed modules with associated Lie algebra crossed modules
$\mathfrak{m}$, $\mathfrak{m}'$. Suppose that $\mathsans{M}'$ is a submodule of $\mathsans{M}$ and that,
consequently, $\mathfrak{m}'$ is a submodule of $\mathfrak{m}$ (cf. subsect. \cref{subsec:crosumo}). 
As $\DD\mathsans{M}'$ is a Lie subgroup of $\DD\mathsans{M}$, $\Map(T[1]X,\DD\mathsans{M}')$ 
is a Lie subgroup of $\Map(T[1]X,\DD\mathsans{M})$. Similarly, as $\DD\mathfrak{m}'$ is a Lie subalgebra
of $\DD\mathfrak{m}$, $\Map(T[1]X,\DD\mathfrak{m})$ is a Lie subalgebra of $\Map(T[1]X,\DD\mathfrak{m})$.
What is more, $\Map(T[1]X,\ZZ\DD\mathfrak{m}')$ is a differential graded Lie subalgebra of
$\Map(T[1]X,\ZZ\DD\mathfrak{m})$, since it is invariant under the action of $\dd$ as is evident from 
\ceqref{superfield10}. If $\mathsans{M}$ is also equipped
with invariant pairing $\langle\cdot,\cdot\rangle$ with respect to which $\mathsans{M}'$
is isotropic (cf. subsect. \cref{subsec:crosumo}),
then $\mathfrak{m}'$ is isotropic and thus
the Lie algebra $\Map(T[1]X,\ZZ\DD\mathfrak{m}')$ is isotropic, that is
$(\Phi,\Psi)=0$ for $\Phi\in\Map(T[1]X,\DD\mathfrak{m}'[p])$, $\Psi\in\Map(T[1]X,\DD\mathfrak{m}'[q])$. \vfil


\subsection{\textcolor{blue}{\sffamily Higher gauge theory in the derived formulation}}\label{subsec:higau}

In this subsection, we present a formulation of higher gauge theory based on the derived superfield
formalism of subsect. \cref{subsec:superfield}. The framework we are going to devise has the virtue of
showing the close relationship of higher to ordinary gauge theory and allows so to import many
ideas and techniques of the latter to the former. 
The benefits of this approach will become evident in sect. \cref{sec:4dchern} below, where
4--dimensional CS theory is worked out.

In higher gauge theory, one should specify to begin with a Lie group crossed module $\mathsans{M}$
and a higher principal $\mathsans{M}$--bundle $P$ on some base manifold $X$. In this general
setting, higher gauge fields and gauge transformations consist in collections of local
Lie valued map and form data organized respectively as non Abelian differential cocycles
and cocycle morphisms \ccite{Baez:2004in,Baez:2005qu}. For the scope of the present paper,
this level of generality is not necessary. It is enough that $P$ be the trivial
$\mathsans{M}$--bundle for which gauge fields and gauge transformations turn out to be 
maps and forms globally defined on $X$. We shall however
come back to this topic in greater detail in subsect. \cref{subsec:2bund} below. 

The basic field of higher gauge theory is the gauge field, which in the derived framework
is a map
$\Omega\in\Map(T[1]X,\DD\mathfrak{m}[1])$. In components, this reads as 
\begin{equation}
\Omega(\alpha)=\omega-\alpha\varOmega, 
\label{higau1}
\end{equation}
where $\omega\in\Map(T[1]X,\mathfrak{g}[1])$, $\varOmega\in\Map(T[1]X,\mathfrak{e}[2])$
(cf. eq. \ceqref{superfield5}). $\omega$, $\varOmega$ are nothing but the familiar $1$-- and $2$--form
gauge fields of 
higher gauge theory.

The higher gauge field $\Omega$ is characterized by its curvature $\Phi$ defined 
\begin{equation}
\Phi=\dd\Omega+\tfrac{1}{2}[\Omega,\Omega], 
\label{higau2}
\end{equation}
where the Lie bracket $[\cdot,\cdot]$ and the differential $\dd$ are defined by \ceqref{superfield6}
and \ceqref{superfield10}, respectively. The expression of $\Phi$ is otherwise formally identical to that
the curvature of a gauge field in ordinary gauge theory. By construction,
$\Phi\in\Map(T[1]X,\DD\mathfrak{m}[2])$. Expressed in components, $\Phi$ reads as 
\begin{equation}
\Phi(\alpha)=\phi+\alpha\varPhi,  \vphantom{\ul{\ul{\ul{g}}}}
\label{higau3}
\end{equation}
where $\phi\in\Map(T[1]X,\mathfrak{g}[2])$, $\varPhi\in\Map(T[1]X,\mathfrak{e}[3])$. $\phi$, $\varPhi$
are just the usual higher gauge theoretic $2$-- and $3$--form curvatures. They 
are expressible in terms of $\omega$, $\varOmega$ through the familiar relations 
\begin{align}
&\phi=d\omega+\tfrac{1}{2}[\omega,\omega]-\dot\tau(\varOmega),
\vphantom{\Big]}
\label{higau4}
\\
&\varPhi=d\varOmega+\sdot\mu\sdot(\omega,\varOmega). 
\vphantom{\Big]}
\label{higau5}
\end{align}

The higher curvature $\Phi$ satisfies the higher Bianchi identity
\begin{equation}
\dd\Phi+[\Omega,\Phi]=0
\label{higau6}
\end{equation}
which follows from \ceqref{higau2} in the usual way. This turns into a pair of Bianchi identities for the
curvature components $\phi$, $\varPhi$, viz
\begin{align}
&d\phi+[\omega,\phi]+\dot\tau(\varPhi)=0,
\vphantom{\Big]}
\label{higau7}
\\
&d\varPhi+\sdot\mu\sdot(\omega,\varPhi)-\sdot\mu\sdot(\phi,\varOmega)=0.
\vphantom{\Big]}
\label{higau8}
\end{align}

A higher gauge transformation is codified in a derived Lie group valued map $\mathrm{U}\in\Map(T[1]X,\DD\mathsans{M})$. 
$\mathrm{U}$ acts on the higher gauge field $\Omega$ as
\begin{equation}
\Omega^{\mathrm{U}}=\Ad\mathrm{U}^{-1}(\Omega)+\mathrm{U}^{-1}\mathrm{dU}
\label{higau9}
\end{equation}
where the adjoint action and pulled--back Maurer--Cartan form of $U$ in the right hand side are defined 
in eqs. \ceqref{superfield8} and \ceqref{superfield14}, respectively. Again, in the derived formulation
the higher gauge transformation action is formally identical to that of ordinary gauge theory.
The higher curvature transforms as
\begin{equation}
\Phi^{\mathrm{U}}=\Ad\mathrm{U}^{-1}(\Phi), 
\label{higau10}
\end{equation}
as expected. The gauge transformation $\mathrm{U}$ can be expressed in components as
\begin{equation}
\mathrm{U}(\alpha)=\ee^{\alpha U}u
\label{higau11}
\end{equation}
with $u\in\Map(T[1]X,\mathsans{G})$, $U\in\Map(T[1]X,\mathfrak{e}[1])$  
according to \ceqref{superfield1}. In terms of these,
using systematically relations \ceqref{superfield8}, \ceqref{superfield14},
it is possible to write down the gauge transform of the higher gauge field components
$\omega$, $\varOmega$,  $\vphantom{\ul{\ul{g}}}$ \pagebreak 
\begin{align}
&\omega^{u,U}=\Ad u^{-1}\!\left(\hspace{.5pt}\omega+duu^{-1}+\dot\tau(U)\right),
\vphantom{\Big]}
\label{higau12}
\\
&\varOmega^{u,U}=\mu\sdot\!\left(u^{-1},\varOmega+\sdot\mu\sdot(\omega,U)+dU+\tfrac{1}{2}[U,U]\right),
\vphantom{\Big]}
\label{higau13}
\end{align}
as well as those of the higher curvature components $\phi$, $\varPhi$,
\begin{align}
&\phi^{u,U}=\Ad u^{-1}\!\left(\hspace{.5pt}\phi\right),
\vphantom{\Big]}
\label{higau14}
\\
&\varPhi^{u,U}=\mu\sdot\!\left(u^{-1},\varPhi+\sdot\mu\sdot(\phi,U)\right).
\vphantom{\Big]}
\label{higau15}
\end{align}
These relations are the well--known gauge transformation expressions of 
the 1-- and 2--form gauge fields and the 2-- and 3--form curvatures of  higher gauge theory.

An infinitesimal higher gauge transformation is a derived Lie algebra valued map
$\Theta\in\Map(T[1]X,\DD\mathfrak{m})$. The $\Theta$ variation of  the higher gauge field $\varOmega$ is
\begin{equation}
\delta_\Theta\Omega=\dd\varTheta+[\Omega,\Theta],
\label{higau20}
\end{equation}
where as before the Lie bracket $[\cdot,\cdot]$ and the differential $\dd$ are given by \ceqref{superfield6}
and \ceqref{superfield10}, respectively. In the derived formulation, the infinitesimal higher gauge transformation
action is again formally identical to that of ordinary gauge theory, in particular it still is the linearized form
of its finite counterpart. As expected, so, the $\Theta$ variation of the higher curvature reads as 
\begin{equation}
\delta_\Theta\Phi=[\Phi,\Theta]. 
\label{higau21}
\end{equation}
The gauge transformation $\Theta$ can be expressed in components as
\begin{equation}
\Theta(\alpha)=\theta+\alpha\varTheta
\label{higau22}
\end{equation}
with $\theta\in\Map(T[1]X,\mathfrak{g})$, $\varTheta\in\Map(T[1]X,\mathfrak{e}[1])$  
according to \ceqref{superfield1}. In terms of these,
exploiting relations \ceqref{superfield8}, \ceqref{superfield14},
we can write down the infinitesimal gauge transform of the higher gauge field components
$\omega$, $\varOmega$,  
\begin{align}
&\delta_{\theta,\varTheta}\omega=d\theta+[\omega,\theta]+\dot\tau(\varTheta),
\vphantom{\Big]}
\label{higau23}
\\
&\delta_{\theta,\varTheta}\varOmega=d\varTheta+\sdot\mu\sdot(\omega,\varTheta)-\sdot\mu\sdot\left(\theta,\varOmega\right),
\vphantom{\Big]}
\label{higau24}
\end{align}
as well as those of the higher curvature components $\phi$, $\varPhi$,$\vphantom{\ul{\ul{\ul{\ul{\ul{g}}}}}}$ \pagebreak 
\begin{align}
&\delta_{\theta,\varTheta}\phi=[\phi,\theta],
\vphantom{\Big]}
\label{higau25}
\\
&\delta_{\theta,\varTheta}\varPhi=\sdot\mu\sdot\left(\phi,\varTheta\right)-\sdot\mu\sdot\left(\theta,\varPhi\right).
\vphantom{\Big]}
\label{higau26}
\end{align}
Again, these are the infinitesimal gauge variation expressions
of the 1-- and 2--form gauge fields and the 2-- and 3--form curvatures in higher gauge theory.



A gauge transformation $\mathrm{U}$ is special if its components $u$, $U$ have the form 
\begin{align}
&u=\tau(A),
\vphantom{\Big]}
\label{higau16}
\\
&U=-dAA^{-1}-\sdot\mu(\omega,A)
\vphantom{\Big]}
\label{higau17}
\end{align}
where $A\in\Map(T[1]X,\mathsans{E})$. Note the dependence on the underlying gauge field $\Omega$. 
Its action on the higher gauge field components $\omega$, $\varOmega$ is particularly simple, 
\begin{align}
&\omega^{u,U}=\omega.
\vphantom{\Big]}
\label{higau18}
\\
&\varOmega^{u,U}=\varOmega+\sdot\mu(\phi,A^{-1}).
\vphantom{\Big]}
\label{higau19}
\end{align}
Therefore, $\omega$ is invariant. $\varOmega$ is not except for when the curvature component $\phi$ vanishes.
The requirement $\phi=0$ is known in higher gauge theory as vanishing fake curvature condition.
A special gauge transformation is one related to the trivial gauge transformation $1$ by a gauge for gauge
transformation. The latter is codified in the group valued map $A$. 

A special infinitesimal gauge transformation $\Theta$ has components $\theta$, $\varTheta$ given by 
\begin{align}
&\theta=\dot\tau(\varXi),
\vphantom{\Big]}
\label{higau27}
\\
&\varTheta=-d\varXi-\sdot\mu\sdot\left(\omega,\varXi\right)
\vphantom{\Big]}
\label{higau28}
\end{align}
where $\varXi\in\Map(T[1]X,\mathfrak{e})$. In keeping with \ceqref{higau18}, \ceqref{higau19},
the corresponding variations of the higher gauge field components $\omega$, $\varOmega$ are 
\begin{align}
&\delta_{\theta,\varTheta}\omega=0, 
\vphantom{\Big]}
\label{higau29}
\\
&\delta_{\theta,\varTheta}\varOmega=-\sdot\mu\sdot\left(\phi,\varXi\right),
\vphantom{\Big]}
\label{higau30}
\end{align}
with $\varOmega$ invariant for $\phi=0$. 

We introduce$\vphantom{\ul{\ul{\ul{g}}}}$ \pagebreak
some notations that will be used frequently in the following. 
The field space of higher gauge theory consists of the higher gauge field manifold
$\mathcal{C}_{\mathsans{M}}(X)=\Map(T[1]X,\DD\mathfrak{m}[1])$. The higher gauge transformations
constitute an infinite dimensional Lie group $\mathcal{G}_{\mathsans{M}}(X)=\Map(T[1]X,\DD\mathsans{M})$ acting on
$\mathcal{C}_{\mathsans{M}}(X)$ according to \ceqref{higau9}. Similarly the infinitesimal higher gauge transformations
make up an infinite dimensional Lie algebra $\mathfrak{G}_{\mathsans{M}}(X)=\Map(T[1]X,\DD\mathfrak{m})$
acting variationally on $\mathcal{C}_{\mathsans{M}}(X)$ through \ceqref{higau20}. 
The special gauge transformations form a subgroup $\mathcal{G}_{\mathsans{M}}(X,\omega)$
of $\mathcal{G}_{\mathsans{M}}(X)$ depending on an assigned gauge field $\Omega$ through its component $\omega$
with associated Lie algebra $\mathfrak{G}_{\mathsans{M}}(X,\omega)$.


\subsection{\textcolor{blue}{\sffamily The derived functional framework
of higher gauge theory }}\label{subsec:funcal}

In field theoretic analysis, one deals with functionals of the relevant higher gauge field 
on some compact manifold $X$. These are given as integrals on the shifted tangent bundle $T[1]X$ of $X$
of certain functions of $\Fun(T[1]X)$ constructed using the gauge field.
Integration is carried out using the Berezinian $\varrho_X$ of $X$.

In higher gauge theory, the relevant field manifold is the higher gauge field space 
$\mathcal{C}_{\mathsans{M}}(X)=\Map(T[1]X,\DD\mathfrak{m}[1])$ introduced in subsect.
\cref{subsec:higau}. The field functionals we will consider belong to 
the graded algebra $\matheul{O}^*{}_{\mathsans{M}}(X)=\Fun(T[1]\mathcal{C}_{\mathsans{M}}(X))$.
$\matheul{O}^*{}_{\mathsans{M}}(X)$ is a complex, its differential being the canonical
homological vector field $\sfdelta$ of $T[1]\mathcal{C}_{\mathsans{M}}(X)$. 
In more conventional terms, the field functionals algebra $\matheul{O}^*{}_{\mathsans{M}}(X)$
envisaged here is that of non homogeneous differential forms on the space 
$\mathcal{C}_{\mathsans{M}}(X)$ and $\sfdelta$ is the corresponding de Rham
differential. Below, we shall set $\matheul{F}{}_{\mathsans{M}}(X)=\matheul{O}^0{}_{\mathsans{M}}(X)$
for convenience.

The graded algebra $\mathcal{F}^*(X)=\Fun(T[1]X)$ is also involved in the considerations below
via the spaces $\Map(T[1]X,\DD\mathfrak{m}[p])$. 
It must be kept in mind here that the grading of $\matheul{O}^*{}_{\mathsans{M}}(X)$ is distinct
from that of $\mathcal{F}^*(X)$. We adopt the convention by which
$\sfPhi\sfPsi=(-1)^{pq+jk}\sfPsi\sfPhi$ for $\sfPhi\in\mathcal{F}^p(X)\otimes\matheul{O}^j{}_{\mathsans{M}}(X)$,  
$\sfPsi\in\mathcal{F}^q(X)\otimes\matheul{O}^k_{\mathsans{M}}(X)$. 

The differential of the higher gauge field $\varOmega\in\mathcal{C}_{\mathsans{M}}(X))$
can be formally viewed as a special element  
$\sfdelta\Omega\in\Map(T[1]X,\DD\mathfrak{m}[1])\otimes\matheul{O}^1{}_{\mathsans{M}}(X)$. 
If $\sfF\in\matheul{F}_{\mathsans{M}}(X)$ is a given field functional, its differential $\sfdelta\sfF$
can be written as 
\begin{equation}
\sfdelta\sfF=\int_{T[1]X}\varrho_X\left(\sfdelta\Omega,\frac{\delta\sfF}{\delta\Omega}\right)
\label{holchern1}
\end{equation}
where $\delta\sfF/\delta\Omega\in\Map(T[d_X-2]X,\DD\mathfrak{m}[1])\otimes\matheul{F}_{\mathsans{M}}(X)$ 
with $d_X=\dim X$, because of the non 
singularity of the field pairing $(\cdot,\cdot)$. This relation defines
the functional derivative $\delta \sfF/\delta\Omega$. We can write relation \ceqref{holchern1} formally as 
\begin{equation}
\sfdelta \sfF=\int_{T[1]X}\varrho_X\left(\sfdelta\Omega,\frac{\delta }{\delta\Omega}\right)\sfF.
\label{holchern3}
\end{equation}
This relation defines in turn
$\delta/\delta\Omega$ as a formally $\DD\mathfrak{m}[d_X-2]$ valued functional
differentiation operator. It is not difficult to verify that $\delta/\delta\Omega$ 
satisfies the Leibniz property as required.

Let $\matheul{V}_{\mathsans{M}}(X)=\Vect(\mathcal{C}_{\mathsans{M}}(X))$ denote the Lie algebra 
of the functional vector fields of $\mathcal{C}_{\mathsans{M}}(X)$. 
A vector field $\sfV\in\matheul{V}_{\mathsans{M}}(X)$ can then be expressed as 
\begin{equation}
\sfV=\int_{T[1]X}\varrho_X\left(\mathrm{V},\frac{\delta }{\delta\Omega}\right), 
\label{holchern5}
\end{equation}
where $\mathrm{V}\in\Map(T[1]X,\DD\mathfrak{m}[1])\otimes\matheul{F}_{\mathsans{M}}(X)$. 
The contraction operator associated with $\sfV$, $\sfiota_{\sfV}$, is 
characterized by the property that 
\begin{equation}
\sfiota_{\sfV}\int_{T[1]X}\varrho_X(\sfdelta\Omega,\Xi)=\int_{T[1]X}\varrho_X(\mathrm{V},\Xi).
\label{holchern7}
\end{equation}
for $\Xi\in\Map(T[1]X,\DD\mathfrak{m}[d_X-2])$. 

On account of \ceqref{higau1}, the differential $\sfdelta\Omega$ of the higher gauge field
$\Omega$ introduced above has the component expression  \hphantom{xxxxxxx} %
\begin{equation}
\sfdelta\Omega(\alpha)=\sfdelta\omega-\alpha\sfdelta\varOmega
\label{holchern2/1}
\end{equation}
Correspondingly, the functional differentiation operator $\delta/\delta\Omega$
defined through relation \ceqref{holchern1} can be written in components as 
\begin{equation}
\frac{\delta}{\delta\Omega}(\alpha)=\frac{\delta}{\delta\varOmega}+(-1)^{d_X}\alpha\frac{\delta}{\delta\omega}
\label{holchern2}
\end{equation}
where $\delta/\delta\varOmega$, $\delta/\delta\omega$ are $\mathfrak{g}[d_X-2]$, $\mathfrak{e}[d_X-1]$
valued functional differentiation operators, respectively. Using these expressions
and employing relations \ceqref{superfield15}, \ceqref{holchern1},
\ceqref{holchern5}, it is possible
to obtain component expressions of the functional derivative $\delta \sfF/\delta\Omega$
of a functional $\sfF\in\matheul{F}_{\mathsans{M}}(X)$ as well as that of a vector field
$\sfV\in\matheul{V}_{\mathsans{M}}(X)$.

Above, we did not \pagebreak define the precise content of the functional algebra
$\matheul{O}^*{}_{\mathsans{M}}(X)$. Doing so is a technical task beyond the scope of this paper
and moreover is not required by the formal developments of later sections. There are however variations 
of the derived functional framework expounded above based on modifications of that content.
First, one could replace the smooth function space $\mathcal{F}^*(X)=\Fun(T[1]X)$ by its distributional extension
$\mathcal{F}'^*(X)=\Fun'(T[1]X)$. This would lead a larger functional algebra $\matheul{O}'^*{}_{\mathsans{M}}(X)$
of functionals whose integral expressions allow for distributions in addition to smooth functions.
Second, one could add the field theoretic constraint of locality, obtaining the local versions
$\matheul{O}^*{}_{\mathsans{M}\mathrm{loc}}(X)$ and $\matheul{O}'^*{}_{\mathsans{M}\mathrm{loc}}(X)$ 
of the previous two functional algebras. Let us recall how these are defined.
By pointwise local smooth functional $\sfpsi$ we mean an element of $\mathcal{F}^*(X)\otimes\matheul{O}^*{}_{\mathsans{M}}(X)$
that can be expressed at each point of $T[1]X$ as a polynomial in the higher gauge field $\Omega$
and its differential $\sfdelta\Omega$ and a finite number of derivatives of $\Omega$ but none of $\sfdelta\Omega$.
Similarly, by pointwise local distributional functional
$\sfpsi$ we mean an element of $\mathcal{F}'^*(X)\otimes\matheul{O}'^*{}_{\mathsans{M}}(X)$ that can be expressed
as $\sfpsi=\sum_i\sfpsi_i\delta_{X_i}$, where the $\sfpsi_i$ are pointwise local smooth functionals and the $\delta_{X_i}$
are Dirac distributions supported on certain submanifolds $X_i$ of $X$.  
An element $\sfPsi\in\matheul{O}^*{}_{\mathsans{M}\mathrm{loc}}(X)$ is one that can be expressed as an integral over
$T[1]X$ of some pointwise local functional $\sfpsi$. Likewise, an element $\sfPsi\in\matheul{O}'^*{}_{\mathsans{M}\mathrm{loc}}(X)$
is one that can be expressed as an integral over $T[1]X$ of some pointwise local distributional functional $\sfpsi$.






\subsection{\textcolor{blue}{\sffamily Derived description of non trivial higher principal bundles}}\label{subsec:2bund}

In this subsection, we shall show that the derived set--up can be used to describe higher gauge theory
on a non trivial higher principal $\mathsans{M}$--bundle $P$ for 
any Lie group crossed module $\mathsans{M}=(\mathsans{E},\mathsans{G},\tau,\mu)$. 
Albeit we shall not encounter this
situation, it is still important to examine this issue in order to ascertain 
whether the derived formulation is capable to efficiently handle this more general case. 

In what follows, we shall describe higher gauge fields and gauge transformations by certain local 
data and their global definedness through matching data. For this reason, we pick an open covering
$\{O_i\}$ of the base manifold $X$ of the bundle $P$. For conciseness, we shall denote by
$O_{ij}=O_i\cap O_j$, $O_{ijk}=O_i\cap O_j\cap O_k$, etc. the non empty intersections
of the covering's opens. 

At a most basic level, a higher gauge field $\Omega$ is a collection $\{\Omega_i\}$ of local
maps $\Omega_i\in\Map(T[1]O_i,\DD\mathfrak{m}[1])$. $\Omega_i$ can expanded in components
\begin{equation}
\Omega_i(\alpha)=\omega_i-\alpha\varOmega_i,
\label{2bund1}
\end{equation}
in keeping with \ceqref{superfield5}, 
where $\omega_i\in\Map(T[1]O_i,\mathfrak{g}[1])$, $\varOmega_i\in\Map(T[1]O_i,\mathfrak{e}[2])$.

In order the local gauge field data $\Omega_i$ to describe a globally defined entity,
the data must match on any double intersection $O_{ij}$  in a way governed by
a collection $\mathrm{F}$
of local matching data $\{\mathrm{F}_{ij}\}$, where $\mathrm{F}_{ij}\in\Map(T[1]O_{ij},\DD\mathsans{M})$.
In accordance with \ceqref{superfield1}, $\mathrm{F}_{ij}$ has the component structure
\begin{equation}
\mathrm{F}_{ij}(\alpha)=\ee^{\alpha F_{ij}}f_{ij},
\label{2bund2}
\end{equation}
where $f_{ij}\in\Map(T[1]O_{ij},\mathsans{G})$, $F_{ij}\in\Map(T[1]O_{ij},\mathfrak{e}[1])$. 
The matching of gauge field data $\Omega_i$ then read as 
\begin{equation}
\Omega_i={}^{F_{ij}}\Omega_j=\Ad\mathrm{F}_{ij}(\Omega_j)-\dd\mathrm{F}_{ij}\mathrm{F}_{ij}{}^{-1}.
\label{2bund3}
\end{equation}
Note the formal analogy of these relations to the corresponding one of ordinary gauge theory.
Using \ceqref{2bund1}, \ceqref{2bund2}, 
eq. \ceqref{2bund3} takes the component form
\begin{align}
\omega_i&=\Ad f_{ij}(\omega_j)-df_{ij}f_{ij}{}^{-1}-\dot\tau(F_{ij}),
\vphantom{\Big]}
\label{2bund4}
\\
\varOmega_i&=\mu\sdot(f_{ij},\varOmega_j)-dF_{ij}-\tfrac{1}{2}[F_{ij},F_{ij}]-\sdot\mu\sdot(\omega_i,F_{ij}).
\vphantom{\Big]}
\label{2bund5}
\end{align}
We recover in this way the well--known gluing relations of higher gauge field components
in higher gauge theory \ccite{Baez:2004in,Baez:2005qu}.

Consistency of the matching of local gauge field data $\Omega_i$ on the triple intersections $O_{ijk}$ does
not require  simply that $\mathrm{F}_{ik}=\mathrm{F}_{ij}\mathrm{F}_{jk}$,
as is the case in ordinary gauge theory, but more generally that
\begin{equation}
\mathrm{F}_{ik}=\mathrm{H}_{ijk}\mathrm{F}_{ij}\mathrm{F}_{jk},
\label{2bund6}
\end{equation}
where $\mathrm{H}_{ijk}\in\Map(T[1]O_{ijk},\DD\mathsans{M})$ such that
\begin{equation}
{}^{\mathrm{H}_{ijk}}\Omega_i=\Ad\mathrm{H}_{ijk}(\Omega_i)-\dd\mathrm{H}_{ijk}\mathrm{H}_{ijk}{}^{-1}=\Omega_i.
\label{2bund7}
\end{equation}
Again, these identities read  more explicit in components. Let 
\begin{equation}
\mathrm{H}_{ijk}(\alpha)=\ee^{\alpha H_{ijk}}h_{ijk},
\label{2bund8}
\end{equation}
where $h_{ijk}\in\Map(T[1]O_{ijk},\mathsans{G})$, $H_{ijk}\in\Map(T[1]O_{ijk},\mathfrak{e}[1])$. 
Using \ceqref{2bund2}, \ceqref{2bund8}, relations \ceqref{2bund6} take the form
\begin{align}
&f_{ik}=h_{ijk}f_{ij}f_{jk},
\vphantom{\Big]}
\label{2bund9}
\\
&F_{ik}=H_{ijk}+\mu\sdot(h_{ijk},F_{ij}+\mu\sdot(f_{ij},F_{jk})).
\vphantom{\Big]}
\label{2bund10}
\end{align}
Using \ceqref{2bund8} once more, conditions \ceqref{2bund7} become
\begin{align}
&{}^{h_{ijk},H_{ijk}}\omega_i=\Ad h_{ijk}(\omega_i)-dh_{ijk}h_{ijk}{}^{-1}-\dot\tau(H_{ijk})=\omega_i,
\vphantom{\Big]}
\label{2bund11}
\\
&{}^{h_{ijk},H_{ijk}}\varOmega_i=\mu\sdot(h_{ijk},\varOmega_i)-dH_{ijk}
-\tfrac{1}{2}[H_{ijk},H_{ijk}]-\sdot\mu\sdot(\omega_i,H_{ijk})=\varOmega_i.
\vphantom{\Big]}
\label{2bund12}
\end{align}
In refs. \ccite{Baez:2004in,Baez:2005qu}, it is shown that the higher gauge field $\Omega$ being fake flat, 
\begin{equation}
d\omega_i+\tfrac{1}{2}[\omega_i,\omega_i]-\dot\tau(\varOmega_i)=0,
\label{2bund13}
\end{equation}
is a necessary and sufficient condition for the well--definedness of higher holonomies. 
It can be checked that this requirement is compatible with the matching relations
\ceqref{2bund4}, \ceqref{2bund5}. In this case, the data $h_{ijk}$, $H_{ijk}$ obeying \ceqref{2bund11},
\ceqref{2bund12} are of the special gauge transformation form of eqs. \ceqref{higau16}, \ceqref{higau17},
\begin{align}
&h_{ijk}=\tau(T_{ijk}),
\vphantom{\Big]}
\label{2bund14}
\\
&H_{ijk}=-\sdot\mu(\omega_i,T_{ijk})-dT_{ijk}T_{ijk}{}^{-1},
\vphantom{\Big]}
\label{2bund15}
\end{align}
where $T_{ijk}\in\Map(T[1]O_{ijk},\mathsans{E})$. 
Relations \ceqref{2bund9}, \ceqref{2bund10} then get
\begin{align}
&f_{ik}=\tau(T_{ijk})f_{ij}f_{jk},
\vphantom{\Big]}
\label{2bund16}
\\
&F_{ik}=\Ad T_{ijk}\left(F_{ij}+\mu\sdot(f_{ij},F_{jk})\right)-\sdot\mu(\omega_i,T_{ijk})-dT_{ijk}T_{ijk}{}^{-1}.
\vphantom{\Big]}
\label{2bund17}
\end{align}

The collection $\mathrm{H}=\{\mathrm{H}_{ijk}\}$ of Lie valued data must itself satisfy a set
of consistency conditions on the quadruple intersections $O_{ijkl}$, 
\begin{equation}
\mathrm{H}_{ikl}\mathrm{H}_{ijk}=\mathrm{H}_{ijl}\mathrm{F}_{ij}\mathrm{H}_{jkl}\mathrm{F}_{ij}{}^{-1}.
\label{2bund18}
\end{equation}
On account of \ceqref{2bund2}, \ceqref{2bund8}, the component form of \ceqref{2bund18} read as
\begin{align}
&h_{ikl}h_{ijk}=h_{ijl}f_{ij}h_{jkl}f_{ij}{}^{-1},
\vphantom{\Big]}
\label{2bund19}
\\
&H_{ikl}+\mu\sdot(h_{ikl},H_{ijk})=H_{ijl}+\mu\sdot(h_{ijl}f_{ij},H_{jkl})
\vphantom{\Big]}
\label{2bund20}
\\
&\hspace{5cm}+\mu\sdot(h_{ijl},F_{ij})-\mu\sdot(h_{ikl}h_{ijk},F_{ij}).
\vphantom{\Big]}
\nonumber
\end{align}
A straightforward calculation shows that when the data $h_{ijk}$, $H_{ijk}$ have the special
form \ceqref{2bund14}, \ceqref{2bund15} of refs. \ccite{Baez:2004in,Baez:2005qu},
conditions \ceqref{2bund19}, \ceqref{2bund20} reduce into 
\begin{align}
&\tau(T_{ikl}T_{ijk})=\tau(T_{ijl}\mu(f_{ij},T_{jkl})),
\vphantom{\Big]}
\label{2bund21}
\\
&\sdot\mu(\omega_i,T_{ikl}T_{ijk})=\sdot\mu(\omega_i,T_{ijl}\mu(f_{ij},T_{jkl})).
\vphantom{\Big]}
\label{2bund22}
\end{align}
Both of these are fulfilled if 
\begin{equation}
T_{ikl}T_{ijk}=T_{ijl}\mu(f_{ij},T_{jkl}). 
\label{2bund23}
\end{equation}

The collection of Lie valued data $f=\{f_{ij}\}$, $T=\{T_{ijk}\}$ obeying \ceqref{2bund16}, 
\ceqref{2bund23} defines a non Abelian cocycle \ccite{Giraud:1971cna}. It describes
the background higher principal $\mathsans{M}$--bundle $P$ on $X$ supporting the higher gauge fields.

The collection of Lie valued data $\omega=\{\omega_i\}$, $\varOmega=\{\varOmega_i\}$, 
$f=\{f_{ij}\}$, $F=\{F_{ij}\}$, $T=\{T_{ijk}\}$ obeying \ceqref{2bund4}, \ceqref{2bund5}, 
\ceqref{2bund16}, \ceqref{2bund17}, \ceqref{2bund23},  
constitutes a non Abelian differential cocycle \ccite{Breen:2001ie}. It encodes a higher gauge field
as a globally defined 2--connection of the bundle $P$. 

At a basic level, a higher gauge transformation $\mathrm{U}$ is a collection $\{\mathrm{U}_i\}$ of Lie valued mappings
$\mathrm{U}_i\in\Map(T[1]O_i,\DD\mathsans{M})$. In components, these maps read as 
\begin{equation}
\mathrm{U}_i(\alpha)=\ee^{\alpha U_i}u_i,
\label{2bund24}
\end{equation}
by \ceqref{superfield1}, where $u_i\in\Map(T[1]O_i,\mathsans{G})$, $U_i\in\Map(T[1]O_i,\mathfrak{e}[1])$. 
$\mathrm{U}$ acts on a higher gauge field $\Omega$ yielding a gauge field ${}^{\mathrm{U}}\Omega$ locally given by  
\begin{equation}
{}^{\mathrm{U}}\Omega_i={}^{\mathrm{U}_i}\Omega_i
=\Ad\mathrm{U}_i(\Omega_i)-\dd\mathrm{U}_i\mathrm{U}_i{}^{-1}.
\label{2bund25}
\end{equation}
Note that this gauge transformation action is related to the one defined in \ceqref{higau9}
by ${}^{\mathrm{U}}\Omega=\Omega^{\mathrm{U}^{-1}}$. We use here the left form of the action
to comply with the most common convention. In components, \ceqref{2bund25} reads 
\begin{align}
&{}^{u,U}\omega_i=\Ad u_i(\omega_i)-du_iu_i{}^{-1}-\dot\tau(U_i), 
\vphantom{\Big]}
\label{2bund26}
\\
&{}^{u,U}\varOmega_i=\mu\sdot(u_i,\varOmega_i)-dU_i-\tfrac{1}{2}[U_i,U_i]
-\sdot\mu\sdot({}^{u.U}\omega_i,U_i),
\label{2bund27}
\end{align}
the local component form of higher gauge transformation 
of refs. \ccite{Baez:2004in,Baez:2005qu}. 

The local data of a higher gauge transformation $\mathrm{U}$
must satisfy certain matching relations implied by the gauge transform ${}^{\mathrm{U}}\Omega$
of a higher gauge field $\Omega$ being itself a gauge field. In this regard, one must keep in mind that
the matching data collection $\mathrm{F}$ depends on the underlying gauge field $\Omega$, as $\mathrm{F}$
is constrained by relations \ceqref{2bund6} in which the data collection $\mathrm{H}$ depending on $\Omega$
via condition \ceqref{2bund7} appears. 
So, the matching data collections ${}^{\mathrm{U}}\mathrm{F}$, ${}^{\mathrm{U}}\mathrm{H}$
of the transformed gauge field ${}^{\mathrm{U}}\Omega$ are generally different
from the corresponding collections $\mathrm{F}$,
$\mathrm{H}$ of the given gauge field $\Omega$. The fact that ${}^{\mathrm{U}}\Omega$
obeys the same kind of matching relations as $\Omega$, viz \ceqref{2bund3}, entails that the local data $\mathrm{U}_i$
satisfy relations of the form
\begin{equation}
\mathrm{U}_i=\mathrm{A}_{ij}{}^{\mathrm{U}}\mathrm{F}_{ij}\mathrm{U}_j\mathrm{F}_{ij}{}^{-1}
\label{2bund28}
\end{equation}
on the double intersections $O_{ij}$, where $\mathrm{A}_{ij}\in\Map(T[1]O_{ij},\DD\mathsans{M})$ such that
\begin{equation}
{}^{\mathrm{A}_{ij}\mathrm{U}}\Omega_i=\Ad\mathrm{A}_{ij}({}^{\mathrm{U}}\Omega_i)
-\dd\mathrm{A}_{ij}\mathrm{A}_{ij}{}^{-1}={}^{\mathrm{U}}\Omega_i.
\label{2bund29}
\end{equation}
Again, these identities read more explicit in components. Let 
\begin{equation}
\mathrm{A}_{ij}(\alpha)=\ee^{\alpha A_{ij}}a_{ij},
\label{2bund30}
\end{equation}
where $a_{ij}\in\Map(T[1]O_{ij},\mathsans{G})$, $A_{ij}\in\Map(T[1]O_{ij},\mathfrak{e}[1])$. 
Using \ceqref{2bund2}, \ceqref{2bund24}, \ceqref{2bund30}, conditions \ceqref{2bund29} lead to 
\begin{align}
&u_i=a_{ij}\,{}^{u,U}\!f_{ij}u_jf_{ij}{}^{-1},
\vphantom{\Big]}
\label{2bund31}
\\
&U_i=\mu\sdot(a_{ij}\,{}^{u,U}\!f_{ij},U_j)+A_{ij}-\mu\sdot(u_i,F_{ij})+\mu\sdot(a_{ij},{}^{u,U}\!F_{ij}).
\vphantom{\Big]}
\label{2bund32}
\end{align}
Using \ceqref{2bund30} again, conditions \ceqref{2bund29} get 
\begin{align}
&{}^{a_{ij},A_{ij}}{}^{u,U}\!\omega_i=\Ad a_{ij}(\,{}^{u,U}\!\omega_i)-da_{ij}a_{ij}{}^{-1}-\dot\tau(A_{ij})={}^{u,U}\!\omega_i,
\vphantom{\Big]}
\label{2bund33}
\\
&{}^{a_{ij},A_{ij}}{}^{u,U}\!\varOmega_i=\mu\sdot(a_{ij},{}^{u,U}\!\varOmega_i)-dA_{ij}
-\tfrac{1}{2}[A_{ij},A_{ij}]-\sdot\mu\sdot(\,{}^{u,U}\!\omega_i,A_{ij})={}^{u,U}\!\varOmega_i.
\vphantom{\Big]}
\label{2bund34}
\end{align}
In the formulation of refs. \ccite{Baez:2004in,Baez:2005qu} in which the gauge field $\Omega$
is fake flat, the data $a_{ij}$, $A_{ij}$ obeying \ceqref{2bund33},
\ceqref{2bund34} are of special gauge transformation type of eqs. \ceqref{higau16}, \ceqref{higau17}) 
\begin{align}
&a_{ij}=\tau(B_{ij}),
\vphantom{\Big]}
\label{2bund35}
\\
&A_{ij}=-\sdot\mu(\,{}^{u,U}\!\omega_i,B_{ij})-dB_{ij}B_{ij}{}^{-1}.
\vphantom{\Big]}
\label{2bund36}
\end{align}
The matching relations \ceqref{2bund31}, \ceqref{2bund32} then read as 
\begin{align}
&u_i=\tau(B_{ij}){}^{u,U}\!f_{ij}u_jf_{ij}{}^{-1},
\vphantom{\Big]}
\label{2bund37}
\\
&U_i=\Ad B_{ij}(\mu\sdot(\,{}^{u,U}\!f_{ij},U_j)+{}^{u,U}\!F_{ij})-\mu\sdot(u_i,F_{ij})
\vphantom{\Big]}
\label{2bund38}
\\
&\hspace{5.5cm}
-\sdot\mu(\,{}^{u,U}\!\omega_i,B_{ij})-dB_{ij}B_{ij}{}^{-1}.
\vphantom{\Big]}
\nonumber
\end{align}

Relations \ceqref{2bund28} can be used to express the gauge transformed matching data ${}^{\mathrm{U}}\mathrm{F}_{ij}$
in terms of the original data $\mathrm{F}_{ij}$ and the gauge transformation data $\mathrm{U}_i$.
Similarly, relations \ceqref{2bund31}, \ceqref{2bund32} 
allow to write the gauge transformed matching
data components ${}^{u,U}f_{ij}$, ${}^{u,U}F_{ij}$ through the matching data 
components $f_{ij}$, $F_{ij}$ and the gauge transformation data components $u_i$, $U_i$.
This is a simple exercise that we leave to the reader.

The collection $\mathrm{A}=\{\mathrm{A}_{ijk}\}$ of Lie valued data must itself satisfy a set
of consistency conditions on the triple intersections $O_{ijk}$, 
\begin{equation}
\mathrm{A}_{ik}=\mathrm{U}_i\mathrm{H}_{ijk}\mathrm{U}_i{}^{-1}
\mathrm{A}_{ij}{}^{\mathrm{U}}\mathrm{F}_{ij}\mathrm{A}_{jk}{}^{\mathrm{U}}\mathrm{F}_{ij}{}^{-1}
{}^{\mathrm{U}}\mathrm{H}_{ijk}{}^{-1}.
\label{2bund39}
\end{equation}
By \ceqref{2bund2}, \ceqref{2bund8}, \ceqref{2bund24}, \ceqref{2bund30},
the component form of \ceqref{2bund39} read as
\begin{align}
&a_{ik}=u_i\hfpt h_{ijk}\hfpt u_i{}^{-1}\hfpt a_{ij}\hfpt {}^{u,U}\!f_{ij}\hfpt
a_{jk}\hfpt {}^{u,U}\!f_{ij}{}^{-1}\hfpt {}^{u,U}\!h_{ijk}{}^{-1},
\vphantom{\Big]}
\label{2bund40}
\\
&A_{ik}=U_i-\mu\sdot(u_i\hfpt h_{ijk}\hfpt u_i{}^{-1},U_i)+\mu\sdot(u_i\hfpt h_{ijk}
\hfpt u_i{}^{-1}\hfpt a_{ij},{}^{u,U}\!F_{ij})
\vphantom{\Big]}
\label{2bund41}
\\
&\hspace{.6cm}-\mu\sdot(u_i\hfpt h_{ijk}\hfpt u_i{}^{-1}\hfpt a_{ij}\hfpt
{}^{u,U}\!f_{ij}\hfpt a_{jk}\hfpt {}^{u,U}\!f_{ij}{}^{-1},{}^{u,U}\!F_{ij})
+\mu\sdot(u_i\hfpt h_{ijk}\hfpt u_i{}^{-1},A_{ij})
\vphantom{\Big]}
\nonumber
\\
&\hspace{.6cm}+\mu\sdot(u_i\hfpt h_{ijk}\hfpt u_i{}^{-1}\hfpt a_{ij}\hfpt {}^{u,U}\!f_{ij},A_{jk})+\mu\sdot(u_i,H_{ijk})
\vphantom{\Big]}
\nonumber
\\
&\hspace{.6cm}-\mu\sdot(u_i\hfpt h_{ijk}\hfpt u_i{}^{-1}\hfpt a_{ij}\hfpt {}^{u,U}\!f_{ij}\hfpt
a_{jk}\hfpt {}^{u,U}\!f_{ij}{}^{-1}\hfpt {}^{u,U}\!h_{ijk}{}^{-1},{}^{u,U}\!H_{ijk}). 
\vphantom{\Big]}
\nonumber
\end{align}
These relations are rather messy, but it is still possible to find a simple solutions
in the framework of refs. \ccite{Baez:2004in,Baez:2005qu}. 
So, we assume again that the underlying gauge field $\Omega$ is fake flat and that 
the data $h_{ijk}$, $H_{ijk}$ and $a_{ij}$, $A_{ij}$ obeying \ceqref{2bund11}, \ceqref{2bund12} 
and \ceqref{2bund33}, \ceqref{2bund34} are of special gauge transformation type
\ceqref{2bund14}, \ceqref{2bund15} and \ceqref{2bund35}, \ceqref{2bund36}, respectively. 
Then, a straightforward calculation shows that conditions \ceqref{2bund40}, \ceqref{2bund41} reduce to 
\begin{align}
&\tau(B_{ik})=\tau(\mu(u_i,T_{ijk})B_{ij}\hfpt\mu({}^{u,U}f_{ij},B_{jk})\hfpt {}^{u,U}T_{ijk}{}^{-1}), 
\vphantom{\Big]}
\label{2bund42}
\\
&\sdot\mu({}^{u,U}\omega_i,B_{ik})+dB_{ik}B_{ik}{}^{-1}
\vphantom{\Big]}
\label{2bund43}
\\
&\hspace{2cm}=\sdot\mu({}^{u,U}\omega_i,\mu(u_i,T_{ijk})B_{ij}\hfpt\mu({}^{u,U}f_{ij},B_{jk})\hfpt {}^{u,U}T_{ijk}{}^{-1})
\vphantom{\Big]}
\nonumber
\\
&\hspace{2cm}+d(\mu(u_i,T_{ijk})B_{ij}\hfpt\mu({}^{u,U}f_{ij},B_{jk})\hfpt {}^{u,U}T_{ijk}{}^{-1})
\vphantom{\Big]}
\nonumber
\\
&\hspace{3cm}\times
(\mu(u_i,T_{ijk})B_{ij}\hfpt\mu({}^{u,U}f_{ij},B_{jk})\hfpt {}^{u,U}T_{ijk}{}^{-1})^{-1}
\vphantom{\Big]}
\nonumber
\end{align}
Both of these are fulfilled if 
\begin{equation}
B_{ik}=\mu(u_i,T_{ijk})B_{ij}\hfpt\mu({}^{u,U}f_{ij},B_{jk})\hfpt {}^{u,U}T_{ijk}{}^{-1}
\label{2bund44}
\end{equation}
This is almost the property such data are required to have
in the formulation of refs. \ccite{Baez:2004in,Baez:2005qu}. We shall come back to this point
momentarily.

Relations \ceqref{2bund39} combined with the \ceqref{2bund22}
can be used to express the gauge transformed matching data ${}^{\mathrm{U}}\mathrm{H}_{ijk}$
in terms of the original data $\mathrm{H}_{ijk}$ and the gauge transformation data $\mathrm{U}_i$.
Similarly, relations \ceqref{2bund40}, \ceqref{2bund41} together with the \ceqref{2bund31}, \ceqref{2bund32} 
allow to write the gauge transformed matching
data components ${}^{u,U}h_{ijk}$, ${}^{u,U}H_{ijk}$ through the given matching data 
components $h_{ijk}$, $H_{ijk}$, and the gauge transformation data components $u_i$, $U_i$.
This is again 
left to the reader.

The collection of Lie valued data $u=\{U_i\}$, $U=\{U_i\}$, 
$B=\{B_{ij}\}$ obeying \ceqref{2bund37}, \ceqref{2bund38}, \ceqref{2bund44}
is an equivalence of the non Abelian differential cocycle pair 
$\omega=\{\omega_i\}$, $\varOmega=\{\varOmega_i\}$, 
$f=\{f_{ij}\}$, $F=\{F_{ij}\}$, $T=\{T_{ijk}\}$, ${}^{u,U}\omega=\{{}^{u,U}\omega_i\}$,
${}^{u,U}\varOmega=\{{}^{u,U}\varOmega_i\}$, 
${}^{u,U}f=\{{}^{u,U}f_{ij}\}$, ${}^{u,U}F=\{{}^{u,U}F_{ij}\}$, ${}^{u,U}T=\{{}^{u,U}T_{ijk}\}$ \ccite{Breen:2001ie}.

As we have seen above, the collection of Lie valued data $f=\{f_{ij}\}$, $T=\{T_{ijk}\}$ obeying \ceqref{2bund16},
\ceqref{2bund23} define a non Abelian cocycle describing the background higher principal $\mathsans{M}$--bundle
structure for the gauge fields. \pagebreak It is natural to require that such data be gauge invariant
\begin{align}
&{}^{u,U}\!f_{ij}=f_{ij},
\vphantom{\Big]}
\label{2bund45}
\\
&{}^{u,U}\!T_{ijk}=T_{ijk}.
\vphantom{\Big]}
\label{2bund46}
\end{align}
This is in keeping with the analogous requirement imposed on the matching data
$f_{ij}$ in ordinary principal bundle theory. It ensures that the $\mathsans{M}$--bundle
structure  constitutes the background for the gauge transformations too. 
With \ceqref{2bund45}, \ceqref{2bund46} holding, conditions \ceqref{2bund44} 
take the form they have in the framework of refs. \ccite{Baez:2004in,Baez:2005qu}.

In conclusion, the derived set--up can be used to describe rather compactly
a higher principal $\mathsans{M}$--bundle and the gauge fields and gauge transformations
it supports. It further renders manifest the formal analogy of higher to ordinary principal
bundles, gauge fields and gauge transformations. However, to make certain implicit constraints
explicit, it is unavoidable to resort to a component analysis. 

We end this subsection with a discussion shedding light upon a basic difference
existing between the nature of higher gauge field and gauge transformations in a non trivial
higher principal bundle background and of their ordinary counterparts. This diversity is responsible 
for making the construction of 4--dimensional CS theory in such a background problematic.
(More on this in subsect. \cref{subsec:chernglob}.)

For a full global description of the higher gauge field $\Omega$, its local data $\Omega_i$ are not
sufficient. The data $\Omega_i$ match through 
certain data $\mathrm{F}_{ij}$ obeying consistency conditions involving further data $\mathrm{H}_{ijk}$ 
depending in turn on the $\Omega_i$ (cf. eqs. \ceqref{2bund3}, \ceqref{2bund6}, \ceqref{2bund7}, \ceqref{2bund18}). 
Thus, it is not possible to organize all these data in a hierarchy and consider in particular
the $\mathrm{F}_{ij}$ as those encoding a fixed higher principal bundle structure independent and preexisting
any gauge field superimposed to it, as is the case in ordinary gauge theory. Only the components $f_{ij}$ and $h_{ijk}$ of
$\mathrm{F}_{ij}$, $\mathrm{H}_{ijk}$ can 
be assumed to be independent from 
the gauge field data $\Omega_i$ and therefore be ascribed to a fixed bundle
background. The components $F_{ij}$ and $H_{ijk}$ conversely cannot and might be considered as part of
the data of a 2--connection on the same footing as the components $\omega_i$, $\varOmega_i$ of the $\Omega_i$. 

Similarly, for a full global description of a higher gauge transformation $\mathrm{U}$, its local data
$\mathrm{U}_i$ are not enough. The data $\mathrm{U}_i$ match through the data $\mathrm{F}_{ij}$ associated with
some assigned gauge field data $\Omega_i$ and further data $\mathrm{A}_{ij}$ obeying compatibility conditions
involving the $\Omega_i$, $\mathrm{U}_i$ and also $\mathrm{H}_{ijk}$ (cf. eqs. \ceqref{2bund28}, \ceqref{2bund30},
\ceqref{2bund39}). Again, it is not possible to organize all these data in a hierarchy and consider in particular
the $\mathrm{U}_i$ as those representing a gauge transformation standing 
independently from the gauge fields it acts on, as in ordinary gauge theory. Only the components
$a_{ij}$ of $\mathrm{A}_{ij}$ can 
be assumed to be independent from the
gauge field and gauge transformation data $\Omega_i$ and $\mathrm{U}_i$. The components $A_{ij}$ instead
cannot and might considered part of the data defining an equivalence of two 2--connections on a par 
as the components $u_i$, $U_i$ of the $\mathrm{U}_i$. 

To allow for the well--definedness of higher holonomies,
the gauge field components $\omega_i$, $\varOmega_i$ are required to obey the vanishing fake
curvature condition \ceqref{2bund13}. The components $h_{ijk}$, $H_{ijk}$ and $a_{ij}$, $A_{ij}$
then have the structure shown in \ceqref{2bund14}, \ceqref{2bund15} and \ceqref{2bund42}, \ceqref{2bund43}
with conditions \ceqref{2bund23} and \ceqref{2bund44} satisfied. Conditions \ceqref{2bund45}, \ceqref{2bund46}
are further imposed. However, again, the matching data are or obey condition depending on the gauge field
data.

\vfil\eject

\section{\textcolor{blue}{\sffamily 4--Dimensional Chern--Simons theory}}\label{sec:4dchern}

In this section, we introduce and study the $4$--dimensional higher CS model,
which is the main topic of this paper, focusing in particular on its 
gauge symmetries. We illustrate further its Hamiltonian formulation. 

As we shall see, 4--dimensional CS theory exhibits its most
interesting features when the underlying $4$--fold has a boundary. This fact highlights its rich
holographic properties, which we shall describe in great detail. 

As already anticipated in sect. \cref{sec:liecrmod}, we shall work in a graded geometric setting 
where forms are ordinary maps from the shifted tangent bundle $T[1]X$ of a relevant manifold $X$
to some target graded manifold. Integration will be implemented through the Berezinian
$\varrho_X$ of $X$. 

The basic algebraic structures the model are a Lie group crossed module
$\mathsans{M}=(\mathsans{E},\mathsans{G},\tau,\mu)$ and the associated Lie algebra crossed module
$\mathfrak{m}=(\mathfrak{e},\mathfrak{g},\dot\tau,\sdot\mu{}\sdot\hfpt)$
(cf. subsect. \cref{subsec:liecrmod}). We assume that $\mathsans{M}$ is equipped with an invariant pairing
$\langle\cdot,\cdot\rangle$ and that $\mathsans{M}$ is fine and that the conditions sufficient for 
the direct sum decomposition \ceqref{crmodinv12} to hold are verified (cf. subsect. \cref{subsec:crmodinv}),
though some of our results do not hinge on this restriction. 

All the fields occurring in the theory are valued either in the derived Lie group
$\DD\mathsans{M}$ of $\mathsans{M}$ or in the derived Lie algebra $\DD\mathfrak{m}$
of $\mathfrak{m}$ (cf.  subsect. \cref{subsec:dergralg}). 
The derived superfield formalism of subsect.  \cref{subsec:superfield} is employed throughout.
The field pairing $(\cdot,\cdot)$ induced by $\langle\cdot,\cdot\rangle$
(cf. eq. \ceqref{superfield15}) is used systematically in the construction. 
The higher gauge theoretic framework of subsect \cref{subsec:higau} 
conjoined with the derived functional framework of subsect.
\cref{subsec:funcal} allow for a particularly geometrically intuitive
formulation highlighting the close relationship of the higher CS theory to the ordinary one.




\subsection{\textcolor{blue}{\sffamily 4-d Chern--Simons theory}}\label{subsec:4dchern}

In this subsection, we present the $4$--dimensional higher CS model, which is the main topic of this paper.
In component form, this model first appeared in \ccite{Zucchini:2011aa} and was 
further studied in \ccite{Soncini:2014ara,Zucchini:2015ohw} on $4$--folds without boundary. 

Below, we assume that $M$ is an oriented, compact $4$--fold, possibly with boundary. No further
restrictions are imposed. 

The action of $4$--dimensional CS theory is 
\begin{equation}
\sfC\hspace{-.75pt}\sfS(\Omega)
=\frac{k}{4\pi}\int_{T[1]M}\varrho_M\!\left(\Omega,\dd\Omega+\tfrac{1}{3}[\Omega,\Omega]\right)\!,
\label{4dchern1}
\end{equation}
where $\Omega\in\mathcal{C}_{\mathsans{M}}(M)$ is a higher gauge field (cf. subsect. \cref{subsec:higau})
and $k$ is a constant, the CS level. 
Formally, the expression of the higher CS action put forth here is identical to that of familiar CS
theory with the pairing $(\cdot,\cdot)$ in place of the usual Lie algebraic trace. However, since the latter
has degree $1$ rather than $0$, the Lagrangian has degree $4$ instead than $3$.
It is precisely for this reason that the present higher CS theory works in $4$ dimensions. 

Expressed through the components $\omega$, $\varOmega$ of the higher gauge field $\Omega$, the
$4$--dimensional CS action $\sfC\hspace{-.75pt}\sfS$ reads as 
\begin{multline}
\sfC\hspace{-.75pt}\sfS(\omega,\varOmega)=\frac{k}{2\pi}\int_{T[1]M}\varrho_M
\left\langle d\omega+\tfrac{1}{2}[\omega,\omega]-\tfrac{1}{2}\dot\tau(\varOmega),\varOmega\right\rangle
\\
-\frac{k}{4\pi}\int_{T[1]\partial M}\varrho_{\partial M} \left\langle\omega,\varOmega\right\rangle.
\label{4dchern2}
\end{multline}
The boundary contribution,
absent in \ceqref{4dchern1}, is yielded by an integration by parts. 
So, higher CS theory can be described as as a generalized BF theory with boundary term
and cosmological term determined by the Lie differential $\dot\tau$ of the target map $\tau$. 
This way of regarding it is however somewhat reductive. $4$--dimensional CS theory 
is characterized by a higher gauge symmetry which places it safely in the realm of higher gauge theory.
We shall analyze this matter in greater detail in subsect. \cref{subsec:levelchern} below. 

The variation of the 4--dimensional CS action $\sfC\hspace{-.75pt}\sfS$
under a variation of the higher gauge field $\Omega$ is given by 
\begin{equation}
\sfdelta\sfC\hspace{-.75pt}\sfS 
=\frac{k}{2\pi}\int_{T[1]M}\varrho_M\!\left(\sfdelta\Omega,\Phi\right)
+\frac{k}{4\pi}\int_{T[1]\partial M}\varrho_{\partial M}\!\left(\sfdelta\Omega,\Omega\right),
\label{holchern0}
\end{equation}
where $\Phi$ is the higher gauge curvature defined in eq. \ceqref{higau2}.
(Here and below, the variational operator $\sfdelta$ is defined as in 
subsect. \cref{subsec:funcal}.) If a suitable boundary condition is imposed on $\Omega$
which makes the boundary term in \ceqref{holchern0} vanish, 
rendering $\sfC\hspace{-.75pt}\sfS$
differentiable in the sense of refs. \ccite{Regge:1974zd,Benguria:1976in}, 
the field equations read 
\begin{equation}
\Phi=0. 
\label {holchern0/1}
\end{equation}
These can be written in terms of the components $\omega$, $\varOmega$ using relations \ceqref{higau4}, \ceqref{higau5}.
In this way, as in ordinary CS theory, the higher CS field equations enforce the flatness of $\Omega$.

If no boundary condition is imposed, the 4--dimensional 
CS action $\sfC\hspace{-.75pt}\sfS$ belongs 
to the distributional extension $\matheul{F}'{}_{\mathsans{M}\mathrm{loc}}(M)$
of the local smooth functional space $\matheul{F}{}_{\mathsans{M}\mathrm{loc}}(M)$,
as the variation $\sfdelta\sfC\hspace{-.75pt}\sfS$ of $\sfC\hspace{-.75pt}\sfS$ 
given in \ceqref{holchern0} contains a boundary term 
\begin{equation}
\sfGamma=-\frac{k}{4\pi}\int_{T[1]\partial M}\varrho_{\partial M}
\left(\sfdelta\Omega,\Omega\right), 
\label{holchern-7}
\end{equation}
that cannot be turned into a legitimate bulk one using Stokes' theorem (cf. subsect. \cref{subsec:funcal}). 
Imposing an appropriate boundary condition eliminates the offending boundary contribution 
and makes $\sfC\hspace{-.75pt}\sfS$ belong to $\matheul{F}{}_{\mathsans{M}\mathrm{loc}}(M)$
with a well defined variational problem leading to the field equations \ceqref{holchern0/1}. 

The choice of the appropriate boundary condition to be prescribed to the higher gauge field $\Omega$
depends on the type of physics the higher CS theory is meant to describe. 
To analyze this matter in full generality within the scope of local field theory, we proceed as follows.

First, to have available the broadest possible range of boundary conditions, we allow for
the addition to the action $\sfC\hspace{-.75pt}\sfS$ of a 
local boundary term $\varDelta\sfC\hspace{-.75pt}\sfS$ independent from any boundary background field.
The resulting modified CS action is then 
\begin{equation}
\sfC\hspace{-.75pt}\sfS'=\sfC\hspace{-.75pt}\sfS+\varDelta\sfC\hspace{-.75pt}\sfS. 
\label{holchern-4}
\end{equation}
The inclusion of $\varDelta\sfC\hspace{-.75pt}\sfS$
provides the variation $\sfdelta\sfC\hspace{-.75pt}\sfS'$
of $\sfC\hspace{-.75pt}\sfS'$ with a boundary contribution that is added to
the problematic boundary contribution $\sfGamma$ yielded by $\sfdelta\sfC\hspace{-.75pt}\sfS$.
Note that the two boundary contributions cannot cancel out since the former is $\sfdelta$-exact
in $\matheul{O}'^*{}_{\mathsans{M}\mathrm{loc}}(M)$ while the latter is not. 


Second, we impose a local boundary condition on the higher gauge field $\Omega$. 
The most general such condition is specified by a local functional submanifold
$\mathcal{L}$ of the boundary higher gauge field space $\mathcal{C}_{\mathsans{M}}(\partial M)$, 
that is one defined by means of local constraints in $\mathcal{C}_{\mathsans{M}}(\partial M)$, and takes the form 
\begin{equation}
\Omega|_{T[1]\partial M}\in\mathcal{L}.
\label{holchern-8}
\end{equation}
The boundary condition must by such to completely cancel the boundary contribution to the variation
$\sfdelta\sfC\hspace{-.75pt}\sfS'$ of $\sfC\hspace{-.75pt}\sfS'$.

The above two step procedure ensures that $\sfC\hspace{-.75pt}\sfS'$
does indeed belong to local smooth field functional space
$\matheul{F}{}_{\mathsans{M}\mathrm{loc}}(M)$, making the associated
variational problem well defined, as we now show.
The assumed qualifications of the boundary term $\varDelta\sfC\hspace{-.75pt}\sfS$ guarantee 
the existence of a boundary local smooth functional $\varDelta\sfC\hspace{-.75pt}\sfS_\partial
\in\matheul{F}{}_{\mathsans{M}\mathrm{loc}}(\partial M)$ independent from any boundary background field
such that $\varDelta\sfC\hspace{-.75pt}\sfS(\Omega)
=\varDelta\sfC\hspace{-.75pt}\sfS_\partial(\Omega_\partial)|_{\Omega_\partial=\Omega|_{T[1]\partial M}}$.
(Here and below we denote boundary fields and field functionals thereof by a subscript
$\partial$ for 
clarity.)  The boundary contribution to the variation
$\sfdelta\sfC\hspace{-.75pt}\sfS$ of $\sfC\hspace{-.75pt}\sfS$ in \ceqref{holchern0}, 
$\sfGamma$, is similarly related to 
the boundary local functional $1$--form $\sfGamma_\partial\in\matheul{O}^1{}_{\mathsans{M}\mathrm{loc}}(\partial M)$,
\begin{equation}
\sfGamma_\partial=-\frac{k}{4\pi}\int_{T[1]\partial M}\varrho_{\partial M}
\left(\sfdelta_\partial\Omega_\partial,\Omega_\partial\right), 
\label{holchern-1}
\end{equation}
as $\sfGamma(\Omega)=\sfGamma_\partial(\Omega_\partial)|_{\Omega_\partial=\Omega|_{T[1]\partial M}}$.
Hence, the total boundary contribution to $\sfdelta\sfC\hspace{-.75pt}\sfS'$ is going to vanish
if $\varDelta\sfC\hspace{-.75pt}\sfS$ and $\mathcal{L}$ are such that 
\begin{equation}
\sfGamma_\partial-\sfdelta_\partial\varDelta\sfC\hspace{-.75pt}\sfS_\partial=0 \qquad \text{in $\mathcal{L}$}.
\label{holchern-5}
\end{equation}
This can be achieved by suitably adjusting either $\varDelta\sfC\hspace{-.75pt}\sfS_\partial$ or $\mathcal{L}$
or both. 

A functional submanifold $\mathcal{L}$ of $\mathcal{C}_{\mathsans{M}}(\partial M)$
will be called admissible if it can be employed to define a viable boundary condition. 
The boundary term $\varDelta\sfC\hspace{-.75pt}\sfS$ as well as the associated modified action
$\sfC\hspace{-.75pt}\sfS'$ are fixed once a choice of one such submanifold $\mathcal{L}$
is made. We shall denote them as $\varDelta\sfC\hspace{-.75pt}\sfS_{\mathcal{L}}$ and
$\sfC\hspace{-.75pt}\sfS'{}_{\mathcal{L}}$ when it is necessary to indicate such dependence. 
The problem of classifying the possible choices of boundary conditions
reduces in this way to that of classifying the admissible submanifolds. 

In the spirit of the covariant canonical approach (see e. g. ref. \ccite{Henneaux:1992ig} for a standard 
review), the boundary functional 1--form $\sfGamma_\partial$ given in eq.
\ceqref{holchern-1} provides the expression of the appropriate
symplectic potential of the higher CS model field space $\mathcal{C}_{\mathsans{M}}(\partial M)$. 
The associated symplectic form $\sfUpsilon_\partial\in\matheul{O}^2{}_{\mathsans{M}\mathrm{loc}}(\partial M)$ thus is 
\begin{equation}
\sfUpsilon_\partial=\sfdelta_\partial\sfGamma_\partial
=\frac{k}{4\pi}\int_{T[1]\partial M}\varrho_{\partial M}
\left(\sfdelta_\partial\Omega_\partial,\sfdelta_\partial\Omega_\partial\right).
\label{holchern-2}
\end{equation}
The admissible submanifolds $\mathcal{L}$ of $\mathcal{C}_{\mathsans{M}}(\partial M)$ 
which describe the higher CS theory boundary conditions then constitute a distinguish subset
of isotropic submanifolds to $\sfUpsilon$, that is the submanifolds $\mathcal{L}$ such that 
\begin{equation}
\sfUpsilon_\partial=0 \qquad \text{in $\mathcal{L}$}
\label{holchern-6}
\end{equation}

In terms of the gauge field components $\omega$, $\varOmega$ the symplectic form reads
\begin{equation}
\sfUpsilon_\partial=\frac{k}{2\pi}\int_{T[1]\partial M}\varrho_{\partial M}
\left\langle\sfdelta_\partial\omega_\partial,\sfdelta_\partial\varOmega_\partial\right\rangle. 
\label{holchern-3}
\end{equation}
The local boundary condition classification problem is therefore similar enough to that
of the description of isotropic submanifolds in ordinary Hamiltonian mechanics. 
We shall not tackle this issue in full generality and for the time being content ourselves
with a basic class of such conditions. 

With any isotropic submodule $\mathsans{M}'$ of the Lie group crossed module $\mathsans{M}$, 
(cf. subsect. \cref{subsec:crosumo}), there is associated the submanifold
$\mathcal{C}_{\mathsans{M}'}(\partial M)$ of $\mathcal{C}_{\mathsans{M}}(\partial M)$. 
By virtue of the isotropy $\mathsans{M}'$, 
$\mathcal{C}_{\mathsans{M}'}(\partial M)$ is a local submanifold of $\mathcal{C}_{\mathsans{M}}(\partial M)$
such that $\sfGamma_\partial=0$ on $\mathcal{C}_{\mathsans{M}'}(\partial M)$, hence an
admissible submanifold of $\mathcal{C}_{\mathsans{M}}(\partial M)$
with $\varDelta\sfC\hspace{-.75pt}\sfS_{\mathcal{C}_{\mathsans{M}'}(\partial M)}=0$
and $\sfC\hspace{-.75pt}\sfS_{\mathcal{C}_{\mathsans{M}'}(\partial M)}=\sfC\hspace{-.75pt}\sfS$. 
$\mathcal{C}_{\mathsans{M}'}(\partial M)$ thus defines
a special choice of boundary condition for 4--dimensional CS theory. In the following we shall refer mostly
to this kind of boundary condition, which we shall call isotropic linear of type $\mathsans{M}'$ for reference. 
The most permissive isotropic linear boundary condition is that for which $\mathsans{M}'$ is Lagrangian.
This will be called  Lagrangian linear of type $\mathsans{M}'$.

\subsection{\textcolor{blue}{\sffamily Gauge invariance of the 4--d Chern--Simons model
}}\label{subsec:levelchern}

In this subsection, we analyze in some detail the gauge symmetry of the 4--dimensional CS model introduced in
subsect. \cref{subsec:4dchern}. In spite of the formal resemblance of higher to ordinary CS theory when the derived
formulation is used, the invariance properties of the 4--dimensional CS model differ in several important aspects
from those of the 3--dimensional one, especially in relation to the effect of a boundary in the base manifold.

For a higher gauge transformation $\mathrm{U}\in\mathcal{G}_{\mathsans{M}}(M)$ (cf. subsect. \cref{subsec:higau}), 
the 4--dimensional CS action \ceqref{4dchern1} varies as 
\begin{equation}
\sfC\hspace{-.75pt}\sfS(\Omega^{\mathrm{U}})=\sfC\hspace{-.75pt}\sfS(\Omega)+\sfA(\Omega;\mathrm{U})
\label{4dchern3}
\end{equation}
for $\Omega\in\mathcal{C}_{\mathsans{M}}(M)$,  where $\sfA(\Omega;\mathrm{U})$ is given by 
\begin{multline}
\sfA(\Omega;\mathrm{U})=
-\frac{k}{24\pi}\int_{T[1]M}\varrho_M\!\left(\mathrm{dUU}^{-1},[\mathrm{dUU}^{-1},\mathrm{dUU}^{-1}]\right)
\\
+\frac{k}{4\pi}\int_{T[1]\partial M}\varrho_{\partial M}\!\left(\Omega,\mathrm{dUU}^{-1}\right)
\label{4dchern4}
\end{multline}
The gauge variation term $\sfA(\Omega;\mathrm{U})$ is formally identical to that of ordinary CS
theory: a bulk WZNW--like term plus a boundary term.

The real nature of the gauge variation term \ceqref{4dchern4} emerges when it is expressed
through the components $\omega$, $\varOmega$ of the higher gauge field $\Omega$ 
and $u$, $U$ of the higher gauge transformation $\mathrm{U}$. Using \ceqref{superfield13}, it can be verified that
the bulk term is exact and hence reduces to a boundary term, yielding the expression \vspace{-1.5mm}
\begin{multline}
\sfA(\omega,\varOmega;u,U)
\\
=\frac{k}{4\pi}\int_{T[1]\partial M}\varrho_{\partial M}\Big[
\left\langle\dot\tau(U), dU+\tfrac{1}{3}\left[U,U\right]\right\rangle 
-\left\langle duu^{-1}+\dot\tau(U),dU+\tfrac{1}{2}[U,U]\right\rangle
\\
+\left\langle\omega,dU+\tfrac{1}{2}[U,U]-\sdot\mu\sdot(duu^{-1}+\dot\tau(U),U)\right\rangle
-\left\langle duu^{-1}+\dot\tau(U),\varOmega\right\rangle\Big].
\label{4dchern5}
\end{multline}
In particular, $\sfA=0$ identically if $\partial M=\slash\hspace{-5pt}0$.
In this sense, in the higher theory, the gauge non invariance of the CS action
is 'holographic' in nature. 
This property distinguishes 4--dimensional CS theory from its 3--dimensional counterpart. 

When a boundary term $\varDelta\sfC\hspace{-.75pt}\sfS$ is added to the basic higher CS action $\sfC\hspace{-.75pt}\sfS$,
a modified action $\sfC\hspace{-.75pt}\sfS'$ is obtained (cf. eq. \ceqref{holchern-4}). $\varDelta\sfC\hspace{-.75pt}\sfS$ is
generally non invariant under gauge transformation. One hence has 
\begin{equation}
\varDelta\sfC\hspace{-.75pt}\sfS(\Omega^{\mathrm{U}})
=\varDelta\sfC\hspace{-.75pt}\sfS(\Omega)
+\varDelta\sfA(\Omega,\mathrm{U})
\label{4dchern-4}  
\end{equation}
where $\varDelta\sfA$ is a boundary gauge variation. 
The modified gauge variation $\sfA'$, the variation of the modified action $\sfC\hspace{-.75pt}\sfS'$,
is therefore given by 
\begin{equation}
\sfA'=\sfA+\varDelta\sfA. 
\label{4dchern-5}  
\end{equation}
Depending on the form of $\varDelta\sfC\hspace{-.75pt}\sfS$, $\sfA'{}$ may differ considerably from $\sfA$.

The gauge invariance properties of 4--dimensional CS theory depend to a large 
extent on the kind of boundary condition one imposes on the higher gauge fields
$\Omega$ to render the CS variational problem well--defined.

As we have explained in subsect. \cref{subsec:4dchern}, in 4--dimensional CS theory a choice of boundary condition 
is specified by an admissible submanifold $\mathcal{L}$ of $\mathcal{C}_{\mathsans{M}}(\partial M)$. The boundary
condition requires that $\Omega$ 
satisfies $\Omega|_{T[1]\partial M}\in\mathcal{L}$. With the boundary condition, further, there is associated
a boundary term $\varDelta\sfC\hspace{-.75pt}\sfS_{\mathcal{L}}$ that is to be added to the basic CS action 
$\sfC\hspace{-.75pt}\sfS$ yielding the appropriate variationally
well--behaved modified CS action $\sfC\hspace{-.75pt}\sfS'{}_{\mathcal{L}}$
(cf. eq. \ceqref{holchern-4}).

When a certain boundary condition is prescribed for the higher gauge fields $\Omega$ 
a corresponding boundary condition must be imposed to the higher gauge transformations $\mathrm{U}$:
they must preserve $\mathcal{L}$. 
The boundary condition can therefore be expressed as the 
requirement that 
\begin{equation}
\mathrm{U}|_{T[1]\partial M}\in\mathcal{I}_{\mathcal{L}}
\label{4dchern-0}  
\end{equation}
where $\mathcal{I}_{\mathcal{L}}$ 
is the invariance subgroup of $\mathcal{L}$
in $\mathcal{G}_{\mathsans{M}}(\partial M)$, the subgroup formed by the boundary gauge transformations
$\mathrm{U}_\partial\in\mathcal{G}_{\mathsans{M}}(\partial M)$ such that
$\mathcal{L}^{\mathrm{U}_\partial}=\mathcal{L}$. 

Since the modified action $\sfC\hspace{-.75pt}\sfS'{}_{\mathcal{L}}$ results from adding 
the boundary term $\varDelta\sfC\hspace{-.75pt}\sfS_{\mathcal{L}}$ to the basic
CS action $\sfC\hspace{-.75pt}\sfS$ according to
\ceqref{holchern-4}, the boundary gauge variation $\varDelta\sfA_{\mathcal{L}}$ is added to
the basic gauge variation term $\sfA$ to yield the modified gauge variation term $\sfA'{}_{\mathcal{L}}$ given by
\ceqref{4dchern-5}. As we shall see momentarily, the expression $\sfA'{}_{\mathcal{L}}$ may take a simpler form when
the boundary conditions obeyed by both the gauge fields and transformations
are taken into account. 

For the isotropic linear boundary condition of type $\mathsans{M}'$ (cf. subsect. \cref{subsec:4dchern}),
where $\mathsans{M}'$ is an isotropic crossed submodule of $\mathsans{M}$, more detailed information can be
provided. In this case,$\vphantom{\ul{\ul{\ul{g}}}}$
\pagebreak the condition is specified by the admissible submanifold $\mathcal{C}_{\mathsans{M}'}(\partial M)$
of $\mathcal{C}_{\mathsans{M}}(\partial M)$. The precise content of the invariance subgroup 
$\mathcal{I}_{\mathcal{C}_{\mathsans{M}'}(\partial M)}$ of $\mathcal{C}_{\mathsans{M}'}(\partial M)$
is not straightforward to describe in simple terms, but it is not difficult to
identify a broad distinguished subgroup $\mathcal{I}^{\NN}{}_{\mathcal{C}_{\mathsans{M}'}(\partial M)}$
of $\mathcal{I}_{\mathcal{C}_{\mathsans{M}'}(\partial M)}$. $\mathcal{I}^{\NN}{}_{\mathcal{C}_{\mathsans{M}'}(\partial M)}$
consists of the  boundary gauge transformations  
$\mathrm{U}_\partial\in\mathcal{G}_{\NN\mathsans{M}'}(\partial M)$ satisfying 
\begin{align}
&d_\partial u_\partial u_\partial{}^{-1}+\dot\tau(U_\partial)=0 \qquad \text{mod $\Map(T[1]\partial M,\mathfrak{g}'[1])$},
\vphantom{\Big]}
\label{4dchern+1}
\\
&d_\partial U_\partial+\tfrac{1}{2}[U_\partial,U_\partial]=0 \qquad \text{mod $\Map(T[1]\partial M,\mathfrak{e}'[1])$},
\vphantom{\Big]}
\label{4dchern+2}
\end{align}
where $\NN\mathsans{M}'$ is the normalizer crossed module of $\mathsans{M}'$ (cf. subsect. \cref{subsec:crosumo}).
$\mathcal{I}^{\NN}{}_{\mathcal{C}_{\mathsans{M}'}(\partial M)}$ being contained in $\mathcal{I}_{\mathcal{C}_{\mathsans{M}'}(\partial M)}$
follows from \ceqref{higau12}, \ceqref{higau13} and 
the defining properties of $\mathcal{G}_{\NN\mathsans{M}'}(\partial M)$. 
It can be further shown that $\mathcal{I}^{\NN}{}_{\mathcal{C}_{\mathsans{M}'}(\partial M)}$ contains
$\mathcal{G}_{\mathsans{M}'}(\partial M)$ as a subgroup and that 
$\mathcal{I}^{\NN}{}_{\mathcal{C}_{\mathsans{M}'}(\partial M)}=\mathcal{I}_{\mathcal{C}_{\mathsans{M}'}(\partial M)}$
if the groups $\mathsans{G}'$, $\mathsans{E}'$ are connected.

Since the boundary term $\varDelta\sfC\hspace{-.75pt}\sfS_{\mathcal{C}_{\mathsans{M}'}(\partial M)}=0$ identically
for the isotropic linear boundary condition, the modified action
$\sfC\hspace{-.75pt}\sfS'{}_{\mathcal{C}_{\mathsans{M}'}(\partial M)}$ and the associated gauge variation term
$\sfA'{}_{\mathcal{C}_{\mathsans{M}'}(\partial M)}$ are equal to their basic counterparts $\sfC\hspace{-.75pt}\sfS$ and
$\sfA$, respectively. If $\mathrm{U}$ is a higher gauge transformation obeying the boundary condition 
$\mathrm{U}|_{T[1]\partial M}\in\mathcal{I}^{\NN}{}_{\mathcal{C}_{\mathsans{M}'}(\partial M)}$, the gauge variation
\ceqref{4dchern5} takes the CS form 
\begin{equation}
\sfA(\omega,\varOmega;u,U)
=\frac{k}{4\pi}\int_{T[1]\partial M}\varrho_{\partial M}
\left\langle\dot\tau(U), dU+\tfrac{1}{3}\left[U,U\right]\right\rangle,
\label{4dchern+3}
\end{equation}
by the isotropy of $\mathsans{M}'$. $\sfA(\omega,\varOmega;u,U)$ is so independent from
the higher gauge field components $\omega$, $\varOmega$. 
Note that $\sfA(\omega,\varOmega;u,U)=0$ when $\mathrm{U}|_{T[1]\partial M}\in\mathcal{G}_{\mathsans{M}'}(\partial M)$. 



\subsection{\textcolor{blue}{\sffamily Level quantization }}\label{subsec:levelquant}

To quantize 4--dimensional CS theory, one should allow for the widest gauge symmetry
leaving the Boltzmann exponential $\exp(i\hfpt\sfC\hspace{-.75pt}\sfS)$ invariant possibly
restricting the value of the CS level $k$. 
In ordinary CS theory, this permits the incorporation of large gauge transformation
in the symmetry, when the CS level is suitably quantized.  One wonders if something similar 
happens in our higher setting.

By \ceqref{4dchern5}, when the boundary $\partial M$ of $M$ is empty, the higher CS theory enjoys full higher
gauge symmetry and there is no problem. When $\partial M$ is non empty, one should impose
on the relevant higher gauge fields $\Omega$ and transformations $\mathrm{U}$ the weakest possible boundary
conditions capable to render the gauge variation term $\sfA$ an integer multiple of $2\pi$.
Given the varied form such conditions can take, here we can only examine basic examples. 


If an isotropic linear boundary condition of type $\mathsans{M}'$ is implemented,
where $\mathsans{M}'$ is an isotropic crossed submodule of $\mathsans{M}$, 
the higher gauge fields $\Omega\in\mathcal{C}_{\mathsans{M}}(M)$ and gauge transformations
$\mathrm{U}\in\mathcal{G}_{\mathsans{M}}(M)$ must satisfy
$\Omega|_{T[1]\partial M}\in\mathcal{C}_{\mathsans{M}'}(\partial M)$ and 
$\mathrm{U}|_{T[1]\partial M}\in\mathcal{I}_{\mathcal{C}_{\mathsans{M}'}(\partial M)}$
(cf. subsects. \cref{subsec:4dchern}, \cref{subsec:levelchern}). We identified a subgroup
$\mathcal{I}^{\NN}{}_{\mathcal{C}_{\mathsans{M}'}(\partial M)}$
of $\mathcal{I}_{\mathcal{C}_{\mathsans{M}'}(\partial M)}$ essentially exhausting it formed by 
the  boundary gauge transformations  
$\mathrm{U}_\partial\in\mathcal{G}_{\NN\mathsans{M}'}(\partial M)$ obeying
\ceqref{4dchern+1}, \ceqref{4dchern+2}. For the gauge transformations $\mathrm{U}$ such that 
$\mathrm{U}|_{T[1]\partial M}\in\mathcal{I}^{\NN}{}_{\mathcal{C}_{\mathsans{M}'}(\partial M)}$, the gauge variations
$\sfA$ has the simple CS form \ceqref{4dchern+3}. This however neither vanishes nor enjoys any quantization
property a priori. We are thus forced to consider a more restrictive boundary condition
for the $\mathrm{U}$. An option is
replacing the invariance subgroup $\mathcal{I}^{\NN}{}_{\mathcal{C}_{\mathsans{M}'}(\partial M)}$
by its orthogonal subgroup $\mathcal{I}^{\ON}{}_{\mathcal{C}_{\mathsans{M}'}(\partial M)}$,
where $\ON\mathsans{M}'$ is the orthogonal normalizer crossed module of $\mathsans{M}'$
(cf. subsect. \cref{subsec:crosumo}). $\mathcal{I}^{\ON}{}_{\mathcal{C}_{\mathsans{M}'}(\partial M)}$
is constituted by the  boundary gauge transformations  
$\mathrm{U}_\partial\in\mathcal{G}_{\ON\mathsans{M}'}(\partial M)$ satisfying
\ceqref{4dchern+1}, \ceqref{4dchern+2}. For the gauge transformations
$\mathrm{U}$ such that 
$\mathrm{U}|_{T[1]\partial M}\in\mathcal{I}^{\ON}{}_{\mathcal{C}_{\mathsans{M}'}(\partial M)}$, the gauge variation
$\sfA$ takes the form 
\begin{equation}
\sfA(\omega,\varOmega;u,U)
=-\frac{k}{24\pi}\int_{T[1]\partial M}\varrho_{\partial M}\left\langle\dot\tau(U),\left[U,U\right]\right\rangle.
\label{4dchern+-1}
\end{equation}
This can be roughly viewed as a kind of winding number of a Lie group valued map,
since by \ceqref{4dchern+1}
$\dot\tau(U)$ can be identified with $duu^{-1}$ on the boundary $T[1]\partial M$ up to
a term belonging to $\Map(T[1]\partial M,\mathfrak{g}'[1])$.

As explained in subsect. \cref{subsec:crmodinv}, under weak assumptions the Lie algebra
crossed module with invariant pairing $\mathfrak{m}$ is isomorphic to the direct sum
of the Lie algebra crossed module $\INN\ran\dot\tau$ with a suitable invariant pairing and
$\AD^*(\mathfrak{g}/\ran\dot\tau)$ with canonical invariant pairing. 
Having this in mind, we are going to find out which form the expression \ceqref{4dchern+-1}
of the gauge variation $\sfA$ takes in the cases where 
the Lie group crossed module $\mathsans{M}$ is either $\INN\mathsans{G}$
with a suitable invariant pairing or $\AD^*\mathsans{G}$ with the canonical duality pairing, where 
$\mathsans{G}$ is a Lie group,  
and $\mathsans{M}'$ is an isotropic crossed submodule of these for which the orthogonal
normalizer crossed module $\ON\mathsans{M}'$ exists,  such as the submodules
$\INN_{\mathsans{H}}\mathsans{K}$ or $\AD_{\mathsans{H}}{}^*\mathsans{K}$ studied in subsect.
\cref{subsec:crosumo} with $\mathsans{H}$, $\mathsans{K}$ suitable connected Lie
subgroups of $\mathsans{G}$. 

We consider first the case where $\mathsans{M}=\INN\mathsans{G}$ and $\mathsans{M}'=\INN_{\mathsans{H}}\mathsans{K}$. 
We have then $\mathsans{E}=\mathsans{G}$ and $\dot\tau(X)=X$ for $X\in\mathfrak{g}$.
The less  restrictive boundary conditions on the gauge
transformation $\mathrm{U}$ 
are those for which the
gauge variation $\sfA$ is an integer times $2\pi$. The weakest conditions one can envisage are as follows.
The boundary gauge transformations $\mathrm{U}_\partial\in
\mathcal{I}^{\ON}{}_{\mathcal{C}_{\INN_{\mathsans{H}}\mathsans{K}}(\partial M)}$
such that$\vphantom{\ul{\ul{\ul{a}}}}$
there is a boundary gauge transformation $\mathrm{V}_\partial\in
\mathcal{G}_{\INN_{\mathsans{H}}\mathsans{K}}(\partial M)$ obeying 
\begin{equation}
d_\partial v_\partial v_\partial{}^{-1}+V_\partial=-\Ad u_\partial{}^{-1}(d_\partial u_\partial u_\partial{}^{-1}+U_\partial)
\label{4dchern+-4}
\end{equation}
form a distinguished subgroup $\mathcal{I}^{\ON*}{}_{\mathcal{C}_{\INN_{\mathsans{H}}\mathsans{K}}(\partial M)}$
of $\mathcal{I}^{\ON}{}_{\mathcal{C}_{\INN_{\mathsans{H}}\mathsans{K}}(\partial M)}$. For a transformation
$\mathrm{U}_\partial\in
\mathcal{I}^{\ON}{}_{\mathcal{C}_{\INN_{\mathsans{H}}\mathsans{K}}(\partial M)}$, one has
\begin{multline}
-\frac{1}{24\pi}\int_{T[1]\partial M}\varrho_{\partial M}\left\langle U_\partial,[U_\partial,U_\partial]\right\rangle
\\=\int_{T[1]\partial M}\varrho_{\partial M}\frac{1}{24\pi}\left\langle d_\partial \tilde u_\partial \tilde u_\partial{}^{-1},
[d_\partial \tilde u_\partial \tilde u_\partial{}^{-1},d_\partial \tilde u_\partial \tilde u_\partial{}^{-1}]\right\rangle
\vphantom{\Big]}
\label{4dchern+-5}
\end{multline}
where $\tilde u_\partial=u_\partial v_\partial$. If the closed form 
$\frac{1}{48\pi^2}\left\langle \kappa,[\kappa,\kappa]\right\rangle$ of $\mathsans{G}$,
where $\kappa$ is the Maurer--Cartan form, 
is a representative of an integer cohomology class, the above expression
takes integer values time $2\pi$. Let us assume this is indeed the case.
By virtue of \ceqref{4dchern+-1}, if $\mathrm{U}$ is gauge transformation obeying the boundary condition 
$\mathrm{U}|_{T[1]\partial M}\in\mathcal{I}^{\ON*}{}{}_{\mathcal{C}_{\INN_{\mathsans{H}}\mathsans{K}}(\partial M)}$, 
$\sfA$ takes the form 
\begin{equation}
\sfA
=\frac{k}{24\pi}\int_{T[1]\partial M}\varrho_{\partial M}
\left\langle d\tilde u\tilde u^{-1},\left[d\tilde u\tilde u^{-1},d\tilde u\tilde u^{-1}\right]\right\rangle. 
\label{4dchern+-2}
\end{equation}
where $\tilde u=uv$, $v$ being an extension to a neighborhood of $\partial M$
of the component $v_\partial$ a boundary gauge transformation $\mathrm{V}_\partial\in
\mathcal{G}_{\INN_{\mathsans{H}}\mathsans{K}}(\partial M)$ satisfying condition \ceqref{4dchern+-4}
with $\mathrm{U}_\partial=\mathrm{U}|_{T[1]\partial M}$. If the level $k$ is an integer, 
$\sfA$ is integer values time $2\pi$ as desired, much as in ordinary CS theory. Level quantization so occurs.





We consider next the case where $\mathsans{M}=\AD^*\mathsans{G}$. We have then $\mathsans{E}=\mathfrak{g}^*$,
viewed as an Abelian group and $\dot\tau(X)=0$ for $X\in\mathfrak{g}^*$.
By \ceqref{4dchern+-1}, $\sfA$ then vanishes, 
\begin{equation}
\sfA
=0.
\label{4dchern+-3}
\end{equation}
In this case, level quantization of course does not occur.

The isotropic linear boundary conditions considered above
serve the purpose of rendering the CS variational problem well--defined and gauge covariant.
By virtue of their origin, 
they suit the perturbative semiclassical limit $k\to\infty$ in which $k$ can be considered
as a continuous parameter regardless its integrality. Below, we envisage
other types of boundary conditions$\vphantom{\ul{\ul{g}}}$
are appropriate for the opposite non perturbative quantum finite
$k$ regime.

The boundary condition we shall study is best expressed in components. 
We require that the higher gauge fields $\Omega\in\mathcal{C}_{\mathsans{M}}(M)$ to be fake flat on the boundary  
\begin{equation}
\phi=d\omega+\tfrac{1}{2}[\omega,\omega]-\dot\tau(\varOmega)=0 \qquad \text{on $T[1]\partial M$}
\label{4dchern11}
\end{equation}
(cf. subsect. \ceqref{subsec:higau}). 
We require further that the allowed gauge transformations $\mathrm{U}\in\mathcal{G}_{\mathsans{M}}(M)$
are special on the boundary, that is
of the form
\begin{align}
&u=\tau(B), 
\vphantom{\Big]}
\label{4dchern6}
\\
&U=-dBB^{-1}-\sdot\mu(\omega,B) \qquad \text{on $T[1]\partial M$}, 
\vphantom{\Big]}
\vspace{-1cm}
\label{4dchern7}
\end{align}
where $B\in\Map(T[1]\partial M,\mathsans{E})$ (cf. eqs. \ceqref{higau16}, \ceqref{higau17}).
Note that the $\Omega$ are not fake flat in general, as the fake flatness condition
\ceqref{4dchern11} is required to hold only on the boundary $\partial M$ of $M$. Likewise, the  
$\mathrm{U}$ are not special in general, as they are required to be of the form \ceqref{higau16}, \ceqref{higau17}
only on $\partial M$. Finally notice that by 
\ceqref{higau18}, \ceqref{higau19}
we have $\omega^{u,U}=\omega$, $\varOmega^{u,U}=\varOmega$ on $T[1]\partial M$. These gauge transformations 
so leave the boundary values of the gauge field components fixed.

The boundary fake flatness condition \ceqref{4dchern11} can be enforced by adding to CS action $\sfC\hspace{-.75pt}\sfS$
a boundary term of the form 
\begin{equation}
\varDelta\sfC\hspace{-.75pt}\sfS(\Omega,\varLambda)
=\frac{k}{4\pi}\int_{T[1]\partial M}\varrho_{\partial M}\left\langle\phi,\varLambda\right\rangle,
\label{4dchern11/1}
\end{equation}
where $\varLambda\in\Map(T[1]\partial M,\mathfrak{e}[1])$ is an auxiliary boundary field.
By \ceqref{crmodinv14/rep}, \ceqref{higau14} 
and \ceqref{4dchern6}, $\varDelta\sfC\hspace{-.75pt}\sfS$ is invariant under any gauge transformation $\mathrm{U}$
with the boundary form \ceqref{4dchern6}, \ceqref{4dchern7} provided $\varLambda$ transforms as 
\begin{equation}
\varLambda^{\mathrm{U}}=\Ad B^{-1}(\varLambda). 
\label{4dchern11/2}
\end{equation}

When the relevant higher gauge \pagebreak field $\omega$, $\varOmega$ and the gauge transformation $u$, $U$ obey the boundary
conditions \ceqref{4dchern11} and \ceqref{4dchern6},\ceqref{4dchern7},
the gauge variation term \ceqref{4dchern5} takes the form 
\begin{equation}
\sfA 
=\frac{k}{24\pi}\int_{T[1]\partial M}\varrho_{\partial M}
\left\langle\dot\tau(dBB^{-1}),\left[dBB^{-1},dBB^{-1}\right]\right\rangle.
\label{4dchern8}
\end{equation}
\footnote{$\vphantom{\dot{\dot{\dot{g}}}}$ The full expression of $\sfA$
contains in the integrand a term $-6\left\langle\phi,\sdot\mu(\omega,B)\right\rangle$ which vanishes
by \ceqref{4dchern11} and a further term $\left\langle\left[\omega,\omega\right],\sdot\mu(\omega,B)\right\rangle
-\left\langle\omega,\sdot\mu\left(\left[\omega,\omega\right],B^{-1}\right)\right\rangle$ which vanish 
by \ceqref{crmodinv15} which in turn holds by the assumed fineness of the crossed module $\mathsans{M}$.}. 
This can again viewed as a kind of the winding number of a Lie group valued map.
$\sfA$ is a homotopy invariant, as one might expect. Indeed, under a variation $\delta B$ of $B$, the
variation of the integrand in the right hand side of \ceqref{4dchern8} is
$3d\left\langle\dot\tau(\delta BB^{-1}),[dBB^{-1},dBB^{-1}]\right\rangle$ entailing that $\delta\sfA=0$.


For reasons we explained earlier,  
we are going to obtain the form the expression \ceqref{4dchern8} 
of the gauge variation $\sfA$ takes when 
the Lie group crossed module $\mathsans{M}$ is either $\INN\mathsans{G}$
with a suitable invariant pairing or $\AD^*\mathsans{G}$ with the canonical duality pairing, where 
$\mathsans{G}$ is a Lie group. 

We consider first the case where $\mathsans{M}=\INN\mathsans{G}$ for which $\mathsans{E}=\mathsans{G}$ and 
$\dot\tau(X)$ $=X$ $X\in\mathfrak{g}$.
By \ceqref{4dchern8}, $\sfA$ takes the form 
\begin{equation}
\sfA
=\frac{k}{24\pi}\int_{T[1]\partial M}\varrho_{\partial M}
\left\langle dBB^{-1},\left[dBB^{-1},dBB^{-1}\right]\right\rangle.
\label{4dchern9}
\end{equation}
Again, if the closed form 
$\frac{1}{48\pi^2}\left\langle \kappa,[\kappa,\kappa]\right\rangle$ of $\mathsans{G}$
with $\kappa$ the Maurer--Cartan form is a representative of an integer cohomology class,
as we assume presently, $\sfA$ takes integer values time $2\pi$ provided the level $k$ is integer.
Level quantization once more occurs. 



We consider next the case where $\mathsans{M}=\AD^*\mathsans{G}$. We have then $\mathsans{E}=\mathfrak{g}^*$,
viewed as an Abelian group and $\dot\tau(X)=0$ for $X\in\mathfrak{g}^*$.
By \ceqref{4dchern8}, $\sfA$ then vanishes
\begin{equation}
\sfA
=0.
\label{4dchern10}
\end{equation}
In this case, again level quantization does not occur.

The calculations carried out above show that level quantization, when it occurs, is a boundary effect. 
This remarkable property markedly distinguishes higher CS theory from its ordinary counterpart.


\subsection{\textcolor{blue}{\sffamily Global issues in 4--d Chern--Simons theory}}\label{subsec:chernglob}

In this subsection, we examine  the issue whether it is possible to give a reasonable definition of
4--dimensional CS theory on a non trivial higher principal bundle. We refer the reader to subsect.
\cref{subsec:2bund} for a preliminary discussion of this matter.

The components $\omega$, $\varOmega$ of the higher gauge field $\Omega$ are only locally defined
when the underlying higher principal bundle is non trivial. Consequently also the 4--dimensional CS Lagrangian density
is only locally defined and formula \ceqref{4dchern1} giving the CS action is unusable. 
This is only the first of a number of subtle points which must be settled before attempting
a definition of 4--dimensional CS theory on a non trivial background. We leave a more thorough analysis
of these issues for future work and here we shall limit ourselves to tackle the problem from a
different more elementary perspective. 

We look for an expression of the 4--dimensional CS action $\sfC\hspace{-.75pt}\sfS$ on a trivial
higher gauge principal bundle that can be sensibly extended also on a non trivial one.  
To this end, we try to adapt a strategy that has shown itself to be successful in the familiar 3--dimensional
case. We write the gauge field $\Omega$ as the sum of a background gauge field $\overline{\Omega}$
and a deviation $\mathrm{W}$, viz \hphantom{xxxxxxxxxxxx}
\begin{equation}
\Omega=\overline{\Omega}+\mathrm{W}. 
\label{chernglob1}
\end{equation}
We assume furthermore that $\overline{\Omega}$ is flat 
\begin{equation}
\overline{\Phi}=\dd\overline{\Omega}+\tfrac{1}{2}\big[\overline{\Omega},\overline{\Omega}\big]=0.
\label{chernglob2}
\end{equation}
This is not done only for mathematical convenience, but also because it allows for a more precise
characterization of the CS action $\sfC\hspace{-.75pt}\sfS(\overline{\Omega})$, as shown momentarily.
We also assume that $\overline{\Omega}$ obeys an isotropic linear boundary condition
(cf. subsect. \cref{subsec:4dchern}) and require that $\mathrm{W}$ also does. In this way, $\Omega$
will satisfy it too. 

The Lagrangian of 4--dimensional CS theory can now expressed as 
\begin{multline}
\left(\Omega,\dd\Omega+\tfrac{1}{3}[\Omega,\Omega]\right)
=\left(\overline{\Omega},\dd\overline{\Omega}+\tfrac{1}{3}[\overline{\Omega},\overline{\Omega}]\right)
\vphantom{\Big]}
\\
+\left(\mathrm{W},\overline{\mathrm{D}}\mathrm{W}+\tfrac{1}{3}[\mathrm{W},\mathrm{W}]\right)
-d\left(\overline{\Omega},\mathrm{W}\right),\vphantom{\Big]}
  \vphantom{\ul{\ul{\ul{\ul{g}}}}}
\label{chernglob3}
\end{multline}
where 
we have conventionally set
\begin{equation}
\overline{\mathrm{D}}=\dd+\ad\overline{\Omega}. 
\label{chernglob4}
\end{equation}
Upon integration on $T[1]M$, the last term in the right hand side of \ceqref{chernglob3}
gives a vanishing contribution because of the isotropic boundary conditions obeyed by both
$\overline{\Omega}$ and $\mathrm{W}$. From \ceqref{4dchern1}, we find so that 
\begin{equation}
\sfC\hspace{-.75pt}\sfS(\Omega)
=\sfC\hspace{-.75pt}\sfS(\overline{\Omega})
+\frac{k}{4\pi}\int_{T[1]M}\varrho_M\! 
\left(\mathrm{W},\overline{\mathrm{D}}\mathrm{W}+\tfrac{1}{3}[\mathrm{W},\mathrm{W}]\right).
\label{chernglob5}
\end{equation}
We now concentrate on the background CS action $\sfC\hspace{-.75pt}\sfS(\overline{\Omega})$. 

Denote by $\overline{\sfdelta}$ a variation with respect to the background gauge field $\overline{\Omega}$
respecting both the flatness requirement \ceqref{chernglob2}
and the given isotropic linear boundary condition. 
Then, since the flatness condition \ceqref{chernglob2} coincides with the CS field equation
(cf. subsect. \cref{subsec:4dchern}), we have  
\begin{equation}
\overline{\sfdelta}\sfC\hspace{-.75pt}\sfS(\overline{\Omega})=0. 
\label{chernglob6}
\end{equation}
$\sfC\hspace{-.75pt}\sfS(\overline{\Omega})$ is therefore constant on each connected component
of the space of flat background gauge fields $\overline{\Omega}$. If $M$ has no boundary, 
$\sfC\hspace{-.75pt}\sfS(\overline{\Omega})$ is also fully gauge invariant. In such a case,  
$\sfC\hspace{-.75pt}\sfS(\overline{\Omega})$ represent a locally constant function on the moduli space
of flat gauge fields $\overline{\Omega}$. If conversely $M$ has a boundary, then $\sfC\hspace{-.75pt}\sfS(\overline{\Omega})$, 
or more precisely $\ee^{i\sfC\hspace{-.75pt}\sfS(\overline{\Omega})}$, is a section of a flat
unitary line bundle on the moduli space whose matching data are defined by the exponentiated gauge variation
\ceqref{4dchern+3}. 

When $M$ has no boundary, $\sfC\hspace{-.75pt}\sfS(\overline{\Omega})$ can be evaluated by a method
borrowed once more from the 3--dimensional case. Suppose that the 4--fold $M$ is the boundary of a 5--fold
$\widetilde{M}$. 
We extend the background gauge field 
$\overline{\Omega}$ to a gauge field $\widetilde{\Omega}$ on $\widetilde{M}$ such that
$\widetilde{\Omega}|_{T[1]\overline{M}}=\overline{\Omega}$. Since
\begin{equation}
\widetilde{d}\big(\widetilde{\Omega},\widetilde{\dd}\widetilde{\Omega}
+\tfrac{1}{3}\big[\widetilde{\Omega},\widetilde{\Omega}\big]\big)
=\big(\widetilde{\Phi},\widetilde{\Phi}\big), 
\label{chernglob7}
\end{equation}
where $\widetilde{\Phi}$ is the curvature of $\widetilde{\Omega}$ defined according to \ceqref{higau2}, we have   
\begin{equation}
\sfC\hspace{-.75pt}\sfS(\overline{\Omega})=k\sfI_{\widetilde{M}}(\widetilde{\Omega}),
\label{chernglob8}
\end{equation}
where $\sfI_{\widetilde{M}}(\widetilde{\Omega})$ is given by 
\begin{equation}
\sfI_{\widetilde{M}}(\widetilde{\Omega})=\frac{1}{4\pi}\int_{T[1]\widetilde{M}}\varrho_{\widetilde{M}}
\big(\widetilde{\Phi},\widetilde{\Phi}\big).
\label{chernglob9}
\end{equation}
The value of $\sfI_{\widetilde{M}}(\widetilde{\Omega})$ does not depend on the choice of $\widetilde{\Omega}$ as
\begin{equation}
\widetilde{\sfdelta}\sfI_{\widetilde{M}}(\widetilde{\Omega})
=\frac{1}{2\pi}\int_{T[1]\partial\widetilde{M}}\varrho_{\widetilde{M}}\,
\big(\widetilde{\sfdelta}\hfpt\widetilde{\Omega},\widetilde{\Phi}\big)
=\frac{1}{2\pi}\int_{T[1]M}\varrho_M\,
\big(\overline{\sfdelta}\hfpt\overline{\Omega},\overline{\Phi}\big)=0
\label{chernglob10}
\end{equation}
by \ceqref{chernglob2}. The value of $\sfI_{\widetilde{M}}(\widetilde{\Omega})$ does not depend also
on the choice of $\widetilde{M}$ because $\big(\widetilde{\Phi},\widetilde{\Phi}\big)$ is exact
by \ceqref{chernglob7}. It is interesting to notice here that quadratic curvature polynomial
$\frac{1}{8\pi^2}\big(\widetilde{\Phi},\widetilde{\Phi}\big)$ is formally analogous the familiar
Chern 4--form. It has however degree 5. Expressed through the curvature components
$\widetilde{\phi}$, $\widetilde{\varPhi}$,
$\sfI_{\widetilde{M}}(\widetilde{\omega},\widetilde{\varOmega})$ reads in fact as \hphantom{xxxxxxxxx}
\begin{equation}
\sfI_{\widetilde{M}}(\widetilde{\omega},\widetilde{\varOmega})
=\frac{1}{2\pi}\int_{T[1]\widetilde{M}}\varrho_{\widetilde{M}}\,
\big\langle\widetilde{\phi},\widetilde{\varPhi}\big\rangle.
\label{chernglo11}
\end{equation}

Suppose now that the background principal bundle is non trivial. We pick again a background gauge field
specified, as explained in subsect. \ceqref{subsec:2bund}, by a collection
of local data $\overline{\Omega}_i$. The $\overline{\Omega}_i$ relate via matching data
$\overline{\mathrm{F}}_{ij}$ as in \ceqref{2bund3}. The $\overline{\mathrm{F}}_{ij}$ adapt in turn via 
the consistency data $\overline{\mathrm{H}}_{ijk}$ as in \ceqref{2bund6} with the $\overline{\mathrm{H}}_{ijk}$
obeying conditions \ceqref{2bund7} and \ceqref{2bund18}. Because of \ceqref{2bund3}, the integrand of the second
term in the right hand side of \ceqref{chernglob5}
will be globally defined if the local data $\mathrm{W}_i$ of the deviation field 
match according to
\begin{equation}
\mathrm{W}_i=\Ad\overline{\mathrm{F}}_{ij}(\mathrm{W}_j).
\label{chernglo12}
\end{equation}
Integration on $T[1]M$ is then possible. In this respect, the context is formally similar to that of the 3--dimensional
theory. If \ceqref{chernglo12} holds, the local data of the gauge field 
are $\Omega_i=\overline{\Omega}_i+\mathrm{W}_i$. The corresponding matching data $\mathrm{F}_{ij}$
and consistency data $\mathrm{H}_{ijk}$ so equal their background counterparts $\overline{\mathrm{F}}_{ij}$ and
$\overline{\mathrm{H}}_{ijk}$. However, the data $\mathrm{H}_{ijk}$ will obey 
\ceqref{2bund7} only if 
\begin{equation}
\Ad\overline{\mathrm{H}}_{ijk}(\mathrm{W}_i)=\mathrm{W}_i.
\label{chernglo13}
\end{equation}
This is a constraint \pagebreak on the deviation data $\mathrm{W}_i$ whose implementation in the classical as well
quantum theory is problematic. Alternatively, we can disregard \ceqref{chernglo13} giving up \ceqref{2bund7}, but then the
gauge field data $\Omega_i$ no longer can be considered as specifying a 2--connection.

Leaving aside these issues, when the background principal bundle is non trivial
other problems arise with regard to the proper definition of the background
CS action $\sfC\hspace{-.75pt}\sfS(\overline{\Omega})$
using the procedure outlined above valid for a base 4--fold $M$ with no boundary. To begin with, we have to
 extend the background bundle and 2--connection structure on $M$,
given by the data $\overline{\Omega}_i$, $\overline{\mathrm{F}}_{ij}$, $\overline{\mathrm{H}}_{ijk}$, 
to one on the chosen 5--fold $\widetilde{M}$, given by the data
$\widetilde{\Omega}_i$, $\widetilde{\mathrm{F}}_{ij}$, $\widetilde{\mathrm{H}}_{ijk}$. Assuming that this is indeed
possible, the extended higher curvature data $\widetilde{\Phi}_i$ match as
\begin{equation}
\widetilde{\Phi}_i=\Ad\widetilde{\mathrm{F}}_{ij}(\widetilde{\Phi}_j).
\label{chernglo14}
\end{equation}
By virtue of this, the integrand in the right hand side \ceqref{chernglob9} is globally defined
and its integration over $T[1]\widetilde{M}$ can be carried out. The problem arising here is that
the quadratic curvature polynomial $\frac{1}{8\pi^2}\big(\widetilde{\Phi},\widetilde{\Phi}\big)$ 
has no a priori integrality properties and so the value of
$\sfI_{\widetilde{M}}(\widetilde{\Omega})$ depends
in principle on the choice of the extending 5--fold $\widetilde{M}$ by an amount that does not vanish modulo
$2\pi\mathbb{Z}$.  The quantization of the level $k$ as integer is of no avail here in sharp contrast
with what happens in the corresponding 3--dimensional setting.

We conclude this subsection with one more remark pointing to a further problem. 
To allow for the well--definedness of higher holonomies, in turn necessary for the incorporation of Wilson surfaces
in 4--dimensional CS theory, the gauge field components $\omega_i$, $\varOmega_i$ are required to satisfy the vanishing
fake curvature condition \ceqref{2bund13}.  
The fake flatness condition is however one the field equations of the CS model.
So, it should emerge from the classical variational problem and should not be assumed from the onset.


\subsection{\textcolor{blue}{\sffamily Canonical formulation}}\label{subsec:holchern}

In this subsection, we shall illustrate the canonical analysis of the 4--dimensional CS model
introduced and studied in the previous subsections. 
The close relationship of the canonical formulations of 4-- and 3--dimensional CS theory is again
especially evident in the derived framework. 
We shall describe the phase space of the model in the derived set--up and obtain compact derived expressions of its Poisson bracket.
We shall further identify the model's phase space constraint manifold as the vanishing higher curvature locus 
and describe the reduced phase space and its Poisson bracket. 
The results of the ordinary theory generalize however to the higher one only up to a certain extent, which we shall make
precise in due course. 

To carry out the canonical analysis, we assume that $M=\mathbb{R}^1\times S$,
where $S$ is an oriented compact $3$--fold possibly with boundary, viewing the Cartesian factors
$\mathbb{R}^1$ and $S$ respectively as a time axis and a space manifold. 
$M$ of course is not compact, as we assumed earlier,
making it necessary imposing integrability conditions on fields to have a finite action integral. 
Alternatively, when $S$ is compact, one may compactify $\mathbb{R}^1$ into the circle $\mathbb{S}^1$ 
requiring fields to be periodic. 


In the canonical formulation, it is natural to rely on a hybrid geometrical  framework whereby
the function algebra $\Fun(T[1]M)$ of $M$ is viewed as the algebra $\Map(T[1]\mathbb{R}^1,\bfs{{\Fun}}(T[1]S))$
of maps from the shifted tangent bundle $T[1]\mathbb{R}^1$ of $\mathbb{R}^1$ 
into the internal function
algebra $\bfs{{\Fun}}(T[1]S)$ of $S$. Proceeding in this way,
a generic derived superfield 
field $\Psi\in\Map(T[1]M,\DD\mathfrak{m}[p]))$ decomposes as
\begin{equation}
\Psi=dt\Psi_t+\Psi_S,
\label{cnholchern1}
\end{equation}
where $\Psi_t\in\Map(\mathbb{R}^1,\Map(T[1]S,\DD\mathfrak{m}[p-1]))$,
$\Psi_S\in\Map(\mathbb{R}^1,\Map(T[1]S,\DD\mathfrak{m}[p]))$ and $t$ and $dt$ denote conventionally
the base and fiber coordinates of $T[1]\mathbb{R}^1$.
Similarly, the differential $\dd$ of $T[1]M$ decomposes as 
\begin{equation}
\dd=dtd_t+\dd_S
\label{cnholchern2}
\end{equation}
in terms of the differential $\dd_S$ of $T[1]S$, where $d_t=d/dt$ 
and both $\dd$ and $\dd_S$ are defined according to \ceqref{superfield10}. 

A higher gauge field $\Omega\in\mathcal{C}_{\mathsans{M}}(M)$ can so be expressed in terms of components
$\Omega_t\in\Map(\mathbb{R}^1,\Map(T[1]S,\DD\mathfrak{m}[0]))$,
$\Omega_S\in\Map(\mathbb{R}^1,\Map(T[1]S,\DD\mathfrak{m}[1]))$ in accordance 
with \ceqref{cnholchern1}. Its curvature $\Phi$ can be similarly decomposed in components 
$\Phi_t\in\Map(\mathbb{R}^1,\Map(T[1]S,\DD\mathfrak{m}[1]))$, 
$\Phi_S\in\Map(\mathbb{R}^1,\Map(T[1]S,\DD\mathfrak{m}[2]))$. \pagebreak 
Geome\-trically, $\Omega_S\in\Map(\mathbb{R}^1,\mathcal{C}_{\mathsans{M}}(S))$ is to be regarded as a time dependent
higher gauge field on $S$.   
$\Phi_S$ is then identified with the curvature of $\Omega_S$, since $\Phi_S$ is given by \ceqref{higau2}
in terms of $\dd_S$ and $\Omega_S$. 


Expressed in terms of the higher gauge field components $\Omega_t$, $\Omega_S$,
the 4--dimensional CS action \ceqref{4dchern1} takes the form 
\begin{multline}
\sfC\hspace{-.75pt}\sfS(\Omega_t,\Omega_S)
=\frac{k}{4\pi}\int_{\mathbb{R}^1}dt\int_{T[1]S}\left[(d_t\Omega_S,\Omega_S)+2(\Omega_t,\Phi_S)\right]
\\
+\frac{k}{4\pi}\int_{\mathbb{R}^1}dt\int_{T[1]\partial S}(\Omega_t,\Omega_S).
\label{cnholchern4}
\end{multline}
It is natural to interpret the component $\Omega_t$ as a Lagrange multiplier
implementing the vanishing curvature constraint \hphantom{xxxxxxx}
\begin{equation}
\Phi_S=0.
\label{cnholchern6}
\end{equation}
upon variation of the action $\sfC\hspace{-.75pt}\sfS$. However, $\sfC\hspace{-.75pt}\sfS$ is not differentiable with
respect to $\Omega_t$ in the sense established in refs. 
\ccite{Regge:1974zd,Benguria:1976in} because of the presence of the boundary term.

Naively, it would seem that the problem could be solved by requiring that 
\begin{equation}
\Omega_t|_{T[1]\partial S}=0.
\label{cnholchern5}
\end{equation}
A similar boundary condition was imposed in ref. \ccite{Elitzur:1989nr} to cope with the
analogous issue arising in the canonical formulation of 3--dimensional CS theory. The question
of the stability of a boundary condition of this sort under gauge transformation is however
quite different in the ordinary and higher cases. In the ordinary theory, the condition is preserved
by gauge transformations which are time independent on the boundary, which constitute a tractable subgroup
of the full gauge group. In the higher theory, the condition \ceqref{cnholchern5}
is preserved by gauge transformations obeying a complicated boundary condition involving also
$\Omega_S$, as emerges by inspection of the component expressions the transformations
of eqs. \ceqref{higau12}, \ceqref{higau13}, leaving doubts about the eventual viability of
the whole approach.

It seems more natural to resort \pagebreak to a boundary condition of the
isotropic linear kind introduced in subsect. \cref{subsec:4dchern}. 
We thus demand that the higher gauge field $\Omega\in\mathcal{C}_{\mathsans{M}}(M)$ satisfies
the requirement that 
$\Omega|_{T[1]\partial M}\in\mathcal{C}_{\mathsans{M}'}(\partial M)$,
where $\mathsans{M}'$ is an isotropic submodule of $\mathsans{M}$. 
When $\Omega$ is expressed in terms of the components $\Omega_t$, $\Omega_S$, the
boundary condition constrains$\vphantom{\ul{\ul{\ul{g}}}}$
$\Omega_t|_{T[1]\partial S}$, $\Omega_S|_{T[1]\partial S}$ to be
$\DD\mathfrak{m}'[0]$, $\DD\mathfrak{m}'[1]$ valued respectively, 
making the problematic boundary term in the right hand side of 
\ceqref{cnholchern4} vanish. For $\Omega_S$, the condition can be cast transparently as  
\begin{equation}
\Omega_S|_{T[1]\partial S}\in\Map(\mathbb{R}^1,\mathcal{C}_{\mathsans{M}'}(\partial S)).
\label{cnholchern8}
\end{equation}

Next, we examine the issue of higher gauge symmetry. 
In the hybrid  geometrical framework we are employing here, 
a higher gauge transformation
$\mathrm{U}\in\mathcal{G}_{\mathsans{M}}(M)$ factorizes as $\mathrm{U}=\mathrm{U}_t\mathrm{U}_S$, 
where $\mathrm{U}_t\in\Map(T[1]\mathbb{R}^1,\bfs{{\Map}}(T[1]S,\DD\mathsans{M}))$ is a 
gauge transformation of the form $\mathrm{U}(\alpha)=\ee^{\alpha dtU_t}$
with $U_t\in\Map(\mathbb{R}^1,\Map(T[1]S,\mathfrak{e}[0]))$
and $\mathrm{U}_S\in\Map(\mathbb{R}^1,\Map(T[1]S,\DD\mathsans{M}))$.

The isotropic linear boundary condition which we have imposed on the higher gauge field $\Omega$, viz 
$\Omega|_{T[1]\partial M}\in\mathcal{C}_{\mathsans{M}'}(\partial M)$, 
is stable under the gauge transformations
$\mathrm{U}\in\mathcal{G}_{\mathsans{M}}(M)$ which satisfy the boundary condition
$U|_{T[1]\partial M}\in\mathcal{I}^{\NN}{}_{\mathcal{C}_{\mathsans{M}'}(\partial M)}$,
where $\mathcal{I}^{\NN}{}_{\mathcal{C}_{\mathsans{M}'}(\partial M)}$
is the invariance subgroup of $\mathcal{C}_{\mathsans{M}'}(\partial M)$
introduced and studied in subsect. \cref{subsec:levelchern}.
When $\mathrm{U}$ is expressed in terms of its components $\mathrm{U}_t$, $\mathrm{U}_S$ as indicated above,
this condition can be written suggestively as  
\begin{equation}
U_S|_{T[1]\partial S}\in\Map(\mathbb{R}^1,\mathcal{I}^{\NN}{}_{\mathcal{C}_{\mathsans{M}'}(\partial S)}),
\label{cnholchern10}
\end{equation}
There is however a further restriction involving both $\mathrm{U}_t$ and $\mathrm{U}_S$ following from
\ceqref{4dchern+1}, \ceqref{4dchern+2}. It ensures that the boundary condition obeyed by $\Omega_t$
is stable under gauge transformation. For fixed $\mathrm{U}_S$, this restriction may fail to be
satisfied by any $\mathrm{U}_t$ unless $\mathrm{U}_S$ is further delimited. 
For this reason, in \ceqref{cnholchern10}
it may be necessary to replace the invariance subgroup $\mathcal{I}^{\NN}{}_{\mathcal{C}_{\mathsans{M}'}(\partial S)}$ 
with a proper subgroup of it. However, in the canonical set--up illustrated below both
$\Omega_t$ and $\mathrm{U}_t$ do not appear and we may therefore disregard this extra limitation. 

In canonical theory, we replace the higher gauge field component
$\Omega_S$ with a time independent gauge field $\Omega$ viewed as a point 
of an ambient functional phase space $\mathcal{C}_{\mathsans{M}}(S)$, where we suppress
the subscript $S$ for notational simplicity.  
Similarly, we replace the higher gauge transformation component $\mathrm{U}_S$ with a time independent gauge
transformation $\mathrm{U}$ of an ambient phase space gauge group $\mathcal{G}_{\mathsans{M}}(S)$
acting on $\Omega$ according the familiar prescription \ceqref{higau9}.
No boundary conditions on either $\Omega$ or $\mathrm{U}$ are imposed at this stage.

The physical phase space $\mathcal{C}_{\mathsans{M}\mathrm{ph}}(S)$ is the functional subspace of the ambient
phase space $\mathcal{C}_{\mathsans{M}}(S)$ defined by the flatness constraint corresponding to \ceqref{cnholchern6}
\begin{equation}
\Phi\approx 0, 
\label{cnholchern11}
\end{equation}
where $\Phi$ is the curvature of $\Omega$ defined according \ceqref{higau2}.
$\mathcal{C}_{\mathsans{M}}(S)$ is invariant under the action of the ambient gauge
transformation group $\mathcal{G}_{\mathsans{M}}(S)$. The reduced physical phase space
$\widetilde{\mathcal{C}}_{\mathsans{M}\mathrm{ph}}(S)$ 
is the quotient of $\mathcal{C}_{\mathsans{M}\mathrm{ph}}(S)$ by $\mathcal{G}_{\mathsans{M}}(S)$, 
\begin{equation}
\widetilde{\mathcal{C}}_{\mathsans{M}\mathrm{ph}}(S)
=\mathcal{C}_{\mathsans{M}\mathrm{ph}}(S)/\mathcal{G}_{\mathsans{M}}(S).
\label{cnholchern11/0}
\end{equation}
All this is rather formal, since the above quotient turns out to be singular. 
It therefore calls for a more precise formulation of the
symplectic structure of $\mathcal{C}_{\mathsans{M}}(S)$, which we provide below.

The short action term in the right hand side
of \ceqref{cnholchern4} indicates the appropriate expression of the symplectic potential
$\sfGamma\in\matheul{O}^1{}_{\mathsans{M}}(S)$ as the $1$--form 
\begin{equation}
\sfGamma=\frac{k}{4\pi}\int_{T[1]S}\varrho_S\left(\Omega,\sfdelta\Omega\right)
\label{holchern10}
\end{equation}
(cf. eq. \ceqref{holchern-1}). 
The symplectic $2$--form $\sfUpsilon\in\matheul{O}^2{}_{\mathsans{M}}(S)$ yielded by $\sfGamma$ is 
\begin{equation}
\sfUpsilon=\sfdelta\sfGamma=\frac{k}{4\pi}\int_{T[1]S}\varrho_S\left(\sfdelta\Omega,\sfdelta\Omega\right)
\label{holchern8}
\end{equation}
(cf. eq. \ceqref{holchern-2}). 
The non singularity of $\sfUpsilon$ follows from that of the field pairing $(\cdot,\cdot)$.
The higher gauge field manifold $\mathcal{C}_{\mathsans{M}}(S)$ is in this way  equipped with the appropriate
symplectic structure. Our task now is expressing the associated Poisson bracket.


For any functional $\sfF\in\matheul{F}{}_{\mathsans{M}}(S)$, the Hamiltonian vector field
$\sfH_\sfF\in\matheul{V}_{\mathsans{M}}(S)$ of $\sfF$ is characterized by the property that 
\begin{equation}
\sfiota_{\sfH_\sfF}\sfUpsilon+\sfdelta\sfF=0. \vphantom{\int}
\label{holchern11}
\end{equation}
From \ceqref{holchern1}, 
$\sfH_\sfF$ is given by relation \ceqref{holchern5} with $\mathrm{V}$ replaced by \hphantom{xxxxxxxxxx}
\begin{equation}
\mathrm{H}_\sfF=\frac{2\pi}{k}\frac{\delta\sfF}{\delta\Omega}.
\label{holchern12}
\end{equation}
The field functional algebra $\matheul{F}_{\mathsans{M}}(S)$ is so equipped with the Poisson bracket 
\begin{equation}
\{\sfF,\sfG\}=\sfiota_{\sfH_{\sfF}}\sfdelta\sfG=\frac{2\pi}{k}\int_{T[1]S}\varrho_S
\left(\frac{\delta\sfF}{\delta\Omega},\frac{\delta\sfG}{\delta\Omega}\right)
\label{holchern13}
\end{equation}
for $\sfF,\sfG\in\matheul{F}_{\mathsans{M}}(S)$. The basic Poisson bracket of the theory is in this way 
\begin{equation}
\left\{\int_{T[1]S}\varrho_S(\Omega,\Sigma),\int_{T[1]S}\varrho_S(\Omega,\Sigma')\right\}
=\frac{2\pi}{k}\int_{T[1]S}\varrho_S(\Sigma,\Sigma')
\label{holchern14}
\end{equation}
with $\Sigma,\Sigma'\in\Map(T[1]S,\DD\mathfrak{m}[1])$. 

We consider next the field functionals 
\begin{equation}
\sfQ(\Theta)
=\frac{k}{2\pi}\left[\int_{T[1]S}\varrho_S(\Phi,\Theta)-\int_{T[1]\partial S}\varrho_{\partial S}(\Omega,\Theta)\right].
\label{holchern15}
\end{equation}
where $\Theta\in\mathfrak{G}_{\mathsans{M}}(S)$ is an infinitesimal gauge transformation and
$\Phi$ is the curvature of higher gauge field $\Omega$ given by \ceqref{higau2} as before.
The boundary term is added to render $\sfQ(\Theta)$
differentiable in the sense established in ref. 
\ccite{Regge:1974zd,Benguria:1976in}, as is evident by writing $\sfQ(\Theta)$ in the form 
\begin{equation}
\sfQ(\Theta)=\frac{k}{2\pi}\int_{T[1]S}\varrho_S\left[(\Omega,\dd\Theta)+\tfrac{1}{2}([\Omega,\Omega],\Theta)\right]
\label{holcher16}
\end{equation}
$\sfQ(\Theta)$ is the Hamiltonian of $\Theta$, 
\begin{equation}
\left\{\sfQ(\Theta),\int_{T[1]S}\varrho_S(\Omega,\Sigma)\right\}=\int_{T[1]S}\varrho_S(\delta_\Theta\Omega,\Sigma).
\label{holchern20}
\end{equation}
for $\Sigma\in\Map(T[1]S,\DD\mathfrak{m}[1])$, where the gauge variation $\delta_\Theta\Omega$
is given  by \ceqref{higau20}. Under Poisson bracketing, the Hamiltonians $\sfQ(\Theta)$ form a
centrally extended representation of the gauge transformation Lie algebra. Specifically, we have 
\begin{equation}
\{\sfQ(\Theta),\sfQ(\Theta')\}=\sfQ([\Theta,\Theta'])+\frac{k}{2\pi}\sfc(\Theta,\Theta')
\label{holchern21}
\end{equation}
with $\Theta,\Theta'\in\mathfrak{G}_{\mathsans{M}}(S)$, where $\sfc$ is the $2$--cocycle
\begin{equation}
\sfc(\Theta,\Theta')=\int_{T[1]\partial S}\varrho_{\partial S}(\Theta,\dd\Theta'). 
\label{holchern22}
\end{equation}
The Poisson bracket relation \ceqref{holchern21} describes a higher 3--dimensional current algebra
analogous to the 2--dimensional current algebra appearing in the canonical formulation of ordinary
CS theory. 
More on this in the next subsection. 

We now write the above results in terms of the components $\omega$, $\varOmega$ of the higher gauge field $\Omega$
for the sake of concreteness. 
The symplectic form $\sfUpsilon$ 
defined in eq. \ceqref{holchern8}, has a simple component expression, 
\begin{equation}
\sfUpsilon=\frac{k}{2\pi}\int_{T[1]S}\varrho_S\left\langle\sfdelta\omega,\sfdelta\varOmega\right\rangle,
\label{holchern16}
\end{equation}
which shows that $\omega$, $\varOmega$ are canonical conjugate fields. 
The component expression of the Poisson bracket \ceqref{holchern13} takes so the familiar canonical form 
\begin{equation}
\{\sfF,\sfG\}=\frac{2\pi}{k}\int_{T[1]S}\varrho_S\left[
\left\langle\frac{\delta\sfF}{\delta\varOmega},\frac{\delta\sfG}{\delta\omega}\right\rangle
-\left\langle\frac{\delta\sfG}{\delta\varOmega},\frac{\delta\sfF}{\delta\omega}\right\rangle
\right].
\label{holchern18}
\end{equation}
The basic Poisson bracket \ceqref{holchern14} reads in this way as 
\begin{equation}
\left\{\int_{T[1]S}\varrho_S\,\langle\omega,\varSigma\rangle,\int_{T[1]S}\varrho_S\,\langle\sigma,\varOmega\rangle\right\}
=\frac{2\pi}{k}\int_{T[1]S}\varrho_S\,\langle\sigma,\varSigma\rangle
\label{holchern19}
\end{equation}
for $\sigma\in\Map(T[1]S,\mathfrak{g}[1])$, $\varSigma\in\Map(T[1]S,\mathfrak{e}[2])$.

To write down the component expressions of the gauge transformation Hamiltonians, we need the components
$\phi$, $\varPhi$ of the higher gauge curvature $\Phi$ given by \ceqref{higau4}, \ceqref{higau5}.  
From \ceqref{holchern15}, the Hamiltonian of an infinitesimal higher
gauge transformation of components $\theta$, $\varTheta$ is 
\begin{multline}
\sfQ(\theta,\varTheta)=\frac{k}{2\pi}\left\{\int_{T[1]S}\varrho_S\left[
\left\langle\phi,\varTheta\right\rangle+\left\langle\theta,\varPhi\right\rangle\right]\right.
\label{holchern19/cmp}
\\
-\left.\int_{T[1]\partial S}\varrho_{\partial S}\left[
  \left\langle\omega,\varTheta\right\rangle+\left\langle\theta,\varOmega\right\rangle\right]\right\}.
\end{multline}
The component form of the Hamiltonian relation \ceqref{holchern20} is then 
\begin{align}
&\left\{\sfQ(\theta,\varTheta),\int_{T[1]S}\varrho_S\,\langle\omega,\varSigma\rangle\right\}
=\int_{T[1]S}\varrho_S\,\langle\delta_{\theta,\varTheta}\omega,\varSigma\rangle,
\vphantom{\Big]}
\label{holchern23}
\\
&\left\{\sfQ(\theta,\varTheta),\int_{T[1]S}\varrho_S\,\langle\sigma,\varOmega\rangle\right\}
=\int_{T[1]S}\varrho_S\,\langle\sigma,\delta_{\theta,\varTheta}\varOmega\rangle,
\vphantom{\Big]}
\label{holchern24}
\end{align}
where the gauge variations $\delta_{\theta,\varTheta}\omega$, $\delta_{\theta,\varTheta}\varOmega$ are given by
\ceqref{higau23}, \ceqref{higau24}. 

In components, the $\sfQ$ generator Poisson bracket \ceqref{holchern21} reads as 
\begin{equation}
\{\sfQ(\theta,\varTheta),\sfQ(\theta',\varTheta')\}
=\sfQ([\theta,\theta'], \sdot\mu\sdot(\theta,\varTheta')-\sdot\mu\sdot(\theta',\varTheta))
+\frac{k}{2\pi}\sfc(\theta,\varTheta;\theta',\varTheta')
\label{holchern21/c}
\end{equation}
and that of the occurring $2$--cocycle \ceqref{holchern22} as 
\begin{equation}
\sfc(\theta,\varTheta;\theta',\varTheta')=-\int_{T[1]\partial S}\varrho_{\partial S}\left[
\langle d\theta,\varTheta'\rangle-\langle d\theta',\varTheta\rangle+\langle\dot\tau(\varTheta),\varTheta'\rangle\right]. 
\label{holchern25}
\end{equation}

As we have already stated, the physical phase space $\mathcal{C}_{\mathsans{M}\mathrm{ph}}(S)$
of higher CS theory is the functional hypersurface in the ambient
phase space $\mathcal{C}_{\mathsans{M}}(S)$ defined by the flatness condition $\Phi\approx 0$
(cf. eq. \ceqref{cnholchern11}). As the bulk contribution to the Hamiltonian functionals $\sfQ(\Theta)$
is proportional to $\Phi$ (cf. eq. \ceqref{holchern15}), it seems plausible that the constraint may be expressed
through the weak constraints
\begin{equation}
\sfQ(\Theta)\approx 0
\label{holchern26}
\end{equation}
with $\Theta\in\mathfrak{G}_{\mathsans{M}}(S)$, in analogy to ordinary CS theory. 
There are a number of problems with this approach.
First, the $\sfQ(\Theta)$ contain also a boundary proportional to $\Omega$,
making the use of \ceqref{holchern26} as definition of the physical phase space doubtful. Second,
since the $\sfQ(\Theta)$ generate infinitesimal gauge transformations (cf. eq. \ceqref{holchern20}), they should
be first class functionals, while they are not because of the 2--cocycle $\sfc(\Theta,\Theta')$ appearing
in the Poisson bracket relations \ceqref{holchern21}.

Both the term proportional to $\Omega$ in the $\sfQ(\Theta)$ and the 2--cocycle $\sfc(\Theta,\Theta')$
are supported on the boundary $\partial S$ of $S$. They could be removed by imposing appropriate boundary conditions
on $\Omega$ and $\Theta$. Requiring that 
$\Omega\in\mathcal{C}_{\mathsans{M}}(S)$ obeys 
\begin{equation}
\Omega|_{T[1]\partial S}\in\mathcal{C}_{\mathsans{M}'}(\partial S)
\label{holchern27}
\end{equation}
and that $\Theta\in\mathfrak{G}_{\mathsans{M}'}(\partial S)$ satisfies 
\begin{equation}
\Theta|_{T[1]\partial S}\in\mathfrak{G}_{\mathsans{M}'}(\partial S),
\label{holchern28}
\end{equation}
where $\mathsans{M}'$ is some isotropic submodule of the crossed module $\mathsans{M}$, 
eliminates at once the unwanted term in $\sfQ(\Theta)$ and the 2--cocycle $\sfc(\Theta,\Theta')$, rendering
the $\sfQ(\Theta)$ honest first class functionals defining the physical phase space via \ceqref{holchern26} as desired.
Note that \ceqref{holchern27} precisely answers to the boundary condition \ceqref{cnholchern8} discussed earlier, while
\ceqref{holchern28} is compatible with the boundary condition \ceqref{cnholchern10},
since $\mathcal{G}_{\mathsans{M}'}(\partial M)\subseteq\mathcal{I}^{\NN}{}_{\mathcal{C}_{\mathsans{M}'}(\partial M)}$.
Below,  we shall so refer to the boundary conditioned phase space
$\mathcal{C}_{\mathsans{M},\mathsans{M}'}(S)\subset\mathcal{C}_{\mathsans{M}}(S)$ formed
by those gauge fields $\Omega$ which satisfy \ceqref{holchern27}
and similarly to the  boundary conditioned gauge algebra
$\mathfrak{G}_{\mathsans{M},\mathsans{M}'}(S)\subset\mathfrak{G}_{\mathsans{M}}(S)$ formed
by those infinitesimal gauge transformations $\Theta$ which satisfy \ceqref{holchern28}.
Before proceeding further, we notice that while the boundary conditions \ceqref{holchern28}
is essentially mandated by the requirement of first classness of the functionals $\sfQ(\Theta)$,
the boundary condition \ceqref{holchern27} could be weakened by requiring less restrictively that
$\Omega|_{T[1]\partial S}\in\mathcal{C}_{\ON\mathsans{M}'}(\partial S)$, where $\ON\mathsans{M}'$ is the
orthogonal normalizer of the isotropic crossed submodule $\mathsans{M}'$
(cf. subsect. \cref{subsec:crosumo}). We shall come back to this point in
subsect. \cref{subsec:surfchern}. 

All field functionals $\sfF\in\matheul{F}_{\mathsans{M}}(S)$ we consider
are defined on the full phase space $\mathcal{C}_{\mathsans{M}}(S)$
containing all higher gauge fields $\Omega$ obeying no preassigned boundary condition.
The boundary condition \ceqref{holchern27} 
is implemented by restricting the functionals to the conditioned phase space
$\mathcal{C}_{\mathsans{M},\mathsans{M}'}(S)$.  
The calculation of the relevant Poisson brackets is correspondingly performed employing
the unrestricted phase space canonical framework described above. The boundary condition \ceqref{holchern27} is
imposed at the end of the calculation. Doing so before that may lead to in\-consistencies.

On the basis of the above analysis of boundary conditions of gauge fields and
gauge transformations, it appears that the physical phase space may be described as 
the submanifold $\mathcal{C}_{\mathsans{M},\mathsans{M}'{\rm phys}}(S)$
of $\mathcal{C}_{\mathsans{M},\mathsans{M}'}(S)$ defined by the weak constraints 
\begin{equation}
\sfL\approx 0
\label{holchern32}
\end{equation}
with $\sfL\in\matheul{F}_{\mathsans{M},\mathsans{M}'{\rm triv}}(S)$, $\matheul{F}_{\mathsans{M},\mathsans{M}'{\rm triv}}(S)$
being the ideal of $\matheul{F}_{\mathsans{M}}(S)$
generated by the Hamiltonians $\sfQ(\Theta)$ with $\Theta\in\mathfrak{G}_{\mathsans{M},\mathsans{M}'}(S)$.
$\matheul{F}_{\mathsans{M},\mathsans{M}'{\rm triv}}(S)$ codifies the infinitesimal 
higher gauge symmetry action associated with the Lie subalgebra $\mathfrak{G}_{\mathsans{M},\mathsans{M}'}(S)$.
The reduced physical phase space $\widetilde{\mathcal{C}}_{\mathsans{M},\mathsans{M}'{\rm phys}}(S)$
is the quotient of $\mathcal{C}_{\mathsans{M},\mathsans{M}'{\rm phys}}(S)$ by this gauge symmetry.
The physical field functional algebra is the algebra of field functionals on
$\widetilde{\mathcal{C}}_{\mathsans{M},\mathsans{M}'{\rm phys}}(S)$.
As is well known,$\vphantom{\ul{\ul{\ul{g}}}}$
$\widetilde{\mathcal{C}}_{\mathsans{M},\mathsans{M}'{\rm phys}}(S)$
is a complicated non local object that is problematic to describe in local field theory. Moreover, 
by \ceqref{higau29}, \ceqref{higau30}, special gauge transformations in $\mathfrak{G}_{\mathsans{M},\mathsans{M}'}(S)$
are inert on the flat gauge fields $\Omega$ which constitute $\mathcal{C}_{\mathsans{M},\mathsans{M}'{\rm phys}}(S)$.
Thus, $\widetilde{\mathcal{C}}_{\mathsans{M},\mathsans{M}'\mathrm{ph}}(S)$
is also a singular manifold. Consequently, also the field
functional algebra of $\widetilde{\mathcal{C}}_{\mathsans{M},\mathsans{M}'\mathrm{ph}}(S)$
is problematic to describe. 
A proper treatment of $\widetilde{\mathcal{C}}_{\mathsans{M},\mathsans{M}'\mathrm{ph}}(S)$ and its field functionals
in local field theory requires the full apparatus of BRST--BV theory for reducible gauge symmetries.
For the time being, it is enough to adopt a more modest stance and proceed as follows. 

Since these Hamiltonians $\sfQ(\Theta)$ with $\Theta\in\mathfrak{G}_{\mathsans{M},\mathsans{M}'}(S)$
obey a Poisson algebra of the form \ceqref{holchern21} with vanishing 
central extension by the isotropy of $\mathsans{M}'$, the constraints \ceqref{holchern32} are first class,
as indeed 
\begin{equation}
\{\sfL,\sfM\}\approx 0
\label{holchern33}
\end{equation}
for $\sfL,\sfM\in\matheul{F}_{\mathsans{M},\mathsans{M}'{\rm triv}}(S)$. 
A physical field functional $\widetilde{\sfF}$ is represented by a gauge invariant functional
$\sfF\in\matheul{F}_{\mathsans{M}}(S)$, that is one such that 
\begin{equation}
\{\sfF,\sfL\}\approx 0
\label{holchern34}
\end{equation}
for $\sfL\in\matheul{F}_{\mathsans{M},\mathsans{M}'{\rm triv}}(S)$. The representative $\sfF$ is not unique however
being modifiable by the addition of any functional $\varDelta\sfF\in\matheul{F}_{\mathsans{M},\mathsans{M}'{\rm triv}}(S)$,
so that $\sfF$, $\sfF+\varDelta\sfF$ can be considered as physically equivalent. 
Let us denote by $\matheul{F}_{\mathsans{M},\mathsans{M}'{\rm inv}}(S)$ the subalgebra of $\matheul{F}_{\mathsans{M}}(S)$
of the functionals $\sfF$ satisfying \ceqref{holchern34}. Then, the algebra of physical functionals
is the quotient
\begin{equation}
  \widetilde{\matheul{F}}_{\mathsans{M},\mathsans{M}'{\rm phys}}(S)
  =\matheul{F}_{\mathsans{M},\mathsans{M}'{\rm inv}}(S)/\matheul{F}_{\mathsans{M},\mathsans{M}'{\rm triv}}(S).
\label{holchern35}
\end{equation}
$\widetilde{\matheul{F}}_{\mathsans{M},\mathsans{M}'{\rm phys}}(S)$ supports the induced Poisson bracket. 
If $\widetilde{\sfF},\widetilde{\sfG}\in\widetilde{\matheul{F}}_{\mathsans{M},\mathsans{M}'{\rm phys}}(S)$ are
physical functionals represented by gauge invariant functionals $\sfF,\sfG\in\matheul{F}_{\mathsans{M},\mathsans{M}'{\rm inv}}(S)$,
then $\{\widetilde{\sfF},\widetilde{\sfG}\}$ is represented by the gauge invariant functional
$\{\sfF,\sfG\}$.

The above has a mathematical formalization. 
Let $\matheul{I}_{\mathsans{M},\mathsans{M}'}(S)$ be the ideal of 
$\matheul{F}_{\mathsans{M}}(S)$ generated by the Hamiltonians $\sfQ(\Theta)$ with
$\Theta\in\mathfrak{G}_{\mathsans{M},\mathsans{M}'}(S)$. 
$\matheul{I}_{\mathsans{M},\mathsans{M}'}(S)$ is a Poisson subalgebra 
of $\matheul{F}_{\mathsans{M}}(S)$, but not a Poisson ideal. We consider so 
the Poisson normalizer $\NN_{\rm P\!}\matheul{I}_{\mathsans{M},\mathsans{M}'}(S)$ of
$\matheul{I}_{\mathsans{M},\mathsans{M}'}(S)$, the set of all functionals
$\sfF\in\matheul{F}_{\mathsans{M}}(S)$ such that $\{\sfF,\sfL\}\in\matheul{I}_{\mathsans{M},\mathsans{M}'}(S)$
for all $\sfL\in\matheul{I}_{\mathsans{M},\mathsans{M}'}(S)$.
$\NN_{\rm P\!}\matheul{I}_{\mathsans{M},\mathsans{M}'}(S)$ is both a subalgebra and a Poisson subalgebra
of $\matheul{F}_{\mathsans{M}}(S)$. 
The Poisson Weyl algebra of
$\matheul{I}_{\mathsans{M},\mathsans{M}'}(S)$
\begin{equation}
\WW_{\rm P\!}\matheul{I}_{\mathsans{M},\mathsans{M}'}(S)
=\NN_{\rm P\!}\matheul{I}_{\mathsans{M},\mathsans{M}'}(S)/\matheul{I}_{\mathsans{M},\mathsans{M}'}(S)
\label{holchern30}
\end{equation}
is then defined.  $\WW_{\rm P\!}\matheul{I}_{\mathsans{M},\mathsans{M}'}(S)$ is both an algebra and a Poisson algebra
with the induced Poisson bracket
\begin{equation}
\{\sfF+\matheul{I}_{\mathsans{M},\mathsans{M}'}(S),\sfG+\matheul{I}_{\mathsans{M},\mathsans{M}'}(S)\}
=\{\sfF,\sfG\}+\matheul{I}_{\mathsans{M},\mathsans{M}'}(S). 
\label{holchern31}
\end{equation}
It should be apparent that $\matheul{I}_{\mathsans{M},\mathsans{M}'}(S)$, $\NN_{\rm P\!}\matheul{I}_{\mathsans{M},\mathsans{M}'}(S)$
and $\WW_{\rm P\!}\matheul{I}_{\mathsans{M},\mathsans{M}'}(S)$ correspond respectively
to $\matheul{F}_{\mathsans{M},\mathsans{M}'{\rm triv}}(S)$,
$\matheul{F}_{\mathsans{M},\mathsans{M}'{\rm inv}}(S)$ and $\widetilde{\matheul{F}}_{\mathsans{M},\mathsans{M}'{\rm phys}}(S)$ in the previous
more conventional characterization. 




The constraints \ceqref{holchern26} are not independent though, as we show next. 
Recall that a special infinitesimal gauge transformation $\Theta^*$ is an infinitesimal gauge transformation 
depending on the underlying gauge field $\Omega$ that in component form reads as in \ceqref{higau27}, \ceqref{higau29}
for some map $\varXi\in\Map(T[1]S,\mathfrak{e})$. Such a $\Theta^*$ must hence be regarded as a functional of
$\Omega$ and $\Xi$. $\Theta^*$ will obey the boundary condition \ceqref{holchern28},
if $\Omega$ and $\varXi$ satisfy respectively the boundary condition
\ceqref{holchern27}, reading in components as $\omega|_{T[1]\partial S}\in\Map(T[1]\partial S,\mathfrak{g}'[1])$,
$\varOmega|_{T[1]\partial S}\in\Map(T[1]\partial S,\mathfrak{e}'[2])$,
and $\varXi|_{T[1]\partial S}\in\Map(T[1]\partial S,\mathfrak{e}')$. 
Using the component expression \ceqref{holchern19/cmp} of $\sfQ(\Theta^*)$ and the 
Bianchi identity \ceqref{higau7}, it is found that 
\begin{equation}
\sfQ(\Theta^*)=0
\label{holchern29}
\end{equation}
when $\Omega$ and $\Xi$ are restricted as indicated above. Consequently, $\sfQ(\Theta^*)=0$
strongly. For varying $\Xi$, the 
\ceqref{holchern29} represent a set of relations obeyed by the $\sfQ(\Theta)$
for general infinitesimal gauge transformations $\Theta$ showing their non independence. 
The higher gauge symmetry is so reducible signaling a higher gauge for gauge symmetry
of the theory.

\subsection{\textcolor{blue}{\sffamily Surface charges and holography}}\label{subsec:surfchern}

In this subsection, we analyze one of the most interesting holographic properties of higher CS theory: the 
existence of surface charges obeying under suitable conditions a non trivial Poisson bracket algebra
that is a higher counterpart of the familiar Kac--Moody current algebra. 

The canonical framework of subsect. \cref{subsec:holchern} turns out to be particularly suited for this purpose. 
For any infinitesimal gauge transformation $\Theta'\in\mathfrak{G}_{\mathsans{M},\mathsans{M}'}(S)$,
the Hamiltonian $\sfQ(\Theta')\in\matheul{I}_{\mathsans{M},\mathsans{M}'}(S)$, where $\matheul{I}_{\mathsans{M},\mathsans{M}'}(S)$
is the constraint ideal. 
For a generic infinitesimal gauge transformation $\Theta\in\mathfrak{G}_{\mathsans{M}}(S)$, though, 
the Hamiltonian $\sfQ(\Theta)\not\in\NN_{\rm P\!}\matheul{I}_{\mathsans{M},\mathsans{M}'}(S)$, 
$\NN_{\rm P\!}\matheul{I}_{\mathsans{M},\mathsans{M}'}(S)$ being
the Poisson normalizer of $\matheul{I}_{\mathsans{M},\mathsans{M}'}(S)$,
as we have $\{\sfQ(\Theta),\sfQ(\Theta')\}\not\in\matheul{I}_{\mathsans{M},\mathsans{M}'}(S)$ 
for $\Theta'\in\mathfrak{G}_{\mathsans{M},\mathsans{M}'}(S)$ in general.
In order $\sfQ(\Theta)\in\NN_{\rm P\!}\matheul{I}_{\mathsans{M},\mathsans{M}'}(S)$, the 
gauge transformation $\Theta$ must be suitably restricted and since what distinguishes 
the gauge transformations $\Theta'\in\mathfrak{G}_{\mathsans{M},\mathsans{M}'}(S)$ is only their
obeying the boundary condition \ceqref{holchern28}, it is a boundary condition that presumably has
to be imposed on $\Theta$. A detailed analysis shows indeed that
$\sfQ(\Theta)\in\NN_{\rm P\!}\matheul{I}_{\mathsans{M},\mathsans{M}'}(S)$
if $\Theta\in\mathfrak{G}_{\mathsans{M},\ON\mathsans{M}'}(S)$, 
where $\ON\mathsans{M}'$ is the orthogonal normalizer of the crossed submodule $\mathsans{M}'$
(cf. subsect. \cref{subsec:crosumo}) and $\mathfrak{G}_{\mathsans{M},\ON\mathsans{M}'}(S)$
is the subalgebra of $\mathfrak{G}_{\mathsans{M}}(S)$ 
of the infinitesimal gauge transformations $\Theta$ satisfying the boundary condition 
\begin{equation}
\Theta|_{T[1]\partial S}\in\mathfrak{G}_{\ON\mathsans{M}'}(\partial S)
\label{surfchern1}
\end{equation}
analogously to the subalgebra $\mathfrak{G}_{\mathsans{M},\mathsans{M}'}(S)$. To see this, 
we note that $\ON\mathfrak{m}'$ being a crossed submodule of $\NN\mathfrak{m}'$ ensures that for
$\Theta'\in\mathfrak{G}_{\mathsans{M},\mathsans{M}'}(S)$,  
$[\Theta,\Theta']\in\mathfrak{G}_{\mathsans{M},\mathsans{M}'}(S)$ by \ceqref{superfield6}. Further,
by the expression \ceqref{holchern22} of the 2--cocycle $\sfc$, $\ON\mathfrak{m}'$ being the orthogonal complement
of $\mathfrak{m}'$ in $\NN\mathfrak{m}'$ and \ceqref{superfield10} imply that
for $\Theta'\in\mathfrak{G}_{\mathsans{M},\mathsans{M}'}(S)$,  
$\sfc(\Theta,\Theta')=0$. It follows by virtue of\ceqref{holchern21} that 
$\{\sfQ(\Theta),\sfQ(\Theta')\}\in\matheul{I}_{\mathsans{M},\mathsans{M}'}(S)$, 
as required.


Since $\mathfrak{m}'$ is a crossed submodule of $\ON\mathfrak{m}'$, for
$\Theta\in\mathfrak{G}_{\mathsans{M},\ON\mathsans{M}'}(S)$, $\Theta'\in\mathfrak{G}_{\mathsans{M},\mathsans{M}'}(S)$
one has $\Theta+\Theta'\in\mathfrak{G}_{\mathsans{M},\ON\mathsans{M}'}(S)$. The identity 
$\sfQ(\Theta+\Theta')=\sfQ(\Theta)+\sfQ(\Theta')$ shows then that $\sfQ(\Theta+\Theta')$ and $\sfQ(\Theta)$
are equivalent modulo $\matheul{I}_{\mathsans{M},\mathsans{M}'}(S)$ and so define the same element
$\sfQ(\Theta+\mathfrak{G}_{\mathsans{M},\mathsans{M}'}(S))\in\WW_{\rm P\!}\matheul{I}_{\mathsans{M},\mathsans{M}'}(S)$,
the reduced gauge invariant functional algebra, as explained in subsect. \cref{subsec:holchern}. 

Remarkably, the reduced Hamiltonians 
$\sfQ(\Theta+\mathfrak{G}_{\mathsans{M},\mathsans{M}'}(S))\in\WW_{\rm P\!}\matheul{I}_{\mathsans{M},\mathsans{M}'}(S)$
with $\Theta\in\mathfrak{G}_{\mathsans{M},\ON\mathsans{M}'}(S)$ form a subalgebra of the reduced Poisson algebra 
$\WW_{\rm P\!}\matheul{I}_{\mathsans{M},\mathsans{M}'}(S)$, as we now show.
Pick $\Theta_1,\Theta_2\in\mathfrak{G}_{\mathsans{M},\ON\mathsans{M}'}(S)$.
$\ON\mathfrak{m}'$ being a crossed submodule of $\NN\mathfrak{m}'$ implies by \ceqref{superfield6}
that $\mathfrak{G}_{\mathsans{M},\mathsans{M}'}(S)$ is an ideal of the Lie algebra
$\mathfrak{G}_{\mathsans{M},\ON\mathsans{M}'}(S)$. So, the Lie bracket $[\Theta_1,\Theta_2]$ depends
on the choice of $\Theta_1$, $\Theta_2$ mod $\mathfrak{G}_{\mathsans{M},\mathsans{M}'}(S)$
only mod $\mathfrak{G}_{\mathsans{M},\mathsans{M}'}(S)$. Further, as $\ON\mathfrak{m}'$ is the orthogonal complement
of $\mathfrak{m}'$ in $\NN\mathfrak{m}'$, $\sfc(\Theta_1,\Theta_2)$ is independent from 
the choice of $\Theta_1$, $\Theta_2$ mod $\mathfrak{G}_{\mathsans{M},\mathsans{M}'}(S)$ and may be denoted as 
$\sfc(\Theta_1+\mathfrak{G}_{\mathsans{M},\mathsans{M}'}(S),\Theta_2+\mathfrak{G}_{\mathsans{M},\mathsans{M}'}(S))$.
$\vphantom{\ul{\ul{\ul{\ul{g}}}}}$By virtue of \ceqref{holchern21}, 
the Poisson bracket of $\sfQ(\Theta_1+\mathfrak{G}_{\mathsans{M},\mathsans{M}'}(S))$,
$\sfQ(\Theta_2+\mathfrak{G}_{\mathsans{M},\mathsans{M}'}(S))$ thus read 
\begin{align}
&\{\sfQ(\Theta_1+\mathfrak{G}_{\mathsans{M},\mathsans{M}'}(S)),\sfQ(\Theta_2+\mathfrak{G}_{\mathsans{M},\mathsans{M}'}(S))\}
\label{surfchern2}
\\
&=\sfQ([\Theta_1+\mathfrak{G}_{\mathsans{M},\mathsans{M}'}(S),\Theta_2+\mathfrak{G}_{\mathsans{M},\mathsans{M}'}(S)])
+\frac{k}{2\pi}\sfc(\Theta_1+\mathfrak{G}_{\mathsans{M},\mathsans{M}'}(S),\Theta_2+\mathfrak{G}_{\mathsans{M},\mathsans{M}'}(S)).
\nonumber    
\end{align}
This is the Poisson subalgebra of $\WW_{\rm P\!}\matheul{I}_{\mathsans{M},\mathsans{M}'}(S)$ sought for. 

As $\mathfrak{G}_{\mathsans{M},\mathsans{M}'}(S)$ is an ideal of the Lie algebra
$\mathfrak{G}_{\mathsans{M},\ON\mathsans{M}'}(S)$, the quotient Lie algebra
$\mathfrak{G}_{\mathsans{M},\ON\mathsans{M}'}(S)/\mathfrak{G}_{\mathsans{M},\mathsans{M}'}(S)$
is defined. The reduced Hamiltonians $\sfQ(\Theta+\mathfrak{G}_{\mathsans{M},\mathsans{M}'}(S))$
are parametrized by the cosets $\Theta+\mathfrak{G}_{\mathsans{M},\mathsans{M}'}(S)
\in\mathfrak{G}_{\mathsans{M},\ON\mathsans{M}'}(S)/\mathfrak{G}_{\mathsans{M},\mathsans{M}'}(S)$. By \ceqref{surfchern2}, 
the resulting map $\sfQ:\mathfrak{G}_{\mathsans{M},\ON\mathsans{M}'}(S)/\mathfrak{G}_{\mathsans{M},\mathsans{M}'}(S)\rightarrow
\WW_{\rm P\!}\matheul{I}_{\mathsans{M},\mathsans{M}'}(S)$ is a Lie algebra morphism. This morphism
is projective because of the central extension term. 

The Lie algebra $\mathfrak{G}_{\mathsans{M},\ON\mathsans{M}'}(S)/\mathfrak{G}_{\mathsans{M},\mathsans{M}'}(S)$
is non trivial only if $\mathsans{M}'$ is isotropic but not Lagrangian.
In fact, when $\mathsans{M}'$ is Lagrangian, one has $\ON\mathfrak{m}'=\mathfrak{m}'$ and hence 
$\mathfrak{G}_{\mathsans{M},\ON\mathsans{M}'}(S)=\mathfrak{G}_{\mathsans{M},\mathsans{M}'}(S)$. The same thus holds for the
reduced Hamiltonians $\sfQ(\Theta+\mathfrak{G}_{\mathsans{M},\mathsans{M}'}(S))$ and their Poisson bracket algebra 
\ceqref{surfchern2}, 

From \ceqref{holcher16}, it is apparent that on the constraint submanifold, where $\Phi\approx 0$, 
the Hamiltonians $\sfQ(\Theta+\mathfrak{G}_{\mathsans{M},\mathsans{M}'}(S))$ reduce to the
surface term supported on $\partial S$. For this reason, the
$\sfQ(\Theta+\mathfrak{G}_{\mathsans{M},\mathsans{M}'}(S))$ are identified with the surface charges of higher CS theory.
The nature of these charges, in particular their non triviality, depends on the boundary conditions imposed 
on the higher gauge field $\Omega$. 

If we required the gauge field $\Omega$ to obey the boundary condition \ceqref{holchern27}, the surface charges
$\sfQ(\Theta+\mathfrak{G}_{\mathsans{M},\mathsans{M}'}(S))$ would vanish since the crossed submodule
$\ON\mathfrak{m}'$ is the orthogonal complement of $\mathfrak{m}'$ in $\NN\mathfrak{m}'$. 
If we want as we do the $\sfQ(\Theta+\mathfrak{G}_{\mathsans{M},\mathsans{M}'}(S))$ to be non trivial
a less severe boundary condition is required. We have already anticipated in subsect. \cref{subsec:holchern}
that in order the Hamiltonians $\sfQ(\Theta')$ with $\Theta'\in\mathfrak{G}_{\mathsans{M},\mathsans{M}'}(S)$
to define through the weak constraints $\sfQ(\Theta')\approx 0$ the flat higher gauge functional submanifold
$\Phi\approx 0$, it is enough to require that
\begin{equation}
\Omega|_{T[1]\partial S}\in\mathcal{C}_{\ON\mathsans{M}'}(\partial S)
\label{surgchern3}
\end{equation}
This boundary condition is weaker than \ceqref{holchern27} and subsumes it. Further, it is invariant under
the infinitesimal gauge transformation action of $\mathfrak{G}_{\mathsans{M},\mathsans{M}'}(S)$ (cf. eqs.
\eqref{higau23}, \ceqref{higau24}) and when the Lie group $\mathsans{G}'$ in $\mathsans{M}'$ is connected 
also under the finite gauge transformation action of
$\mathcal{G}_{\mathsans{M},\mathsans{M}'}(S)$ (cf. eqs. \ceqref{higau12}, \ceqref{higau13}),
where $\mathcal{G}_{\mathsans{M},\mathsans{M}'}(S)$ is the subgroup of
$\mathcal{G}_{\mathsans{M}}(S)$ of the gauge transformations $\mathrm{U}$ such that
$\mathrm{U}|_{T[1]\partial S}\in\mathcal{G}_{\mathsans{M}'}(\partial S)$. Finally, it makes the boundary term
of the $\sfQ(\Theta')$ with $\Theta'\in\mathfrak{G}_{\mathsans{M},\mathsans{M}'}(S)$ vanish identically ensuring that
the constraints $\sfQ(\Theta')\approx 0$ are equivalent to $\Phi\approx 0$ as required. 




The Poisson algebra \ceqref{surfchern2} bears striking formal similarities to the 2--dimen\-sional current algebra
known also as Kac--Moody algebra in mathematics \ccite{Kac:1990gs}, which occurs also in ordinary
CS theory in an analogous context \ccite{Banados:1994tn,Banados:1998ta,Banados:1998gg,Troessaert:2013fba}.
The structure of 2--cocycle $\sfc$ given in eq. \ceqref{holchern22}
shows this rather clearly. So, \ceqref{surfchern2} can be considered a higher current algebra hinged on a
Lie algebra crossed module rather than an ordinary Lie algebra.

The occurrence of a non trivial Poisson algebra of surface charges is a holographic feature that 4--dimensional CS theory
shares not only with its familiar 3--dimensional counterpart but also with other 4--dimensional theories, 
notably electrodynamics and general relativity. (See ref. \ccite{Compere:2018aar} for a review.)
4--dimensional CS theory might so provide an ideal testing ground for studying holography in 4 dimensions. 



\subsection{\textcolor{blue}{\sffamily Toward the edge field theory of 4--d Chern--Simons theory}}\label{subsec:canedge}

Gauge theories, including topological ones, 
on manifolds with boundaries normally exhibit emergent boundary degrees of
freedom called edge fields. In this subsection, we outline a canonical theory of the edge modes of 4--dimensional
CS theory and their physical symmetries, extending the corresponding analysis of the 3-dimensional theory
\ccite{Wen:1992vi,Carlip:2005zn,Afshar:2017okz,Donnelly:2016auv,Geiller:2019bti}.
Although there still remain basic issues to be clarified, \pagebreak as discussed shortly, 
it is already possible to
shed light on some of its main features. 
A more in--depth analysis will be provided elsewhere \ccite{Zucchini:2021inp}. 


We follow the method originally
worked out in ref. \ccite{Donnelly:2016auv}. 
In the canonical framework of  4--dimensional
CS theory, where the underlying 4--fold $M=\mathbb{R}^1\times S$
with $S$ a 3--fold, it is possible to construct an extended
phase space $\mathcal{P}_{\mathsans{M}}(S)$, which comprises extra degrees
of freedom localized at the boundary besides the interior ones.
Edge fields are in this way added to the original bulk gauge fields. 
Gauge invariance dictates the nature of the edge modes and the
form of their Poisson brackets by the requirement that
$\mathcal{P}_{\mathsans{M}}(S)$ be invariant under the group
$\mathcal{G}_{\mathsans{M}}(S)$ gauge transformations
of the original system, inclusive of those which are effective at
the boundary, on the physical shell. In addition, the edge
fields are acted upon by another set of Hamiltonian 
transformations. 
This form an infinite dimensional boundary symmetry group $\mathcal{K}_{\mathsans{M}}(\partial S)$
emerging as a consequence of the original gauge invariance. The
boundary symmetry and gauge transformations reciprocally commute.
Therefore, the charges generating $\mathcal{K}_{\mathsans{M}}(\partial S)$ are 
gauge invariant, i.e. physical boundary observables.

The way gauge invariance is implemented in the extended phase space is a bit subtle.
The bulk and edge symplectic 2--forms, $\sfUpsilon$ and $\sfUpsilon_\partial$, 
are not separately invariant under bulk gauge transformations. The gauge variation
of the former is however a boundary term which is cancelled by that of the latter.
For a mechanism like this to work out for a given bulk field content,
gauge transformation prescription and symplectic structure, the edge field content and its
gauge transformation properties and symplectic structure must be suitably adjusted.
There is no a priori guarantee that this is possible at all, but happily it is in our case. 

We shall now describe the above construction in more precise terms.
We utilize the derived formalism of subsect. \cref{subsec:higau}
for convenience and frame our analysis in the covariant canonical theory \ccite{Henneaux:1992ig}.
On the 3--fold $S$ with boundary $\partial S$, the interior fields are just
the bulk gauge fields $\Omega\in\mathcal{C}_{\mathsans{M}}(S)$ already considered. 
The boundary fields 
comprise the edge gauge fields $\Omega_\partial\in\mathcal{C}_{\mathsans{M}}(\partial S)$
and Stueckelberg fields $\mathrm{H}_\partial\in\mathcal{G}_{\mathsans{M}}(\partial S)$,
boundary gauge transformations promoted to dynamical edge fields.
The extended phase space
$\mathcal{P}_{\mathsans{M}}(S)$ is the submanifold of the product field manifold
$\mathcal{C}_{\mathsans{M}}(S)\times\mathcal{C}_{\mathsans{M}}(\partial S)
\times\mathcal{G}_{\mathsans{M}}(\partial S)$ defined by the condition
$\Omega|_{T[1]\partial S}=\Omega_\partial$ imposing the compatibility of the
bulk and edge gauge fields.

Under a bulk gauge transformation $\mathrm{U}\in\mathcal{G}_{\mathsans{M}}(S)$,
a bulk gauge field $\Omega$ transforms as in \ceqref{higau9/intr}, while
a edge gauge field $\Omega_\partial$ and Stueckelberg field $\mathrm{H}_\partial$
transform as 
\begin{align}
&\Omega_\partial{}^{\mathrm{U}}
=\Ad\mathrm{U}{}^{-1}(\Omega_\partial)+\mathrm{U}{}^{-1}\mathrm{dU},
\vphantom{\Big]}
\label{canedge1}
\\
&\mathrm{H}_\partial{}^{\mathrm{U}}=\mathrm{U}{}^{-1}\mathrm{H}_\partial,
\vphantom{\Big]}
\label{canedge2}
\end{align} 
where $\mathrm{U}$ is tacitly restricted $T[1]\partial S$. Notice that the expression of
$\Omega_\partial{}^{\mathrm{U}}$ is dictated by the gauge covariance of the compatibility requirement
$\Omega|_{T[1]\partial S}=\Omega_\partial$.

As we found in subsect. \cref{subsec:holchern},
the bulk field symplectic 2--form $\sfUpsilon$ and Poisson brackets $\{\cdot,\cdot\}$ are 
given by expressions \ceqref{holchern8} and \ceqref{holchern14}. The edge field
symplectic form $\sfUpsilon_\partial$ and Poisson brackets $\{\cdot,\cdot\}_\partial$ 
cannot be assigned independently 
being determined by the requirement of the full gauge invariance of the 
total symplectic 2--form $\sfUpsilon_{\mathrm{tot}}=\sfUpsilon+\sfUpsilon_\partial$
and Poisson brackets $\{\cdot,\cdot\}_{\mathrm{tot}}$ thereof  
on the shell \ceqref{cnholchern11}. This can be stated more precisely as follows. 

Let $\mathrm{gau}:\mathcal{P}_{\mathsans{M}}(S)\times\mathcal{G}_{\mathsans{M}}(S)\rightarrow\mathcal{P}_{\mathsans{M}}(S)$, 
$\mathrm{pr}:\mathcal{P}_{\mathsans{M}}(S)\times\mathcal{G}_{\mathsans{M}}(S)\rightarrow\mathcal{P}_{\mathsans{M}}(S)$ be 
the gau\-ge transformation action and projection maps, respectively.
On the shell \ceqref{cnholchern11}, the pull--backs $\mathrm{gau}^*\sfUpsilon$,
$\mathrm{pr}^*\sfUpsilon$ of the bulk symplectic 2--form $\sfUpsilon$ are found to differ by a boundary term. 
A straightforward analysis shows however that the edge symplectic 2--form $\sfUpsilon_\partial$ can be
defined such that
$\mathrm{gau}^*\sfUpsilon-\mathrm{pr}^*\sfUpsilon\approx
-\mathrm{gau}^*\sfUpsilon_\partial+\mathrm{pr}^*\sfUpsilon_\partial$. Setting
$\sfUpsilon_{\mathrm{tot}}=\sfUpsilon+\sfUpsilon_\partial$, one has therefore
\begin{equation}
\mathrm{gau}^*\sfUpsilon_{\mathrm{tot}}\approx\mathrm{pr}^*\sfUpsilon_{\mathrm{tot}}. 
\label{canedge3}
\end{equation}
In this sense, $\sfUpsilon_{\mathrm{tot}}$ is gauge invariant.

The above result can be understood intuitively as follows. A Stueckelberg field
$\mathrm{H}_\partial\in\mathcal{G}_{\mathsans{M}}(\partial S)$ can be extended non uniquely in
the interior of $S$ to a field $\mathrm{H}\in\mathcal{G}_{\mathsans{M}}(S)$.
If $\mathrm{H}$ is an extension of $\mathrm{H}_\partial$ and $\mathrm{U}\in\mathcal{G}_{\mathsans{M}}(S)$
is a gauge transformation, then $\mathrm{H}^{\mathrm{U}}=\mathrm{U}^{-1}\mathrm{H}$
is an extension of the gauge transform $\mathrm{H}_\partial{}^{\mathrm{U}}$ of $\mathrm{H}_\partial$.
We view $\mathrm{H}^{\mathrm{U}}$ as the gauge transform of $\mathrm{H}$,
in keeping with \ceqref{canedge2}.
If $\Omega\in\mathcal{C}_{\mathsans{M}}(S)$ is a bulk gauge field,
its gauge transform $\Omega^{\mathrm{H}}$ depends on the extension $\mathrm{H}$ used, but for a fixed
choice of $\mathrm{H}$ it is gauge invariant. The 2--form $\sfUpsilon_{\mathrm{tot}}$ obtained from
$\sfUpsilon$ by replacing $\Omega$ with $\Omega^{\mathrm{H}}$ in \ceqref{holchern8}
is by construction gauge invariant. On the shell \ceqref{cnholchern11}, 
$\sfUpsilon_{\mathrm{tot}}$ turns out to equal the sum of $\sfUpsilon$ and a boundary term $\sfUpsilon_\partial$,
depending on the edge fields $\Omega_\partial$, $\mathrm{H}_\partial$ but not on the chosen extension $\mathrm{H}$
of $\mathrm{H}_\partial$, which is precisely the edge symplectic 2--form.

The procedure described in the previous paragraph provides a practical way of computing
the edge symplectic 2--form $\sfUpsilon_\partial$. 
The expression of $\sfUpsilon_\partial$ that we find reads as 
\begin{align}
\sfUpsilon_\partial
&=\frac{k}{2\pi}\int_{T[1]\partial S}\varrho_{\partial S}
\left[\left(\sfdelta_\partial\mathrm{H}_\partial\mathrm{H}_\partial{}^{-1},\sfdelta_\partial\Omega_\partial\right)
\right.
\vphantom{\Big]}
\label{canedge4}
\\
&\left.
\hspace{1.75cm}-\tfrac{1}{2}\left(\dd(\sfdelta_\partial\mathrm{H}_\partial\mathrm{H}_\partial{}^{-1}),
\sfdelta_\partial\mathrm{H}_\partial\mathrm{H}_\partial{}^{-1}\right)
-\tfrac{1}{2}\left(\Omega_\partial,[\sfdelta_\partial\mathrm{H}_\partial\mathrm{H}_\partial{}^{-1},
\sfdelta_\partial\mathrm{H}_\partial\mathrm{H}_\partial{}^{-1}]\right)\right].
\vphantom{\Big]}
\nonumber
\end{align}
From this expression, the edge Poisson brackets $\{\cdot,\cdot\}_\partial$ can be determined.
In particular, it is found that edge Poisson brackets of functionals of the Stueckelberg field
$\mathrm{H}_\partial$ only Poisson commute. 

As we found in subsect. \cref{subsec:holchern}, the bulk gauge transformation action on bulk fields
is Hamiltonian. For an infinitesimal gauge transformation $\Theta\in\mathfrak{G}_{\mathsans{M}}(S)$,
the associated bulk Hamiltonian functional $\sfQ(\Theta)$ is given by eq. \ceqref{holchern15}
and the gauge transformation action on bulk fields can be expressed through the Poisson
bracket \ceqref{holchern20}. 
Remarkably, the bulk gauge transformation action on edge fields is Hamiltonian as well.
For the gauge transformation $\Theta$, 
the associated edge Hamiltonian functional $\sfQ_\partial(\Theta)$ reads as  
\begin{equation}
\sfQ_\partial(\Theta)
=\frac{k}{2\pi}\int_{T[1]\partial S}\varrho_{\partial S}\left(\Omega_\partial,\Theta\right).
\label{canedge5}
\end{equation}
and the gauge transformation action on edge fields can be cast as 
\begin{equation}
\left\{\sfQ_\partial(\Theta),\sfF_\partial\right\}_\partial=\delta_{\Theta}\sfF_\partial,
\label{canedge6}
\end{equation}
where $\sfF_\partial$ is an edge field functionals and $\delta_\Theta$ denotes variation with respect the infinitesimal form
of the gauge transformation \ceqref{canedge1}, \ceqref{canedge2}.

The bulk Hamiltonian functionals $\sfQ(\Theta)$ the Poisson bracket algebra \ceqref{holchern21}
featuring a central extension with $\mathfrak{G}_{\mathsans{M}}(S)$--2--cocycle
$\sfc$ given by \ceqref{holchern22}. 
The edge Hamiltonian functionals $\sfQ_\partial(\Theta)$ obey a totally 
similar Poisson bracket algebra, 
\begin{equation}
\left\{\sfQ_\partial(\Theta),\sfQ_\partial(\Theta')\right\}_\partial
=\sfQ_\partial([\Theta,\Theta'])-\frac{k}{2\pi}\sfc(\Theta,\Theta'), 
\label{canedge7}
\end{equation}
with the same central extension up to sign. 

For an infinitesimal gauge transformation $\Theta$, the total bulk plus edge Hamiltonian
functional is 
\begin{equation}
\sfQ_{\mathrm{tot}}(\Theta)=\sfQ(\Theta)+\sfQ_\partial(\Theta)
=\int_{T[1]S}\varrho_S\left(\Phi,\Theta\right).
\label{canedge8}
\end{equation}
By virtue of \ceqref{holchern21}, \ceqref{canedge7}, the Hamiltonian functionals $\sfQ_{\mathrm{tot}}(\Theta)$
obey a centerless Poisson bracket algebra, 
\begin{equation}
\left\{\sfQ_{\mathrm{tot}}(\Theta),\sfQ_{\mathrm{tot}}(\Theta)\right\}_{\mathrm{tot}}
=\sfQ_{\mathrm{tot}}([\Theta,\Theta']). 
\label{canedge9}
\end{equation}
Thanks to \ceqref{canedge8} and \ceqref{canedge9}, the physical on--shell condition \ceqref{cnholchern11} can
hence be consistently cast in the form 
\begin{equation}
\sfQ_{\mathrm{tot}}(\Theta)\approx 0
\label{canedge10}
\end{equation}
with non need to impose any boundary conditions on either the bulk gauge field $\Omega$ or $\Theta$,
as we were forced to in subsect. \cref{subsec:holchern}. 

As we outlined at the beginning of this subsection, the extended phase space enjoys a second
physical non gauge surface symmetry. We now describe it in greater detail. A surface transformation is
specified by an element $\mathrm{T}_\partial\in\mathcal{G}_{\mathsans{M}}(\partial S)$. It acts on the
bulk gauge fields $\Omega$ trivially
and on the edge fields as follows, 
\begin{align}
&\Omega_\partial{}^{\mathrm{T}_\partial}=\Omega_\partial,
\vphantom{\Big]}
\label{canedge11}
\\
&\mathrm{H}_\partial{}^{\mathrm{T}_\partial}=\mathrm{H}_\partial\mathrm{T}_\partial.
\vphantom{\Big]}
\label{canedge12}
\end{align}
As for the bulk gauge symmetry, the form of $\Omega_\partial{}^{\mathrm{T}_\partial}$ is dictated by the
surface symmetry covariance of the compatibility requirement $\Omega|_{T[1]\partial S}=\Omega_\partial$.

The surface symmetry turns out to be Hamiltonian.
For an infinitesimal surface transformation $\Lambda_\partial\in\mathfrak{G}_{\mathsans{M}}(\partial S)$,
the associated edge Hamiltonian functional $\sfC_\partial(\Lambda_\partial)$ has the simple form
\begin{equation}
\sfC_\partial(\Lambda_\partial)=-\frac{k}{2\pi}\int_{T[1]\partial S}\varrho_{\partial S}
\left(\Omega_\partial{}^{\mathrm{H}_\partial},\Lambda_\partial\right)
\label{canedge13}
\end{equation}
and the surface transformation action on edge fields takes the expected form 
\begin{equation}
\left\{\sfC_\partial(\Lambda_\partial),\sfF_\partial\right\}_\partial=\delta_{\Lambda_\partial}\sfF_\partial,
\label{canedge14}
\end{equation}
where $\sfF_\partial$ is an edge field functionals and $\delta_{\Lambda_\partial}$ denotes variation with
respect the infinitesimal form of the transformations \ceqref{canedge11}, \ceqref{canedge12}. 

The surface edge Hamiltonian functionals $\sfC_\partial(\Lambda_\partial)$ obey a Poisson bracket algebra
formally identical to that of the gauge edge Hamiltonian functionals $\sfQ_\partial(\Theta)$ shown in
eq. \ceqref{canedge7}. One has indeed, 
\begin{equation}
\left\{\sfC_\partial(\Lambda_\partial),\sfC_\partial(\Lambda_\partial{}')\right\}_\partial
=\sfC_\partial([\Lambda_\partial,\Lambda_\partial{}'])-\frac{k}{2\pi}\sfc(\Lambda_\partial,\Lambda_\partial{}')
\label{canedge15}
\end{equation}
where the $\mathfrak{G}_{\mathsans{M}}(\partial S)$--2--cocycle $\sfc$ is given again
by expression \ceqref{holchern22} with $\Theta,~\Theta'$ replaced by $
\Lambda_\partial,~\Lambda_\partial{}'$ throughout.

An important property of the surface Hamiltonians $\sfC_\partial(\Lambda_\partial)$
is their Poisson commuting with the gauge Hamiltonians $\sfQ_\partial(\Theta)$, 
\begin{equation}
\left\{\sfQ_\partial(\Theta),\sfC_\partial(\Lambda_\partial)\right\}_\partial=0.
\label{canedge16}
\end{equation}
That a relation like the above must hold is evident also from the fact that $\sfC_\partial(\Lambda_\partial)$
is defined in \ceqref{canedge13} through the gauge invariant combination $\Omega_\partial{}^{\mathrm{H}_\partial}$.
Relation \ceqref{canedge16} proves further the physical nature of the surface symmetry. The 
$\sfC_\partial(\Lambda_\partial)$ are the associated charges. Surface symmetry is therefore infinite dimensional.

By \ceqref{canedge15}, the surface charges are in involution if the
infinitesimal surface transformations $\Lambda_\partial\in\mathfrak{G}_{\mathsans{M}}(\partial S)$
are restricted in such a way to make the central term vanish.
An inspection of \ceqref{holchern22} shows readily
that there are several ways in which this can be achieved. 
For instance, we may require that $\Lambda_\partial\in\mathfrak{G}_{\mathsans{M}'}(\partial S)$, 
where $\mathsans{M}'$ is an isotropic crossed submodule of $\mathsans{M}$ (cf. subsect. \cref{subsec:crosumo}),
or that $\Lambda_\partial\in\mathfrak{G}_{\mathsans{M}\mathrm{cl}}(\partial S)$, the Lie subalgebra of
$\mathfrak{G}_{\mathsans{M}}(\partial S)$ spanned by those elements $\Lambda_\partial$ which satisfy
the equation $\dd\Lambda_\partial=0$. 

The above account of the edge sector of 4--dimensional CS theory
is still in\-complete.  There remains a basic problem 
to be solved: the lack of a Lagrangian and Hamiltonian
formulation describing the dynamics of edge fields, if any. 
In fact, the corresponding analysis for 3--dimensional CS theory
(see e.g. \ccite{Geiller:2019bti} for a discussion of this point)
shows that the edge dynamics of topological gauge theories may be non
trivial as that of non topological ones. Whether this is the case also
for out 4--dimensional model is an issue deserving further investigation.
\vfil


\subsection{\textcolor{blue}{\sffamily Covariant Schroedinger quantization }}\label{subsec:edge}

In this final subsection, we study the covariant Schroedinger quantization of 4--dimen\-sional CS theory.
Although there still are points requiring clarification and
a more in--depth analysis, in particular in connection to the
edge field theory of the model discussed in subsect. \cref{subsec:canedge}, 
it is still possible to elucidate its outlines to a considerable extent.  

The covariant Schroedinger quantization scheme of 4--dimensional CS theory is based on its covariant phase space.  
This is the space $\mathcal{C}_{\mathsans{M}}(\partial M)$ of boundary gauge field configurations 
$\omega_\partial$, $\varOmega_\partial$. 
A straightforward analysis totally analogous to that of subsect. \cref{subsec:surfchern} shows that 
the symplectic form of $\sfUpsilon_\partial$, the associated Poisson bracket $\{\cdot,\cdot\}_\partial$,
the Hamiltonians $\sfQ_\partial(\theta_\partial, \varTheta_\partial)$
of the boundary infinitesimal gauge transformations $\theta_\partial$, $\varTheta_\partial$,
their Poisson action and Poisson bracket algebra are given respectively 
by relations \ceqref{holchern16}, \ceqref{holchern18}, \ceqref{holchern19},
\ceqref{holchern19/cmp}, \ceqref{holchern23}, \ceqref{holchern24} and \ceqref{holchern21/c}
with $S=\partial M$, $\partial S=\emptyset$, $\omega$, $\varOmega$ replaced by $\omega_\partial$, $\varOmega_\partial$,
$\theta$, $\varTheta$ by $\theta_\partial$, $\varTheta_\partial$ and $\sfc=0$. 
In this case, so, the $\sfQ_\partial$ Poisson algebra features no central extension.  

The Hilbert space of 4--dimensional CS theory consists of 
complex wave functionals on $\mathcal{C}_{\mathsans{M}}(\partial M)$
obeying a polarization condition, that is annihilated by the vector fields
on $\mathcal{C}_{\mathsans{M}}(\partial M)$ belonging to an integrable Lagrangian distribution
of $T\mathcal{C}_{\mathsans{M}}(\partial M)$. 
There are two obvious choices of the distribution. 
The first is spanned by the vector fields $\delta/\delta\varOmega_\partial$ 
and produces wave functionals $\Psi(\omega_\partial)$ of $\omega_\partial$.
The second is generated by the vector fields $\delta/\delta\omega_\partial$ 
and leads to wave functionals $\Psi(\varOmega_\partial)$ of $\varOmega_\partial$. 
Although these two alternatives are defined in a seemingly symmetrical manner, only the first one is viable 
once gauge transformation is implemented. In fact, inspection of \ceqref{higau12}, \ceqref{higau13}
shows that under gauge transformation $\omega_\partial$ does not mix with $\varOmega_\partial$
whilst $\varOmega_\partial$ does with $\omega_\partial$.

In the following, we therefore consider only the first choice of polarization. In this 
canonical quantum set--up, the Hilbert space inner product reads as 
\begin{equation}
\langle\varPsi_1,\varPsi_2\rangle=\int\mathcal{D}\omega_\partial\,\varPsi_1(\omega_\partial)^*\,\varPsi_2(\omega_\partial),
\label{edge1}
\end{equation}
where $\mathcal{D}\omega_\partial$ is a suitable formal functional measure. Further,
the operators $\widehat{\omega}_\partial$, $\widehat{\varOmega}_\omega$ quantizing
$\omega_\partial$, $\varOmega_\omega$ take the familiar form
\begin{align}
&\widehat{\omega}_\partial=\omega_\partial\,\cdot\,,
\vphantom{\Big]}
\label{edge2}
\\
&\widehat{\varOmega}_\partial=-\frac{2\pi i}{k}\frac{\delta}{\delta\omega_\partial}.
\vphantom{\Big]}
\label{edge3}
\end{align}
They are formally selfadjoint with respect to the inner product structure \eqref{edge1}.

The infinitesimal gauge transformation 
Hamiltonians $\sfQ_\partial(\theta_\partial, \varTheta_\partial)$ constitute a set of first class 
covariant phase space functionals. Since the $\sfQ_\partial(\theta_\partial, \varTheta_\partial)$ are linear in
the curvature components $\phi_\partial$, $\varPhi_\partial$, the physical phase space is defined by the constraints
$\sfQ_\partial(\theta_\partial, \varTheta_\partial)\approx 0$. At the quantum level, the constraints translate
into a set of linear conditions the wave functionals must satisfy,  
\begin{equation}
\widehat{\sfQ}_\partial(\theta_\partial, \varTheta_\partial)\Psi=0.
\label{edge4}
\end{equation}
Here, the $\widehat{\sfQ}_\partial(\theta_\partial, \varTheta_\partial)$ are operators quantizing
the phase functionals $\sfQ_\partial(\theta_\partial, \varTheta_\partial)$. The quantization must
be such that the commutator algebra 
\begin{equation}
\big[\widehat{\sfQ}_\partial(\theta_\partial,\varTheta_\partial),
\widehat{\sfQ}_\partial(\theta_\partial{}',\varTheta_\partial{}')\big]
=i\hfpt\widehat{\sfQ}_\partial([\theta_\partial,\theta_\partial{}'],
\sdot\mu\sdot(\theta_\partial,\varTheta_\partial{}')-\sdot\mu\sdot(\theta_\partial{}',\varTheta_\partial))
\label{edge5}
\end{equation}
is obeyed in conformity with \ceqref{holchern21/c}. This guarantees in particular 
the consistency of the conditions \ceqref{edge1}.

Conditions \ceqref{edge1} imply that the wave functional $\varPsi$ 
satisfy a pair of functional differential equations,  
\begin{align}
&\int_{T[1]\partial M}\varrho_{\partial M}\left\langle d\omega_\partial+\frac{1}{2}[\omega_\partial,\omega_\partial]
+\frac{2\pi i}{k}\dot\tau\Big(\frac{\delta}{\delta \omega_\partial}\Big),
\varTheta_\partial\right\rangle\varPsi(\omega_\partial)=0,
\vphantom{\Big]}
\label{edge6}
\\
&\int_{T[1]\partial M}\varrho_{\partial M}\left\langle\theta_\partial,
-\frac{2\pi i}{k}\left[d\frac{\delta}{\delta \omega_\partial}
+\sdot\mu\sdot\Big(\omega_\partial,\frac{\delta}{\delta \omega_\partial}\Big)\right]
\right\rangle\varPsi(\omega_\partial)=0.
\vphantom{\Big]}
\label{edge7}
\end{align}
On account of \ceqref{higau23}, these identities imply that 
\begin{equation}
\delta_{\theta_\partial,\varTheta_\partial}\varPsi(\omega_\partial)
=\frac{ik}{2\pi}\int_{T[1]\partial M}\varrho_{\partial M}\left\langle d\omega_\partial
+\tfrac{1}{2}[\omega_\partial,\omega_\partial],\varTheta_\partial\right\rangle\varPsi(\omega_\partial).
\label{edge8}
\end{equation}
Therefore, the variation of $\varPsi$   
under a finite boundary gauge transformation
$u_\partial$, $U_\partial$ is given by a multiplicative factor
\begin{equation}
\varPsi(\omega_\partial{}^{u_\partial,U_\partial})=
\exp(i\sfW\mhfpt\sfZ_\partial(u_\partial,U_\partial;\omega_\partial))\varPsi(\omega_\partial).
\label{edge9}
\end{equation}
By the very structure of this relation, the functional $\sfW\mhfpt\sfZ_\partial(u_\partial,U_\partial;\omega_\partial)$ 
appearing in it is a $\mathsans{U}(1)$--valued cocycle for the boundary gauge transformation action on the degree 1
boundary gauge field component, 
\begin{multline}
\sfW\mhfpt\sfZ_\partial(u_\partial v_\partial,U_\partial+\mu\sdot(u_\partial,V_\partial);\omega_\partial)
\\
=\sfW\mhfpt\sfZ_\partial(u_\partial,U_\partial;\omega_\partial)
+\sfW\mhfpt\sfZ_\partial(v_\partial,V_\partial;\omega_\partial{}^{u_\partial,U_\partial})\quad \text{mod $2\pi\mathbb{Z}$}.
\label{edge10}
\end{multline}
To reproduce the infinitesimal variation \eqref{edge8}, $\sfW\mhfpt\sfZ_\partial(u_\partial,U_\partial;\omega_\partial)$ must
further satisfy the normalization condition 
\begin{equation}
\delta_{\theta_\partial,\varTheta_\partial}\sfW\mhfpt\sfZ_\partial(u_\partial,U_\partial;\widetilde{\omega}_\partial)
=\frac{k}{2\pi}\int_{T[1]\partial M}\varrho_{\partial M}
\left\langle d\omega_\partial+\tfrac{1}{2}[\omega_\partial,\omega_\partial],\varTheta_\partial\right\rangle,
\label{edge11}
\end{equation}
where the tilde indicates that $\delta_{\theta_\partial,\varTheta_\partial}$ is inert on $\omega_\partial$.
Properties \ceqref{edge10}, \ceqref{edge11} determine the cocycle $\sfW\mhfpt\sfZ_\partial$ up to a trivial cocycle,
\begin{multline}
\sfW\mhfpt\sfZ_\partial(u_\partial,U_\partial;\omega_\partial)
=\frac{k}{4\pi}\int_{T[1]\partial M}\varrho_{\partial M}\Big[
\left\langle\dot\tau(U_\partial), d U_\partial+\tfrac{1}{3}\left[U_\partial,U_\partial\right]\right\rangle 
\\
+\left\langle\omega_\partial,[U_\partial,U_\partial]\right\rangle
+2\left\langle d\omega_\partial+\tfrac{1}{2}[\omega_\partial,\omega_\partial],U_\partial\right\rangle\Big]
+\sfK_\partial(\omega_\partial{}^{u_\partial,U_\partial})
-\sfK_\partial(\omega_\partial),
\label{edge12}
\end{multline}
where $\sfK_\partial(\omega_\partial)$ is a local boundary functional
which cannot be determined in the present method. We expect $\sfK_\partial$ to be generated by
quantum effects as we shall discuss in greater detail momentarily.

At this point, it is important to remark that the above Schroedinger quantization scheme
of 4--dimensional CS theory mirrors closely the Bargmann one
used in ordinary 3--dimensional CS theory \ccite{Axelrod:1989xt}.
In particular, relations \ceqref{edge6}, \ceqref{edge7} are the higher counterpart of the familiar
WZNW Ward identities, the cocycle $\sfW\mhfpt\sfZ_\partial$ appearing in \ceqref{edge9}
is a higher gauged WZNW functional and the cocycle relation \ceqref{edge10} is just a higher version of the
Polyakov--Wiegmann identity \ccite{Polyakov:1984et}.
However, unlike its ordinary counterpart, the WZNW functional $\sfW\mhfpt\sfZ_\partial$ is fully topological,
being independent on any background metric structure. Further, it depends only on the second component of the 
underlying gauge transformation, $U_\partial$, but not on the first one, $u_\partial$.

The issue of the cohomological triviality of the cocycle functional $\sfW\mhfpt\sfZ_\partial$,
the property that $\sfW\mhfpt\sfZ_\partial(\omega_\partial)=\sfS_\partial(\omega_\partial{}^{u_\partial,U_\partial})
-\sfS_\partial(\omega_\partial)$ mod $2\pi\mathbb{Z}$ \pagebreak 
for a local boundary functional $\sfS_\partial(\omega_\partial)$,  
is relevant. When it occurs, the modified wave functional 
\begin{equation}
\Psi'(\omega_\partial)=\ee^{i\sfS_\partial(\omega_\partial)}\Psi(\omega_\partial)
\label{edge13}
\end{equation}
is fully gauge invariant so that
\begin{equation}
\sfW\mhfpt\sfZ_\partial{}'=0 \quad \text{mod $2\pi\mathbb{Z}$}.
\label{edge14}
\end{equation}
It is interesting to illustrate this point by some examples.
Consider the case where $\mathsans{M}=\INN\mathsans{G}$ is the
inner automorphism crossed module of a Lie group $\mathsans{G}$
with an invariant symmetric non singular bilinear form $\langle\cdot,\cdot\rangle$
on $\mathfrak{g}$. Then, a simple calculation 
shows that the WZNW functional $\sfW\mhfpt\sfZ_\partial$ can be cast as 
\begin{multline}
\sfW\mhfpt\sfZ_\partial(u_\partial,U_\partial;\omega_\partial)
=\frac{k}{24\pi}\int_{T[1]\partial M}\varrho_{\partial M}
\left\langle d u_\partial u_\partial{}^{-1},
\left[d u_\partial u_\partial{}^{-1},d u_\partial u_\partial{}^{-1}\right]\right\rangle 
\\
+\sfC\mhfpt\sfS_\partial(\omega_\partial{}^{u_\partial,U_\partial})-\sfC\mhfpt\sfS_\partial(\omega_\partial)
+\sfK_\partial(\omega_\partial{}^{u_\partial,U_\partial})
-\sfK_\partial(\omega_\partial), 
\label{edge15}
\end{multline}
where $\sfC\mhfpt\sfS_\partial(\omega_\partial)$ is the boundary CS action 
\begin{equation}
\sfC\mhfpt\sfS_\partial(\omega_\partial)
=\frac{k}{4\pi}\int_{T[1]\partial M}\varrho_{\partial M}
\left\langle \omega_\partial, d \omega_\partial
+\tfrac{1}{3}\left[\omega_\partial,\omega_\partial\right]\right\rangle.
\vphantom{\ul{\ul{\ul{\ul{g}}}}}
\label{edge16}
\end{equation}
If $\mathsans{G}$ is a compact semisimple Lie group, $\langle\cdot,\cdot\rangle$ is the suitably normalized 
Killing form of $\mathfrak{g}$ and $k$ is an integer, then the first term in the right hand side of \ceqref{edge15}
vanishes mod $2\pi\mathbb{Z}$ and so $\sfW\mhfpt\sfZ_\partial$ is cohomologically trivial.  
When $\mathsans{M}=\AD^*\mathsans{G}$ is the coadjoint action crossed module of $\mathsans{G}$
with the canonical duality pairing
or a generic crossed module, $\sfW\mhfpt\sfZ_\partial$ is cohomologically non trivial.

In 4--dimensional CS theory on a 4--fold $M$, a wave functional $\Psi_M(\omega_\partial)$ is yielded 
by path integration over all gauge field configuration $\omega$, $\varOmega$ such that
$\omega|_{T[1]\partial M}=\omega_\partial$. Formally, one has 
\begin{equation}
\Psi_M(\omega_\partial)=\int_{\omega|_{T[1]\partial M}=\omega_\partial}\mathcal{D}\omega\mathcal{D}\varOmega
\ee^{i\sfC\hspace{-.75pt}\sfS'(\omega,\varOmega)},
\label{edge17}
\end{equation}
leaving aside such relevant issues such as normalization and gauge fixing. 
The consistent quantization of the theory requires however
that the CS action $\sfC\hspace{-.75pt}\sfS'$
employed be differentiable in the sense of refs. \ccite{Regge:1974zd,Benguria:1976in}
under the boundary con\-dition enforced. \vspace{2mm}


As explained in subsect. \cref{subsec:4dchern}, 
the variation $\sfdelta\sfC\hspace{-.75pt}\sfS$ of the CS action $\sfC\hspace{-.75pt}\sfS$,
given by eq. \ceqref{holchern0}, exhibits a boundary contribution
showing that $\sfC\hspace{-.75pt}\sfS$ is not differentiable as it is. To obtain a differentiable CS action,
it is necessary to $i)$ replace $\sfC\hspace{-.75pt}\sfS$ by a modified the CS action 
$\sfC\hspace{-.75pt}\sfS'$ obtained by adding to $\sfC\hspace{-.75pt}\sfS$ a
suitable boundary term $\varDelta\sfC\hspace{-.75pt}\sfS$
as in  eq. \ceqref{holchern-4} and $ii)$ impose a boundary condition on the gauge field
components $\omega$, $\varOmega$ such that the variation $\sfdelta\sfC\hspace{-.75pt}\sfS'$
of $\sfC\hspace{-.75pt}\sfS'$ is given by the bulk contribution in the right hand side of eq.
\ceqref{holchern0} only, once the boundary condition is enforced.
For the boundary condition used in \ceqref{edge17}, 
the expression of the appropriate boundary term $\varDelta\sfC\hspace{-.75pt}\sfS$
is readily found, 
\begin{equation}
\varDelta\sfC\hspace{-.75pt}\sfS(\omega,\varOmega)
=\frac{k}{4\pi}\int_{T[1]\partial M}\varrho_{\partial M} \left\langle\omega,\varOmega\right\rangle.
\label{edge18}
\end{equation}
With this choice, the modified CS action $\sfC\hspace{-.75pt}\sfS'$ is given by
the right hand side of \ceqref{4dchern2} with the boundary term removed. 
The boundary part of the variation $\sfdelta\sfC\hspace{-.75pt}\sfS'$ of
$\sfC\hspace{-.75pt}\sfS'$ then turns out to be 
$\sfdelta\sfC\hspace{-.75pt}\sfS'{}_{\mathrm{boundary}}=\frac{k}{2\pi}
\int_{T[1]\partial M}\varrho_{\partial M}\left\langle\sfdelta\omega,\varOmega\right\rangle$.
This vanishes when the boundary condition $\omega|_{T[1]\partial M}=\omega_\partial$ with
assigned $\omega_\partial$ is impos\-ed,
rendering $\sfC\hspace{-.75pt}\sfS'$ differentiable
as required. Here, it is appropriate to recall that the above procedure has a
well--known counterpart in 3--dimensional CS theory. In that case, however, the boundary
term depends on the choice of a conformal structure on the 2--dimensional boundary, since the two
boundary gauge field 1--form components are canonically conjugated \ccite{Axelrod:1989xt}.
In the present case, conversely, since $\omega$ is canonically conjugate to $\varOmega$, there is no need
to introduce a new structure in the theory and so the boundary term is fully topological.


In sect. \cref{subsec:levelchern},  we found that 
the gauge variation $\sfA'$ of the modified CS action $\sfC\hspace{-.75pt}\sfS'$ is given by \ceqref{4dchern-5} in
terms of the gauge variations $\sfA$ and $\varDelta\sfA$ of the CS action $\sfC\hspace{-.75pt}\sfS$ and the boundary
term $\varDelta\sfC\hspace{-.75pt}\sfS$. Using relations \ceqref{4dchern5} for $\sfA$ and computing
$\varDelta\sfC\hspace{-.75pt}\sfS$ from \ceqref{edge18} employing \ceqref{higau12}, \ceqref{higau13}, it is
straightforward to obtain 
\begin{multline}
\sfA'(\omega;u,U)
=\frac{k}{4\pi}\int_{T[1]\partial M}\varrho_{\partial M}\Big[
\left\langle\dot\tau(U), dU+\tfrac{1}{3}\left[U,U\right]\right\rangle 
\\
+\left\langle\omega,[U,U]\right\rangle
+2\left\langle d\omega+\tfrac{1}{2}[\omega,\omega],U\right\rangle\Big]. 
\label{edge19}
\end{multline} 
Comparing \ceqref{edge12} and \ceqref{edge19}, we find that $\sfA'(\omega;u,U)$ reproduces the first term in the
right hand side of \ceqref{edge12} when $u|_{T[1]\partial M}=u_\partial$, $U|_{T[1]\partial M}=U_\partial$,
$\omega|_{T[1]\partial M}=\omega_\partial$. On account of \ceqref{edge17}, this shows that such term 
is the full classical contribution to $\sfW\mhfpt\sfZ_\partial$.
The remaining terms, therefore, if they arise at all, are of a quantum nature.

\vfil\eject

\section{\textcolor{blue}{\sffamily Sample applications}}\label{sec:aplc}

In this section, we illustrate a few field theoretic models which are interesting instances of 
4--dimensional CS theory: the toric and the Abelian projection models.
Here, our aim is showing by direct construction that 
4--dimensional CS theory can find explicit realizations related to
various areas of theoretical research on one hand and prepare the ground
for a more systematic study of the models presented to appear 
in future work on the other. So, this last section should also provide an outlook 
for perspective applications of the theory which we have developed.


\subsection{\textcolor{blue}{\sffamily The toric 4--dimensional CS model}}\label{subsec:toricth}

Dijkgraaf--Witten theory  \cite{Dijkgraaf:1989pz} is known to classify symmetry protected topological
phases without fermions in low dimension. 
4--dimensional Dijkgraaf--Witten theory in turn has a continuum description in terms of 
toric 4--dimensional CS theory \ccite{Kapustin:2014gua,Gaiotto:2014kfa,Delcamp:2019fdp}.

The Lie group crossed module of the toric model is the toric crossed module 
$\mathsans{M}=(\mathsans{T},\mathsans{T},\varsigma,\varpi)$, where $\mathsans{T}$ is a torus,
that is a ompact connected Abelian group, 
$\varsigma:\mathsans{T}\rightarrow\mathsans{T}$ is an endomorphism
and $\varpi:\mathsans{T}\times\mathsans{T}\rightarrow\mathsans{T}$ is the trivial action of 
$\mathsans{T}$ on itself. The associated Lie algebra crossed module is
$\mathfrak{m}=(\mathfrak{t},\mathfrak{t},\dot\varsigma,\sdot\varpi\sdot)$.

The torus $\mathsans{T}$ can be represented as the quotient $\mathfrak{t}/\varLambda$,
where $\varLambda$ is the integral lattice of $\mathsans{T}$ defined by the property that
$\ee^l=1_{\mathsans{T}}$ for $l\in\mathfrak{t}$. $\varsigma$ being an endomorphism of $\mathsans{T}$
entails that $\dot\varsigma:\varLambda\rightarrow\varLambda$ is a lattice endomorphism.

Let $\langle\cdot,\cdot\rangle:\mathfrak{t}\times\mathfrak{t}\rightarrow\mathbb{R}$ be a
symmetric non singular bilinear form on $\mathfrak{t}$. Since the Lie algebra $\mathfrak{t}$
is Abelian and the action $\sdot\varpi\sdot$ is trivial, the form $\langle\cdot,\cdot\rangle$
satisfies the invariance property \ceqref{crmodinv2/rep} trivially. $\langle\cdot,\cdot\rangle$
satisfies the symmetry property \ceqref{crmodinv1/rep} if the endomorphism $\dot\varsigma$ is
symmetric with respect to $\langle\cdot,\cdot\rangle$, which we assume henceforth to be the case. 

It is natural to suppose that the form $\langle\cdot,\cdot\rangle$ restricts to a lattice bilinear form
$\langle\cdot,\cdot\rangle:\varLambda\times\varLambda\rightarrow\mathbb{Z}$, which we denote by the same
symbol for the sake of simplicity. $\dot\varsigma:\varLambda\rightarrow\varLambda$ is then a symmetric
lattice endomorphism. 

The toric higher gauge field components are $\omega\in\Omega^1(M,\mathfrak{t})$,
$\varOmega\in\Omega^2(M,\mathfrak{t})$. The toric CS action is  
\begin{equation}
\sfC\hspace{-.75pt}\sfS(\omega,\varOmega)
=\frac{k}{2\pi}\int_M\left\langle d\omega 
  -\tfrac{1}{2}\dot\varsigma(\varOmega),\varOmega\right\rangle
-\frac{k}{4\pi}\int_{\partial M}\left\langle\omega,\varOmega\right\rangle
\label{toric1}
\end{equation}
A toric gauge transformation consists of a map $u\in\Map(M,\mathsans{T})$, $U\in\Omega^1(M,\mathfrak{t})$. 
Its action on the toric gauge field is according to eqs. \ceqref{higau12}, \ceqref{higau13}, 
\begin{align}
&\omega^{u,U}=\omega+duu^{-1}+\dot\varsigma(U),
\vphantom{\Big]}
\label{toric2}
\\
&\varOmega^{u,U}=\varOmega+dU.
\vphantom{\Big]}
\label{toric3}
\end{align}
These transformations are gauge symmetries only if $\partial M$ is empty as we have seen.

If we identify $\mathfrak{t}\simeq\mathbb{R}^r$ and $\varLambda\simeq\mathbb{Z}^r$ for some integer $r$, then we have
\begin{equation}
\langle x,y\rangle=\sss_{i,j=1}^rK_{ij}x_iy_j,
\label{toric4}
\end{equation}
where $K$ is an $r\times r$ matrix of the form
\begin{equation}
K_{ij}=n_i\delta_{ij}
\label{toric5}
\end{equation}
with $n_i\in\mathbb{Z}$, $n_i>0$. The endomorphism $\dot\varsigma$ is similarly expressed as 
\begin{equation}
\dot\varsigma(x)=\sss_{j=1}^rs_{ij}x_j,
\label{toric6}
\end{equation}
where $s$ is a certain $r\times r$ matrix. Requiring that is a symmetric lattice endomorphism
leads to the property that $s_{ij}\in\mathbb{Z}$ and the condition
\begin{equation}
n_is_{ij}-n_js_{ji}=0. 
\label{toric7}
\end{equation}
In the toric models of ref. \ccite{Kapustin:2014gua,Gaiotto:2014kfa}, one has 
\begin{equation}
n_is_{ij}=-p_{ij}\mathrm{lcm}(n_i,n_j)-\chi_ip_{ii}n_i\delta_{ij},
\label{toric8}
\end{equation}
where $p_{ij}$ is a symmetric integer matrix and $\chi_i=0$ or $1$ according to whether
$n_i$ is even or odd respectively. It is immediately checked that the matrix $t$ furnished
by \ceqref{toric8} is integer and satisfies \ceqref{toric7}. 

The Lie group crossed submodules of $\mathsans{M}$, which as we have seen
in sect. \cref{sec:4dchern} are the basic datum of linear 
boundary conditions, have a simple structure in toric CS theory. The most general one is of the form
$\mathsans{M}'=(\mathsans{V},\mathsans{U},\varsigma|_{\mathsans{V}},\varpi|_{\mathsans{U}\times\mathsans{V}})$
where $\mathsans{U}$, $\mathsans{V}$ are subgroups of $\mathsans{T}$ such that
$\varsigma(\mathsans{V})\subseteq\mathsans{U}$.   
The corresponding Lie algebra crossed submodule is therefore 
$\mathfrak{m}'=(\mathfrak{v},\mathfrak{u},\dot\varsigma|_{\mathfrak{v}},\sdot\varpi\sdot|_{\mathfrak{u}\times\mathfrak{v}})$. 
Thus, $\mathfrak{m}'$ isotropic if $\mathfrak{v}\subseteq\mathfrak{u}^\perp$ and Lagrangian
if $\mathfrak{v}=\mathfrak{u}^\perp$. The associated orthogonal normalizer and Weyl
crossed modules are respectively $\ON\mathfrak{m}'=(\mathfrak{u}^\perp,\mathfrak{v}^\perp,\dot\varsigma|_{\mathfrak{u}^\perp},
\sdot\varpi\sdot|_{\mathfrak{v}^\perp\times\mathfrak{u}^\perp})$ and 
$\OW\mathfrak{m}'=(\mathfrak{u}^\perp/\mathfrak{v},\mathfrak{v}^\perp/\mathfrak{u},
\dot\varsigma|_{\mathfrak{u}^\perp/\mathfrak{v}},
\sdot\varpi\sdot|_{\mathfrak{v}^\perp/\mathfrak{u}\times\mathfrak{u}^\perp/\mathfrak{v}})$.


\subsection{\textcolor{blue}{\sffamily The Abelian projection model}}\label{subsec:cartanth}

Abelian projection \ccite {tHooft:1981bkw} is a theoretical framework for investigating the properties of confining gauge
theories. It consists in a gauge choice reducing the gauge symmetry from
a non Abelian group to a maximal Abelian subgroup. 
Abelian gauge fields emerge then from the non Abelian background and with these magnetic monopoles
presumably responsible for confinement.
In this subsection, we show how a kind of Abelian projection can be implemented in 4--dimensional CS theory,
even though no Higgs field is provisioned by it, 
leaving to future work the exploration of possible physical applications. 
Cartan--Weyl theory of semisimple Lie algebras
is used throughout. See app. \cref{app:cartan} for a brief review of some of basic facts and notation used. 


We begin by showing that we can associate a toric 4--dimensional CS model
(cf. subsect. \cref{subsec:toricth}) to a maximal torus $\mathsans{F}$
of a compact semisimple Lie group $\mathsans{E}$ whose Lie algebra 
$\mathfrak{e}$ is endowed with an invariant symmetric non singular bilinear form.
The model's toric crossed module $\mathsans{M}_{\mathsans{E}}=(\mathsans{T},\mathsans{T},\varsigma,\varpi)$
is defined as follows. The torus $\mathsans{T}$ is just $\mathsans{U}(1)^r$, where $r$ is the rank of $\mathsans{E}$.
Below, we shall view the elements $\mathsans{T}$ as ordered $r$-uples $(\ee^{x_\alpha})_{\alpha\in\Pi_+}$
of $\mathsans{U}(1)$ elements indexed by a set $\Pi_+$ of simple positive roots of $\mathfrak{e}$,
where $x_\alpha\in i\mathbb{R}$. 
The target map $\varsigma$ is given by
\begin{equation}
\varsigma\big((\ee^{x_\alpha})_{\alpha\in\Pi_+}\big)
=\big(
\ee^{\hfpt\sum_{\beta\in\Pi_+}\!C_{\beta\alpha}x_\beta}
\big)_{\alpha\in\Pi_+},
\label{cartanth1}
\end{equation}
where $C_{\alpha\beta}$ is the Cartan matrix of $\mathfrak{e}$
defined by \ceqref{cartan3}. Note that $\varsigma$ is well defined
because $C$ is a matrix with integer entries. The action map $\varpi$ is trivial.
$\mathsans{M}_{\mathsans{E}}$ can be further equipped with an invariant pairing.
Writing the elements of $\mathfrak{t}$ as ordered $r$-uples $(x_\alpha)_{\alpha\in\Pi_+}$
of $\mathfrak{u}(1)$ elements indexed by $\Pi_+$ analogously to the finite case, this reads  
\begin{equation}
\big\langle \left(x_\alpha\right)_{\alpha\in\Pi_+},\left(y_\alpha\right)_{\alpha\in\Pi_+}\big\rangle
 =\sss_{\alpha\in\Pi_+}\kappa_\alpha x_\alpha y_\alpha, 
\label{cartanth2}
\end{equation}
where $\kappa_{\alpha}$ is the inverse half lengths square of the root $\alpha$
defined by \ceqref{cartan4}. 
The symmetry property \ceqref{crmodinv1/rep} is fulfilled as
can be easily checked upon noticing that 
$\dot\varsigma\big(\left(ix_\alpha\right)_{\alpha\in\Pi_+}\big)
=\big(i\sum_{\beta\in\Pi_+}\!C_{\beta\alpha}x_\beta\big)_{\alpha\in\Pi_+}$
and using the identity $\kappa_\alpha C_{\beta\alpha}=\kappa_{\beta}C_{\alpha\beta}$.

In the above formal framework, 
the components of a toric higher gauge field are ordered $r$--tuples
$(\omega_\alpha)_{\alpha\in\Pi_+}$, $(\varOmega_\alpha)_{\alpha\in\Pi_+}$
with $\omega_\alpha\in\Omega^1(M,i\mathbb{R})$, $\varOmega_\alpha\in\Omega^2(M,i\mathbb{R})$.
The CS toric model action \ceqref{toric1} then reads explicitly as 
\begin{multline}
\sfC\hspace{-.75pt}\sfS\big((\omega_\alpha)_{\alpha\in\Pi_+},(\varOmega_\alpha)_{\alpha\in\Pi_+}\big)
=\frac{k}{2\pi}\int_M\sss_{\hfpt\alpha\in\Pi_+}\kappa_\alpha
\left(d\omega_\alpha-\tfrac{1}{2}\sss_{\hfpt\beta\in\Pi_+}C_{\beta\alpha}\varOmega_\beta\right)\varOmega_\alpha
\\
-\frac{k}{4\pi}\int_{\partial M}\sss_{\hfpt\alpha\in\Pi_+}\kappa_\alpha\omega_\alpha\varOmega_\alpha. 
\label{cartanth3}
\end{multline}

The components of a toric higher gauge transformation are similarly ordered $r$--tuples
$\left(\ee^{f_\alpha}\right)_{\alpha\in\Pi_+}$, $\left(U_\alpha\right)_{\alpha\in\Pi_+}$
with $f_\alpha\in\Omega^0(\widetilde{M},i\mathbb{R})$, $U_\alpha\in\Omega^1(M,i\mathbb{R})$.
Here, the functions $f_\alpha$ are generally multivalued and thus properly defined on the
universal covering $\widetilde{M}$ of $M$. The integrality condition
\begin{equation}
\frac{1}{2\pi i}\int_c\,df_\alpha \in\mathbb{Z},
\label{cartanth4}
\end{equation}
where $c$ is a 1--cycle of $M$ must be satisfied in order $\ee^{f_\alpha}$ to be a well
defined element of $\Map(M,\mathsans{U}(1))$. In accordance with \ceqref{toric2}, \ceqref{toric3}.
the gauge transformed gauge field $(\omega^{(\ee^{f}),(U)}{}_\alpha)_{\alpha\in\Pi_+}$,
$(\varOmega^{(\ee^{f}),(U)}{}_\alpha)_{\alpha\in\Pi_+}$ reads as 
\begin{align}
&\omega^{(\ee^{f}),(U)}{}_\alpha=\omega_\alpha+df_\alpha+\sss_{\hfpt\beta\in\Pi_+}C_{\beta\alpha}U_\beta,
\vphantom{\Big]}
\label{cartanth5}
\\
&\varOmega^{(\ee^{f}),(U)}{}_\alpha=\varOmega_\alpha+dU_\alpha.
\vphantom{\Big]}
\label{cartanth6}
\end{align}
Again, these transformations are gauge symmetries only if $\partial M$ is empty.


The Abelian projection associates a CS toric model of the type just described
with a CS model based on a Lie group crossed module 
$\mathsans{M}=(\mathsans{E},\mathsans{G},\tau,\mu)$ whose source group 
$\mathsans{E}$ is a compact semisimple Lie group and a choice of
a maximal torus $\mathsans{F}$ of $\mathsans{E}$. Its explicit construction
goes through a few steps detailed next. 

The source $\mathfrak{e}$ of the Lie algebra crossed module $\mathfrak{m}=(\mathfrak{e},\mathfrak{g},\dot\tau,\sdot\mu\sdot)$  
is a compact semisimple Lie algebra. Hence, $\mathfrak{z}(\mathfrak{e})=0$. 
So, since $\ker\dot\tau$ is a central ideal of $\mathfrak{e}$, we have $\ker\dot\tau=0$. 

Suppose $\mathsans{M}$ is equipped with an invariant pairing $\langle\cdot,\cdot\rangle$.
Then, $\mathfrak{m}$ is balanced so that $\dim\mathfrak{e}=\dim\mathfrak{g}$. 
Since $\ker\dot\tau=0$ as seen above, we have $\ran\dot\tau=\mathfrak{g}$.
$\dot\tau$ is therefore a Lie algebra isomorphism and $\mathfrak{e}\simeq\mathfrak{g}$.
$\mathfrak{g}$ is consequently also a compact semisimple Lie algebra. 

$\dot\tau$ being a Lie algebra isomorphism allows us to define a distinguished
invariant symmetric non singular bilinear of the Lie algebra $\mathfrak{e}$, namely 
\begin{equation}
\langle X,Y\rangle_\tau=\langle\dot\tau(X),Y\rangle,
\label{cartanth7}
\end{equation}
which we shall tacitly employ in what follows. 

A higher gauge field $\omega$, $\varOmega$  induces a
toric higher gauge field $(\omega_\alpha)_{\alpha\in\Pi_+}$, $(\varOmega_\alpha)_{\alpha\in\Pi_+}$
\begin{align}
&\omega_\alpha=\langle\omega, H_\alpha\rangle, 
\vphantom{\Big]}
\label{cartanth8}
\\
&\varOmega_\alpha=\langle\dot\tau(H_{\alpha^*}),\varOmega\rangle,
\vphantom{\Big]}
\label{cartanth9}
\end{align}
where $H_{\alpha},\,H_{\alpha^*}\in i\mathfrak{f}$ are the root and weight Cartan subalgebra
generators associated with the roots and weights $\alpha$, $\alpha^*$. 
We shall call $(\omega_\alpha)_{\alpha\in\Pi_+}$, $(\varOmega_\alpha)_{\alpha\in\Pi_+}$
the Abelian projection of $\omega$, $\varOmega$. 
In turn, with any toric higher gauge field
$(\omega_\alpha)_{\alpha\in\Pi_+}$, $(\varOmega_\alpha)_{\alpha\in\Pi_+}$
there is associated a higher gauge field $\omega$, $\varOmega$ of the form 
\begin{align}
&\omega=\sss_{\hfpt\alpha\in\Pi_+}\omega_\alpha\dot\tau(H_{\alpha^{*\mhfpt\vee}}),
\vphantom{\Big]}
\label{cartanth10}
\\
&\varOmega=\sss_{\hfpt\alpha\in\Pi_+}\varOmega_\alpha H_{\alpha\mhfpt{}^\vee},
\vphantom{\Big]}
\label{cartanth11}
\end{align}
where $H_{\alpha\mhfpt{}^\vee},\,H_{\alpha^{*\vee}}\in i\mathfrak{f}$ are the coroot and coweight Cartan subalgebra
generators associated with the coroots and coweights $\alpha\mhfpt{}^\vee$, $\alpha^{*\vee}$. 
We shall call a gauge field of this form Abelian projected. It is immediately verified that the Abelian projection
of the Abelian projected gauge field corresponding to the toric gauge field 
$(\omega_\alpha)_{\alpha\in\Pi_+}$, $(\varOmega_\alpha)_{\alpha\in\Pi_+}$ equals this latter.
See app. \cref{app:cartan} for some technical details. 

The CS action of an Abelian projected higher gauge field
$\omega$, $\varOmega$ equals precisely the toric CS action of the underlying
toric higher gauge field $(\omega_\alpha)_{\alpha\in\Pi_+}$, $(\varOmega_\alpha)_{\alpha\in\Pi_+}$
given by eq. \ceqref{cartanth3},
\begin{equation}
\sfC\hspace{-.75pt}\sfS(\omega,\varOmega)
=\sfC\hspace{-.75pt}\sfS\big((\omega_\alpha)_{\alpha\in\Pi_+},(\varOmega_\alpha)_{\alpha\in\Pi_+}\big). 
\label{cartanth12}
\end{equation}



The Abelian projection involves a reduction of the higher gauge symmetry similarly to ordinary gauge
theory. The residual gauge symmetry is described by the normalizer crossed module $\NN\mathsans{M}_{\mathsans{F}}$
of a certain crossed submodule $\mathsans{M}_{\mathsans{F}}$ of $\mathsans{M}$
depending on the maximal torus $\mathsans{F}$ (cf. subsect. \cref{subsec:crosumo}). 

The characteristic crossed module of $\mathsans{F}$ is 
$\mathsans{M}_{\mathsans{F}}=(\mathsans{F},\tau(\mathsans{F}),\tau|_{\mathsans{F}},
\mu|_{\tau(\mathsans{F})\times\mathsans{F}})$. $\mathsans{M}_{\mathsans{F}}$ is a crossed submodule of $\mathsans{M}$.
$\mathsans{M}_{\mathsans{F}}$ is toric as $\mathsans{F}$, $\tau(\mathsans{F})$ are
maximal tori of $\mathsans{E}$, $\mathsans{G}$, respectively, and $\mu|_{\tau(\mathsans{F})\times\mathsans{F}}$ is trivial. 
It is simple to show that the normalizer crossed module of $\mathsans{M}_{\mathsans{F}}$
is $\NN\mathsans{M}_{\mathsans{F}}=(\NN\mathsans{F},\mu\NN\mathsans{F},\tau|_{\NN\mathsans{F}},
\mu|_{\mu\NN\mathsans{F}\times\NN\mathsans{F}})$. The Weyl crossed module of $\mathsans{M}_{\mathsans{F}}$
so turns out to be $\WW\mathsans{M}_{\mathsans{F}}=(\NN\mathsans{F}/\mathsans{F},
\mu\NN\mathsans{F}/\tau(\mathsans{F}),\tau|_{\NN\mathsans{F}/\mathsans{F}},
\mu|_{\mu\NN\mathsans{F}/\tau(\mathsans{F})\times\NN\mathsans{F}/\mathsans{F}})$.
$\WW\mathsans{M}_{\mathsans{F}}$ is a finite discrete crossed module. Indeed,
$\NN\mathsans{F}/\mathsans{F}=\WW\mathsans{E}$, the familiar Lie theoretic Weyl group
of $\mathsans{E}$. Furthermore, since it turns out that $\mu\NN\mathsans{F}=\NN\tau(\mathsans{F})$,
as is straightforwardly shown, 
$\mu\NN\mathsans{F}/\tau(\mathsans{F})=\WW\mathsans{G}$, the Weyl group of $\mathsans{G}$. 
Notice that $\WW\mathsans{G}\simeq\WW\mathsans{E}$, since $\mathfrak{g}\simeq\mathfrak{e}$.

As already mentioned, $\NN\mathsans{M}_{\mathsans{F}}$ is the crossed module of the residual higher gauge
symmetry left over by the Abelian projection. Therefore, to implement the projection, we
restrict to the subgroup of $\NN\mathsans{M}_{\mathsans{F}}$--valued higher gauge transformations, 
that is, by what found in the previous paragraph, the gauge transformations $u$, $U$,
with $u\in\Map(M,\mu\NN\mathsans{F})$ and $U\in\Omega^1(M,\mathfrak{f})$.
$u$, $U$ have for this reason  a special form. 
Since $\mu\NN\mathsans{F}$ is a disjoint union of finitely many cosets of $\tau(\mathsans{F})$,
on each connected component of $M$, one has 
\begin{equation}
u=u_c\hfpt a,
\label{cartanth13}
\end{equation}
where $u_c\in\Map(M,\tau(\mathsans{F}))$ and $a\in\mu\NN\mathsans{F}$ is constant. 
$u_c$ can thus be expressed as 
\begin{equation}
u_c=\ee^{\hfpt\sum_{\alpha\in\Pi_+}\!f_\alpha \dot\tau(H_{\alpha^{*\mhfpt\vee}})}
\label{cartanth14}
\end{equation}
with $f_\alpha\in\Omega^0(\widetilde{M},i\mathbb{R})$ through
the coweight generators $H_{\alpha^{*\mhfpt\vee}}\in i\mathfrak{f}$. As before, the $f_\alpha$
are generally multivalued functions. The well-definedness of the exponential
in the right hand side of \ceqref{cartanth14} requires that \vfil\eject\noindent
\begin{equation}
\frac{1}{2\pi i}\int_c\sss_{\hfpt\alpha\in\Pi_+}df_\alpha H_{\alpha^{*\mhfpt\vee}}\in\Lambda_{\mathfrak{f}}
\label{cartanth15}
\end{equation}
for any 1--cycle $c$ of $M$, 
where $\Lambda_{\mathfrak{f}}$ is the integral lattice of $\mathfrak{f}$. Condition \ceqref{cartanth15}
is compatible with \ceqref{cartanth4} since the integral lattice $\Lambda_{\mathfrak{f}}$ is a sublattice of the coweight
lattice $\Lambda_{\mathrm{cw}\mathfrak{f}}$. However, the integer values which the periods of $df_\alpha$ can take are restricted
unless $\Lambda_{\mathfrak{f}}=\Lambda_{\mathrm{cw}\mathfrak{f}}$, which happens when the center $\mathsans{Z}(\mathsans{E})$
of $\mathsans{E}$ is trivial, as $\mathsans{Z}(\mathsans{E})=\Lambda_{\mathrm{cw}\mathfrak{f}}/\Lambda_{\mathfrak{f}}$. 
Finally, $U$ can be expanded as
\begin{equation}
U=\sss_{\hfpt\alpha\in\Pi_+}U_\alpha H_{\alpha\mhfpt{}^\vee}.
\label{cartanth16}
\end{equation}
with $U_\alpha\in\Omega^1(M,i\mathbb{R})$ in terms of 
the coroot generators $H_{\alpha\mhfpt{}^\vee}\in i\mathfrak{f}$.
In this way, we can associate with an $\NN\mathsans{M}_{\mathsans{F}}$--valued gauge transformation
$u$, $U$ a toric gauge transformation $\left(\ee^{f_\alpha}\right)_{\alpha\in\Pi_+}$,
$\left(U_\alpha\right)_{\alpha\in\Pi_+}$ satisfying \ceqref{cartanth15}, the Abelian projection of $u$, $U$.
This can be defined alternatively through 
\begin{align}
&df_\alpha=\langle duu^{-1}, H_\alpha\rangle, 
\vphantom{\Big]}
\label{cartanth17}
\\
&U_\alpha=\langle \dot\tau(H_{\alpha^*}),U\rangle,
\vphantom{\Big]}
\label{cartanth18}
\end{align}
analogously to \ceqref{cartanth8}, \ceqref{cartanth9} projecting on the
root and weight generators $H_\alpha$, $H_{\alpha^*}\in i\mathfrak{f}$.
Viceversa, with any toric gauge transformation $\left(\ee^{f_\alpha}\right)_{\alpha\in\Pi_+}$,
$\left(U_\alpha\right)_{\alpha\in\Pi_+}$ satisfying \ceqref{cartanth15}, we can associated
an Abelian projected $\NN\mathsans{M}_{\mathsans{F}}$--gauge
transformation $u_c$, $U$ through \ceqref{cartanth14}, \ceqref{cartanth16}.
Note that in performing the Abelian projection of a $\NN\mathsans{M}_{\mathsans{F}}$--valued gauge transformation
$u$, $U$ all the information about the discrete factor $a\in\mu\NN\mathsans{F}$ appearing in 
\ceqref{cartanth13} is lost. Correspondingly, for the Abelian projected gauge transformation 
$u_c$, $U$ yielded by a toric gauge transformation we have 
$u_c\in\Map(M,\tau(\mathsans{F}))$ only. 


For $a\in\mu\NN\mathsans{F}$, the $\mu\sdot\hfpt(a,\cdot)$ are automorphisms
of the Lie subalgebra $\mathfrak{f}$. The findings of the previous paragraph indicate that 
their action on the root, coroot, weight and coweight lattices $\Lambda_{\mathrm{r}\mathfrak{f}}$,
$\Lambda_{\mathrm{cr}\mathfrak{f}}$, $\Lambda_{\mathrm{w}\mathfrak{f}}$, $\Lambda_{\mathrm{cw}\mathfrak{f}}$
of $\mathfrak{f}$ may be relevant to the analysis of the gauge invariance of Abelian projection. One finds 
\begin{align}
&\mu\sdot\hfpt(a,H_\alpha)=\sss_{\hfpt\beta\in\Pi_+}\chi{}_{\beta\alpha}(a)H_\beta,
\vphantom{\Big]}
\label{cartanth19}
\end{align}
\begin{align}
&\mu\sdot\hfpt(a,H_{\alpha\mhfpt{}{}^\vee})=\sss_{\hfpt\beta\in\Pi_+}\chi\mhfpt{}^\vee{}_{\beta\alpha}(a)H_{\beta\mhfpt{}{}^\vee},
\vphantom{\Big]}
\label{cartanth20}
\\
&\mu\sdot\hfpt(a,H_{\alpha^{*}})=\sss_{\hfpt\beta\in\Pi_+}\chi\mhfpt{}^\vee{}_{\alpha\beta}(a^{-1})H_{\beta^{*}},
\vphantom{\Big]}
\label{cartanth21}
\\
&\mu\sdot\hfpt(a,H_{\alpha^{*\vee}})=\sss_{\hfpt\beta\in\Pi_+}\chi{}_{\alpha\beta}(a^{-1})H_{\beta^{*\vee}},
\vphantom{\Big]}
\label{cartanth22}
\end{align}
where $\chi:\mu\NN\mathsans{F}\rightarrow\GL(r,\mathbb{Q})$,
$\chi\mhfpt{}^\vee{}:\mu\NN\mathsans{F}\rightarrow\GL(r,\mathbb{Q})$ are certain group morphisms.
In fact, $\chi\mhfpt{}^\vee{}$, $\chi$ are simply related \hphantom{xxxxxxxxx}
\begin{equation}
\chi(a)=\kappa\chi\mhfpt{}^\vee{}(a)\kappa^{-1},
\label{cartanth23}
\end{equation}
where $\kappa\in\GL(r,\mathbb{R})$ is the diagonal matrix
\begin{equation}
\kappa_{\alpha\beta}=\kappa_\alpha\delta_{\alpha\beta}, 
\label{cartanth24}
\end{equation}
$\kappa_\alpha$ being defined by eq. \ceqref{cartan4}.
These relations follow form observing that the $\mu\sdot\hfpt(a,\cdot)$ are automorphisms
of both the integral lattice $\Lambda_{\mathfrak{f}}$ of $\mathfrak{f}$ and its dual lattice
$\Lambda_{\mathfrak{f}}{}^*$ and that the root and coroot generators span 
$\Lambda_{\mathfrak{f}}{}^*$, $\Lambda_{\mathfrak{f}}$ over $\mathbb{Q}$,  
since the root and coroot lattices 
$\Lambda_{\mathrm{r}\mathfrak{f}}$, $\Lambda_{\mathrm{cr}\mathfrak{f}}$ are sublattices 
of $\Lambda_{\mathfrak{f}}{}^*$, $\Lambda_{\mathfrak{f}}$ respectively. Note also that, 
as $\Aut(\mathsans{F})$ is a discrete group, 
$\chi\mhfpt{}^\vee{}(a)=\chi(a)=1_{r}$ for $a\in\tau(\mathsans{F})$. So,
recalling that $\mu\NN\mathsans{F}/\tau(\mathsans{F})\simeq\WW\mathsans{E}$,
we have induced group morphisms 
$\chi:\WW\mathsans{E}\rightarrow\GL(r,\mathbb{Q})$,
$\chi\mhfpt{}^\vee{}:\WW\mathsans{E}\rightarrow\GL(r,\mathbb{Q})$, which we
denote by the same symbol for simplicity. 

Let $u$, $U$ be a $\NN\mathsans{M}_{\mathsans{F}}$--valued higher gauge transformation with associated
toric gauge transformation $\left(\ee^{f_\alpha}\right)_{\alpha\in\Pi_+}$, $\left(U_\alpha\right)_{\alpha\in\Pi_+}$
and discrete factor $a$ as defined by eqs. \ceqref{cartanth13}, \ceqref{cartanth14}, \ceqref{cartanth16}. 
If $\omega$, $\varOmega$ is a higher gauge field and $\omega^{u,U}$, $\varOmega^{u,U}$ is the gauge transformed 
gauge field, the toric higher gauge fields $(\omega_\alpha)_{\alpha\in\Pi_+}$, $(\varOmega_\alpha)_{\alpha\in\Pi_+}$
and $(\omega^{u,U}{}_\alpha)_{\alpha\in\Pi_+}$, $(\varOmega^{u,U}{}_\alpha)_{\alpha\in\Pi_+}$ associated to
$\omega$, $\varOmega$ and $\omega^{u,U}$, $\varOmega^{u,U}$ by Abelian projection
according to eqs. \ceqref{cartanth8}, \ceqref{cartanth9} are related as 
\begin{align}
&\omega^{u,U}{}_\alpha=\sss_{\beta\in\Pi_+}\chi_{\beta\alpha}(a)\hfpt\omega^{(\ee^{f}),(U)}{}_\beta,
\vphantom{\Big]}
\label{cartanth25}
\\
&\varOmega^{u,U}{}_\alpha=\sss_{\beta\in\Pi_+}\chi\mhfpt{}^\vee{}_{\alpha\beta}(a^{-1})\hfpt\varOmega^{(\ee^{f}),(U)}{}_\beta,
\vphantom{\Big]}
\label{cartanth26}
\end{align}
where in accordance with eqs. \ceqref{cartanth5}, \ceqref{cartanth6}
$(\omega^{(\ee^{f}),(U)}{}_\alpha)_{\alpha\in\Pi_+}$, $(\varOmega^{(\ee^{f}),(U)}{}_\alpha)_{\alpha\in\Pi_+}$
is the toric gauge transform of $(\omega_\alpha)_{\alpha\in\Pi_+}$, $(\varOmega_\alpha)_{\alpha\in\Pi_+}$.
Correspondingly, if $\omega$, $\varOmega$ is an Abelian projected higher
gauge field and $\omega^{u,U}$, $\varOmega^{u,U}$ is again the gauge
transformed gauge field, then $\omega^{u,U}$, $\varOmega^{u,U}$ is also Abelian projected and the toric gauge fields
$(\omega_\alpha)_{\alpha\in\Pi_+}$, $(\varOmega_\alpha)_{\alpha\in\Pi_+}$
and $(\omega^{u,U}{}_\alpha)_{\alpha\in\Pi_+}$, $(\varOmega^{u,U}{}_\alpha)_{\alpha\in\Pi_+}$ underlying
$\omega$, $\varOmega$ and $\omega^{u,U}$, $\varOmega^{u,U}$ in eqs. \ceqref{cartanth10}, \ceqref{cartanth11}
are again related as in \ceqref{cartanth25}, \ceqref{cartanth26}.

Hence, the toric gauge transformation $\left(\ee^{f_\alpha}\right)_{\alpha\in\Pi_+}$, $\left(U_\alpha\right)_{\alpha\in\Pi_+}$
induced by the gauge transformation $u$, $U$ does not exhaust its action. There is a residual 
finite discrete $\WW\mathsans{E}$ action. This constitutes an extra 
Weyl group symmetry of the toric CS action
\ceqref{cartanth3} for Abelian projected gauge fields $\omega$, $\varOmega$ in addition to the
toric higher gauge symmetry (for $\partial M=\emptyset$).


\vfil\eject

\appendix

\section{\textcolor{blue}{\sffamily Appendixes}}\label{app:app}

The following appendixes collect basic results and identities used repeatedly in the main body of the paper
and provided also the proofs of a few basic statements relevant in our analysis, which are original to the
best of out knowledge. 


\subsection{\textcolor{blue}{\sffamily Basic definitions and identities of crossed module theory}}\label{app:def}

In this appendix, we collect a number of basic definitions and relations which are assumed and
used throughout the main text of the paper. This will also allow us to set our notation. A part of
this material is fairly standard \ccite{Baez5}, the rest is original to the best of our knowledge. 

\vspace{2.5mm}

\noindent
{\it  Lie group crossed modules and module morphisms}

\noindent
A Lie group crossed module $\mathsans{M}$ consists of two Lie groups $\mathsans{E}$ and $\mathsans{G}$ 
together with Lie group morphisms $\tau:\mathsans{E}\rightarrow\mathsans{G}$ and
$\mu:\mathsans{G}\rightarrow\Aut(\mathsans{E})$ such that 
\begin{align}
&\tau(\mu(a,A))=a\tau(A)a^{-1},
\vphantom{\Big]}
\label{liecrmod1}
\\
&\mu(\tau(A),B)=ABA^{-1}
\vphantom{\Big]}
\label{liecrmod2}
\end{align}
for $a\in\mathsans{G}$, $A,B\in\mathsans{E}$, where here and below we view
$\mu:\mathsans{G}\times\mathsans{E}\rightarrow\mathsans{E}$ for convenience.
$\tau$ and $\mu$ are called the target and action maps and 
\ceqref{liecrmod1}, \ceqref{liecrmod2}
are called equivariance and Peiffer properties, respectively.
As a rule, we write $\mathsans{M}=(\mathsans{E},\mathsans{G},\tau,\mu)$
to specify the crossed module through its constituent data. 


A morphism $\beta:\mathsans{M}'\rightarrow\mathsans{M}$ of Lie group crossed modules
consists of two Lie group morphisms $\phi:\mathsans{G}'\rightarrow\mathsans{G}$  and 
$\varPhi:\mathsans{E}'\rightarrow\mathsans{E}$ with the property that 
\begin{align}
&\tau(\varPhi(A))=\phi(\tau'(A)).
\vphantom{\Big]}
\label{liecrmod3}
\\
&\varPhi(\mu'(a,A))=\mu(\phi(a),\varPhi(A))
\vphantom{\Big]}
\label{liecrmod4}
\end{align}
for $a\in\mathsans{G}'$, $A\in\mathsans{E}'$.
The morphism $\beta$ is an isomorphism precisely when $\varPhi$, $\phi$ are both isomorphisms.
We normally write $\beta:\mathsans{M}'\rightarrow\mathsans{M}=(\varPhi,\phi)$
to indicate the constituent morphisms of the crossed module morphism. 


There are obvious notions \pagebreak of direct product $\mathsans{M}_1\times\mathsans{M}_2$ 
of two Lie group crossed modules $\mathsans{M}_1$, $\mathsans{M}_2$ and direct product $\beta_1\times\beta_2$
of two Lie group crossed module morphisms $\beta_1$, $\beta_2$
consisting in taking the direct product of the corresponding constituent data in the Lie group category.


Lie group crossed modules and morphisms thereof with the direct product operation
constitute a monoidal category. 

\vspace{2.5mm}

\noindent
{\it Lie group crossed submodules}

\noindent
Let $\mathsans{M}$, $\mathsans{M}'$
be Lie group crossed modules. $\mathsans{M}'$ is a submodule of $\mathsans{M}$ if 
$\mathsans{E}'$, $\mathsans{G}'$ are Lie subgroups of $\mathsans{E}$, $\mathsans{G}$ and $\tau'$, $\mu'$
are restrictions of $\tau$, $\mu$, respectively or, equivalently, if there are inclusion Lie group morphisms
$\mathsans{E}'\subseteq\mathsans{E}$, $\mathsans{G}'\subseteq\mathsans{G}$ which are the components
of an inclusion Lie group crossed module morphism $\mathsans{M}'\subseteq\mathsans{M}$.

Let $\mathsans{M}'$, $\mathsans{M}''$ be crossed submodules of a Lie group crossed module $\mathsans{M}$
with $\mathsans{M}'$ a submodule of $\mathsans{M}''$. $\mathsans{M}''$ is said to normalize $\mathsans{M}'$ 
if the following conditions are met.
For $a\in\mathsans{G}''$, $b\in\mathsans{G}'$, one has $aba^{-1}\in\mathsans{G}'$. 
For $a\in\mathsans{G}''$, $B\in\mathsans{E}'$, one has $\mu(a,B)\in\mathsans{E}'$. 
Finally, for $b\in\mathsans{G}'$, $A\in\mathsans{E}''$, one has $\mu(b,A)A^{-1}\in\mathsans{E}'$.

If $\mathsans{M}''$ normalize $\mathsans{M}'$, it is possible to define the quotient crossed
module $\mathsans{M}''/\mathsans{M}'$. By the condition listed in the previous paragraph,
$\mathsans{G}'$, $\mathsans{E}'$ are normal Lie subgroups of
$\mathsans{G}''$, $\mathsans{E}''$ respectively, making it possible to construct the quotient Lie groups
$\mathsans{G}''/\mathsans{G}'$, $\mathsans{E}''/\mathsans{E}'$. Then,
$\mathsans{M}''/\mathsans{M}'=(\mathsans{E}''/\mathsans{E}',\mathsans{G}''/\mathsans{G}',
\tau''{}_{\mathsans{M}'},\mu''{}_{\mathsans{M}'})$, where 
$\tau''{}_{\mathsans{M}'}$, $\mu''{}_{\mathsans{M}'}$ are the structures maps defined by
\begin{align}
&\tau''{}_{\mathsans{M}'}(A\mathsans{E}')=\tau(A)\mathsans{G}',
\vphantom{\Big]}
\label{}
\\
&\mu''{}_{\mathsans{M}'}(a\mathsans{G}',A\mathsans{E}')=\mu(a,A)\mathsans{E}'
\vphantom{\Big]}
\label{}
\end{align}
for $a\in\mathsans{G}''$, $A\in\mathsans{E}''$.
It can be verified that $\tau''{}_{\mathsans{M}'}$, $\mu''{}_{\mathsans{M}'}$ 
are well defined and obey relations \ceqref{liecrmod1}, \ceqref{liecrmod2}.


Just as the notions of Lie algebra and algebra morphism and Lie subalgebra
are the infinitesimal counterpart of those of Lie group and group morphism and Lie subgroup, so 
the concepts of Lie algebra crossed module and module morphism and Lie algebra crossed submodule
are the infinitesimal counterpart of those of Lie group crossed module and module morphism and Lie group
crossed submodule. 

\vspace{2.5mm}

\noindent
{\it  Lie algebra crossed modules and module morphisms}

\noindent
A Lie algebra crossed module $\mathfrak{m}$ consists of two Lie algebras $\mathfrak{e}$ and $\mathfrak{g}$
together with Lie algebra morphisms $t:\mathfrak{e}\rightarrow\mathfrak{g}$ and
$m:\mathfrak{g}\rightarrow\Der(\mathfrak{e})$ such that 
\begin{align}
&t(m(u,U))=[u,t(U)],
\vphantom{\Big]}
\label{liecrmod5}
\\
&m(t(U),V)=[U,V]
\vphantom{\Big]}
\label{liecrmod6}
\end{align}
for $u\in\mathfrak{g}$, $U,V\in\mathfrak{e}$, where here and below
we view $m:\mathfrak{g}\times\mathfrak{e}\rightarrow\mathfrak{e}$ for convenience.
$t$ and $m$ are called the target and action maps and 
\ceqref{liecrmod5}, \ceqref{liecrmod6} are called equivariance and Peiffer properties, respectively, in
analogy to the group case. Again, 
we write $\mathfrak{m}=(\mathfrak{e},\mathfrak{g},t,m)$
to identify the crossed module through its defining elements. 

A morphism $p:\mathfrak{m}'\rightarrow\mathfrak{m}$ of Lie algebra crossed modules 
consists of two Lie algebra morphisms $H:\mathfrak{e}'\rightarrow\mathfrak{e}$ and
$h:\mathfrak{g}'\rightarrow\mathfrak{g}$ with the property that 
\begin{align}
&t(H(U))=h(t'(U)).
\vphantom{\Big]}
\label{liecrmod7}
\\
&H(m'(u,U))=m(h(u),H(U)).
\vphantom{\Big]}
\label{liecrmod8}
\end{align}
for $u\in\mathfrak{g}'$, $U\in\mathfrak{e}'$.
The morphism $p$ is said to be an isomorphism if and only if $H$, $h$ are both isomorphisms. 
Again, we write as a rule $p:\mathfrak{m}'\rightarrow\mathfrak{m}=(H,h)$
to specify the crossed module morphism by means of its defining morphisms. 

Similarly to the Lie group case, there are obvious notions of direct sum $\mathfrak{m}_1\oplus\mathfrak{m}_2$ 
of two Lie algebra crossed modules $\mathfrak{m}_1$, $\mathfrak{m}_2$ and direct sum $p_1\oplus p_2$
of two Lie algebra crossed module morphisms $p_1$, $p_2$
consisting in taking the direct sum of the corresponding constituent data in the Lie algebra category.

Lie algebra crossed modules and morphisms thereof with the direct sum operation
constitute a monoidal category.

\vspace{2.5mm}

\noindent
{\it Lie algebra crossed submodules}

\noindent
Let $\mathfrak{m}$, $\mathfrak{m}'$
be Lie algebra crossed modules. $\mathfrak{m}'$ is a submodule of $\mathfrak{m}$ if 
$\mathfrak{e}'$, $\mathfrak{g}'$ are Lie subalgebras of $\mathfrak{e}$, $\mathfrak{g}$ and $t'$, $m'$
are restrictions of $t$, $m$, respectively or, equivalently, if there are inclusion Lie algebra morphisms
$\mathfrak{e}'\subseteq\mathfrak{e}$, $\mathfrak{g}'\subseteq\mathfrak{g}$ which are the components
of an inclusion Lie algebra crossed module morphism $\mathfrak{m}'\subseteq\mathfrak{m}$.

Let $\mathfrak{m}'$, $\mathfrak{m}''$ be crossed \pagebreak submodules of a Lie algebra crossed module $\mathfrak{m}$
with $\mathfrak{m}'$ a submodule of $\mathfrak{m}''$. $\mathfrak{m}''$ is said to normalize $\mathfrak{m}'$ 
if the following conditions are met.
For $u\in\mathfrak{g}''$, $v\in\mathfrak{g}'$, one has $[u,v]\in\mathfrak{g}'$. 
For $u\in\mathfrak{g}''$, $V\in\mathfrak{e}'$, one has $m(u,V)\in\mathfrak{e}'$. 
Finally, for $v\in\mathfrak{g}'$, $U\in\mathfrak{e}''$, one has $m(v,U)\in\mathfrak{e}'$.

If $\mathfrak{m}''$ normalize $\mathfrak{m}'$, it is possible to define the quotient crossed
module $\mathfrak{m}''/\mathfrak{m}'$. By the condition listed in the previous paragraph,
$\mathfrak{g}'$, $\mathfrak{e}'$ are Lie ideals of
$\mathfrak{g}''$, $\mathfrak{e}''$ respectively, 
making it possible to construct the quotient Lie algebras
$\mathfrak{g}''/\mathfrak{g}'$, $\mathfrak{e}''/\mathfrak{e}'$. Then,
$\mathfrak{m}''/\mathfrak{m}'=(\mathfrak{e}''/\mathfrak{e}',\mathfrak{g}''/\mathfrak{g}',
t''{}_{\mathfrak{m}'},m''{}_{\mathfrak{m}'})$, where 
$t''{}_{\mathfrak{m}'}$, $m''{}_{\mathfrak{m}'}$ are the structures maps 
\begin{align}
&t''{}_{\mathfrak{m}'}(U+\mathfrak{e}')=t(U)+\mathfrak{g}',
\vphantom{\Big]}
\label{}
\\
&m''{}_{\mathfrak{m}'}(u+\mathfrak{g}',U+\mathfrak{e}')=m(u,U)+\mathfrak{e}'
\vphantom{\Big]}
\label{}
\end{align}
for $u\in\mathfrak{g}''$, $U\in\mathfrak{e}''$.
It can be verified that $t''{}_{\mathfrak{m}'}$, $m''{}_{\mathfrak{m}'}$ 
are well defined and obey relations \ceqref{liecrmod1}, \ceqref{liecrmod2}.


\subsection{\textcolor{blue}{\sffamily Lie differentiation of crossed modules}}\label{app:ident}

From what shown in app.\cref{app:def}, 
it is apparent that the Lie algebra crossed module category is the infinitesimal
counterpart of the Lie group crossed module one. As it might be expected, they are related by Lie differentiation.

As a convention, whenever a Lie group theoretic structure and $\mathsans{S}$ and a Lie algebra
theoretic structure $\mathfrak{s}$ denoted by the same letter appear in a given context, it is tacitly assumed that
$\mathfrak{s}$ is yielded by $\mathsans{S}$ via Lie differentiation, unless otherwise stated.

\vspace{2.5mm}

\noindent
{\it Lie differentiation}

\noindent
Let $\mathsans{M}=(\mathsans{E},\mathsans{G},\tau,\mu)$ be a Lie group crossed module. 
With the structure map $\tau:\mathsans{E}\rightarrow\mathsans{G}$,
there is associated its Lie differential
$\dot\tau:\mathfrak{e}\rightarrow\mathfrak{g}$. Likewise, with the structure
map $\mu:\mathsans{G}\times\mathsans{E}\rightarrow\mathsans{E}$, there are associated three
distinct Lie differentials, namely
$\mu\sdot:\mathsans{G}\times\mathfrak{e}\rightarrow\mathfrak{e}$,
$\sdot\mu:\mathfrak{g}\times\mathsans{E}\rightarrow\mathfrak{e}$ and
$\sdot\mu\sdot:\mathfrak{g}\times\mathfrak{e}\rightarrow\mathfrak{e}$.
Since $\tau$ is a Lie group morphism, $\dot\tau$ is a Lie algebra morphism. 
Similarly, since $\mu$ encodes a Lie group morphism $\mu:\mathsans{G}\rightarrow\Aut(\mathsans{E})$, 
$\mu\sdot$ and $\sdot\mu\sdot$ encode respectively a Lie group morphism
$\mu\sdot:\mathsans{G}\rightarrow\Aut(\mathfrak{e})$ 
and Lie algebra morphism $\sdot\mu\sdot:\mathfrak{g}\rightarrow\Der(\mathfrak{e})$.
The interpretation of $\sdot\mu$ is less obvious: as it turns out, 
$\sdot\mu:\mathfrak{g}\rightarrow C_{\Ad}{}^1(\mathsans{E},\mathfrak{e})$
is a linear morphism of $\mathfrak{g}$ into the linear space of adjoint action $1$--cocycles of $\mathsans{E}$
on $\mathfrak{e}$. 
The precise definition and main properties of these objects are provided 
below.

Let $\beta:\mathsans{M}'\rightarrow\mathsans{M}=(\varPhi,\phi)$
be a Lie group crossed module morphism.
The Lie differentials of the Lie group morphisms 
$\varPhi:\mathsans{E}'\rightarrow\mathsans{E}$, $\phi:\mathsans{G}'\rightarrow\mathsans{G}$
are then Lie algebra morphisms $\dot\varPhi:\mathfrak{e}'\rightarrow\mathfrak{e}$,
$\dot\phi:\mathfrak{g}'\rightarrow\mathfrak{g}$.

Given a Lie group crossed module $\mathsans{M}=(\mathsans{E},\mathsans{G},\tau,\mu)$, 
the data $\mathfrak{m}=(\mathfrak{e},\mathfrak{g},\dot\tau,\sdot\mu{}\sdot\hfpt)$
define a Lie algebra crossed module.
$\mathfrak{m}$ is in this way associated with $\mathsans{M}$
much as a Lie algebra is associated with a Lie group.
Similarly,  given a Lie group crossed module morphism
$\beta:\mathsans{M}'\rightarrow\mathsans{M}=(\varPhi,\phi)$, the data
$\dot\beta:\mathfrak{m}'\rightarrow\mathfrak{m}=(\dot\varPhi,\dot\phi)$
define a Lie algebra crossed module morphism. Again, $\dot\beta$ is associated with $\beta$
just as a Lie algebra morphism is associated with a Lie group morphism.
The Lie algebra crossed module associated with the direct product $\mathsans{M}_1\times\mathsans{M}_2$
of two Lie group crossed modules $\mathsans{M}_1$, $\mathsans{M}_2$ is the direct sum 
$\mathfrak{m}_1\oplus\mathfrak{m}_2$ of the associated Lie algebra crossed modules $\mathfrak{m}_1$, $\mathfrak{m}_2$. 
Similarly, the Lie algebra crossed module morphism associated with the direct product $\beta_1\times\beta_2$
of two Lie group crossed module morphisms $\beta_1$, $\beta_2$ is the direct sum 
$\dot\beta_1\oplus\dot\beta_2$ of the associated Lie algebra crossed module morphisms
$\dot\beta_1$, $\dot\beta_2$. 

The map that associates with each Lie group crossed module $\mathsans{M}$ its Lie algebra
crossed module $\mathfrak{m}$ and with each Lie group crossed module morphism 
$\beta:\mathsans{M}'\rightarrow\mathsans{M}$
its Lie algebra crossed module morphism $\dot\beta:\mathfrak{m}'\rightarrow\mathfrak{m}$ is a functor
of the Lie group into the  Lie algebra crossed module monoidal category. 

Let $\mathsans{M}$, $\mathsans{M}'$, be Lie group crossed modules with associated Lie algebra crossed modules
$\mathfrak{m}$, $\mathfrak{m}'$. If $\mathsans{M}'$ is a crossed submodule of $\mathsans{M}$, then 
$\mathfrak{m}'$ is a crossed submodule of $\mathfrak{m}$. Next, let $\mathsans{M}''$ be
a Lie group crossed module with Lie algebra crossed module $\mathfrak{m}''$. If $\mathsans{M}''$ is a 
crossed submodule of $\mathsans{M}$ normalizing $\mathsans{M}'$, then $\mathfrak{m}''$ is
crossed submodule of $\mathfrak{m}$ normalizing $\mathfrak{m}'$. Moreover, the Lie algebra crossed module
of the quotient crossed module $\mathsans{M}''/\mathsans{M}'$ is precisely $\mathfrak{m}''/\mathfrak{m}'$.

\vspace{2.5mm}

\noindent
{\it Basic Lie theoretic identities}

\noindent
The relevant differentiated structure mappings $\dot\tau:\mathfrak{e}\rightarrow\mathfrak{g}$,
$\mu\sdot\,:\mathsans{G}\times\mathfrak{e}\rightarrow\mathfrak{e}$,
$\sdot\mu:\mathfrak{g}\times\mathsans{E}\rightarrow\mathfrak{e}$ and 
$\sdot\mu\sdot\,:\mathfrak{g}\times\mathfrak{e}\rightarrow\mathfrak{e}$
of a Lie group crossed module $\mathsans{M}=(\mathsans{E},\mathsans{G},\tau,\mu)$
which we introduced above satisfy a host of identities often used in detailed calculations
and analyses. 

$\mu\sdot$ obeys the following algebraic identities:
\begin{align}
&\mu\sdot(ab,X)=\mu\sdot(a,\mu\sdot(b,X)),
\vphantom{\Big]}
\label{ident10/1}
\\
&\mu\sdot(a,\sdot\mu\sdot(x,X))
=\sdot\mu\sdot(\Ad a(x),\mu\sdot(a,X)),
\vphantom{\Big]}
\label{ident10}
\end{align}
where $a,b\in\mathsans{G}$, $x\in\mathfrak{g}$, $X\in\mathfrak{e}$.
$\sdot\mu$ in turn satisfies the following relations:
\begin{align}
&\dot\tau(\,\,\sdot\mu(x,A))=x-\Ad\tau(A)(x), 
\vphantom{\Big]}
\label{ident5}
\\
&\,\sdot\mu(\dot\tau(X),A)=X-\Ad A(X), 
\vphantom{\Big]}
\label{ident6}
\\
&\,\sdot\mu([x,y],A)
=\sdot\mu\sdot\,(x,\,\sdot\mu(y,A))
-\sdot\mu\sdot\,(y,\,\sdot\mu(x,A))
-[\,\,\sdot\mu(x,A),\,\sdot\mu(y,A)],
\vphantom{\Big]}
\label{ident7}
\\
&\,\sdot\mu(x,AB)=\sdot\mu(x,A)+\Ad A(\,\,\sdot\mu(x,B)), 
\vphantom{\Big]}
\label{ident8}
\\
&\,\mu\sdot\,(a,\,\sdot\mu(x,A))=\sdot\mu(\Ad a(x),\mu(a,A)),
\vphantom{\Big]}
\label{ident9}
\end{align}
where $a,b\in\mathsans{G}$, $A,B\in\mathsans{E}$, $x,y\in\mathfrak{g}$, $X\in\mathfrak{e}$.

The following variational identities hold: 
\begin{align}
&\delta\mu(a,A)\mu(a,A)^{-1}
=\mu\sdot\,(a,\,\sdot\mu(a^{-1}\delta a,A)+\delta AA^{-1}), 
\vphantom{\Big]}
\label{ident11}
\\
&\delta\mu\sdot\,(a,X)=\mu\sdot\,(a,\sdot\mu\sdot\,(a^{-1}\delta a,X)+\delta X), 
\label{ident12}
\\
&\delta\,\,\sdot\mu(x,A)
=\sdot\mu(\delta x,A)
+\sdot\mu\sdot\,(x,\delta AA^{-1})-[\,\,\sdot\mu(x,A),\delta AA^{-1}],
\vphantom{\Big]}
\label{ident13}
\end{align}
where $a\in\mathsans{G}$, $A\in\mathsans{E}$, $x\in\mathfrak{g}$, $X\in\mathfrak{e}$.


\subsection{\textcolor{blue}{\sffamily Crossed modules with invariant pairing}}\label{app:crmodinv}

In this appendix, we provide the definition and main properties of Lie group and algebra crossed modules
with invariant pairing used in the main text. We also provide details on isotropic crossed submodules.

\vspace{2.5mm}

\noindent
{\it Lie algebra crossed modules with invariant pairing}

\noindent
A Lie algebra crossed module with invariant pairing is a Lie algebra crossed module
$\mathfrak{m}=(\mathfrak{e},\mathfrak{g},t,m)$ endowed with 
a non singular bilinear map $\langle\cdot,\cdot\rangle:\mathfrak{g}\times\mathfrak{e}\rightarrow\mathbb{R}$
enjoying the properties that \hphantom{xxxxxxxxxx}
\begin{equation}
\langle\ad z(x),X\rangle+\langle x,m(z,X)\rangle=0
\label{crmodinv2}
\end{equation}
for $z,x\in\mathfrak{g}$, $X\in\mathfrak{e}$ and that \hphantom{xxxxxxxxxxxxxxxx}
\begin{equation}
\langle t(X),Y\rangle=\langle t(Y),X\rangle
\label{crmodinv1}
\end{equation}
for $X,Y\in\mathfrak{e}$. The non singularity of $\langle\cdot,\cdot\rangle$ implies that
$\mathfrak{m}$ is balanced, $\dim\mathfrak{e}=\dim\mathfrak{g}$. 

A morphism $p:\mathfrak{m}'\rightarrow\mathfrak{m}=(H,h)$ of Lie algebra crossed modules with invariant pairing
is a crossed module morphism that respects the pairing, that is 
\begin{equation}
\langle h(x),H(X)\rangle=\langle x,X\rangle'
\label{crmodinv3}
\end{equation}
for $x\in\mathfrak{g}'$, $X\in\mathfrak{e}'$.

If $\mathfrak{m}_1$, $\mathfrak{m}_2$ are Lie algebra crossed modules with invariant pairing, then their direct sum 
$\mathfrak{m}=\mathfrak{m}_1\oplus\mathfrak{m}_2$ is a crossed module with the invariant pairing
\begin{equation}
\langle x_1\oplus x_2,X_1\oplus X_2\rangle=\langle x_1,X_1\rangle_1+\langle x_2,X_2\rangle_2,
\label{crmodinv4}
\end{equation}
where $x_1\in\mathfrak{g}_1$, $X_1\in\mathfrak{e}_1$, $x_2\in\mathfrak{g}_2$, $X_2\in\mathfrak{e}_2$.

Lie algebra crossed modules with invariant pairing and morphisms thereof with the direct sum operation
constitute a monoidal category that is a subcategory of the monoidal category of Lie algebra
crossed modules and module morphisms.

\vspace{2.5mm}

\noindent
{\it Lie group crossed modules with invariant pairing} 

\noindent

A Lie group crossed module with invariant pairing is a 
crossed module 
$\mathsans{M}=(\mathsans{E},\mathsans{G},\tau,\mu)$ such that the associated Lie algebra crossed module
$\mathfrak{m}=(\mathfrak{e},\mathfrak{g},\dot\tau,\sdot\mu\sdot)$
(cf. app. \cref{app:ident}) is a 
crossed module with invariant pairing $\langle\cdot,\cdot\rangle$ 
satisfying 
\begin{equation}
\langle\Ad a(x),\mu\sdot(a,X)\rangle=\langle x,X\rangle
\label{crmodinv14}
\end{equation}
for $a\in\mathsans{G}$, $x\in\mathfrak{g}$, $X\in\mathfrak{e}$. 
Note that \ceqref{crmodinv14} implies \ceqref{crmodinv2} with $m=\sdot\mu\sdot$ 
through Lie differentiation with respect to $a$. Again, the non singularity of $\langle\cdot,\cdot\rangle$ implies that
$\mathsans{M}$ is balanced, $\dim\mathsans{E}=\dim\mathsans{G}$. 


A morphism $\beta:\mathsans{M}'\rightarrow\mathsans{M}$  
of Lie group crossed modules with invariant pairing
is a morphism of the underlying 
crossed modules such that the induced Lie algebra crossed module
morphism $\dot\beta:\mathfrak{m}'\rightarrow\mathfrak{m}$ is a morphism of \pagebreak 
crossed modules with invariant pairing as defined earlier (cf. eq. \ceqref{crmodinv3}). 

If $\mathsans{M}_1$, $\mathsans{M}_2$ are Lie group crossed modules with invariant pairing, then their direct product 
$\mathsans{M}=\mathsans{M}_1\times\mathsans{M}_2$ is a crossed module with the invariant pairing,
since the associated Lie algebra crossed module $\mathfrak{m}=\mathfrak{m}_1\oplus\mathfrak{m}_2$ is endowed
with the invariant pairing \ceqref{crmodinv4} satisfying \ceqref{crmodinv14}.

Lie group crossed modules with invariant pairing and morphisms thereof
with the direct product operation constitute a monoidal category that is a
subcategory of the monoidal category of Lie group crossed modules and module morphisms.

\vspace{3mm}  

\noindent
{\it Fine crossed modules}


\noindent
Let $\mathsans{M}=(\mathsans{E},\mathsans{G},\tau,\mu)$ 
be a Lie group crossed module with invariant pairing $\langle\cdot,\cdot\rangle$. 
$\mathsans{M}$ is said to be fine if for $x,y\in\mathfrak{g}$ and $A\in\mathsans{E}$ one has 
\begin{equation}
\langle x,\sdot\mu(y,A)\rangle=\langle y,\sdot\mu(x,A^{-1})\rangle.
\label{invident5}
\end{equation}
This property is a sense dual to \ceqref{crmodinv14}.
$\mathsans{M}$ is fine under mild assumptions on the Lie group
$\mathsans{E}$. In particular, $\mathsans{M}$ is fine when $\mathsans{E}$
is connected and also when  $\mathsans{E}$ is not connected
in the connected component of the identity of $\mathsans{E}$ and in any connected component
of $\mathsans{E}$ where it holds for at least one element. $\mathsans{M}$ is fine also
when $\tau$ is invertible with no restrictions on $\mathsans{E}$.




\subsection{\textcolor{blue}{\sffamily Proof of the decomposition theorem}}\label{app:coreres}

In this appendix, we provide a sketch of the proof of the decomposition theorem
\ceqref{crmodinv12} of a Lie algebra crossed module with invariant pairing
$\mathfrak{m}$ satisfying the hypothesis \ceqref{crmodinv11}. The theorem
states the isomorphism 
\begin{equation}
\mathfrak{m}\simeq\CC\mathfrak{m}\oplus\RR\mathfrak{m},
\label{coreres1}
\end{equation}
where $\CC\mathfrak{m}$, $\RR\mathfrak{m}$ are the thee core and residue of $\mathfrak{m}$, 
the Lie algebra crossed modules with invariant pairing defined by
\ceqref{crmodinv5}, \ceqref{crmodinv6}, \ceqref{crmodinv9}
and \ceqref{crmodinv7}, \ceqref{crmodinv8}, \ceqref{crmodinv10}, respectively. 

By the assumption \ceqref{crmodinv11}, we have the Lie algebra direct sum decomposition 
\begin{equation}
\mathfrak{g}=\ran t\oplus\mathfrak{h},
\label{coreres2}
\end{equation}
where $\mathfrak{h}$ is an ideal of $\mathfrak{g}$ with $\ran t\cap \mathfrak{h}=0$.
The projector $\pi:\mathfrak{g}\rightarrow\ran t$ associated with the decomposition
is a Lie algebra morphism.

The duality pairing of $\mathfrak{g}$ and $\mathfrak{e}$ established by the invariant pairing
$\langle\cdot,\cdot\rangle$ entails the Lie algebra direct sum decomposition 
\begin{equation}
\mathfrak{e}=\ker t\oplus\mathfrak{h}^\perp,
\label{coreres3}
\end{equation}
where $\mathfrak{h}^\perp$, the orthogonal complement of $\mathfrak{h}$
with respect to the pairing, an ideal of $\mathfrak{e}$ with $\ker t\cap \mathfrak{h}^\perp=0$.
The projector $\varPi:\mathfrak{g}\rightarrow\ker t$ associated with the decomposition
is again a Lie algebra morphism. 

We define mappings $h:\mathfrak{g}\rightarrow\ran t\oplus(\mathfrak{g}/\ran t)$ and 
$H:\mathfrak{e}\rightarrow(\mathfrak{e}/\ker t)\oplus\ker t$ by 
\begin{align}
&h(x)=\pi(x)\oplus((1-\pi)(x)+\ran t),
\vphantom{\Big]}
\label{coreres4}
\\
&H(X)=((1-\varPi)(X)+\ker t)\oplus \varPi(X)
\vphantom{\Big]}
\label{coreres5}
\end{align}
with $x\in\mathfrak{g}$, $X\in\mathfrak{e}$. $h$, $H$ are the components of Lie
of a Lie algebra crossed module isomorphism $p:\mathfrak{m}\rightarrow\CC\mathfrak{m}\oplus\RR\mathfrak{m}$
Indeed, as it is straightforward to verify, $h$, $H$ are Lie algebra isomorphisms.  
Further, by virtue of the relations, 
\begin{align}
&\pi(t(X))=t((1-\varPi)(X)),
\vphantom{\Big]}
\label{coreres6}
\\
&(1-\pi)(t(X))=0,
\vphantom{\Big]}
\label{coreres7}
\\
&\varPi(m(x,X))=m((1-\pi)(x),\varPi(X)),
\vphantom{\Big]}
\label{coreres8}
\\
&(1-\varPi)(m(x,X))=m(\pi(x),(1-\varPi)(X)),
\vphantom{\Big]}
\label{coreres9}
\end{align}
$h$, $H$ obey the required conditions \ceqref{liecrmod7}, \ceqref{liecrmod8}.
Property \ceqref{crmodinv3} is immediately checked.


\subsection{\textcolor{blue}{\sffamily Basic results of Cartan--Weyl theory}}\label{app:cartan}

In this appendix, we review the basic notions of the Cartan--Weyl theory of lie algebras
used in sect. \cref{sec:aplc}. A standard reference is \ccite{Humphreys:1980dw}. 

Let $\mathsans{E}$ be a compact semisimple Lie group and
$\mathsans{F}$ a maximal torus of $\mathsans{E}$. 
Then, $\mathfrak{e}$ is a compact semisimple Lie algebra and 
$\mathfrak{f}$ is a maximal toroidal Lie subalgebra of $\mathfrak{e}$.
The integer $r=\dim\mathfrak{f}$ is the rank of $\mathsans{E}$. 

The structure of the Lie algebra $\mathfrak{e}$ is best analyzed by complexification. 
We let $\mathfrak{e}_{\mathbb{C}}=\mathbb{C}\otimes\mathfrak{e}$ 
and $\mathfrak{f}_{\mathbb{C}}=\mathbb{C}\otimes\mathfrak{f}$, the Cartan subalgebra of $\mathfrak{e}_{\mathbb{C}}$. 
Then, $\mathfrak{e}_{\mathbb{C}}$ has the vector space direct sum decomposition
\begin{equation}
\mathfrak{e}_{\mathbb{C}}=\ddd_{\alpha\in\Delta}\mathfrak{e}_\alpha\oplus\mathfrak{f}_{\mathbb{C}},
\label{cartan1}
\end{equation}
where $\Delta\subset\mathfrak{f}_{\mathbb{C}}\mhfpt{}^{*}$ is set of roots of $\mathfrak{e}_{\mathbb{C}}$,
the eigenvalues of $\ad\mathfrak{f}_{\mathbb{C}}$,  
and the $\mathfrak{e}_\alpha$ are the root subspaces, the associated eigenspaces
of $\ad\mathfrak{f}_{\mathbb{C}}$. The $\mathfrak{e}_\alpha$ can be shown to be all 1--dimensional. 

$\mathfrak{e}_{\mathbb{C}}$ admits an invariant symmetric non singular
bilinear pairing $\langle\cdot,\cdot\rangle_K$
unique up to normalization in each simple component of $\mathfrak{e}_{\mathbb{C}}$. 
$\langle\cdot,\cdot\rangle_K$ restricts to a non singular pairing on $\mathfrak{f}_{\mathbb{C}}$.
Through $\langle\cdot,\cdot\rangle_K$,
each element $\kappa\in\mathfrak{f}_{\mathbb{C}}\mhfpt{}^{*}$ is then identified with a unique generator 
$H_\kappa\in\mathfrak{f}_{\mathbb{C}}$. 
An symmetric non singular bilinear pairing $\langle\cdot,\cdot\rangle_K$ on
$\mathfrak{f}_{\mathbb{C}}\mhfpt{}^{*}$,
defined by $\langle\kappa,\lambda\rangle_K=\langle H_\kappa,H_\lambda\rangle_K$
for $\kappa,\lambda\in\mathfrak{f}_{\mathbb{C}}\mhfpt{}^{*}$, is so induced.

The roots $\alpha\in\Delta$ are therefore identified with generators
$H_\alpha$ of $\mathfrak{f}_{\mathbb{C}}$. With the
$\alpha$, there are further associated the coroots $\alpha\mhfpt{}^\vee\in\mathfrak{f}_{\mathbb{C}}$
given by $\alpha\mhfpt{}^\vee=2\alpha/\langle\alpha,\alpha\rangle_K$, which in turns are
identified with generators $H_{\alpha\mhfpt{}^\vee}$ of $\mathfrak{f}_{\mathbb{C}}$.  It can be shown that 
$H_\alpha,\,H_{\alpha\mhfpt{}^\vee}\in i\mathfrak{f}$.
Unlike the $H_\alpha$, however, the $H_{\alpha\mhfpt{}^\vee}$ do not depend on the normalization of
$\langle\cdot,\cdot\rangle_K$. Besides these, there exist normalized generators 
$X_\alpha\in\mathfrak{e}_\alpha$ such that the basic Lie brackets of $\mathfrak{e}_{\mathbb{C}}$ read as 
\begin{equation}
[H_{\alpha\mhfpt{}^\vee},X_{\pm\alpha}]=\pm 2X_{\pm\alpha}, \qquad [X_\alpha,X_{-\alpha}]=H_{\alpha\mhfpt{}^\vee}
\label{cartan2}
\end{equation}

The root set $\Delta$ is spanned over $\mathbb{Z}$ by a set of positive simple roots
$\Pi_+\subset\Delta$. Note that $|\Pi_+|=r$. The Cartan matrix
\begin{equation}
C_{\alpha\beta}     
=\frac{2\langle\alpha,\beta\rangle_K}
{\langle\alpha,\alpha\rangle_K}
=\frac{2\langle \alpha\mhfpt{}^\vee,\beta\mhfpt{}^\vee\rangle_K}
{\langle\beta\mhfpt{}^\vee,\beta\mhfpt{}^\vee\rangle_K}, \qquad \alpha,\beta\in\Pi_+
\label{cartan3}
\end{equation}
is independent from the normalization of the invariant pairing $\langle\cdot,\cdot\rangle_K$.
$C$ is an invertible $r\times r$ matrix with integer entries
in the range $-3,-2,-1,0,2$ completely codifying $\mathfrak{e}$ as a Lie algebra. In particular, $C$ determines
essentially all the ratios
of the normalization dependent simple root inverse half lengths squares  
\begin{equation}
\kappa_{\alpha}=\frac{2}{\langle\alpha,\alpha\rangle_K}
=\frac{\langle\alpha\mhfpt{}^\vee,\alpha\mhfpt{}^\vee\rangle_K}{2},\qquad \alpha\in\Pi_+.
\vphantom{\ul{\ul{\ul{\ul{\ul{g}}}}}}
\label{cartan4}
\end{equation}
For non orthogonal roots $\alpha,\beta\in\Pi_+$, 
$\kappa_\alpha/\kappa_\beta=C_{\alpha\beta}/C_{\alpha\beta}=1,2,3,1/2,1/3$ depending on cases. 

The real subspace $i\mathfrak{f}\subset\mathfrak{f}_{\mathbb{C}}$ is characterized by six lattices: 
\begin{itemize}

\item the root lattice $\Lambda_{\mathrm{r}\mathfrak{f}}$, the lattice of $i\mathfrak{f}$ generated by $H_\alpha$
with $\alpha\in\Pi_+$,

\item the coroot lattice $\Lambda_{\mathrm{cr}\mathfrak{f}}$, the lattice of $i\mathfrak{f}$ generated by $H_{\alpha\mhfpt{}^\vee}$
with $\alpha\in\Pi_+$,

\item the integral lattice $\Lambda_{\mathfrak{f}}$, the set of all $X\in i\mathfrak{f}$
such that $\ee^{2\pi iX}=1_{\mathsans{E}}$,
\end{itemize}

and their dual lattices with respect to $\langle\cdot,\cdot\rangle_K$:
\begin{itemize}

\item the weight lattice $\Lambda_{\mathrm{w}\mathfrak{f}}=\Lambda_{\mathrm{cr}\mathfrak{f}}\mhfpt{}^{*}$,

\item the coweight lattice $\Lambda_{\mathrm{cw}\mathfrak{f}}=\Lambda_{\mathrm{r}\mathfrak{f}}\mhfpt{}^{*}$,

\item the dual integral lattice $\Lambda_{\mathfrak{f}}\mhfpt{}^{*}$.

\end{itemize}
The weight and coweight lattices can also be defined by through generators. 
The simple roots $\alpha\in\Pi_+$ can be paired with the fundamental weights and coweights
$\alpha^{*},\alpha^{*\vee}\in\mathfrak{f}_{\mathbb{C}}{}^*$ defined by 
\begin{equation}
\langle\alpha^{*},\beta\mhfpt{}^\vee\rangle_K
=\langle\alpha^{*\vee},\beta\rangle_K
=\delta_{\alpha\beta}.
\label{cartan5}
\end{equation}
The weight and coweight lattices $\Lambda_{\mathrm{w}\mathfrak{f}}$,
$\Lambda_{\mathrm{cw}\mathfrak{f}}$ are then the lattices of $i\mathfrak{f}$ generated by $H_{\alpha^{*}}$,
$H_{\alpha^{*\vee}}$ with $\alpha\in\Pi_+$, respectively.  
Explicitly, we have $H_{\alpha^{*}}=\sum_{\hfpt\beta\in\Pi_+}C^{-1}{}_{\beta\alpha}H_\beta$,
$H_{\alpha^{*\vee}}=\sum_{\hfpt\beta\in\Pi_+}C^{-1}{}_{\alpha\beta}H_{\beta\mhfpt{}^\vee}$.
It is known that 
$\Lambda_{\mathrm{cr}\mathfrak{f}}\subseteq\Lambda_{\mathfrak{f}}\subseteq\Lambda_{\mathrm{cw}\mathfrak{f}}$
and $\Lambda_{\mathrm{r}\mathfrak{f}}\subseteq\Lambda_{\mathfrak{f}}\mhfpt{}^{*}\subseteq\Lambda_{\mathrm{w}\mathfrak{f}}$.
While the lattices $\Lambda_{\mathrm{cr}\mathfrak{f}}$, $\Lambda_{\mathrm{cw}\mathfrak{f}}$
$\Lambda_{\mathrm{r}\mathfrak{f}}$, $\Lambda_{\mathrm{w}\mathfrak{f}}$ depend only on the Lie algebra
$\mathfrak{e}$ and are therefore the same for all Lie groups $\mathsans{E}$ which share 
$\mathfrak{e}$ as their Lie algebra, the lattices $\Lambda_{\mathfrak{f}}$, $\Lambda_{\mathfrak{f}}\mhfpt{}^{*}$
do depend on $\mathsans{E}$. In this regard, one has 
$\mathsans{Z}(\mathsans{E})\simeq\Lambda_{\mathrm{cw}\mathfrak{f}}/\Lambda_{\mathfrak{f}}\simeq
\Lambda_{\mathfrak{f}}\mhfpt{}^{*}/\Lambda_{\mathrm{r}\mathfrak{f}}$ and $\pi_1(\mathsans{E})
\simeq\Lambda_{\mathfrak{f}}/\Lambda_{\mathrm{cr}\mathfrak{f}}\simeq
\Lambda_{\mathrm{w}\mathfrak{f}}/\Lambda_{\mathfrak{f}}\mhfpt{}^{*}$.


\vfil\eject

\noindent
\textcolor{blue}{Acknowledgements.} 
The author thanks the organizers of the Erwin Schroedinger Institute
Program ``Higher Structures and Field Theory''
whose seminars and discussion sessions have been source of much inspiration for him.
He thanks in particular Thomas Strobl for inviting him to the Program and for the interest shown
in his work. The author further acknowledges financial support from INFN Research Agency
under the provisions of the agreement between University of Bologna and INFN.

\vfil\eject


\begin{thebibliography}{99}

  \begin{small}

\bibitem{MacDowell:1977jt}
S.~W.~MacDowell and F.~Mansouri,
{\it Unified geometric theory of gravity and supergravity}, \\
\textcolor{blue}
{\href{https://journals.aps.org/prl/abstract/10.1103/PhysRevLett.38.739}
  {Phys.\ Rev.\ Lett.\  {\bf 38} (1977) 739}}, Erratum:
\textcolor{blue}
{\href{https://journals.aps.org/prl/abstract/10.1103/PhysRevLett.38.1376}
{Phys.\ Rev.\ Lett.\  {\bf 38} (1977) 1376}}.

\bibitem{Plebanski:1977zz}
J.~F.~Plebanski,
{\it On the separation of Einsteinian substructures},\\
\textcolor{blue}
{\href{https://aip.scitation.org/doi/10.1063/1.523215}
{J.\ Math.\ Phys.\  {\bf 18} (1977) 2511}}.


\bibitem{Morales:2017zjw}
I.~Morales, B.~Neves, Z.~Oporto and O.~Piguet
{\it Chern–Simons gravity in four dimensions}, \\
\textcolor{blue}
{\href{https://link.springer.com/article/10.1140/epjc/s10052-017-4653-8}
{Eur.\ Phys.\ J.\ C {\bf 77} (2017) no.2,  87}},
[\textcolor{blue}
{\href{https://arxiv.org/abs/1701.03642}
{\sffamily arXiv:1701.03642 [gr-qc]}}].

\bibitem{Carroll:1989vb}
S.~M.~Carroll, G.~B.~Field and R.~Jackiw,
{\it Limits on a Lorentz and parity violating modification of electrodynamics},\\
\textcolor{blue}
{\href{https://journals.aps.org/prd/abstract/10.1103/PhysRevD.41.1231}
{Phys.\ Rev.\ D {\bf 41} (1990) 1231}}.


\bibitem{Grumiller:2007rv}
D.~Grumiller and N.~Yunes,
{\it How do black holes spin in Chern-Simons modified gravity?},\\
\textcolor{blue}
{\href{https://journals.aps.org/prd/abstract/10.1103/PhysRevD.77.044015}
{Phys.\ Rev.\ D {\bf 77} (2008) 044015}},
[\textcolor{blue}
{\href{https://arxiv.org/abs/0711.1868}
{\sffamily arXiv:0711.1868 [gr-qc]}}].

\bibitem{Shiu:2015xda}
G.~Shiu, W.~Staessens and F.~Ye,
{\it Large field inflation from axion mixing},\\
\textcolor{blue}
{\href{https://link.springer.com/article/10.1007/JHEP06(2015)026}
{JHEP {\bf 1506} (2015) 026}},
[\textcolor{blue}
{\href{https://arxiv.org/abs/1503.02965}
{\sffamily arXiv:1503.02965 [hep-th]}}].

\bibitem{Costello:2013zra}
K.~Costello,
{\it Supersymmetric gauge theory and the Yangian},\\
\textcolor{blue}
{\href{https://arxiv.org/abs/1303.2632}
{\sffamily arXiv:1303.2632 [hep-th]}}.

\bibitem{Costello:2013sla}
K.~Costello,
{\it Integrable lattice models from four-dimensional field theories},\\
\textcolor{blue}
{\href{https://www.ams.org/books/pspum/088/}
{Proc.\ Symp.\ Pure Math.\  {\bf 88} (2014) 3}},
[\textcolor{blue}
{\href{https://arxiv.org/abs/1308.0370}
{\sffamily arXiv:1308.0370 [hep-th]}}].

\bibitem{Costello:2017dso}
K.~Costello, E.~Witten and M.~Yamazaki,
{\it Gauge theory and integrability, I},\\
\textcolor{blue}
{\href{https://www.intlpress.com/site/pub/pages/journals/items/iccm/content/vols/0006/0001/a006/index.php}
{ICCM Not. 6, 46-191 (2018)}},
[\textcolor{blue}
{\href{https://arxiv.org/abs/1709.09993}
{\sffamily arXiv:1709.09993 [hep-th]}}]. 

\bibitem{Costello:2018gyb}
K.~Costello, E.~Witten and M.~Yamazaki,
{\it Gauge theory and integrability, II},\\
\textcolor{blue}
{\href{https://www.intlpress.com/site/pub/pages/journals/items/iccm/content/vols/0006/0001/a007/index.php}
{ICCM Not. 6, 120-149 (2018)}},
[\textcolor{blue}
{\href{https://arxiv.org/abs/1802.01579}
{\sffamily arXiv:1802.01579 [hep-th]}}].

\bibitem{Turaev:1992hq}
  V.~G.~Turaev and O.~Y.~Viro,
{\it State sum invariants of 3 manifolds and quantum 6j symbols}, \\
 \textcolor{blue}
{\href{https://www.sciencedirect.com/science/article/pii/004093839290015A}
{Topology {\bf 31} (1992) 865}}.

\bibitem{Barrett:1993ab}
J.~W.~Barrett and B.~W.~Westbury,
{\it Invariants of piecewise linear three manifolds},\\
\textcolor{blue}
{\href{https://www.ams.org/journals/tran/1996-348-10/S0002-9947-96-01660-1/}
{Trans.\ Am.\ Math.\ Soc.\  {\bf 348} (1996) 3997}},
[\textcolor{blue}
{\href{https://arxiv.org/abs/hep-th/9311155}
{\sffamily hep-th/9311155}}].

\bibitem{Horowitz:1989ng}
G.~T.~Horowitz,
{\it Exactly soluble diffeomorphism invariant theories},\\
 \textcolor{blue}
{\href{https://link.springer.com/article/10.1007/BF01218410}
{Commun.\ Math.\ Phys.\  {\bf 125} (1989) 417}}.

\bibitem{Blau:1989bq}
M.~Blau and G.~Thompson,
{\it Topological gauge theories of antisymmetric tensor fields},\\
\textcolor{blue}
{\href{https://www.sciencedirect.com/science/article/abs/pii/0003491691902409?via\%3Dihub}
{Annals \ Phys.\ {\bf 205} (1991) 130}}.
  
\bibitem{Cattaneo:1995tw}
A.~S.~Cattaneo, P.~Cotta-Ramusino, J.~Frohlich and M.~Martellini,
{\it Topological BF theories in three-dimensions and four-dimensions},\\
\textcolor{blue}
{\href{https://aip.scitation.org/doi/10.1063/1.531238}
{J.\ Math.\ Phys.\  {\bf 36} (1995) 6137}},
[\textcolor{blue}
{\href{https://arxiv.org/abs/hep-th/9505027}
{\sffamily hep-th/9505027}}].

\bibitem{Reshetikhin:1991tc}
N.~Reshetikhin and V.~G.~Turaev,
{\it Invariants of three manifolds via link polynomials and quantum groups},\\
\textcolor{blue}
{\href{https://link.springer.com/article/10.1007\%2FBF01239527}
{Invent.\ Math.\  {\bf 103} (1991) 547}}.
  
\bibitem{Witten:1988hf}
E.~Witten,
{\it Quantum field theory and the Jones polynomial},\\
\textcolor{blue}
{\href{https://link.springer.com/article/10.1007/BF01217730}
{Commun.\ Math.\ Phys.\  {\bf 121} (1989) 351}}.

\bibitem{Frohlich:1989gr}
J.~Frohlich and C.~King,
{\it The Chern-Simons theory and knot polynomials},\\
\textcolor{blue}
{\href{https://link.springer.com/article/10.1007/BF02124336}
{Commun.\ Math.\ Phys.\  {\bf 126} (1989) 167}}.

\bibitem{Dijkgraaf:1989pz}
R.~Dijkgraaf and E.~Witten,
{\it Topological gauge theories and group cohomology},\\
 \textcolor{blue}
{\href{https://link.springer.com/article/10.1007/BF02096988}
  {Commun.\ Math.\ Phys.\  {\bf 129} (1990) 393}}.

\bibitem{Crane:1993cra}
L.~Crane and D.~N.~Yetter,
{\it A categorical construction of 4d topological quantum field theories},\\
in {\it Series on Knots and Everything}, vol. 3, 
\textcolor{blue}
{\href{https://www.worldscientific.com/doi/10.1142/9789812796387_0005}
{Quantum Topology (1993) 120}}, eds. L.~Kauffman and R.~Baadhio,
World Scientific, 
[\textcolor{blue}
{\href{https://arxiv.org/abs/hep-th/9301062}
{\sffamily hep-th/9301062}}].

\bibitem{Broda:1993bu}
B.~Broda,
{\it Surgical invariants of four manifolds},\\
proceedings  
\textcolor{blue}
{\href{https://www.worldscientific.com/doi/10.1142/9789814534505}
{Quantum Topology (1994) 45}}, ed. D.~N.~Yetter, World Scientific, 
[\textcolor{blue}
{\href{https://arxiv.org/abs/hep-th/9302092}
{\sffamily hep-th/9302092}}].

\bibitem{Crane:1994ji}
L.~Crane, L.~H.~Kauffman and D.~N.~Yetter,
{\it State sum invariants of four manifolds. 1},\\
\textcolor{blue}
{\href{https://arxiv.org/abs/hep-th/9409167}
{\sffamily hep-th/9409167}}.


\bibitem{Baez:1995ph}
J.~C.~Baez,
{\it Four-dimensional BF theory with cosmological term as a topological quantum field theory},\\
\textcolor{blue}
{\href{https://link.springer.com/article/10.1007/BF00398315}
{Lett.\ Math.\ Phys.\  {\bf 38} (1996) 129}},
[\textcolor{blue}
{\href{https://arxiv.org/abs/q-alg/9507006}
{\sffamily arXiv:q-alg/9507006}}].

\bibitem{Yetter:1993dh}
D.~N.~Yetter,
{\it TQFTs from homotopy 2-types}, \\
\textcolor{blue}
{\href{https://www.worldscientific.com/doi/abs/10.1142/S0218216593000076}
  {J. \ Knot \ Theor. Ramifications {\bf 2} (1993) 113}}.

\bibitem{Kitaev:2003fta}
A.~Yu.~Kitaev
{\it Fault-tolerant quantum computation by anyons},\\
\textcolor{blue}
{\href{https://www.sciencedirect.com/science/article/abs/pii/S0003491602000180}
{Annals \ Phys. \ {\bf 303} (2003) 2}},
[\textcolor{blue}
{\href{https://arxiv.org/abs/quant-ph/9707021}
{\sffamily quant-ph/9707021}}].

\bibitem{Levin:2004mi}
M.~A.~Levin and X.~G.~Wen,
{\it String net condensation: a physical mechanism for topological phases},\\
\textcolor{blue}
{\href{https://journals.aps.org/prb/abstract/10.1103/PhysRevB.71.045110}
{Phys.\ Rev.\ B {\bf 71} (2005) 045110}},
[\textcolor{blue}
{\href{https://arxiv.org/abs/cond-mat/0404617}
{\sffamily cond-mat/0404617}}].

\bibitem{Walker:2011mda}
K.~Walker and Z.~Wang,
{\it (3+1)-TQFTs and topological insulators},\\
\textcolor{blue}
{\href{https://link.springer.com/article/10.1007/s11467-011-0194-z}
{Front. Phys. {\bf 7} (2012) 150}},
[\textcolor{blue}
{\href{https://arxiv.org/abs/1104.2632}
{\sffamily arXiv:1104.2632 [cond-mat.str-el]}}].

\bibitem{Balachandran:1992qg}
A.~P.~Balachandran and P.~Teotonio-Sobrinho,
{\it The edge states of the BF system and the London equations},\\
\textcolor{blue}
{\href{https://www.worldscientific.com/doi/abs/10.1142/S0217751X9300028X}
{Int.\ J.\ Mod.\ Phys.\ A {\bf 8} (1993) 723}},
[\textcolor{blue}
{\href{https://arxiv.org/abs/hep-th/9205116}
  {\sffamily hep-th/9205116}}].

\bibitem{Bergeron:1994ym}
M.~Bergeron, G.~W.~Semenoff and R.~J.~Szabo,
{\it Canonical BF type topological field theory and fractional statistics of strings},\\
\textcolor{blue}
{\href{https://arxiv.org/abs/hep-th/9407020}
{Nucl.\ Phys.\ B {\bf 437} (1995) 695}},
[\textcolor{blue}
{\href{https://arxiv.org/abs/hep-th/9407020}
{\sffamily hep-th/9407020}}].

\bibitem{Szabo:1998ej}
R.~J.~Szabo,
{\it String holonomy and extrinsic geometry in four-dimensional topological gauge theory},\\
\textcolor{blue}
{\href{https://www.sciencedirect.com/science/article/abs/pii/S0550321398005860?via\%3Dihub}
{Nucl.\ Phys.\ B {\bf 531} (1998) 525}},
[\textcolor{blue}
{\href{https://arxiv.org/abs/hep-th/9804150}
{\sffamily hep-th/9804150}}].

\bibitem{Wilson:1974sk}
K.~G.~Wilson,
{\it Confinement of quarks},\\
\textcolor{blue}
{\href{http://dx.doi.org/10.1103/PhysRevD.10.2445}
{Phys. Rev. D {\bf 10}  (1974) 2445}}.

\bibitem{CottaRamusino:1994ez}
P.~Cotta-Ramusino and M.~Martellini,
{\it BF theories and 2--knots},\\
in \textcolor{blue}
{\href{https://global.oup.com/academic/product/knots-and-quantum-gravity-9780198534907?q=Knots\%20and\%20Quantum\%20Gravity&lang=en&cc=it}
{Knots and Quantum Gravity (1994) 169}}, ed. J.~C.~Baez, Oxford University Press, 
[\textcolor{blue}
{\href{https://arxiv.org/abs/hep-th/9407097}
{\sffamily hep-th/9407097}}].

\bibitem{Soncini:2014zra}
E.~Soncini and R.~Zucchini,
{\it A new formulation of higher parallel transport in higher gauge theory},\\
\textcolor{blue}
{\href{http://dx.doi.org/10.1016/j.geomphys.2015.04.010}
{J. Geom. Phys. {\bf 95} (2015) 28}},
[\textcolor{blue}
{\href{http://www.arxiv.org/abs/1410.0775}
{\sffamily  arXiv:1410.0775 [hep-th]}}].

\bibitem{Zucchini:2015wba}
  R.~Zucchini,
{\it On higher holonomy invariants in higher gauge theory I},\\
\textcolor{blue}{\href{http://www.worldscientific.com/doi/10.1142/S0219887816500900} 
{Int.\ J.\ Geom.\ Meth.\ Mod.\ Phys.\  {\bf 13} 07 (2016) 1650090}},
\hfill \hfill \hfill \hfill \hfill \hfill \hfill \hfill

\vspace{-1.75mm}\noindent
[\textcolor{blue}{\href{https://arxiv.org/abs/1505.02121}{\sffamily arXiv:1505.02121 [hep-th]}}].

\bibitem{Zucchini:2015xba}
  R.~Zucchini,
{\it On higher holonomy invariants in higher gauge theory II},\\
\textcolor{blue}{\href{http://www.worldscientific.com/doi/10.1142/S0219887816500912} 
{Int.\ J.\ Geom.\ Meth.\ Mod.\ Phys.\  {\bf 13} 07 (2016) 1650091}},
\hfill \hfill \hfill \hfill \hfill \hfill \hfill \hfill

\vspace{-1.75mm}\noindent
[\textcolor{blue}{\href{https://arxiv.org/abs/1505.02122}{\sffamily arXiv:1505.02122 [hep-th]}}].

\bibitem{Alekseev:2015hda}
A.~Alekseev, O.~Chekeres and P.~Mnev,
{\it Wilson surface observables from equi- variant cohomology}, \\
\textcolor{blue}
{\href{https://link.springer.com/article/10.1007\%2FJHEP11\%282015\%29093}
{JHEP {\bf 1511} (2015) 093}},
[\textcolor{blue}
{\href{https://arxiv.org/abs/1507.06343}
{\sffamily arXiv:1507.06343 [hep-th]}}].

\bibitem{Chekeres:2018kmh}
O.~Chekeres,
{\it Quantum Wilson surfaces and topological interactions}, \\
\textcolor{blue}
{\href{https://link.springer.com/article/10.1007/JHEP02(2019)030}
{JHEP {\bf 1902} (2019) 030}},
[\textcolor{blue}{\href
{https://arxiv.org/abs/1805.10992}
{\sffamily arXiv:1805.10992 [hep-th]}}].

\bibitem{Zucchini:2019mbz}
R.~Zucchini,
{\it Wilson surfaces for surface knots},\\
\textcolor{blue}
{\href{https://onlinelibrary.wiley.com/doi/abs/10.1002/prop.201910026}
{Fortsch.\ Phys.\  {\bf 67} (2019) no.8-9,  1910026}},
[\textcolor{blue}{\href
{https://arxiv.org/abs/1903.02853}
{\sffamily  arXiv:1903.02853 [hep-th]}}].



\bibitem{tHooft:1993dmi}
G.~'t Hooft,
{\it Dimensional reduction in quantum gravity},\\
\textcolor{blue}
{\href{https://inspirehep.net/conferences/968293}
{Conf.\ Proc.\ C {\bf 930308} (1993) 284}},
[\textcolor{blue}
{\href{https://arxiv.org/abs/gr-qc/9310026}
{\sffamily gr-qc/9310026}}].

\bibitem{Susskind:1994vu}
L.~Susskind,
{\it The world as a hologram},\\
\textcolor{blue}
{\href{https://aip.scitation.org/doi/10.1063/1.531249}
{J.\ Math.\ Phys.\  {\bf 36} (1995) 6377}},
[\textcolor{blue}
{\href{https://arxiv.org/abs/hep-th/9409089}
{\sffamily hep-th/9409089}}].

\bibitem{Wess:1971yu}
J.~Wess and B.~Zumino,
{\it Consequences of anomalous Ward identities},\\
\textcolor{blue}
{\href{https://www.sciencedirect.com/science/article/pii/037026937190582X?via\%3Dihub}
{Phys.\ Lett.\ {\bf 37B} (1971) 95}}.

\bibitem{Witten:1983tw}
E.~Witten,
{\it Global aspects of current algebra},\\
\textcolor{blue}
{\href{https://www.sciencedirect.com/science/article/abs/pii/0550321383900639?via\%3Dihub}
{Nucl.\ Phys.\ B {\bf 223} (1983) 422}}.

\bibitem{Fujita:2009kw}
M.~Fujita, W.~Li, S.~Ryu and T.~Takayanagi,
{\it Fractional quantum Hall effect via holography: Chern-Simons, edge states, and hierarchy},\\
\textcolor{blue}
{\href{https://iopscience.iop.org/article/10.1088/1126-6708/2009/06/066}
{JHEP {\bf 0906} (2009) 066}},
[\textcolor{blue}
{\href{https://arxiv.org/abs/0901.0924}
{\sffamily arXiv:0901.0924 [hep-th]}}].

\bibitem{Takayanagi:2013uya}
T.~Takayanagi,
{\it Holographic entanglement entropy, fractional quantum Hall effect and Lifshitz-like fixed point},\\
\textcolor{blue}
{\href{https://iopscience.iop.org/article/10.1088/1742-6596/462/1/012053}
{J.\ Phys.\ Conf.\ Ser.\  {\bf 462} no. 1 (2013) 012053}}.

\bibitem{Baez:2010ya}
J.~C.~Baez and J.~Huerta,
{\it An invitation to higher gauge theory},\\
\textcolor{blue}{\href{http://dx.doi.org/10.1007/s10714-010-1070-9}
{Gen. Relativ. Gravit. {\bf 43} (2011) 2335}},
[\textcolor{blue}{\href{http://www.arxiv.org/abs/1003.4485}{\sffamily 1003.4485[hep-th]}}].

\bibitem{Zucchini:2011aa} 
R.~Zucchini,
{\it AKSZ models of semistrict higher gauge theory},\\
\textcolor{blue}{\href{http://link.springer.com/article/10.1007\%2FJHEP03\%282013\%29014}
{JHEP {\bf 1303} 014 (2013)}},
[\textcolor{blue}{\href{https://arxiv.org/abs/1112.2819}
{\sffamily arXiv:1112.2819 [hep-th]}}].

\bibitem{Soncini:2014ara}
  E.~Soncini and R.~Zucchini,
{\it 4-d semistrict higher Chern-Simons theory I},\\
\textcolor{blue}{\href{http://link.springer.com/article/10.1007\%2FJHEP10\%282014\%29079}
{JHEP {\bf 1410} (2014) 79}},
[\textcolor{blue}{\href{https://arxiv.org/abs/1406.2197}
{\sffamily arXiv:1406.2197 [hep-th]}}].

\bibitem{Zucchini:2015ohw}
  R.~Zucchini,
 {\it A Lie based 4-dimensional higher Chern-Simons theory},\\
 \textcolor{blue}{\href{http://scitation.aip.org/content/aip/journal/jmp/57/5/10.1063/1.4947531}
{J.\ Math.\ Phys.\  {\bf 57} 5 (2016) 052301}},
[\textcolor{blue}{\href{https://arxiv.org/abs/1512.05977}
{\sffamily arXiv:1512.05977 [hep-th]}}].


\bibitem{Baez5}
J.~Baez and A.~Lauda, 
{\it Higher dimensional algebra V: 2-groups}, \\
\textcolor{blue}{\href{http://www.tac.mta.ca/tac/volumes/12/14/12-14abs.html}
{Theor.\ Appl.\ Categor.\ {\bf 12} (2004) 423}},
[\textcolor{blue}{\href{https://arxiv.org/abs/math/0307200}
{\sffamily arXiv:math.0307200}}].

\bibitem{Baez:2003fs}
J.~C.~Baez and A.~S.~Crans,
{\it Higher dimensional algebra VI: Lie $2$--algebras},\\
\textcolor{blue}{\href{http://www.tac.mta.ca/tac/volumes/12/15/12-15abs.html}
{Theor.\ Appl.\ Categor.\  {\bf 12} (2004) 492}},
[\textcolor{blue}{\href{https://arxiv.org/abs/math/0307263}{\sffamily arXiv:math/0307263}}].


\bibitem{Zucchini:2019rpp}
R.~Zucchini,
{\it Operational total space theory of principal 2--bundles I: operational geometric framework},\\
\textcolor{blue}
{\href{https://www.sciencedirect.com/science/article/pii/S0393044020301625?via\%3Dihub}
{J.\ Geom.\ Phys.\  {\bf 156} (2020) 103826}},
[\textcolor{blue}
{\href{https://arxiv.org/abs/1905.10057}
{\sffamily arXiv:1905.10057 [math-ph]}}].

\bibitem{Zucchini:2019pbv}
R.~Zucchini,
{\it Operational total space theory of principal 2--bundles II: 2--connections and 1-- and 2--gauge transformations},\\
\textcolor{blue}
{\href{https://www.sciencedirect.com/science/article/pii/S0393044020301601?via\%3Dihub}
{J.\ Geom.\ Phys.\  {\bf 156} (2020) 103825}},
[\textcolor{blue}
{\href{https://arxiv.org/abs/1907.00155}
{\sffamily arXiv:1907.00155 [math-ph]}}].



\bibitem{Wen:1992vi}
X.~G.~Wen,
{\it Theory of the edge states in fractional quantum Hall effects},\\
\textcolor{blue}
{\href{https://www.worldscientific.com/doi/abs/10.1142/S0217979292000840}
{Int. J. Mod. Phys. B \textbf{6} (1992), 1711}}.

\bibitem{Carlip:2005zn}
S.~Carlip,
{\it Conformal field theory, (2+1)-dimensional gravity, and the BTZ black hole},\\
\textcolor{blue}
{\href{https://iopscience.iop.org/article/10.1088/0264-9381/22/12/R01}
{Class. Quant. Grav. \textbf{22} (2005), R85}},
[\textcolor{blue}
{\href{https://arxiv.org/abs/gr-qc/0503022}
{\sffamily arXiv:gr-qc/0503022 [gr-qc]}}].


\bibitem{Afshar:2017okz}
  H.~Afshar, D.~Grumiller, M.~M.~Sheikh-Jabbari and H.~Yavartanoo,
{\it Horizon fluff, semi-classical black hole microstates -- Log--corrections to BTZ entropy and
black hole/particle correspondence},\\
\textcolor{blue}
{\href{https://link.springer.com/article/10.1007\%2FJHEP08\%282017\%29087}
{JHEP \textbf{08} (2017), 087}},
[\textcolor{blue}
{\href{https://arxiv.org/abs/1705.06257}
{\sffamily arXiv:1705.06257 [hep-th]}}].

\bibitem{Donnelly:2016auv}
W.~Donnelly and L.~Freidel,
{\it Local subsystems in gauge theory and gravity},\\
\textcolor{blue}
{\href{https://link.springer.com/article/10.1007/JHEP09(2016)102}
{JHEP \textbf{09} (2016), 102}},
[\textcolor{blue}
{\href{https://arxiv.org/abs/1601.04744}
{\sffamily arXiv:1601.04744 [hep-th]}}].

  
\bibitem{Geiller:2019bti}
M.~Geiller and P.~Jai-akson,
{\it Extended actions, dynamics of edge modes, and entanglement entropy,},\\
\textcolor{blue}
{\href{https://link.springer.com/article/10.1007\%2FJHEP09\%282020\%29134}
{JHEP \textbf{20} (2020), 134}},
[\textcolor{blue}
{\href{https://arxiv.org/abs/1912.06025}
{\sffamily arXiv:1912.06025 [hep-th]]}}].


\bibitem{Axelrod:1989xt}
S.~Axelrod, S.~Della Pietra and E.~Witten,
{\it Geometric quantization of Chern-Simons gauge theory},\\
\textcolor{blue}
{\href{https://projecteuclid.org/euclid.jdg/1214446565}
{J. Diff. Geom. \textbf{33} (1991) no. 3 787}}.


\bibitem{Balachandran:1977ub}
A.~P.~Balachandran, S.~Borchardt and A.~Stern,
{\it Lagrangian and Hamiltonian descriptions of Yang-Mills particles},\\
\textcolor{blue}
{\href{https://journals.aps.org/prd/abstract/10.1103/PhysRevD.17.3247}
{Phys.\ Rev.\ D {\bf 17} (1978) 3247}}.

\bibitem{Alekseev:1988vx}
A.~Alekseev, L.~D.~Faddeev and S.~L.~Shatashvili,
{\it Quantization of symplectic orbits of compact Lie groups by means of the functional integral},\\
\textcolor{blue}
{\href{https://www.sciencedirect.com/science/article/pii/0393044088900319?via\%3Dihub}
{J.\ Geom.\ Phys.\  {\bf 5} (1988) 391}}.

\bibitem{Diakonov:1989fc}
D.~Diakonov and V.~Y.~Petrov,
{\it A formula for the Wilson loop},\\
\textcolor{blue}
{\href{https://www.sciencedirect.com/science/article/abs/pii/0370269389910629?via\%3Dihub}
{Phys.\ Lett.\ B {\bf 224} (1989) 131}}.

\bibitem{Diakonov:1996zu}
D.~Diakonov and V.~Y.~Petrov,
{\it Non Abelian Stokes theorem and quark--mo\-nopole interaction}, 
talk at the {\it International workshop on non perturbative approaches to QCD}, \\
\textcolor{blue}
{\href{https://inspirehep.net/conferences/969882}
{C95-07-10.10 Proceedings (1995)}},
[\textcolor{blue}
{\href{https://arxiv.org/abs/hep-th/9606104}{\sffamily hep-th/9606104}}].

\bibitem{Beasley:2009mb} 
C.~Beasley,
{\it Localization for Wilson loops in Chern-Simons theory},\\
\textcolor{blue}{\href{https://www.intlpress.com/site/pub/pages/journals/items/atmp/content/vols/0017/0001/a001/}
{Adv.\ Theor.\ Math.\ Phys.\  {\bf 17} (2013) 1}},
[\textcolor{blue}{\href{https://arxiv.org/abs/0911.2687}{\sffamily arXiv:0911.2687 [hep-th]}}].

\bibitem{Zucchini:2021inp}
  R.~Zucchini, in preparation


\bibitem{Baez:2004in}
J.~C.~Baez and U.~Schreiber,
{\it Higher gauge theory: 2-connections on 2-bundles},\\
\textcolor{blue}{\href{http://www.arxiv.org/abs/hep-th/0412325}{\sffamily arXiv:hep-th/0412325}}.

\bibitem{Baez:2005qu}
J.~C.~Baez and U.~Schreiber,
{\it Higher gauge theory},
in {\it Categories in algebra, geometry and mathematical physics}, 
eds. A. Davydov et al., \\
\textcolor{blue}{\href{http://www.ams.org/books/conm/431/}
{Contemp.\ Math.\ {\bf 431} AMS (2007) 7}}
[\textcolor{blue}{\href{https://arxiv.org/abs/math/0511710}{\sffamily arXiv:math/0511710}}].

\bibitem{Giraud:1971cna}
J.~Giraud, 
{\it Cohomologie non-ab\`elienne},\\
\textcolor{blue}{\href{https://www.springer.com/it/book/9783540053071}
{Grundl.\ Math.\ Wiss. {\bf 197}, Springer Verlag (1971)}}. 

\bibitem{Breen:2001ie}
 L.~Breen and W.~Messing,
{\it Differential geometry of gerbes},\\
\textcolor{blue}{\href{http://www.sciencedirect.com/science/article/pii/S0001870805002513}
{Adv.\ Math.\ {\bf 198} (2005) 732}}
[\textcolor{blue}{\href{https://arxiv.org/abs/math/0106083}
{\sffamily arXiv:math/0106083}}].

\bibitem{Batalin:1981jr}
I.~A.~Batalin and G.~A.~Vilkovisky,
{\it Gauge algebra and quantization},\\
\textcolor{blue}{\href{http://www.sciencedirect.com/science/article/pii/0370269381902057}
{Phys.\ Lett.\ B {\bf 102} (1981) 27}}.

\bibitem{Batalin:1984jr}
I.~A.~Batalin and G.~A.~Vilkovisky,
{\it Quantization of gauge theories with linearly dependent generators},\\
\textcolor{blue}{\href{http://journals.aps.org/prd/abstract/10.1103/PhysRevD.28.2567}
{Phys.\ Rev.\ D {\bf 28} (1983) 2567}}
(\textcolor{blue}{\href{http://journals.aps.org/prd/abstract/10.1103/PhysRevD.30.508}
{Erratum-ibid.\ D {\bf 30} (1984) 508)}}.


\bibitem{Henneaux:1992ig}
M.~Henneaux and C.~Teitelboim,
{\it Quantization of gauge systems},\\
\textcolor{blue}
{\href{https://press.princeton.edu/books/paperback/9780691037691/quantization-of-gauge-systems}
{Princeton paperbacks (1992)}}, 
Princeton University Press.

\bibitem{Regge:1974zd}
T.~Regge and C.~Teitelboim,
{\it Role of surface integrals in the Hamiltonian formulation of general relativity},\\
\textcolor{blue}
{\href{https://www.sciencedirect.com/science/article/abs/pii/0003491674904047?via\%3Dihub}
{Annals Phys.\  {\bf 88} (1974) 286}}.

\bibitem{Benguria:1976in}
R.~Benguria, P.~Cordero and C.~Teitelboim,
{\it Aspects of the Hamiltonian dynamics of interacting
gravitational gauge and Higgs fields with applications to spherical symmetry},\\
\textcolor{blue}
{\href{https://www.sciencedirect.com/science/article/abs/pii/0550321377904266?via\%3Dihub}
{Nucl.\ Phys.\ B {\bf 122} (1977) 61}}.
  doi:10.1016/0550-3213(77)90426-6

\bibitem{Elitzur:1989nr}
S.~Elitzur, G.~W.~Moore, A.~Schwimmer and N.~Seiberg,
{\it Remarks on the canonical Quantization of the Chern-Simons-Witten theory},\\
\textcolor{blue}
{\href{https://www.sciencedirect.com/science/article/pii/0550321389904367?via\%3Dihub}
{Nucl.\ Phys.\ B {\bf 326} (1989) 108}}.

\bibitem{Kac:1990gs}
V.~G.~Kac,
{\it Infinite dimensional Lie algebras},\\
\textcolor{blue}
{\href{https://www.cambridge.org/core/books/infinitedimensional-lie-algebras/053FE77E6E9B35C56B5AEF7336FE7306}
  {Cambridge Univ. Pr. website (2010)}}, Cambridge University Press.

\bibitem{Banados:1994tn}
M.~Banados,
{\it Global charges in Chern-Simons field theory and the (2+1) black hole},\\
\textcolor{blue}
{\href{https://journals.aps.org/prd/abstract/10.1103/PhysRevD.52.5816}
{Phys.\ Rev.\ D {\bf 52} (1996) 5816}},
[\textcolor{blue}
{\href{https://arxiv.org/abs/hep-th/9405171}
{\sffamily hep-th/9405171}}].

\bibitem{Banados:1998ta}
M.~Banados, T.~Brotz and M.~E.~Ortiz,
{\it Boundary dynamics and the statistical mechanics of the (2+1)-dimensional black hole},\\
\textcolor{blue}
{\href{https://www.sciencedirect.com/science/article/abs/pii/S0550321399000693?via\%3Dihub}
{Nucl.\ Phys.\ B {\bf 545} (1999) 340}},
[\textcolor{blue}
{\href{https://arxiv.org/abs/hep-th/9802076}
{\sffamily  hep-th/9802076}}].

\bibitem{Banados:1998gg}
 M.~Banados,
{\it Three-dimensional quantum geometry and black holes},\\
\textcolor{blue}
{\href{https://aip.scitation.org/doi/abs/10.1063/1.59661}
{AIP Conf.\ Proc.\ {\bf 484} no.1 (1999) 147}},
[\textcolor{blue}
{\href{https://arxiv.org/abs/hep-th/9901148}
{\sffamily hep-th/9901148}}].

\bibitem{Troessaert:2013fba}
C.~Troessaert,
{\it Canonical structure of field theories with boundaries and applications to gauge theories},\\
\textcolor{blue}
{\href{https://arxiv.org/abs/1312.6427}
{\sffamily arXiv:1312.6427 [hep-th]}}.


\bibitem{Compere:2018aar}
G.~Comp\`ere and A.~Fiorucci,
{\it Advanced lectures on general relativity},\\
\textcolor{blue}
{\href{https://arxiv.org/abs/1801.07064}
{\sffamily  arXiv:1801.07064 [hep-th]}}. 



\bibitem{Polyakov:1984et}
A.~M.~Polyakov and P.~B.~Wiegmann,
{\it Goldstone fields in two-dimensions with multivalued actions},\\
\textcolor{blue}
{\href{https://www.sciencedirect.com/science/article/abs/pii/0370269384902065?via\%3Dihub}
{Phys. Lett. B \textbf{141} (1984), 223}}.
doi:10.1016/0370-2693(84)90206-5

\bibitem{Kapustin:2014gua}
A.~Kapustin and N.~Seiberg,
{\it Coupling a QFT to a TQFT and duality},\\
\textcolor{blue}
{\href{https://link.springer.com/article/10.1007\%2FJHEP04\%282014\%29001}
{JHEP {\bf 1404} (2014) 001}},
[\textcolor{blue}
{\href{https://arxiv.org/abs/1401.0740}
{\sffamily arXiv:1401.0740 [hep-th]}}].

\bibitem{Gaiotto:2014kfa}
D.~Gaiotto, A.~Kapustin, N.~Seiberg and B.~Willett,
{\it Generalized global symmetries},\\
\textcolor{blue}
{\href{https://doi.org/10.1007/JHEP02(2015)172}
{JHEP {\bf 1502} (2015) 172}},
[\textcolor{blue}
{\href{https://arxiv.org/abs/1412.5148}
{\sffamily arXiv:1412.5148 [hep-th}}].

\bibitem{Delcamp:2019fdp}
C.~Delcamp and A.~Tiwari,
{\it On 2-form gauge models of topological phases},\\
\textcolor{blue}
{\href{https://link.springer.com/article/10.1007/JHEP05(2019)064}
{JHEP {\bf 1905} (2019) 064}},
[\textcolor{blue}
{\href{https://arxiv.org/abs/1901.02249}
{\sffamily arXiv:1901.02249 [hep-th]}}].

\bibitem{tHooft:1981bkw}
G.~'t Hooft,
{\it Topology of the gauge condition and new confinement phases in non Abelian gauge theories},\\
\textcolor{blue}
{\href{https://www.sciencedirect.com/science/article/abs/pii/0550321381904429?via\%3Dihub}
{Nucl.\ Phys.\ B {\bf 190} (1981) 455}}.

\bibitem{Humphreys:1980dw}
  J.~E.~Humphreys,
  {\it Introduction to Lie algebras and representation theory}, \\
  {\it Infinite dimensional Lie algebras},\\
\textcolor{blue}
{\href{https://www.springer.com/gp/book/9780387900537}
  {Springer website (1972)}}, Springer. 








\end{small}

\end{thebibliography}
\end{document}